\documentclass[twoside,11pt]{article}

\usepackage{blindtext}
\usepackage{mathrsfs,amsmath,amsthm,amssymb,color}
\usepackage{caption}
\usepackage{subfigure}
\usepackage{algorithm}
\usepackage{algpseudocode}
\usepackage{rotating}
\usepackage{multirow}
\usepackage{lastpage}
\usepackage{jmlr2e}
\usepackage{booktabs}
\usepackage{pifont}
\usepackage{enumitem}
\usepackage[table,xcdraw]{xcolor}
\usepackage{accents}
\usepackage{graphicx}
\usepackage{xparse}
\usepackage{threeparttable}

\allowdisplaybreaks

\usepackage{jmlr2e}

\def\bet{\begin{theorem}}
	\def\eet{\end{theorem}}
\def\bep{\begin{proposition}}
	\def\eep{\end{proposition}}
\def\beq{\begin{equation}}
	\def\eeq{\end{equation}}
\def\bsq{\begin{equation*}}
	\def\esq{\end{equation*}}
\def\bse{\begin{eqnarray*}}
	\def\ese{\end{eqnarray*}}
\def\bel{\begin{lemma}}
	\def\eel{\end{lemma}}
\def\one{{\bf 1}}
\def\bA{{\mathcal A}}
\def\ve{\varepsilon}
\def\pre{\preccurlyeq}
\def\mle{\mbox{mle}}
\def\tr{\mbox{tr}}
\def\obs{\textup{obs}}
\def\lse{\mbox{lse}}
\def\red{\color{red}}
\def\blue{\color{blue}}
\def\cmle{\mbox{cmle}}
\def\n{\nonumber}
\def\ci{\perp\!\!\!\perp}
\def\e{\mathbf{e}}
\def\gap{\textup{gap}}
\def\cle{\preccurlyeq}
\def\V{\mbox{vec}}
\def\wh{\widehat}
\def\wb{\underline}
\def\lmax{\lambda_{\max}}
\def\cc{{\it correct }}
\def\svd{\mS_{\mbox{\sc svd}}}

\def\wY{\widetilde Y}
\def\ol{\overline }
\def\SE{\widehat{\mbox{SE}}}
\def\E{\mathbb{E}}
\def\bB{\mathbf{B}}
\def\mA{\mathcal A}
\def\mC{\mathcal C}
\def\mN{\mathcal{N}}
\def\mJ{\mathcal{J}}
\def\bK{\mathbf{K}}
\def\mP{\mathbb P}
\def\mD{\mathcal D}
\def\mQ{\mathbb Q}
\def\mG{\mathcal G}
\def\calR{{\cal R}}
\def\cP{\mathcal{P}}
\def\mI{\mathcal I}
\def\mL{\mathcal L}
\def\mS{\mathbb S}
\def\mH{\mathcal H}
\def\mM{\mathcal M}
\def\mm{\mathfrak m}
\def\mF{\mathcal F}
\def\F{\mathbf F}
\def\cR{\mathcal{R}}
\def\bR{\mathbf{R}}
\def\bY{\mathcal Y}
\def\mS{\mathcal S}
\def\mE{\mathcal E}
\def\mO{\mathcal O}
\def\mT{\mathcal T}
\def\bH{\mathbb H}
\def\bc{\mathbf c}
\def\mU{\mathbb{U}}
\def\mV{\mathbb{V}}
\def\mW{\mathbb{W}}
\def\cW{\mathcal{W}}
\def\mX{\mathbb{X}}
\def\cX{\mathcal{X}}
\def\wX{{\widetilde{\mathbb{X}}}}
\def\sX{{\mathbb{X}_b^*}}
\def\mY{\mathbb{Y}}
\def\mZ{\mathbb{Z}}
\def\bx{\mathbf{x}}
\def\bZ{\mathbf{Z}}
\def\bU{\mathbf{U}}
\def\bv{\mathbf{v}}
\def\var{\mbox{var}}
\def\supp{\mbox{supp}}
\def\es{E_*}
\def\vs{\mbox{var}_*}
\def\cov{\mbox{cov}}
\def\argmin{\mbox{argmin}}
\def\argmax{\mbox{argmax}}
\def\rank{\mbox{rank}}
\def\diag{\mbox{diag}}

\def\aic{\mbox{AIC}}
\def\bic{\mbox{BIC}}
\def\dbic{\mbox{DBIC}}
\def\rss{\mbox{RSS}}
\def\rmse{\mbox{RMSE}}
\def\vec{\mbox{vec}}
\def\err{\mbox{ERR}}
\def\sis{\mM_{\mbox{\sc sis}}}
\def\pis{\mM_{\mbox{\sc pis}}}
\def\pss{\mM_{\mbox{\sc pss}}}
\def\bgamma{\boldsymbol{\gamma}}
\def\bbeta{\boldsymbol{\beta}}
\def\bpsi{\boldsymbol{\psi}}
\def\bLambda{\boldsymbol{\Lambda}}
\def\bSigma{\boldsymbol{\Sigma}}
\def\bomega{\boldsymbol{\omega}}
\def\bve{\boldsymbol{\varepsilon}}
\def\bX{\mathbf{X}}
\def\bZ{\mathbf{Z}}
\def\bh{\mbox{\boldmath $h$}}
\def\by{\mathbf{y}}
\def\y{\mathbf{y}}
\def\bY{\mathbf{Y}}
\def\bA{\mathbf{A}}
\def\bG{\mathbf{G}}
\def\bD{\mathbf{D}}
\def\bW{\mathbf{W}}
\def\bS{\mathbf{S}}
\def\bH{\mathbf{H}}
\def\b{\mathbf{b}}
\def\bF{\mathbf{F}}
\def\kR{\mathfrak{R}}
\def\bg{\mbox{\boldmath $g$}}
\def\bI{\mathbf{I}}
\def\bu{\mbox{\boldmath $u$}}
\def\lag{\rm lag}
\def\cH{{\mathbb H}}
\def\cB{{\mathbb B}}
\def\cZ{{\mathcal Z}}
\def\cN{{\mathcal N}}

\def\zero{\mathbf{0}}
\def\defeq{\stackrel{\mathrm{def}}{=}}  
\def\sign{\mbox{sign}}
\def\qic{\textup{QIC}}

\def\th{^{th}}
\def\supp{\hbox{supp}}
\def\var{\hbox{var}}
\def\cov{\hbox{cov}}
\def\corr{\hbox{corr}}
\def\trace{\hbox{trace}}
\def\wh{\widehat}
\def\wc{\widecheck}
\def\eff{_{\rm eff}}
\def\sub{{\rm sub}}
\def\cat{{\rm cat}}
\def\th{^{th}}
\def\my{\mathcal Y}
\def\mL{\mathcal L}
\def\mR{\mathbb{R}}
\def\n{\nonumber}
\def\bias{\mbox{bias}}
\def\vecl{\mbox{vecl}}
\def\AIC{\mbox{AIC}}
\def\BIC{\mbox{BIC}}
\def\MSE{\mbox{MSE}}
\def\rank{\mbox{rank}}
\def\cov{\mbox{cov}}
\def\corr{\mbox{corr}}
\def\vec{\mbox{vec}}
\def\argmin{\mbox{argmin}}
\def\argmax{\mbox{argmax}}
\def\diag{\mbox{diag}}
\def\tr{\mbox{tr}}
\def\sir{{\mbox{\tiny SIR}}}
\def\save{{\mbox{\tiny SAVE}}}
\def\phd{{\mbox{\tiny PHD}}}
\def\cume{{\mbox{\tiny CUME}}}
\def\cuve{{\mbox{\tiny CUVE}}}
\def\cuhd{{\mbox{\tiny CUHD}}}
\def\cudr{{\mbox{\tiny CUDR}}}
\def\dr{{\mbox{\tiny DR}}}
\def\sumi{\sum_{i=1}^n}
\def\sumj{\sum_{j=1}^n}
\def\suml{\sum_{l=1}^n}
\def\sumk{\sum_{k=1}^n}
\def\trans{^{\top}}
\def\hDash{\bot\!\!\!\bot}
\def\mS{\mbox{ $\mathcal{S}$}}
\def\bs{\boldsymbol}
\def\bTheta{\boldsymbol\Theta}
\def\ba{\boldsymbol\alpha}
\def\bmu{\boldsymbol\mu}
\def\bnu{\boldsymbol\nu}
\def\beps{\boldsymbol\epsilon}
\def\ha{\widehat{\ba}}
\def\bb{\boldsymbol\beta}
\def\bGamma{\boldsymbol\Gamma}
\def\bOmega{\boldsymbol\Omega}
\def\bdelta{\boldsymbol\delta}
\def\bDelta{\boldsymbol\Delta}
\def\bxi{\boldsymbol\xi}
\def\bphi{\boldsymbol\phi}
\def\btau{\boldsymbol\tau}
\def\bpsi{\boldsymbol\psi}
\def\bzeta{\boldsymbol\zeta}
\def\bmu{\boldsymbol\mu}
\def\bg{\boldsymbol\gamma}
\def\hb{\widehat{\bb}}
\def\he{\widehat{\varepsilon}}
\def\defby{\stackrel{\mbox{\textrm{\tiny def}}}{=}}
\def\0{{\bf 0}}
\def\1{{\bf 1}}
\def\A{{\bf A}}
\def\U{{\bf U}}
\def\V{{\bf V}}
\def\e{{\bf e}}
\def\R{{\bf R}}
\def\G{{\bf G}}
\def\bO{{\bf O}}
\def\a{{\bf a}}
\def\B{{\bf B}}
\def\c{{\bf c}}
\def\D{{\bf D}}
\def\V{{\bf V}}
\def\K{{\bf K}}
\def\g{{\bf g}}
\def\r{{\bf r}}
\def\f{{\bf f}}
\def\h{{\bf h}}
\def\b{{\bf b}}
\def\I{{\bf I}}
\def\mU{\mathcal{U}}
\def\BB{\mbox{ $\mathcal{B}$}}
\def\N{\mbox{ $\mathcal{N}$}}
\def\M{{\bf M}}
\def\bM{{\bf M}}
\def\K{{\bf K}}
\def\t{{\bf t}}
\def\T{{\bf T}}
\def\bd{{\bf d}}
\def\bP{{\bf P}}
\def\bP{{\bf P}}
\def\bJ{{\bf J}}
\def\bV{{\bf V}}
\def\hQ{{\widehat \bQ}}
\def\U{{\bf U}}
\def\S{{\bf S}}
\def\s{{\bf s}}
\def\u{{\bf u}}
\def\m{{\bf m}}
\def\v{{\bf v}}
\def\W{{\bf W}}
\def\T{{\bf T}}
\def\bO{{\bf O}}
\def\w{{\bf w}}
\def\X{{\bf X}}
\def\L{{\bf L}}
\def\x{{\bf x}}
\def\I{{\bf I}}
\def\tx{{\widetilde \x}}
\def\Y{{\bf Y}}
\def\H{{\bf H}}

\def\C{{\bf C}}
\def\tY{{\widetilde Y}}
\def\y{{\bf y}}
\def\Z{{\bf Z}}
\def\z{{\bf z}}
\def\bC{{\bf C}}
\def\Ybar{{\overline{Y}}}
\def\Xbar{{\overline{\X}}}
\def\xbar{{\overline{\x}}}
\def\wbar{{\overline{\W}}}
\def\bSig{{\bf \Sigma}}
\def\bLam{{\bf \Lambda}}
\def\diag{\hbox{diag}}
\def\dfrac#1#2{{\displaystyle{#1\over#2}}}
\def\VS{{\vskip 3mm\noindent}}
\def\refhg{\hangindent=20pt\hangafter=1}
\def\refmark{\par\vskip 2mm\noindent\refhg}
\def\naive{\hbox{naive}}
\def\itemitem{\par\indent \hangindent2\parindent \textindent}
\def\dist{\hbox{dist}}
\def\trace{\hbox{trace}}
\def\refhg{\hangindent=20pt\hangafter=1}
\def\refmark{\par\vskip 2mm\noindent\refhg}
\def\Normal{\hbox{Normal}}
\def\povr{\buildrel p\over\longrightarrow}
\def\ccdot{{\bullet}}
\def\pr{\hbox{pr}}
\def\wh{\widehat}
\def\th{^{th}}
\def\diag{\hbox{diag}}
\def\log{\hbox{log}}
\def\bias{\hbox{bias}}
\def\Siuu{\boldSigma_{i,uu}}
\def\squarebox#1{\hbox to #1{\hfill\vbox to #1{\vfill}}}
\def\btheta{{\boldsymbol \theta}}
\def\bfeta{{\boldsymbol \eta}}
\def\balpha{{\boldsymbol \alpha}}
\def\bOmega{{\boldsymbol \Omega}}
\def\bXi{{\boldsymbol \Xi}}
\def\blambda{{\boldsymbol \lambda}}
\def\bPsi{{\boldsymbol \Psi}}
\def\bzeta{{\boldsymbol \zeta}}
\def\bpi{{\boldsymbol \pi}}
\def\bx{{\bf x}}
\def\bz{{\bf z}}
\def\vec{\mathrm{vec}}
\def\mA{\mathcal{A}}
\def\cA{\mathbb{A}}
\def\bE{\mathbf{E}}
\def\mB{\mathcal{B}}
\def\mC{\mathcal{C}}
\def\mF{\mathcal{F}}
\def\cS{\mathcal{S}}
\def\mH{\mathcal{H}}
\def\mM{\mathcal{M}}
\def\my{\mathcal Y}
\def\cY{\mathcal Y}
\def\bw{\mathbf w}
\def\dfrac#1#2{{\displaystyle{#1\over#2}}}
\def\VS{{\vskip 3mm\noindent}}
\def\refhg{\hangindent=20pt\hangafter=1}
\def\refmark{\par\vskip 2mm\noindent\refhg}
\def\itemitem{\par\indent \hangindent2\parindent \textindent}
\def\var{\hbox{var}}
\def\cov{\hbox{cov}}
\def\corr{\hbox{corr}}
\def\trace{\hbox{trace}}
\def\refhg{\hangindent=20pt\hangafter=1}
\def\Normal{\hbox{Normal}}
\def\povr{\buildrel p\over\longrightarrow}
\def\dovr{\buildrel d\over\longrightarrow}
\def\ccdot{{\bullet}}
\def\pr{\hbox{pr}}
\def\br{{\bf r}}
\def\q{{\bf q}}
\def\wh{\widehat}
\def\wt{\widetilde}
\def\diag{\hbox{diag}}
\def\bias{\hbox{bias}}
\def\Siuu{\boldSigma_{i,uu}}
\def\whT{\widehat{\Theta}}
\def\diag{\hbox{diag}}
\def\th{^{th}}
\def\o{\textup{or}}
\def\logit{{\mbox{logit}}}
\def\bfa{\textbf{a}}
\def\log{\textup{log}}
\definecolor{ForestGreen}{RGB}{34,139,34}

\def\boxit#1{\vbox{\hrule\hbox{\vrule\kern6pt\vbox{\kern6pt#1\kern6pt}\kern6pt\vrule}\hrule}}
\def\yimeng#1{\vskip 2mm\boxit{\vskip 2mm{\color{ForestGreen}\bf#1} {\color{blue}\bf -- Yimeng\vskip 2mm}}\vskip 2mm}
\def\xuening#1{\vskip 2mm\boxit{\vskip 2mm{\color{blue}\bf#1} {\color{blue}\bf -- Xuening\vskip 2mm}}\vskip 2mm}

\renewcommand\thesubfigure{(\alph{subfigure})}

\newtheorem{assumption}{Assumption}
\newtheorem{remark}{Remark}

\setlistdepth{8}
\renewlist{itemize}{itemize}{8}  
\setlist[itemize,5]{leftmargin=2em}

\newcommand{\circled}[1]{\textcircled{\scriptsize #1}} 

\makeatletter
\newcommand{\widecheck}[1]{\mathpalette\widecheck@{#1}}
\newcommand{\widecheck@}[2]{%
	\begingroup
	\sbox\@tempboxa{$#1\widehat{\smash{\phantom{#2}}}$}%
	\accentset{\raisebox{0ex}{$\m@th#1\rotatebox[origin=c]{180}{\copy\@tempboxa}$}}{#2}%
	\endgroup
}
\makeatother

\usepackage{lastpage}
\jmlrheading{27}{2026}{1-\pageref{LastPage}}{7/24; Revised 11/25}{6/26}{24-1191}{Yimeng Ren, Xuening Zhu, Ganggang Xu, and Yanyuan Ma}

\ShortHeadings{Multi-relational Network Autoregression Model with Latent Group Structures}{Ren, Zhu, Xu and Ma}
\firstpageno{1}

\begin{document}

\title{Multi-relational Network Autoregression Model with Latent Group Structures}

\author{\name{Yimeng Ren}$^{1, 2}$ \email{ymren@ust.hk} \\
	\addr $^1$ School of Data Science\\
	Fudan University\\
	Shanghai, China \\
	\addr $^2$ HKUST Business School\\
	The Hong Kong University of Science and Technology\\
	Hong Kong SAR, China
	\AND
	\name Xuening Zhu$^{\ast}$ \email xueningzhu@fudan.edu.cn \\
	\addr School of Management\\
	Fudan University\\
	Shanghai, China
	\AND
	\name Ganggang Xu$^{\ast}$ \email gangxu@bus.miami.edu \\
	\addr Department of Management Science\\
	University of Miami\\
	Coral Gables, FL 33146, USA
	\AND
	\name Yanyuan Ma \email yanyuanma@gmail.com \\
	\addr Department of Statistics\\
	The Pennsylvania State University\\
	University Park, PA 16802, USA
}
\editor{Ji Zhu}

\maketitle
{
	\makeatletter
	\renewcommand{\@makefntext}[1]{\noindent #1} 
	\makeatother
	\footnotetext{$^{\ast}$ Corresponding authors.}
}

\begin{abstract}
	Multi-relational networks among entities are frequently observed in the era of big data.
	Quantifying the effects of multiple networks has attracted significant research interest recently.
	In this work, we model multiple network effects through an autoregressive framework for tensor-valued time series.
	To characterize the potential heterogeneity of the networks
	and handle the high dimensionality of the time series data simultaneously, we assume
	a separate group structure for entities in each network and estimate all group memberships in a data-driven fashion.
	Specifically, we propose a group tensor network autoregression (GTNAR) model,
	which assumes that within each network, entities in the same group share the same set of model parameters, and the parameters differ across networks.
	An iterative algorithm is developed to estimate the model parameters and the latent group memberships simultaneously.
	Theoretically, we show that the group-wise parameters and group
	memberships can be consistently estimated when the group numbers are
	correctly- or possibly over-specified.
	An information criterion for estimating the group number for each network is also provided to consistently select the group numbers.
	Lastly, we apply the GTNAR method to a Yelp dataset to illustrate its usefulness.
\end{abstract}


\section{Introduction}\label{sec:intro}

As the world becomes increasingly connected, studying network effects has become an important research topic across disciplines, including economics, finance, and many others.
The primary focus of our study is to analyze the network effects inherent in high-dimensional time series observed over multiple networks.
The existing literature has seen notable progress in the study of time series within a single network.
For instance, \cite{zhu2017network} introduced a network autoregression model to investigate time series in large social networks.
\cite{armillotta2023nonlinear} extended the framework to the nonlinear network autoregression model, in which the multivariate observations could be both continuous and discrete.
\cite{chen2023community} proposed a community network vector autoregression model for the high-dimensional time series.
\cite{fang2023group} proposed a group network Hawkes process with a single network to model unit event occurrences.
\cite{chen2026group} proposed a group network multivariate GARCH model that incorporates a latent group structure and an observed network adjacency matrix.
Compared with conventional high-dimensional time series models \citep{walden2002wavelet, leng2012sparse, zhou2014gemini, wang2019factor, chen2019constrained, chang2021modelling, wang2023rate, chen2023statistical}, the network autoregressive model distinguishes itself by providing enhanced parameter interpretability, offering deeper insights into the intricacies of network dynamics.

In real-world scenarios, entities within a population commonly establish connections across multiple networks, often referred to as multi-relational or multi-layer networks. Recently, there has been significant interest in investigating these networks collectively, as seen in works like \cite{lei2020consistent}, \cite{ zhang2020flexible}, \cite{jing2021community}, \cite{macdonald2022latent}, and  \cite{ma2023community}.
While the majority of these studies focus on identifying community structures within multi-layer networks, there is also a considerable interest in quantifying the impacts of multi-relational network effects on various research objectives.
For example, \cite{emch2016integration} investigated the joint effects of spatial and social networks on disease transmission.
\cite{chen2017uncovering} proposed the utilization of network metrics from various social networks to predict the adoption of new products in marketing research.
Besides, \cite{corradini2021investigating} investigated the influence of multi-dimensional social networks on negative reviews posted on Yelp.
Although these models are valuable in empirical research, a critical gap remains in the availability of rigorous statistical models capable of providing valid statistical inferences about multiple network effects, and it is our intention to fill this gap.

The objective of our work is to investigate tensor-valued time series indexed across multi-relational networks.
A straightforward way to analyze tensor-valued data is to stack it into vectors or to consider only one of its dimensions.
However, this will destroy the intrinsic multidimensional structure and lack clear interpretations, and provide limited insights into network effects. This drawback has also been recognized in several recent studies on matrix- or tensor-valued time series \citep{zhou2014regularized, wang2019factor, chen2019constrained, chen2021autoregressive, chen2022factor, chen2023statistical, wang2024high, chen2024rank}.
Additionally, although stacking into vectors allows the implementation of techniques for high-dimensional VAR models, this still leads to a much larger number of parameters that need to be estimated, on the order of $\prod_{l=1}^q (N_l)^2$, where $q$ is the dimension of tensor data, and $N_l$ is the number of nodes in the $l$th dimension.
Even with commonly used sparsity regularization, the estimation variance of model parameters is likely to be much higher than that of the network VAR model when the data is generated by the latter \citep{zhu2023simultaneous}.

To address high dimensionality, several recent studies focus on factor models to capture low-rank structures \citep{zhou2023partially, han2023simultaneous, wang2024high}.
For example,
\cite{chen2022factor} introduced a general framework of factor models for tensor-valued time series.
\cite{han2021cp} propose a factor model and a high-order projection estimator to analyze high-dimensional dynamic tensor time series.
Moreover, \cite{wang2022high} used the tensor decomposition technique to deal with the large transition matrices of the high-dimensional VAR model.
\cite{chen2024rank} introduced a pre-averaging procedure, allowing for a spectrum of factor strengths, as well as the weak factors when both cross-sectional and serial correlations are present.
Besides, \cite{wang2023rate} proposed a general robust estimation procedure, which can address many high-dimensional VAR models with low-rank or sparse structure.
However, existing works do not exploit network structure information across tensor dimensions and therefore do not provide direct statistical inference on these network effects.

On the other hand, it is remarkable that nodal heterogeneity widely exists in practice \citep{ke2015homogeneity}.
The utilization of latent group structures to model heterogeneous data has a well-established history in panel data analysis. For instance,
\cite{ke2015homogeneity} introduced a clustering algorithm in regression through data-driven segmentation to identify the groups,
and \cite{vogt2017classification} shed light on the latent grouped structure in the non-parametric regression functions.
\cite{su2016identifying}
introduce a Classifier Lasso (C-Lasso) estimator for panel models.
And more recently, \cite{ando2020quantile} proposed a new estimation method for analyzing the quantile co-movement of large-scale financial time series data, which can identify latent group heterogeneity among the series.
There have also been recent efforts to leverage latent group
structures in modeling time series data within a single network, see, e.g.,
\cite{zhu2023simultaneous}, \cite{chen2023community}, \cite{fang2023group}, \cite{liu2024two}, and \cite{chen2026group}.
However, these works have predominantly focused on time series observed on
a single network; hence, they cannot be directly used in modeling the tensor-valued time series data.
Moreover, addressing the theoretical challenges involved in extending from a single network to multiple networks is non-trivial due to the interactions between different group structures, necessitating the development of new theoretical tools.
To tackle this challenge, we divide the members of the network on each dimension into several latent subgroups. We assume that members within each subgroup share similar characteristics, carrying the same model coefficients.

The main contributions of our work can be summarized as follows.
First, we introduce a highly interpretable network autoregression model for high-dimensional multivariate time series indexed by multiple networks, namely, the Group Tensor-valued Network Autoregression (GTNAR) model. Second, to account for network heterogeneity, we allow for separate group structures on related networks.
Third, we establish estimation consistency for both model parameters and group memberships, not only when the numbers of groups are correct, but also when they are over-specified. Simultaneously estimating group memberships across multiple networks poses significant challenges compared to the existing literature, which typically considers at most one network \citep[e.g.,][]{fang2023group}. We develop new theoretical tools to address this challenge.
Lastly, we develop a selection criterion that consistently selects the true group numbers and establishes asymptotic normality when the group numbers are correctly specified. Our theoretical framework enables rigorous testing of multiple network effects, which are crucial across various research disciplines.

The remainder of this article is structured as follows.
In Section \ref{sec:model},
we first show a motivating example on the Yelp dataset to introduce a simple model form. Then, we introduce the general GTNAR model.
Section \ref{sec:estimate} outlines the model estimation procedure and the selection method for the group numbers.
Theoretical properties concerning parameter estimation, group membership estimation, and group number selection are discussed in Section \ref{sec:theory}.
Two model extensions and corresponding theories are provided in Section \ref{sec:wls_int_theory}.
In Section \ref{sec:simulation}, we present extensive simulation studies to illustrate the finite sample performance of GTNAR.
Section \ref{sec:real} includes an application of GTNAR to the Yelp dataset.
Finally, Section \ref{sec:conclusion} provides concluding remarks.
Additional technical proofs can be found in the Appendix.

\section{Group Tensor-valued Network Autoregression}\label{sec:model}

\subsection{Notations and Tensor Algebra}

Throughout the paper, we use the following notations.
Denote $[n] = \{1,2,\cdots ,n\}$ for an integer $n$.
For a vector $\bv = (v_j: j\in [p])^\top\in \mR^p$, let $\|\bv\| = (\sum_{j = 1}^p v_j^2)^{1/2}$.
For a matrix $\bM = (m_{ij})\in \mR^{n_1\times n_2}$, let
$\bM_{i\cdot}$ be the $i$th row vector and $\bM_{\cdot j}$ as the
$j$th column vector of $\bM$.
In addition, let $\bM^{(\mC,\cdot)} = (m_{ij}: i\in \mC, j\in [n_2])$
and $\bM^{(\cdot,\mC)} = (m_{ij}: i\in [n_1], j\in \mC)$,
where $\mC$ is an index set.
For a symmetric matrix $\bM$, define $\lambda_{\min}(\bM)$ and $\lambda_{\max}(\bM)$ as the smallest and largest eigenvalues, respectively.
For a tensor $\cX \in \mR^{n_1 \times \cdots \times n_q}$, denote $\cX^{(\mC_1, \cdots, \mC_q)}$ as the corresponding subset of the tensor with each dimension selected by the index sets $\mC_l$s.
Denote $\A \otimes \B \in \mR^{n_1 n_3 \times n_2 n_4}$ as the Kronecker product between matrices $\A\in \mR^{n_1 \times n_2}$ and
$\B\in \mR^{n_3 \times n_4}$.
For a matrix series $\{ \A_l: l \in [q] \}$, denote $\B \otimes_{k \neq l} \A_k = \B \otimes \A_1 \otimes \cdots \otimes \A_{l-1} \otimes \A_{l+1} \otimes \cdots \otimes \A_q$.
For a vector, matrix, or tensor $\bM$, let $\|\bM\|_{\max}$
denote its largest absolute entry.
For a set $\cS$, denote $|\cS|$ as the cardinal number of $\cS$.
We denote $\one_p \in \mR^{p}$ as a $p$-dimensional vector with all elements equal to one, and $\I_p$ is an identity matrix.

For a tensor $\cX = (x_{i_1, \cdots, i_q})_{i_1 \in [n_1], \cdots, i_q \in [n_1]} \in \mR^{n_1 \times \cdots \times n_q}$, denote the vectorization as $\bx = \vec(\cX) = (x_{1, 1, \cdots, 1}, x_{2, 1, \cdots, 1}, \cdots,  x_{n_1, 1, \cdots, 1}, x_{1, 2, 1, \cdots, 1},  \cdots, x_{n_1, n_2, \cdots, n_q})^\top \in \mR^{\prod_l n_l}$.
Denote its mode-$l$ matricization as $\cX_{(l)} \in \mR^{n_l \times \prod_{k \neq l} n_k}$, which is calculated by setting the $l$th tensor dimension as the matrix rows, and collapsing all other into its columns, for $l \in [q]$.
For tensor $\cX = (x_{i_1, \cdots, i_q}) \in \mR^{n_1 \times \cdots \times n_q}$ and a matrix $\bM = (m_{i,j}) \in \mR^{q \times n_l}$, denote the mode-$l$ multiplication $\cX \times_l \bM$ as a tensor in $\mR^{n_1 \times \cdots n_{l-1} \times q \times n_{l+1} \times \cdots \times n_q}$, and its $(i_1, \cdots, i_{l-1}, s, i_{l+1}, \cdots, i_q)$th element is calculated by
$
(\cX \times_l \bM)_{i_1, \cdots, i_{l-1}, s, i_{l+1}, \cdots,  i_q} = \sum_{i_l = 1}^{n_l}  x_{i_1, \cdots, i_l, \cdots i_q} m_{s, i_l}.
$
Subsequently, we denote $\cX \times_{l=1}^q \bM_l \defeq (\cX \times_1 \bM_1) \times_2 \cdots \times_q \bM_q$.
For two tensors $\cX \in \mR^{n_1 \times \cdots \times n_q}$ and $\cY \in \mR^{n_1 \times  \cdots \times n_s}$ with $q \ge s$, their generalized inner product $\langle \cX, \cY \rangle$ is a $(q-s)$ dimensional tensor, calculated by $\langle \cX, \cY \rangle_{i_{s+1} \cdots i_q} = \sum_{i_1 = 1}^{n_1} \cdots \sum_{i_s = 1}^{n_s} \cX_{i_1 \cdots i_s i_{s+1} \cdots i_q} \cY_{i_1 \cdots i_s}$.
We use $\mA \odot \cY$ to denote the element-wise product between two tensors ($\mA$ and $\cY$) with the same dimensions.
For a column vector $\x_l \in \mR^{n_l}$, $\bx_l \circ_{k \neq l} \one_{n_k} = \one_{n_1} \circ \cdots \circ \x_l \circ \cdots \circ \one_{n_q}$ is a $q$ dimensional tensor $\mT \in \mR^{n_1 \times \cdots \times n_q}$ with its $(i_1, \cdots ,i_l ,\cdots, i_q)$th element being $x_{i_l}$.

\subsection{Motivating Example and A Simple Model}\label{sec:motivation}

\begin{figure}[htpb!]
	\begin{center}
		\includegraphics[width=0.5\textwidth]{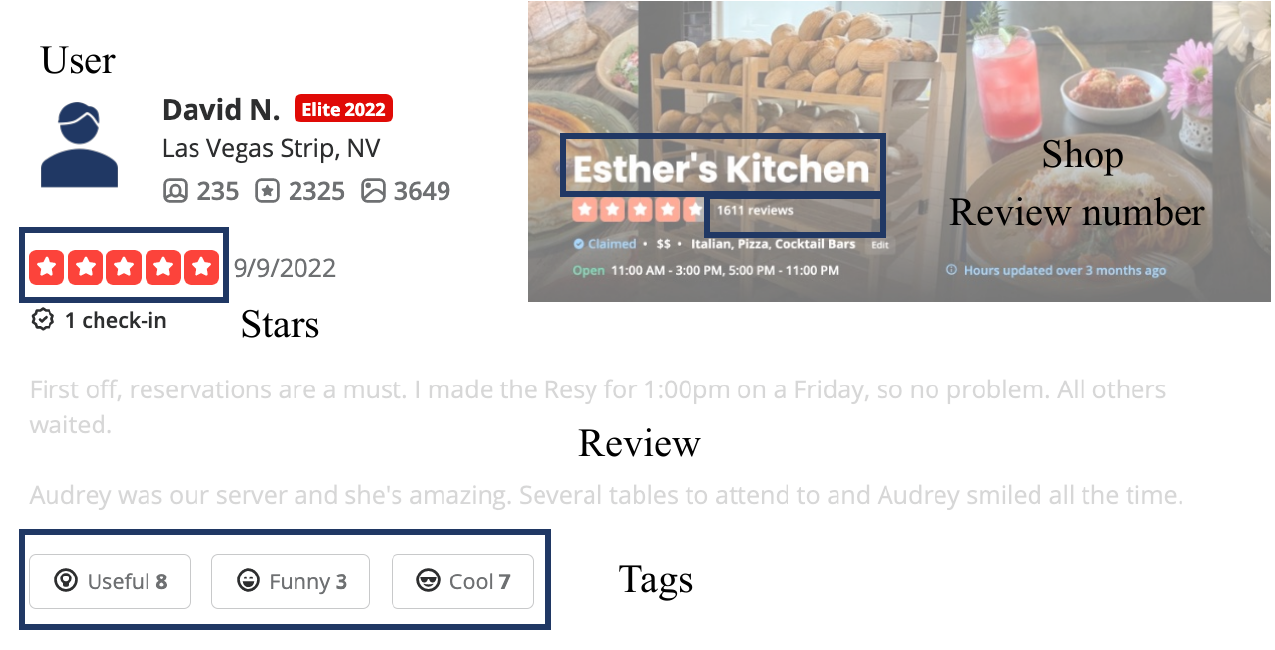}
		\caption{\small A review snapshot of the shop
			``Esther's Kitchen''. It contains the user
			information, shop statistics, review text, and the
			tags assigned to this review.}
		\label{fig:yelp}
	\end{center}
\end{figure}

The objective of our work is to investigate time series indexed across multi-relational networks. To offer a more lucid representation of GTNAR, we demonstrate it through a motivating dataset collected from Yelp (https://www.yelp.com/). Yelp serves as a prominent review platform for various businesses, including restaurants, local retailers, entertainment establishments, and more. It also functions as a social platform where users can share information and their personal experiences.
The dataset covers the period from 2010 to 2018 and is collected from five North American cities (i.e., Charlotte, Las Vegas, Phoenix, Scottsdale and Toronto), and comprises four main categories of information: user data (e.g., user registration time on Yelp), user-friend relationships, business information (including spatial location), and user reviews of businesses.
For example, Figure \ref{fig:yelp} illustrates a user's review of a restaurant named ``Esther's Kitchen'' in Las Vegas. In this case, the user gave the restaurant a five-star rating, and their review received 18 tags from other users, including 8 ``useful'', 3 ``funny'', and 7 ``cool'' tags. Overall, the restaurant has accumulated 1611 reviews.

\begin{figure}[htpb!]
	\centering
	\subfigure[Social network]{\label{fig:social_net}\includegraphics[width=0.3\textwidth]{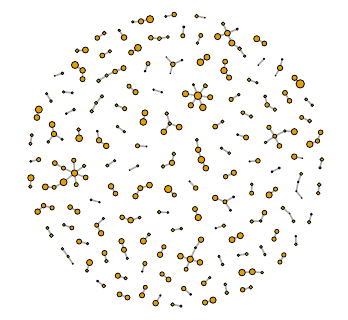}}
	\subfigure[Spatial network]{\label{fig:spat_net}\includegraphics[width=0.3\textwidth]{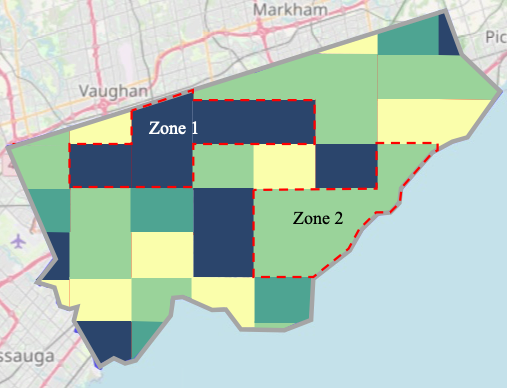}}
	\subfigure[Estimated parameters]{\label{fig:cluster_param}\includegraphics[width=0.3\textwidth]{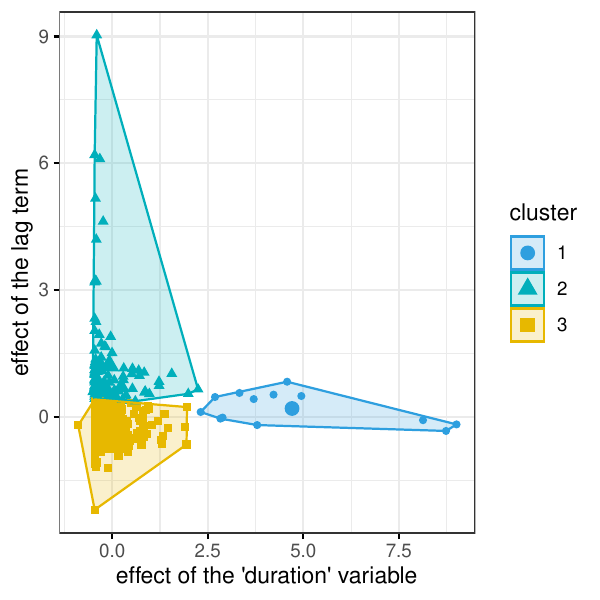}}
	\caption{\small {(a) The social network of Toronto users with degrees greater than one. Node sizes reflect the logarithm of the number of comments made in 2008. (b) The spatial network of Toronto districts during the last quarter of 2018. Colors represent four quantile intervals of the logarithm of the number of comments made in the 4th quarter of 2008. Darker colors indicate districts with higher comment activity. (c) Clustering results based on estimated regression coefficients, categorizing users into three distinct groups. Different colors and shapes visually represent each group.}}
	\label{fig:pre_analysis}
\end{figure}

Within the Toronto segment of the dataset, we have records for $N_1=462$ users who have provided reviews for restaurants in $N_2=56$ distinct locations, spanning $T=36$ quarters.
Our primary focus in this analysis concentrates on the variable denoted as $Y_{ij, t}$, representing the $\log(1+x)$-transformed number of reviews contributed by user $i$ to restaurants in district $j$ during the $t$th quarter. This variable forms a time series indexed by both the user ID ($i$) and the district ID ($j$).
The first challenge we encounter when analyzing this dataset is that neither users nor districts can be considered isolated units. As a result, it becomes imperative to model the $Y_{ij, t}$'s in a collective manner. Specifically, users form a social network, while districts establish a spatial network that fosters substantial interactions among their respective network members. These interactions, in turn, significantly influence the outcome variable $Y_{ij, t}$ when considered jointly. For example, \cite{tiwari2016social} found that peer social networks play a highly effective role in influencing restaurant preferences within social circles of friends. The social network analysis conducted on Yelp data by \cite{fe2023social} suggests that social network friends are 64\% more likely to visit the same restaurant when compared to non-friends. \cite{sun2017spatial} discovered significant spatial effects on ratings across various categories of Yelp venues, and \cite{gan2021spatial} investigated the spatial network effects on the tourism economy. However, these studies are primarily empirical and lack a strong statistical foundation.
Furthermore, they tend to concentrate solely on the impact of a single network, rather than considering multiple network effects jointly.  As a motivating example, Figure \ref{fig:social_net} depicts the social network of Toronto users with at least two friends, highlighting the observation that connected friends often have similar comment volumes. Meanwhile, Figure \ref{fig:spat_net} presents the spatial network of Toronto, indicating that neighboring districts, such as zones 1 and 2, tend to exhibit similar comment volumes. Both network effects contribute jointly to the outcome variable. Therefore, the first challenge we intend to address is how to construct a multivariate time series model that can rigorously quantify the impact of multiple network effects for data similar to the Yelp review dataset.

The second challenge we encounter in the Yelp review dataset relates to the heterogeneity among members in both social and spatial networks.
Previous research in network analysis \citep{newman2006structure} highlighted that users tend to naturally form clusters based on similar behaviors, a phenomenon consistent with Yelp's ecosystem, and users display distinct interaction patterns (e.g., frequent reviewers versus casual browsers) \citep{fe2023social, sun2017spatial}.
There is also empirical research \citep{wedel2000market,fiebig2010generalized}, demonstrating that segmenting users enhances recommendation accuracy on platforms with varied preferences, which is particularly applicable to Yelp's diverse range of restaurants and businesses \citep{jagabathula2018model}.
Similarly, restaurants located in different spatial regions, such as the central business district or other areas, may experience different levels of popularity, leading to varying degrees of spatial spillover effects \citep[see, e.g.,][]{koschinsky2009spatial}.
Recent advances (e.g., \cite{bonhomme2015grouped,su2016identifying, liu2020identification}) in high-dimensional time series modeling
demonstrated that subgrouping mitigates bias in parameter estimation and enhances estimation efficiency given limited sample sizes, which is also a critical feature for the Yelp data.
As a motivating example, we conduct a preliminary regression analysis with aggregated users' comment volumes (log-transformed) as the response variable, denoted as $Y_{it} = \sum_{j} Y_{ij,t}$. Two covariates are included: the lagged response ($Y_{i(t-1)}$) and the user's duration since registration (i.e., the number of months since joining Yelp as of the $(t-1)$th quarter). The estimated regression coefficients obtained from all users are clustered into three groups using the $k$-means algorithm, and these clusters are visualized in Figure \ref{fig:cluster_param}. An evident heterogeneous pattern is observed among the users' coefficients. To tackle this challenge, we adopt the approach proposed in \cite{zhu2023simultaneous}, which involves dividing the members of both social and spatial networks into several subgroups. We assume that members of each subgroup share similar characteristics.

Let $\bA^{(1)} = (a^{(1)}_{i j})\in \mR^{N_1\times N_1}$ and $\bA^{(2)} = (a^{(2)}_{ij})\in \mR^{N_2\times N_2}$ represent the observed exogenous adjacency matrices characterizing the social and spatial networks, respectively.
Specifically, $a_{ij}^{(1)} = 1$ implies that the $i$th user follows the $j$th user, while $a_{ij}^{(1)} = 0$ otherwise. Similarly, $a_{ij}^{(2)} = 1$ indicates that the $i$th district is a spatial neighbor of the $j$th district, while $a_{ij}^{(2)} = 0$ otherwise.
We follow the convention by setting $a_{ii}^{(1)} = 0$ and $a_{jj}^{(2)} = 0$ for $1 \le i \le N_1$ and $1 \le j \le N_2$.
We assume the existence of $G_1$ groups in the social network and $G_2$ groups in the spatial network. The group membership for the $i$th node in the social network is denoted as $g_i^{(1)}$ ($1 \le g_i^{(1)} \le G_1$), and the group membership for the $j$th node in the spatial network is denoted as $g_j^{(2)}$ ($1 \le g_j^{(2)} \le G_2$). We propose the following model with a two-way group structure:
\begin{equation}
	\begin{split}
		Y_{ij, t} &=\underbrace{\lambda_{g_i^{(1)}}^{(1)} \sum_{k = 1}^{N_1} \frac{a^{(1)}_{ik}}{n_{1i}} Y_{kj, (t-1)}}_{\text{\rm Social Network main effect}}
		+
		\underbrace{\lambda_{g_j^{(2)}}^{(2)} \sum_{k = 1}^{N_2} \frac{a^{(2)}_{kj}}{n_{2j}}Y_{ik, (t-1)}}_{\text{\rm Spatial Network main effect}}\\
		&\hspace{13em} + \underbrace{\alpha_{g_i^{(1)} g_j^{(2)}}Y_{ij, (t-1)}}_{\text{\rm Self-momentum}} +\underbrace{\bx_{it}^{(1)\top} \bzeta^{(1)}_{g_i^{(1)}}+
			\bx_{jt}^{(2)\top} \bzeta^{(2)}_{g_j^{(2)}}}_{\text{\rm Covariate effects}}+\ve_{ij, t},\label{eq:model00}
	\end{split}
\end{equation}
where $n_{1i}=\sum_{k=1}^{N_1}a^{(1)}_{ik}$, $n_{2j}=\sum_{k=1}^{N_2} a^{(2)}_{kj}$, $\bx^{(1)}_{it}\in \mR^{p_1}$ and $\bx^{(2)}_{jt}\in \mR^{p_2}$ are exogenous covariate vectors of finite dimensions associated with the $i$th user and $j$th district, respectively. Besides, $\ve_{ij, t}$ represents independent and identically distributed (i.i.d.) white noise with $E(\ve_{ij, t})=0$ and its variance $\var(\ve_{ij, t}) = \sigma^2$. For identifiability, it is required that $\sum_{g^{(1)}=1}^{G_1} \zeta_{g^{(1)},1}^{(1)} = 0$ when both intercepts are included in $\x^{(1)}_{it}$ and $\x^{(2)}_{it}$, where $\zeta^{(1)}_{g_i^{(1)},1}$ (the first element of $\bzeta^{(1)}_{g_i^{(1)}}$) represents the intercept for $\x^{(1)}_{it}$.

The first term in \eqref{eq:model00}, i.e., $\sum_{k = 1}^{N_1}
(a^{(1)}_{ik}/n_{1i})Y_{kj, (t-1)}$, represents the average  number of reviews ($\log(1+x)$-transformed)
by user $i$'s following friends on the restaurants in
district $j$ in the previous quarter. Consequently, $\lambda_{g_i^{(1)}}^{(1)}$
quantifies the influence of following friends on user $i$'s attitude
towards district $j$ and encapsulates a social network main effect. On
the other hand, the second term in \eqref{eq:model00}, i.e., $\sum_{k = 1}^{N_2} (a^{(2)}_{kj}/n_{2j}) Y_{ik, (t-1)}$, calculates the average log
number of reviews by the $i$th user on districts that are connected with the $j$th district in the previous quarter. Thus, $\lambda_{g_j^{(2)}}^{(2)}$ signifies how the user's review count towards district $j$ is influenced by his/her reviews towards the districts connected with district $j$, and can be interpreted as the spatial network main effect.
Additionally, $\alpha_{g_i^{(1)} g_j^{(2)}}$ represents the self-driven time effect
for the $(i,j)$th time series, quantifying the momentum effect of the
review activity by the $i$th user towards the $j$th district in
the previous quarter. A higher value of $\alpha_{g_i^{(1)} g_j^{(2)}}$ suggests a
greater level of loyalty of user $i$ to district $j$. Finally, $\bzeta_{g_i^{(1)}}^{(1)} \in\mR^{p_1}$ and $\bzeta_{g_j^{(2)}}^{(2)} \in\mR^{p_2}$ are external covariate effects at the user and district levels, enhancing the model's capacity to account for user and district heterogeneity in the data. By evaluating the significance of $\lambda_{g^{(1)}}^{(1)}$'s and $\lambda_{g^{(2)}}^{(2)}$'s, one can examine the presence of social and spatial network main effects while controlling for other factors in Model~\eqref{eq:model00}.

Our framework for multiple networks further extends the two-network model~\eqref{eq:model00} to multi-relational networks given by Model~\eqref{eq:model0} in Section~\ref{sec:general} below, which we refer to as the GTNAR model.

\subsection{General Model in Tensor Form}
\label{sec:general}

Suppose there are $q$ observed exogenous networks characterized by adjacency matrices $\bA^{(1)}, \cdots, \bA^{(q)}$, with the $l$th matrix $ \bA^{(l)} = (a_{ij}^{(l)}) \in \mR^{N_l \times N_l}$, where $N_l$ is the number of nodes in the $l$th network, and $q$ is finite.
Specifically, $a_{ij}^{(l)} = 1$ implies that the $i$th node is connected with the $j$th node in the $l$th network, while $a_{ij}^{(l)} = 0$ otherwise. We follow the convention by setting $a_{i_l i_l}^{(l)} = 0$ for $1 \le i_l \le N_l$ and $1 \le l \le q$.
We assume that there exists $G_l$ groups in the $l$th network, and the group membership for the $i$th node in this network is denoted as $g_{i_l}^{(l)}$ ($1\le g_{i_l}^{(l)} \le G_l$). As a result, each network has its own latent group structure among the $N_l$ network nodes, which is denoted by $\mG_l = (g_i^{(l)}: 1 \le i \le N_l)$, $1\le l \le q$.
Under such a model framework, the response variables of interest constitute a tensor-valued time series, denoted by $\cY_t=(Y_{i_1i_2...i_q,t})\in \mR^{N_1\times N_2\times\cdots\times N_q}$, with the following model structure
\begin{align}
	Y_{i_1i_2...i_q,t} =\sum_{l=1}^q
	\underbrace{\lambda_{g_{i_l}^{(l)}}^{(l)}\sum_{k = 1}^{N_l} \frac{a_{i_{l}k}^{(l)}}{n_{li_l}}Y_{i_1...i_{l-1}ki_{l+1}...i_q,(t-1)}}_{\text{\rm The $l$th Network main effect}}
	&+ \underbrace{\alpha_{g_{i_1}^{(1)}...g_{i_q}^{(q)}}Y_{i_1i_2...i_q,(t-1)} }_{\text{\rm Self-momentum}}\nonumber\\
	&+\sum_{l=1}^q
	\underbrace{\bx_{i_lt}^{(l)\top} \bzeta_{g_{i_l}^{(l)}}^{(l)}}_{\text{\rm The $l$th covariate effects}}+\ve_{i_1i_2...i_q,t},\label{eq:model0}
\end{align}
where $n_{li_l}=\sum_{k=1}^{N_1}a_{i_lk}^{(l)}$, $\bx_{i_lt}^{(l)}\in \mR^{p_l}$ represents exogenous covariates associated with the $i_l$th member in the $l$th network, and $\ve_{i_1i_2...i_q,t}$ denotes the white noise with $\var(\ve_{i_1 \cdots i_q, t}) = \sigma^2$.
In this work, we treat the number of layers ($q$) and the covariates dimensions ($p_l, ~l \in [q]$) as fixed constants.
For indentifiability, assume that $\sum_{g^{(l)} =1}^{G_l} \zeta_{g^{(l)}, 1}^{(1)} = 0 ~(l \in [q-1])$.
Define  $\mG_l = (g_{i_l}^{(l)}: 1\le i_l \le N_l)^\top \in \mR^{N_l}$ and denote $\mG = \{\mG_l: 1\le l \le q \}$.
Let $\cR_{g}^{(l)} = \{ i_l: g_{i_l}^{(l)} = g\}$ and further denote $N_{l g} = |\cR_{g}^{(l)}|$.
Denote $g^{-(l)} = (g^{(1)}, \cdots, g^{(l-1)}, g^{(l+1)}, \cdots, g^{(q)})$ as the group membership indices excluding the $l$th network,
and denote the group index $\mI_{g^{(1)}, \cdots, g^{(q)}}$ as a function of $(g^{(1)}, \cdots, g^{(q)})$, which takes values from $\{1,\cdots,\prod_l G_l\}$.

Define $\bW^{(l)} = (a^{(l)}_{ij}/n_{li})$ as the $l$th row-normalized adjacency matrix of $\bA^{(l)}$. Then the GTNAR model \eqref{eq:model0} can be expressed in a tensor form as follows,
\begin{equation}
	\cY_t = \sum_{l = 1}^q (\cY_{t-1} \times_l \bW^{(l)}) \times_l \L^{(l)} + \mA \odot \cY_{t-1} + \sum_{l=1}^q \bbeta_{X_l, t}^{(l)} \circ_{k \neq l} \one_{N_k}
	+ \mE_t,\label{eq:model}
\end{equation}
where
$\L^{(l)} = \diag(\lambda^{(l)}_{g_{i_l}^{(l)}}:1\le i_l \le N_l) \in \mR^{N_l \times N_l}$,
$\mA = (\alpha_{g_{i_1}^{(1)}  \cdots g_{i_q}^{(q)}}: 1\le i_l \le N_l)  \in \mR^{N_1 \times \cdots \times N_q}$,
$\bbeta^{(l)}_{X_l, t} = (\bx_{i_l t}^{(l)\top} \bzeta_{g_{i_l}^{(l)}}^{(l)}:1\le i_l \le N_l)^\top  \in \mR^{N_l}$
and $\mE_t = (\ve_{i_1, \cdots, i_q, t})  \in \mR^{N_1 \times \cdots \times N_q}$.

	\begin{remark}\label{rmk:q2}
		{\bf (Matrix Autoregression When $q =  2$)}
		When $q=2$, the model in \eqref{eq:model} can be simplified as the matrix-valued autoregression
		model in \eqref{eq:model_matrix} below,
		\begin{align}
			\Y_t =  (\L^{(1)}\W^{(1)}) \Y_{t-1} + \Y_{t-1} (\W^{(2)} \L^{(2)})  + \A \circ \Y_{t-1} + \bbeta_{X_1, t} \one_{N_2}^\top + \one_{N_1} \bbeta_{X_2,t}^\top + \bE_t, \label{eq:model_matrix}
		\end{align}
		where $\Y_t \in \mR^{N_1 \times N_2}$ is the response matrix, $\L^{(l)} = \textup{diag}(\lambda^{(l)}_{g_{i_l}^{(l)}}:1\le i_l \le N_l) \in \mR^{N_l \times N_l}$ for $l = 1, 2$,
		$\A = (\alpha_{g_{i_1}^{(1)} g_{i_2}^{(2)}}:1\le i_1\le N_1, 1\le i_2\le N_2)  \in \mR^{N_1 \times N_2}$, and $\bE_t \in \mR^{N_1 \times N_2}$ is the random noise matrix.
	\end{remark}

	\begin{remark}\label{rmk:network_construct}
		We remark that GTNAR is not limited to the motivating example and can be utilized in a wide range of applications.
		Specifically, GTNAR is designed for tensor-valued time series data, provided we can collect network relationships among the units.
		The network under consideration is not restricted to spatial networks, as we use in the motivating example.
		For instance, in financial data analysis, we can use the common shareholding relationship to examine the spillover effects of systematic risk \citep{feng2023systemic}.
		In the international trading market, regions can construct two adjacent networks from the export and import aspects, while the trading products categories form a similarity network \citep{alves2019nested}.
		In the contextual recommendation system, the social network relationships among the users form a network on the first dimension, the similarities based on the item characteristics form the second one, and the proximities of different types of users' activities mode form the third network \citep{adomavicius2005incorporating, wu2016contextual}.
		Other examples include supply chain linkages \citep{serpa2018impact},
		common interests in customer purchase \citep{carroni2020bring}, and others, according to the application scenarios.
	\end{remark}

	\begin{remark}\label{rmk:real_tensor}
		Applications of tensor-valued time series are widespread, spanning numerous fields.
		For instance, 
		a contextual recommender system naturally generates a 3-dimensional tensor of users, items, and contexts, as it recommends items to users under varying contexts (such as the promotion types and the users' locations) over a period of time; 
		each dimension can be further enriched by its corresponding network, such as a user social network, an item similarity network, and a contextual proximity network.
		Similarly, international trade data, capturing the export volumes of different product categories between regions over a period of time, forms another 3-dimensional tensor-valued time series, where geographical and product-similarity networks can be constructed on each dimension. 
		In the domain of neuroimaging, functional Magnetic Resonance Imaging (fMRI) produces a series of 3D brain scans, creating a 3-dimensional tensor-valued time series where networks can be defined by the adjacency of brain regions. 
		A fourth example is found in online live-streaming, where user activities (such as comments and shares) directed at various streamers and different topics form a 4-dimensional tensor. 
		This type of tensor recorded along a period of time naturally forms a 4-dimensional tensor-valued time series.
		The practical and urgent need to analyze this type of complex, high-dimensional data, which is ubiquitous in modern socio-economic and scientific contexts, serves as the primary applied motivation for the GTNAR model.
	\end{remark}
	
	\begin{remark}\label{rmk:reduce_dim}
		{\bf (Parameter Dimension Reduction)}
		We remark that we reduce the parameter dimension by embedding the observed network weighting matrices $\W^{(k)}$
		and a group structure.
		Take the case with $q = 2$ for example.
		In general, one needs to estimate $O(N_1^2+N_2^2)$ parameters in model \eqref{eq:model_matrix}. In contrast, with the imposed model structure, the number of estimated parameters reduces to $O(G_1(p_1+1) + G_2 (p_2+1) + G_1 G_2)$ when the group memberships are given, which is greatly reduced due to incorporating the weighting matrices $\W^{(l)}$ and memberships $\mG_1$ and $\mG_2$.
	\end{remark}

	\subsection{Comparisons with Existing Literature}\label{subsec:review}
	There are numerous works that are closely related to GTNAR.
	The most pertinent works encompass high-dimensional vector autoregression models \citep{davis2016sparse, wang2022high, miao2023high}, tensor-valued time series models \citep{hoff2015multilinear, wang2019factor, chen2021autoregressive, chen2022factor, chen2023statistical, wang2024high}, grouped panel data models \citep{ke2015homogeneity, ando2016panel, su2016identifying}, and the network autoregressive models with heterogeneous effects \citep{zhu2020grouped, zhu2023simultaneous}.
	In this subsection, we illustrate how the GTNAR model differs from the above existing models.

	\subsubsection{High-dimensional Vector Autoregression Models}\label{subsec:sVAR}
	
	Recently, there have been studies focusing on the high-dimensional
	vector autoregression (VAR) models \citep{davis2016sparse, wang2022high, miao2023high}.
	By stacking elements of the tensor
	$\cY_t = (Y_{i_1 \cdots, i_q, t})\in \mR^{N_1\times \cdots \times N_q}$ into a vector, GTNAR model \eqref{eq:model} can be re-written as a VAR model with 
	dimension $N'  =\prod_l N_l$, 
	\begin{align*}
		\y_t = \Big[ \sum_{l=1}^q \Big( (\L^{(l)} \W^{(l)}) \otimes_{k \neq l} \I_{N_k} \Big) + \text{diag}(\mathbf{a}) \Big] \mathbf{y}_{t-1} + \sum_{l=1}^q \Big( \bbeta^{(l)}_{X_l, t} \otimes_{k \neq l} \one_{N_k} \Big) + \bve_t,
	\end{align*}
	where $\a = \vec(\mA) \in \mR^{N_1 \times \cdots \times N_q}$ and $\bve_t = \vec(\mE_t)$.
	This allows for the application of existing techniques of
	high-dimensional VAR models. 
	While this formulation imposes fewer assumptions on the model
	coefficients, it leads to a much larger number of parameters to be
	estimated. Specifically, the number of parameters is
	$O(\prod_l N_l^2)$, in comparison to $O(\sum_l 
	G_l(p_l+1)+\prod_l G_l)$ (when fixing the group memberships) for
	the GTNAR model.
	Even with commonly used sparsity regularization, the estimation variance
	is likely to inflate compared to the network models under certain
	scenarios \citep{zhu2023simultaneous}, which is also investigated
	in our  numerical studies. 
	In addition, the ``stacking'' operation will destroy the
	intrinsic data structure in the original tensor form, 
	hence the coefficients estimated in the general VAR model lack
	clear interpretations and provide limited insights into the network
	effects.
	These drawbacks are also recognized in existing literatures~\citep{chen2021autoregressive, wang2024high}.

	\subsubsection{Tensor-valued Time Series Models}\label{subsec:tensor_ts}
	This line of research contains two main categories, the tensor autoregressive framework \citep{ding2018matrix, chen2021autoregressive, wang2024high} category, and the tensor factor model \citep{hoff2015multilinear, wang2019factor, chen2020constrained, chen2022factor, chen2023statistical} category. 
	Specifically, GTNAR should be classified into the first category.
	For ease of understanding, we compare our modeling approach with existing approaches when $q = 2$, which reduces to the matrix case in \eqref{eq:model_matrix}.

	First, the tensor autoregressive (TAR) models characterize the dynamics of $\Y_t$ using its lagged information $\Y_{t-1}$.
	For example, the TAR model proposed by \cite{wang2024high} takes the following form, i.e.,
	\beq
	\cY_{t} =\langle \mA, \cY_{t-1} \rangle + \mE_t, \label{eq:mar_wang}
	\eeq
	where $\mA \in \mR^{n_1 \times \cdots \times n_q \times n_1 \times \cdots \times n_q}$ is a $2q$ dimensional tensor, which has Tucker ranks $(r_1, \cdots, r_q)$ with $r_l = \rank(\mA_{(l)})$. 
	They assume that the tensor $\mA$ takes a low rank structure as
	$\mA = \mC \times_{l = 1}^{2q} \U_l$, $\mC \in \mR^{r_1 \times \cdots \times r_{2q}}$ is the core low-rank tensor and $\U_l \in \mR^{n_l \times r_l}$ for $l \in [2q]$.
	When $q=2$, the model \eqref{eq:mar_wang} can be rewritten as
	\begin{align}
		\Y_t = \U_3 (\mC \U_1^\top \Y_{t-1} \U_2) \U_4^\top + \bE_t. \label{eq:mar_wang_q2}
	\end{align}
	In model \eqref{eq:mar_wang_q2}, the number of parameters to be estimated is $O\{\prod_{l = 1}^{4} r_l + 2(N_1 + N_2)\}$.
	Compared to the existing TAR modeling approach, the GTNAR model \eqref{eq:model_matrix} has two major differences. 
	First, we embed the network weighting matrices $\W^{(l)}$ into the autoregression matrices to characterize the network dependence structure.
	To further reduce the number of estimated parameters, we consider a group structure with autoregression coefficients in $\L^{(1)}$ and $\L^{(2)}$, respectively.
	As we commented in Remark \ref{rmk:reduce_dim} for $q = 2$, GTNAR contains $O(G_1 (p_1+1) + G_2(p_2+1) + G_1 G_2)$ parameters in the
	network autoregression coefficients when the group memberships are fixed.
	In addition, we provide an interpretable modeling framework with the group-specific autoregression coefficients.
	Furthermore, we would like to remark that the autoregression matrices in model \eqref{eq:model_matrix} cannot be expressed
	with a low rank form unless the network matrices $\W^{(l)}$ take a low rank structure.
	Second, in addition to the network autoregression terms, we use $\A \circ \Y_{t-1}$ to capture the self-momentum effect, which cannot be characterized by \eqref{eq:mar_wang} with the low-rank assumption.
	One can require $\A$ to have a specific low rank structure $\A = \M^{(1)} \G \M^{(2) \top} $, where $\M^{(l)} = (\e_{g_i^{(l)}}: 1\le i\le N_l)^\top \in \mR^{N_l \times G_l}$ is a membership matrix for the $l$-th layer, and $\G = (\alpha_{gg'}: g\in [G_1], g'\in [G_2])\in \mR^{G_1\times G_2}$.
	This relates GTNAR to the traditional low-rank models \citep{chen2021autoregressive, wang2024high}.
	We remark that the self-momentum effect contains cross-layer parameters
	in $\G$, which poses challenges in our theoretical analysis since the memberships from each layer cannot be analyzed separately.

	The second category of tensor-valued time series models includes the tensor factor (TF) models considered by \cite{wang2019factor}, \cite{chen2022factor}, and \cite{chen2023statistical}.
	The TF model assumes that the dynamics of $\cY_t$ can be driven by a low-dimensional latent dynamic tensor factor {$\mF_t$}.
	For example, the TF model proposed by \cite{chen2022factor} takes the form $\cY_t = \mF_t (\prod_k \times_k) \A_k + \mE_t$.
	We refer to \cite{hoff2015multilinear} and \cite{chen2020constrained} for other TF models in various forms.
	Although the TF model can be used to directly conduct a dimension reduction for $\cY_t$, it cannot quantify the historical self-momentum effect for $\cY_t$, and it also lacks a sound interpretation regarding the latent group structure.

	We also compare the numerical performance of GTNAR with that of the multilinear tensor model in \cite{hoff2015multilinear} to illustrate the advantages of our proposed model.

	\subsubsection{Grouped Panel Data Models}

	Utilizing latent group structures to model heterogeneous data
	has a well-established history in statistical models \citep{ke2015homogeneity,bester2016grouped,liu2020identification}.
	For instance,
	\cite{bonhomme2015grouped} and \cite{bester2016grouped} introduced
	grouped linear panel models with time-varying fixed effects and
	individual fixed effects, respectively. \cite{su2016identifying}
	introduced a Classifier Lasso (C-Lasso) estimator for panel models,
	and more recently, \cite{liu2020identification} explored estimation
	and inference in the presence of possible over-specification of the group
	number.
	Specifically, \cite{su2016identifying} considered the grouped linear penal model $Y_{it} = \bbeta_i^\top \x_{it} + \mu_i + \ve_{it}$, where the parameter $\bbeta_i$ takes values from $G_0$ group centers $\{\balpha_1, \cdots, \balpha_{G_0}\}$.
	However, the model format designed for vector-valued responses limits its applicability to complex tensor-valued data, thereby lacking generality.
	Compared to these studies, the GTNAR model further shows the additional advantage in quantifying the multiple heterogeneous network effects by incorporating different types of networks in tensor dimensions.

	\subsubsection{Network Autoregression Model with Group Heterogeneity}\label{subsec:narg_compare}
	There have also been recent efforts to leverage latent group
	structures in modeling time series data within a single network, as
	demonstrated by
	\cite{zhu2020grouped,zhu2023simultaneous, chen2023community}; and etc.
	However,
	these works have predominantly focused on time series observed on
	a single network, and to the best of our knowledge, our work is the first to tackle time series indexed by multiple networks with distinct group structures.
	Specifically, \cite{zhu2023simultaneous} considered an autoregression model for a vector formed response $\y_t$ as
	\begin{align}
		\y_t = \B \y_{t-1} + \bmu_z + \bve_t,\label{eq:gnar_joe}
	\end{align}
	where $\B \in \mR^{N \times N}$ with its $(i,j)$th element being $b_{ij} = w_{ij} \beta_{g_i g_j}$ with $w_{ij}=a_{ij}/n_i$ for $i \neq j$ and $b_{ii} = \nu_{g_i}$ for $i = 1, \cdots, N$.
	In \eqref{eq:gnar_joe}, $\bmu_z = (\x_1^\top \bzeta_{g_1}, \cdots, \x_N^\top \bzeta_{g_N})^\top$ represents the covariates term. 
	The key difference between these two approaches lies in how they model network interactions. The goal of \cite{zhu2023simultaneous} is to capture interactions within a single network as flexibly as possible, where the effect between nodes $i$ and $j$, namely, $w_{ij}\beta_{g_i g_j}$, depends jointly on their group memberships $g_i$ and $g_j$. This makes the consistent estimation of group memberships substantially more challenging than in other existing group panel data models \citep{su2016identifying, liu2020identification}, where group memberships do not interact in the parameter structure. To address this difficulty, \cite{zhu2023simultaneous} proposed a membership refinement procedure based on brute-force enumeration. 
	While a straightforward extension of this approach
	to the multiple network tensor-valued data setting studied might be possible, we adopt a completely different within-network effect model in
	\eqref{eq:model_matrix}, where in each network, the effect between node $i$ and $j$ takes the form $w_{ij}^{(l)}\lambda_{g_{i}^{(l)}}$, depending only on the membership of node $i$ (but not $j$) within the $l$th network, $l=1,\ldots,q$. This approach has the advantage of avoiding the refinement procedure of \cite{zhu2023simultaneous}, while has its own complexity of handling multiple networks simultaneously. Thus, the theoretical challenges in \cite{zhu2023simultaneous} and in our work stem from different sources, and we employ very different strategies to resolve them. 
	The GTNAR model is a new modeling approach that leverages the general grouping idea to address challenges unique to the multi-layer networks setting.

\section{Model Estimation}\label{sec:estimate}

	We first introduce the necessary notations.
	For the network corresponding to the $l$th tensor dimension, there exist $G_l$ groups for the network nodes, denoted by $g^{(l)} \in [G_l]$. For each group, the group-level parameters are denoted by $\btheta_{g^{(l)}}^{(l)} = (\lambda_{g^{(l)}}^{(l)}, \bzeta_{g^{(l)}}^{(l)\top})^\top \in \mR^{p_l+1}$. The collection of layer-specific parameters is then denoted as $\btheta^{(l)} = (\btheta_{g^{(l)}}^{(l)}: g^{(l)} \in [G_l]) \in \mR^{G_l(p_l+1)}$, leading to the collection of $\bxi = (\btheta^{(1)\top}, \cdots, \btheta^{(q)\top}, \vec(\balpha)^\top)^\top \in \mR^{\sum_l G_l(p_l+1) + \prod_l G_l}$ for all $q$ tensor dimensions. Finally, the self-momentum parameter is denoted as $\balpha = (\alpha_{g^{(1)} \cdots g^{(q)}})_{g^{(l)} \in [G_l]} \in \mR^{G_1 \times \cdots \times G_q}$.
	In the following, we discuss estimation for the GTNAR model, which includes the group parameters $\bxi$ and memberships $\mG$.
	We utilize an iterative algorithm to update $\bxi$ and $\mG$.
	We first discuss the estimation of $\bxi$ when $\mG$ is given, and then we introduce the iterative algorithm for joint estimation of $\{\bxi, \mG\}$.
	For easy understanding, we first introduce the estimation procedure for the matrix model in \eqref{eq:model_matrix} when $q = 2$.
	Then we present a general estimation procedure for a general $q$.

	\subsection{Estimation Procedure for Model \eqref{eq:model_matrix} when $q = 2$}\label{subsec:est_matrix}
	We first introduce the estimation of model \eqref{eq:model_matrix} with two networks for clear illustration.
	We aim to minimize the following objective function
	\begin{align}
		Q(\bxi, \mG)  = \sum_{i = 1}^{N_1} \sum_{j = 1}^{N_2} \sum_{t = 1}^T \Big(Y_{ij,t} & - \lambda^{(1)}_{g_i^{(1)}} \sum_{k = 1}^{N_1} \frac{a_{ik}^{(1)}}{n_{1i}} Y_{kj, (t-1)} - \lambda^{(2)}_{g_j^{(2)}} \sum_{k = 1}^{N_2} \frac{a_{kj}^{(2)}}{n_{2j}} Y_{ik, (t-1)} \nonumber \\
		& - \alpha_{g_i^{(1)} g_j^{(2)}} Y_{ij, (t-1)} - \bx_{it}^{(1)\top} \bzeta^{(1)}_{g_i^{(1)}} - \bx_{jt}^{(2)\top} \bzeta^{(2)}_{g_j^{(2)}} \Big)^2. \label{eq:Q_obj_simple}
	\end{align}
	To minimize \eqref{eq:Q_obj_simple}, we solve
	\begin{align*}
		\frac{\partial Q(\bxi, \mG)}{\partial \btheta_{g^{(1)}}^{(1)}} = \zero,~~~
		\frac{\partial Q(\bxi, \mG)}{\partial \btheta_{g^{(2)}}^{(2)}} = \zero, ~~~
		\frac{\partial Q(\bxi, \mG)}{\partial \alpha_{g^{(1)} g^{(2)}}} = 0.
	\end{align*}
	By solving the equations above, we obtain the estimator when $q=2$ as $\wh\btheta = \bM^{-1} \bb$, where
	\begin{align*}
		\mathbf{M}=\left(\begin{array}{ccc}
			\mathbf{M}^{(1)} & \mathbf{M}^{(12)} & \mathbf{M}^{(1 \alpha)} \\
			\mathbf{M}^{(12) \top} & \mathbf{M}^{(2)} & \mathbf{M}^{(2 \alpha)} \\
			\mathbf{M}^{(1\alpha) \top} & \mathbf{M}^{(2 \alpha) \top} & \mathbf{M}^\alpha
		\end{array}\right), \quad \mathbf{b}=\left(\begin{array}{c}
			\mathbf{b}^{(1)} \\
			\mathbf{b}^{(2)} \\
			\mathbf{b}^\alpha
		\end{array}\right),
	\end{align*}
	and the specific expressions are given in Appendix \ref{sec:notation_q2}.

\subsection{Estimation Procedure for General Model with General $q$}

We now discuss the estimation of the GTNAR model \eqref{eq:model} in general.
To simultaneously estimate the model parameters and the group memberships, we aim to minimize the following least squares objective function:
\begin{align}\label{eq:Q_obj}
	Q({\bxi,\mG})  = \sum_{i_1=1}^{N_1} \cdots \sum_{i_q = 1}^{N_q} \sum_{t=1}^T \Big(
	Y_{i_1i_2...i_q,t} & - \sum_{l=1}^q \lambda_{g_{i_l}^{(l)}}^{(l)}\sum_{k = 1}^{N_l} \frac{a^{(l)}_{i_{l}k}}{n_{li_l}}Y_{i_1...i_{l-1}ki_{l+1}...i_q,(t-1)}\nonumber\\
	&
	-\alpha_{g_{i_1}^{(1)}...g_{i_q}^{(q)}}Y_{i_1i_2...i_q,(t-1)}
	-\sum_{l=1}^q \bx_{i_lt}^{(l)\top}\bzeta_{g_{i_l}^{(l)}}^{(l)}
	\Big)^2.
\end{align}

We first discuss the estimation when the group memberships $\mG$ are given. In this case, the minimization of~\eqref{eq:Q_obj} is equivalent to solving
$
\frac{\partial Q(\bxi, \mG)}{\partial \btheta_{g^{(l)}}^{(l)}} = \zero,~~
\frac{\partial Q(\bxi, \mG)}{\partial \alpha_{g^{(1)} \cdots g^{(q)}}} = 0
$
for all $g^{(l)}\in[G_l]$ and $l \in [q]$.
Recall that $N_{l g^{(l)}} = |\cR_{g^{(l)}}^{(l)}|$
and let
\begin{align}
	\hskip -1em	\mX_{g^{(1)} \cdots g^{(q)},t}^{(l)} &= \Big(\vec\big\{(\cY_{t-1} \times_l \bW^{(l)})_{g^{(1)} \cdots g^{(q)}}\big\},
	\one_{N_{1 g^{(1)}}} \otimes \cdots \otimes (\bX_t^{(l)})^{(\cR_{g^{(l)}}^{(l)}, \cdot)} \otimes \cdots \otimes \one_{N_{q g^{(q)}}}   \Big) ,\label{eq:X_gt}
\end{align}
where $\bX_{t}^{(l)} = (\bx_{1t}^{(l)}, \cdots, \bx_{N_lt}^{(l)})^\top \in \mR^{N_l \times p_l}$,
and that
\begin{align*}
	(\cY_{t-1} \times_l \bW^{(l)})_{g^{(1)} \cdots g^{(q)}}  &= \Big(\sum_{i_l = 1}^{N_l} Y_{i_1, \cdots, i_l, \cdots, i_q, t-1} (a^{(l)}_{s, i_l}/n_{ls})  \Big)_{i_1 \in \cR_{g^{(1)}}, \cdots, s \in \cR_{g^{(l)}}, \cdots, i_q \in \cR_{g^{(q)}}}.
\end{align*}
Then one can verify that
\begin{align}
	& \frac{\partial { Q(\bxi, \mG)}}{\partial \btheta_{g^{(l)}}^{(l)}} =
	\Big(\sum_{t, g^{-(l)}}\mX_{g^{(1)} \cdots g^{(q)},t}^{(l)\top} \mX_{g^{(1)} \cdots g^{(q)},t}^{(l)} \Big)
	\btheta_{g^{(l)}}^{(l)}\nonumber\\
	&- \sum_{t, g^{-(l)}}  \Big\{\mX_{g^{(1)} \cdots g^{(q)},t}^{(l)\top} \Big(
	\mY_{g^{(1)} \cdots g^{(q)},t} - \mY_{g^{(1)} \cdots g^{(q)},(t-1)} \alpha_{g^{(1)} \cdots g^{(q)}}  - \sum_{m \neq l} \mX_{g^{(1)} \cdots g^{(q)},t}^{(m)} \btheta_{g^{(m)}}^{(m)}  \Big)\Big\}  ,\label{eq:theta_grad}
\end{align}
where $	\mY_{g^{(1)} \cdots g^{(q)},t} = \vec \Big\{ \cY_t^{\big(\cR_{g^{(1)}}^{(1)}, \cdots, \cR_{g^{(q)}}^{(q)}\big)} \Big\} \in \mR^{N_{1 g^{(1)}}  \cdots  N_{q g^{(q)}}}$,
and the summation $\sum_{g^{-(l)}}$ is the simplified notation for $\sum_{g^{(1)}, \cdots, g^{(l-1)},  g^{(l+1)}, \cdots g^{(q)}}$.
Furthermore, it holds that
\begin{align}
	\frac{\partial {Q(\bxi, \mG)}}{\partial \alpha_{g^{(1)} \cdots g^{(q)}}} & =
	\sum_t \|\mY_{g^{(1)} \cdots g^{(q)},(t-1)} \|^2 \alpha_{g^{(1)} \cdots g^{(q)}}
	\nonumber\\
	& -\sum_t \mY_{g^{(1)} \cdots g^{(q)},(t-1)}^{\top} \Big(
	\mY_{g^{(1)} \cdots g^{(q)},t}  - \sum_{l=1}^q \mX_{g^{(1)} \cdots g^{(q)},t}^{(l)} \btheta_{g^{(l)}}^{(l)} \Big).\label{eq:alpha_grad}
\end{align}
Equations \eqref{eq:theta_grad} and \eqref{eq:alpha_grad} define a system of linear equations, whose solutions have the form ${\wh\bxi = \bM^{-1} \b}$ with $\bM \in \mR^{\{\sum_l G_l(p_l+1) + \prod_l G_l\} \times \{\sum_l G_l(p_l+1) + \prod_l G_l\}}$ and $\b \in \mR^{\sum_l G_l(p_l+1) + \prod_l G_l}$, where
\begin{align}
	&\bM =  \left(
	\begin{array}{cccccc}
		\bM^{(1)}& \bM^{(12)} & \bM^{(13)} & \cdots & \bM^{(1q)} & \bM^{(1\alpha)} \\
		\bM^{(21)} & \bM^{(2)} & \bM^{(23)} & \cdots & \bM^{(2q)} & \bM^{(2 \alpha)} \\
		\cdots & \cdots & \cdots & \cdots & \cdots & \cdots \\
		\bM^{(q1)} & \bM^{(q2)} & \cdots & \cdots & \bM^{(q)}& \bM^{(q \alpha)}\\
		\bM^{(1 \alpha) \top}& \bM^{(2\alpha)\top} & \cdots & \cdots & \bM^{(q \alpha)}& \bM^{(\alpha)}
	\end{array}
	\right), ~~~
	\b =  \left(
	\begin{array}{c}
		\b^{(1)} \\
		\cdots \\
		\b^{(q)}\\
		\b^{\alpha}
	\end{array}
	\right).\label{eq:Mb}
\end{align}
The specific terms in \eqref{eq:Mb} are given in Appendix \ref{sec:M_express}.
Subsequently, given the estimated parameters $\wh\bxi$, we update the
group memberships $\mG_l$ from $l=1$ to $q$ iteratively.
Specifically, given $\bxi$ and $\mG_{-l} \defeq \{ \mG \setminus \mG_l \}$, the $\mG_l$ is updated by
\begin{align}
	\wh g_{i_l}^{(l)}  \in \arg\min_{g_{i_l}^{(l)}  \in [G_l]}
	\sum_{i_{-l}} \sum_{t=1}^T  \Big(
	Y_{i_1i_2...i_q,t}  & - \sum_{l=1}^q \lambda_{g_{i_l}^{(l)}}^{(l)}\sum_{k = 1}^{N_l} \frac{a^{(l)}_{i_{l}k}}{n_{li_l}}Y_{i_1...i_{l-1}ki_{l+1}...i_q,(t-1)}\nonumber\\
	&
	-\alpha_{g_{i_l}^{(l)}...g_{i_q}^{(q)}}Y_{i_1i_2...i_q,(t-1)}
	-\sum_{l=1}^q \bx_{i_lt}^{(l)\top}\bzeta_{g_{i_l}^{(l)}}^{(l)}
	\Big)^2,
	\label{eq:g_i}
\end{align}
where $\sum_{i_{-l}} = \sum_{i_1=1}^{N_1}\cdots \sum_{i_{l-1} = 1}^{N_{l-1}} \sum_{i_{l+1} = 1}^{N_{l+1}} \cdots \sum_{i_q = 1}^{N_q}$.

We summarize the algorithm in Algorithm \ref{alg:gmnar} in Appendix \ref{subsec:alg}, which consists of iterations of two major steps. The first step updates the estimated parameters given the group memberships, while the second step updates group memberships given the parameter estimates.
We note that the update step in \eqref{eq:g_i} requires input of memberships of
	other layers (i.e., $\mG_{-l}$). Therefore, we adopt a sequential updating rule. 
	During the $k$th iteration, for $l'<l$, we substitute $\mG_{l'}$ by $\mG_{l'}^{[k]}$, and for $l'>l$, we substitute $\mG_{l'}$ by $\mG_{l'}^{[k-1]}$ in the previous iteration. See \eqref{eq:update_g} for detailed expression.
Each step can be efficiently computed due to the simple analytical forms.
This algorithm can be validated to converge to a local minimizer, of which the proof is given in Appendix \ref{sec:alg_converge}.
	We remark that whether it can converge to a global optimizer is a challenging question due to the nonconvexity of the objective function \eqref{eq:Q_obj} \citep{murty1987some, lin2020near}.
	To increase the chance of finding the global optimizer, a set of initialization strategies is employed, which are described in detail in Algorithm \ref{alg:init}.
	We try multiple initial values and use the one with the minimum loss function as the best one, which guarantees a stable numerical performance.
	We leave the theoretical investigation for a global optimizer as an interesting future topic.

	\begin{remark}\label{rmk:alg1_reduce_dim}
			In Algorithm \ref{alg:gmnar}, {\sc Step 2} achieves dimension reduction due to the calculation of group memberships.
			Note that in each layer, the update equation \eqref{eq:update_g} only involves subject $i_l$, and does not depend on other inner-layer subjects. 
			Thus, the memberships update equation \eqref{eq:update_g} can be computed in parallel for each inner-layer subject ($i_l$ in the $l$th layer), greatly increasing the computational speed.
			We show our computational cost under each sample size setting in Figure \ref{fig:compute_time}.
			Even at the maximum network size configuration ($N_1 = 300, N_2 = 250$), total computational costs remain no more than 20 seconds across all time horizons.
			More discussion can be found in Appendix \ref{subsec:compute_cost}.
		\end{remark}

	\subsection{Selection of Group Numbers}\label{sec:group_number}

	To simplify the notations, denote $\underline{G} =  (G_1, \cdots, G_q)^\top$, and denote the corresponding true group numbers as $\underline{G}_0$.
	We focus our later theoretical analysis on the case that $G_l$ is finite for $1\le l\le q$.
	Write ${\wh \bxi(\underline{G})}, \wh \mG(\underline{G})$ as the estimators when the group numbers are specified as $\underline{G}$. Then, we estimate $\underline{G}_0$ by utilizing the following information criterion:
	\begin{align}
		\qic(\underline{G} ) = \log\{Q({\wh \bxi(\underline{G})}, \wh \mG(\underline{G} ))\} + \lambda(\underline{G} ),\label{eq:qic}
	\end{align}
	where $Q(\cdot, \cdot)$ is defined in \eqref{eq:Q_obj}, and $\lambda(\underline{G} )$ is a penalty function.
	Then we estimate the group numbers by
	$\wh {\underline{G}}  \in \arg\min_{\underline{G} } \qic(\underline{G})$.
	In practice, we specify $\lambda(\underline{G} ) = \kappa(\sum_l G_l)$, in which $\kappa$ is a tuning parameter.
	Our theoretical analysis shows that if
	$T^{-1/2}(m+ \sum_l \log N_l)\ll \kappa \ll c_\gap c_\pi^q/(\prod_l G_l)$, the QIC can consistently select
	$\underline{G}_0$.
	Here $c_\gap$ and $c_\pi$ are group structures related values depending on the strength of model signals,
	which will be defined in our theoretical analysis in Section~\ref{sec:theory}.
	In our numerical study, we specify $\kappa = \{C (\log T) T^{1/8}\}^{-1}$ with $C = 40$, which achieves reliable finite sample performances in all of our numerical studies.

\section{Theoretical Properties}\label{sec:theory}

\subsection{Estimation Consistency}

Define $\bTheta_{i_1 \cdots i_q} = (\btheta_{g_{i_1}^{(1)}}^{(1)\top}, \cdots,  \btheta_{g_{i_q}^{(q)}}^{(q)\top}, \alpha_{g_{i_1}^{(1)} \cdots g_{i_q}^{(l)}})^\top\in \mR^{\sum_l (p_l+1) + 1}$ and
$\bTheta = (\bTheta_{i_1 \cdots i_q}:i_l \in [N_l], 1\le l\le q)$
as a tensor of dimension $N_1\times \cdots \times N_q \times \{\sum_l (p_l+1) + 1\}$.
With $\bTheta$ we can rewrite the loss function (\ref{eq:Q_obj}) as follows
\begin{align}
	Q(\bTheta) = \sum_{i_1 =1}^{N_1} \cdots \sum_{i_q = 1}^{N_q} \sum_{t = 1}^T
	\big(Y_{i_1 \cdots i_q, t} - \cX_{i_1 \cdots i_q, t}^\top \bTheta_{i_1 \cdots i_q}\big)^2\defeq
	\sum_{i_1 =1}^{N_1} \cdots  \sum_{i_q = 1}^{N_q} Q_{i_1\cdots i_q}(\bTheta_{i_1 \cdots i_q}),\label{def:Qij}
\end{align}
where
\begin{align}
	\cX_{i_1 \cdots i_q, t} \defeq
	\big( & \sum_{k = 1}^{N_1} w^{(1)}_{i_1 k}Y_{k i_2 \cdots i_q, (t-1)}, \bx_{i_1t}^{(1)\top} , \cdots, \nonumber\\
	& \sum_{k = 1}^{N_q} w^{(q)}_{i_q k} Y_{i_1 \cdots i_{(q-1)} k, (t-1)} , \bx_{i_q t}^{(q)\top},
	Y_{i_1 \cdots i_q, (t-1)}\big)^\top\in \mR^{\sum_l (p_l+1) + 1}.\label{eq:X_i1iq}
\end{align}
Denote by $\wh \bTheta = \big(\wh \bTheta_{i_1 \cdots i_q} =
(\wh \btheta_{\wh g_{i_1}^{(1)}}^{(1)\top}, \cdots , \wh \btheta_{\wh g_{i_q}^{(q)}}^{(q)\top}, \wh \alpha_{\wh g_{i_1}^{(1)} \cdots \wh g_{i_q}^{(q)}})^\top \big)$ the global minimizer of $Q(\bTheta)$ with the estimated group memberships $\wh g_{i_l}^{(l)} ~(i_l \in [N_l], l \in [q])$,
we define
a  pseudo distance as follows
\begin{align}
	d(\wh \bTheta, \bTheta) 
	& = \sum_{l=1}^q \frac{1}{N_l}\sum_{i_l = 1}^{N_l}  \|\wh \btheta_{\wh g_{i_l}^{(l)}}^{(l)} - \btheta_{g_{i_l}^{(l)}}^{(l)} \|^2 + \frac{1}{\prod_l N_l }\sum_{i_1 =1}^{N_1} \cdots \sum_{i_q = 1}^{N_q}  |\wh \alpha_{\wh g_{i_1}^{(1)} \cdots \wh g_{i_q}^{(q)}} - \alpha_{g_{i_1}^{(1)} \cdots g_{i_q}^{(q)}}|^2.\label{eq:pseudo_dist}
\end{align}
Intuitively,  $d(\wh \bTheta, \bTheta)$
measures the average distance between $\wh\bTheta$ and $\bTheta$.
Next, we establish the consistency of the global optimizer $\wh \bTheta$ of the loss function \eqref{eq:Q_obj} in this pseudo distance, for which we require the following definition and assumptions.

\begin{definition}\label{def:convex_concen}
	{\sc ($K$-Convex concentration)}
	Let $\bx\in \mR^n$ be a random vector.
	If for every 1-Lipschitz convex function $\varphi: \mR^n\to \mR$, we have
	$E|\varphi(\bx)|<\infty$ and for every $t>0$, it holds that
	$
	P\Big(\Big|\varphi(\bx) - E\{\varphi(\bx)\}\Big|\ge t\Big)\le 2\exp(-t^2/K^2),\nonumber
	$
	then $\bx$ is said to have the $K$-convex concentration property.
\end{definition}

\begin{assumption}\label{assum:para_space}
	{\sc (Parameter Space)} The parameter satisfies that $\|\bTheta\|_{\max} < \infty$.
\end{assumption}

\begin{assumption}\label{assum:tau_min}
	{\sc (Convexity)}
	Let $\bSigma_{i_1 \cdots i_q} = E(\cX_{i_1 \cdots i_q, t}\cX_{i_1 \cdots i_q, t}^\top)$. We assume that
	$\tau_{\min}\defeq \min_{i_1, \cdots,i_q}\lambda_{\min}(\bSigma_{i_1 \cdots i_q})$
	is a positive constant.
\end{assumption}

\begin{assumption}\label{assum:sub_gaussian}
	{\sc (Distribution of Noise Term)}
	Assume $\ve_{i_1 \cdots i_q, t}$ is {\it i.i.d} across $i_l \in [N_l]$ for all $l \in [q]$ and $t\in [T]$.
	In addition, $\ve_{i_1 \cdots i_q, t}$ is a zero-mean sub-Gaussian
	variable with a scale factor $0< \nu <\infty$, i.e.,
	$E\{\exp(u\ve_{i_1 \cdots i_q, t})\}\le \exp(\nu^2 u^2/2)$ for all $u \in \mR$.
	Let $\ve_{i_1 \cdots i_q, t}$ be independent of $\{ \cY_{s-1}, \bX_s^{(1)}, \cdots, \bX_s^{(q)}: s \le t\}$, where $\bX_s^{(l)} = (\bx_{i_l s}^{(l)}: i_l \in [N_l])^\top$.
\end{assumption}

\begin{assumption}\label{assum:mixing}
	{\sc (Distribution of Covariates)}
	Recall that $\bx_{i_lt}^{(l)} \in \mR^{p_l}$ is the covariate vector of the $i_l$th subject in the $l$th layer at time $t$.
	Assume $E(\bx_{i_l t}^{(l)}) = \zero$ for all $i_l \in [N_l]$ and $t \in [T]$.
	Let $\bfeta_l \in \mR^{p_l}$ be a constant vector satisfying $\|\bfeta_l\| \le c$ for $l \in [q]$, where $c$ is a positive constant.
	Define $\bx_t^{(l)\eta} = (\bx_{i_l t}^{(l)\top} \bfeta_{l}: i_l \in [N_l])^\top \in \mR^{N_l}$.
	Assume $\bx^{(l) \eta} =(\bx_t^{(l)\eta\top}: 0\le t\le T)^\top \in \mR^{N_l(T+1)}$
	satisfies the $K$-convex concentration property for some constant $K$
	according to Definition \ref{def:convex_concen}.
\end{assumption}

\begin{assumption}\label{assum:station}
	{\sc (Stability)}
	Denote $\mY_0 \defeq \vec(\cY_0) = \zero$, and assume that
	\begin{align*}
		\max_{g^{(1)} \in [G_{1,0}], \cdots, g^{(q)} \in [G_{q,0}]}  \big|\sum_l \lambda_{g^{(l)}}^0  + \alpha_{g^{(1)} \cdots g^{(q)}}^0 \big| \le \kappa_{\max} <1,
	\end{align*}
	where $G_{l, 0}$ is the true number of groups in the $l$th dimension, and $\kappa_{\max}$ is a positive constant.
\end{assumption}

\begin{assumption}\label{assum:group_diff}
	{\sc (Group Difference)}
	Assume $\min_{g_1^{(l)} \ne g_2^{(l)}}\{\|\btheta_{g_1^{(l)}}^{(l)0}  - \btheta_{g_2^{(l)}}^{(l)0}\|^2 +
	\max_{g^{-(l)} \in [G_{-l}^0]} \\ | \alpha_{g_1^{(l)} g^{-(l)}}^0 - \alpha_{ g_2^{(l)} g^{-(l)}}^0 |^2\}\ge  c_\gap$ holds for all $l \in [q]$,
	where $c_\gap \gg T^{-1} (\sum_l \log N_l)^2$ as $T \to \infty$.
	Here $\{g^{-(l)} \in [G_{-l}^0]\}$ denotes $\{g^{(1)} \in [G_{1,0}], \cdots, g^{(l-1)} \in [G_{l-1,0}], g^{(l+1)} \in [G_{l+1, 0}], \cdots, g^{(q)} \in [G_{q,0}]\}$.
\end{assumption}

\begin{assumption}\label{assum:group_ratio}
	{\sc (Group Proportion)}
	Let $\{g_{i_l}^{(l)0}: i_l\in [N_l]\}$ be non-random true membership sequences.
	Let $\pi_{g^{(l)}, N_l}^{(l)} = \sum_{i_l} I(g_{i_l}^{(l)0} = g^{(l)})/N_l$ for $g^{(l)} \in [G_{l,0}]$.
	Assume that $\min_{l \in [q], g^{(l)} \in [G_{l,0}]} \pi_{g^{(l)}, N_l}^{(l)}  \ge c_{\pi} > 0$ for sufficiently large $N_l$, where $c_\pi$ is a positive constant.
\end{assumption}

Assumption \ref{assum:para_space} requires the parameter space to be bounded.
Assumption \ref{assum:tau_min} ensures the convexity of the element-wise objective function, i.e.,
$Q_{i_1 \cdots i_1}(\bTheta_{i_1 \cdots i_q})$, as a function of $\bTheta_{i_1 \cdots i_q}$ for sufficiently large $T$.
This condition guarantees the unique solution of the local objective function for each node.
and is crucial for establishing the consistency result for the metric \eqref{eq:pseudo_dist}.
These conditions are also commonly used in the literature, see, \cite{self1987asymptotic}, \cite{fan2001variable}, \cite{zou2006adaptive}, etc., for example.

Assumptions \ref{assum:sub_gaussian}--\ref{assum:mixing} concern about the distributions of the error term and covariates, respectively.
Specifically, Assumption \ref{assum:sub_gaussian} requires the error term $\ve_{i_1 \cdots i_q,t}$ to be {\it i.i.d.} sub-Gaussian variables, which is widely used in high-dimensional time series literature \citep{wang2013calibrating,lugosi2019sub,fan2021augmented}.
We also provide a weighted least squares estimation procedure with group-specific variances, i.e., $\var(\ve_{i_1 \cdots i_q, t}) = \sigma_{g_{i_1}^{(1)0} \cdots g_{i_q}^{(q)0}}$, in Section \ref{sec:wlse_est}.
Subsequently, Assumption \ref{assum:mixing} allows the covariates $\{\bx_{i_l t}^{(l)}\}$ to be correlated but satisfying the $K$-convex concentration property according to Definition \ref{def:convex_concen}.
This assumption is employed to establish Hanson-Wright type inequality for dependent variables \citep{adamczak2015note}.
Although this is a high-level condition, there are a variety of random variables satisfying Definition \ref{def:convex_concen}, as discussed in the following Remark \ref{remark.K_convex}.
We further comment that the Assumptions \ref{assum:sub_gaussian}--\ref{assum:mixing} together imply that
$ \bv^\eta \defeq (\bv_t^\eta: 0\le t\le T)^\top$
satisfies the $K$-convex concentration property for some constant $K$.
Here $ \bv_t^\eta = (\bx_t^{(1)\eta\top}, \cdots, \bx_t^{(q)\eta\top},
\E_t^\top)^\top \in \mR^{\sum_l N_l + \prod_l N_l}$, where $\E_t = \vec(\mE_t)$.

\begin{remark}\label{remark.K_convex}
	As discussed by \cite{adamczak2015note},
	there are a variety of random vectors $\bx$ satisfying the $K$-convex concentration property in Definition \ref{def:convex_concen}. For example, (i) Any random vector $\bx$ with its elements $x_i$s independent for all $i$, and $|x_i| \le 1 ~ a.s.$, satisfies Definition \ref{def:convex_concen} \citep{talagrand1988isoperimetric}; (ii) Any random vector $\bx$ with its elements in a bounded interval and geometrically strongly mixing satisfies Definition \ref{def:convex_concen} \citep{samson2000concentration}.
	We refer to \cite{adamczak2015note} for more detailed discussions.
\end{remark}

Next, Assumption \ref{assum:station} ensures the stability of the
tensor-valued time series data as $T$ goes to infinity, as
defined in \cite{lutkepohl2005new}.
Assumptions \ref{assum:group_diff} and \ref{assum:group_ratio} are imposed on certain group properties.
Assumption \ref{assum:group_diff} assumes there is a gap between the true parameters of two different groups within any network.
The condition is an extension of the same type of conditions assumed by the group panel data models with groups assigned on one dimension \citep{su2016identifying,ando2016panel,zhang2019quantile,liu2020identification}.
In Assumption~\ref{assum:group_diff}, special care is paid to the self-momentum parameter $\alpha_{g^{(1)} \cdots g^{(q)}}$, indexed by the tensor dimension-wise group memberships, to ensure the parameter identifiability.
Specifically, we require a min-max type condition for $\alpha_{g^{(1)} \cdots g^{(q)}}^0$ in Assumption \ref{assum:group_diff}.
Furthermore, instead of assuming $c_{\gap}>c>0$ by a positive constant $c$ in existing literature \citep{bonhomme2015grouped, su2016identifying, zhang2019quantile,liu2020identification},
we allow $c_{\gap}\to 0$ to study how this signal strength affects the theoretical properties.
Lastly, Assumption \ref{assum:group_ratio} assumes a lower bound of group proportions.
In the following, we establish the consistency of the pseudo distance.

\bet\label{thm:pseudo_dist}
Suppose $G_l \ge G_{l,0}$ for all $l \in [q]$, where $G_{l,0}$ is the true number of groups.
In addition, assume Assumptions \ref{assum:para_space}--\ref{assum:station} hold.
We have that $d(\wh\bTheta, \bTheta^0) = O_p\{ (\sum_l \log N_l)^2 T^{-1}\}$,
where
$\bTheta^0  = (\bTheta_{i_1 \cdots i_q}^0 = (\btheta_{g_{i_1}^{(1)0}}^{(1)0\top}, \cdots,  \btheta_{g_{i_q}^{(q)0}}^{(q)0\top}, \alpha_{g_{i_1}^{(1)0} \cdots g_{i_q}^{(q)0}}^0)^{\top})$
is the true parameter for $\bTheta$.
\eet

Theorem \ref{thm:pseudo_dist} implies that as long as we have
$T \gg (\sum_l \log N_l)^2$, $\wh\bTheta$ is a consistent estimator
for $\bTheta^0$ in the metric when $G_l$ is possibly over-specified, i.e.,
$G_l\ge G_{l,0}$.
This allows $N_l$ to grow exponentially fast with $\sqrt{T}$, which is a mild condition compared to existing literatures {\citep{su2016identifying}}.
	As a result, it allows for moderately large $N_l$ with respect to $\sqrt{T}$.
Subsequently, we show in Theorem \ref{thm:select_GH} that the QIC can
consistently select the true group numbers.

\bet\label{thm:select_GH}
Under Assumptions \ref{assum:para_space}--\ref{assum:group_ratio},
and assume $\kappa = \lambda(\underline{G})/(\sum_l G_l)$ satisfies
\begin{align}
	T^{-1}(\sum_l \textup{\log} N_l)^2 \ll \kappa \ll c_\gap /(\prod_l G_l). \label{eq:eta_range}
\end{align}
Then we have $P(\wh G_1 = G_{1,0}, \cdots, \wh G_q = G_{q,0})\to 1$ as $\min\{N_1, \cdots, N_q,T\} \to \infty$ and $T^{-1}(\sum_l \textup{\log}  N_l)^2\to 0$.
\eet

Theorem \ref{thm:select_GH} implies that if we set $\kappa$ to satisfy \eqref{eq:eta_range}, then we can consistently estimate the true group numbers.
Denote $G_{-l} = (G_k: k \ne l)^\top \in \mR^{q-1}$,  we need $\kappa\gg T^{-1}(\sum_l \log  N_l)^2$ to ensure that $\qic(\underline{G})>\qic(\underline{G}_0)$ for the over-fitting case, i.e.,
$G_l>G_{l,0}$ and $G_{-l} \ge G_{-l,0}$ for any $l \in [q]$.
Here $G_{-l}\ge G_{-l,0}$ means $G_{k}\ge G_{k,0}$ for any $k\ne l$.
Conversely, we need $\kappa \ll c_\gap /(\prod_l G_l)$ to guarantee that $\qic(\underline{G})>\qic(\underline{G}_0)$ for the under-fitting case, i.e., there exists an $l$ with $G_l < G_{l,0}$.
When both conditions are met, we can obtain $\wh G_l = G_{l,0} $ for all $l \in [q]$ with a probability approaching 1.
In the next subsection, we further discuss the results of node-wise parameter estimation and the strong group membership consistency.

\subsection{Membership Estimation Consistency and Asymptotic Normality}\label{sec:42}

As we stated before, the metric in \eqref{eq:pseudo_dist}
measures the average estimation error between $\wh \bTheta$ and $\bTheta^0$.
Therefore, the result in Theorem \ref{thm:pseudo_dist} is not
sufficient to imply the parameter consistency for each node.
To this end, we derive the following node-wise parameter consistency
result, which will be crucial to building the strong membership estimation
consistency later.

\bep\label{pro:gh_consistency}
Under Assumptions \ref{assum:para_space}--\ref{assum:station},
when $G_l \ge G_{l,0}$ for all $l \in [q]$, we have
\begin{align}
	& \sup_{i_l} \left\{ \big\| \wh\btheta_{\wh g_{i_l}^{(l)}}^{(l)} - \btheta_{g_{i_l}^{(l)0}}^{(l)0}\big \|^2 + \frac{1}{\prod_{m \neq l} N_m}  \sum_{m \neq l} \sum_{i_m = 1}^{N_m} \big|\wh\alpha_{\wh g_{i_1}^{(1)} \cdots \wh g_{i_q}^{(q)}} - \alpha_{g_{i_1}^{(1)0} \cdots g_{i_q}^{(q)0}}^0 \big|^2 \right\} \nonumber\\
	& = O_p\Big\{ T^{-1} (\sum_l \textup{\log} N_l)^2 \Big\}. \label{eq:sup_i_para_diff}
\end{align}
\eep

Recall that $\mG_l = (g_{i_l}^{(l)}: 1 \le i_l \le N_l)^\top$,
and let $\mG_l^0$ denote the corresponding true memberships.
Equation \eqref{eq:sup_i_para_diff} establishes uniform node-wise parameter estimation consistency, which is crucial for achieving the following strong consistency of membership estimation for $\wh \mG_l$ when $G_l \ge G_{l,0}$ for all $l \in [q]$.
Denote the node set $\wh \cR_{\wt g^{(l)}}^{(l)} = \{i_l: \wh g_{i_l}^{(l)} = \wt g^{(l)}\}$ for $\wt g^{(l)} \in [G_l]$ and $\cR_{ g^{(l)}}^{(l)0} = \{ i_l: g_{i_l}^{(l)0} = g^{(l)} \}$ for $g^{(l)} \in [G_{l,0}]$, where $\wh g_{i_l}^{(l)}$ is obtained by \eqref{eq:g_i} when $\wh\bxi$ is specified.

\bet\label{thm:h_consistency2}
(Strong Consistency of Membership Estimation)
Under Assumptions \ref{assum:para_space}--\ref{assum:group_ratio}, and suppose $G_l \ge G_{l,0}$ for all $l \in [q]$.
Then for each estimated group $\wt g^{(l)} \in [G_l]$, there exists a true group $g^{(l)} \in [G_{l,0}]$,
such that $\wh \cR_{\wt g^{(l)}}^{(l)} \subset \cR_{g^{(l)}}^{(l)0}$ with probability
tending to 1.
\eet

As implied by Theorem \ref{thm:h_consistency2}, the true groups are split into subgroups
instead of joining into new groups when $G_l \ge G_{l,0}$ for all $l \in [q]$, as shown in Figure \ref{fig:memb_consist}.
Define $\wh\mG_l = (\wh g_{i_l}^{(l)}: i_l \in [N_l])^\top \in \mR^{N_l}$ as the estimated membership vectors for all $l \in [q]$.
Particularly, when $G_l = G_{l,0}$, we can show that
$\wh \mG_l = \mG_l^0$ holds with probability tending to 1 under certain label permutations.
We would like to remark that establishing the group consistency result is non-trivial due to several challenges under our framework.
As shown in the GTNAR model \eqref{eq:model}, group memberships from different dimensions are involved and tangled in an additive form.
In addition, the network dependence structure creates extra difficulty in establishing the node-wise membership consistency result.
To establish the membership consistency,
for instance, in \cite{zhu2023simultaneous}, a refinement procedure is further required for obtaining this property.
In \cite{su2023identifying}, a two step estimation procedure is used to achieve group membership estimation consistency when the true group number is known.
Specifically, their two-step procedure establishes node-wise parameter convergence and consistent group membership identification sequentially, whereas we allow simultaneous estimation of group membership and corresponding group-wise parameters within a unified framework.
In contrast to their work, a further investigation of our additive model form enables us to address these challenges and establish the group membership consistency result for stochastically estimated $\wh \mG_l$ directly, which is critically important for later statistical inference.
Furthermore, let $\wh\bxi^{\o}$ be the oracle estimator when
the true group memberships $\mG_l^0$s are known for all $l \in [q]$.
Then the oracle property holds that $\wh \bxi = \wh \bxi^{\o}$ with probability tending to 1, where $\wh\bxi$ is the estimator for $\bxi^0 = ( \btheta^{(1)0 \top}, \cdots, \btheta^{(q)0 \top}, \vec(\balpha^0)^\top )^\top$.
The results are presented in the following Corollary.

\begin{corollary}\label{coro:group_consistency}
	Under Assumptions \ref{assum:para_space}--\ref{assum:group_ratio},
	and assume $G_l = G_{l,0}$ for all $l \in [q]$.
	Then under label permutations, we have
	\begin{align}
		&\lim_{\min\{N_1, \cdots, N_q,T\}\to \infty} P\left(\wh \mG_1 = \mG_1^0, \cdots, \wh \mG_q = \mG_q^0 \right)\to 1, \label{eq:membership_consistency}\\
		&  \lim_{\min\{N_1, \cdots, N_q,T\}\to \infty}  P\left(\wh \bxi = \wh \bxi^{\o}
		\right)\to 1. \label{eq:oracle}
	\end{align}
\end{corollary}

The results in Corollary \ref{coro:group_consistency} imply that $\wh \bxi$ is asymptotically equivalent to $\wh \bxi^\o$. Therefore, to derive the asymptotic distribution of $\wh\bxi$, it is sufficient to investigate the case for $\wh \bxi^\o$.
We establish the asymptotic normality result in the following theorem.

\bet\label{thm:normal}
Assume Assumptions \ref{assum:para_space}--\ref{assum:group_ratio}, $G_l = G_{l,0}$ and that there exists $n$, such that $c_1 n \le \min_l N_l \le \max_l N_l \le c_2 n$ for some constants $c_1, c_2 > 0$.
Define $\bM_{nT}^0 = n^{-q} T^{-1} E(\bM)$ and assume $ \bM^0 = \lim_{\min\{n, T\}\to \infty} \bM_{nT}^0$ exists, where $\M$ is given in \eqref{eq:Mb}.
Assume $\lambda_{\min}(\bM^0)\ge \tau>0$ for a positive constant $\tau$.
Let $k = \sum_l \{G_l(p_l+1)\}+\prod_l G_l$.
Then for any $\bfeta\in \mR^{k}$ with $\|\bfeta\| = 1$
we have
\begin{align}
	n^{q/2} T^{1/2} \bfeta^\top (\wh\bxi - \bxi^0) \to_d N\left(\zero, \sigma^2\bfeta^\top  (\bM^0)^{-1}\bfeta\right).\label{eq:normal0}
\end{align}

\eet

Theorem \ref{thm:normal} establishes the asymptotic normality of the estimator.
Specifically, the convergence rates of $\wh\btheta^{(l)}$ is $\sqrt{n^q T}$ for all $l \in [q]$.
Using \eqref{eq:normal0}, we can conduct the statistical inference.

\section{Model Extensions}\label{sec:wls_int_theory}

	The GTNAR model can be flexibly extended to accommodate an interactive model setting and group-specific error variances.
	In this section, we present the corresponding estimation and establish its statistical guarantees.

	\subsection{Mixed GTNAR Model}\label{subsec:int_theory}
	
	GTNAR model can be extended to incorporating the interactive network effects as
	\begin{align}
		\cY_t & = (\cY_{t-1} \times_{l=1}^q \W^{(l)} ) \times_{l=1}^q \bGamma^{(l)} \nonumber\\
		& + \sum_l (\cY_{t-1} \times_l \W^{(l)}) \times_l \L^{(l)} +  \mA \odot \cY_{t-1} + {\sum_{l=1}^q \bbeta_{X_l, t}^{(l)} \circ_{k \neq l} \one_{N_k}}+ \mE_t,\label{eq:model_int}
	\end{align}
	which we refer to as the Mixed GTNAR model.
	The first term represents the interactive network effects with $\bGamma^{(l)} = \textup{diag}\big(\gamma^{(l)}_{g_{i_l}^{(l)}}: i_l \in [N_l]\big) = \diag(\bgamma^{(l)})$.
	Specifically, the interactive network coefficients are involved in \eqref{eq:model_int} in a multiplicative form across different layers.
	Take the case of $q = 2$ as an example, model \eqref{eq:model_int} can be expressed as
	\begin{align}
		\Y_t & =(\bGamma^{(1)} \W^{(1)})  \Y_{t-1} (\W^{(2)} \bGamma^{(2)})+  (\L^{(1)} \W^{(1)} ) \Y_{t-1} + \Y_{t-1} (\W^{(2)} \L^{(2)}) \nonumber\\
		&  + \A \odot \Y_{t-1} + \bbeta_{X_1, t}^{(1)} \one_{N_2}^\top + \one_{N_1} \bbeta_{X_2, t}^{(2)\top} + \bE_t.\label{eq:model_int_2}
	\end{align}
	Note that the multiplication form in $\bGamma^{(1)} \W^{(1)} \Y_{t-1} \W^{(2)} \bGamma^{(2)}$ causes the parameter identification issue.
	Following the convention \citep{chen2021autoregressive}, we set $\| \bGamma^{(1)} \|_F = 1$ to guarantee the interactive network effects $\bGamma^{(1)}$ and $\bGamma^{(2)}$ to be identifiable with sign flips.
	To estimate model \eqref{eq:model_int_2}, we apply an iterative least squares method, and the detailed estimation algorithms are given in Appendix \ref{subsec:int_est_alg}.
	In the following, we present the theoretical properties for the estimator of model \eqref{eq:model_int}.
	Note that one can re-write the first interactive term in \eqref{eq:model_int_2} as $(\bgamma^{(1)} \bgamma^{(2)\top}) \odot (\bW^{(1)} \bY_{t-1} \bW^{(2)}) $,
	hence $\bgamma^{(1)} \bgamma^{(2)\top}$ plays  a similar role as $\A$.
	As a result, we can borrow the proof idea for dealing with the autoregressive matrix $\A$ in our original model \eqref{eq:model_matrix} with $q = 2$ to establish the consistency result for the Mixed GTNAR model.
	We first introduce some necessary conditions.
	For notational simplicity,
	denote $\bTheta_{ij} = (\btheta_{ g_{i}^{(1)}}^{(1)\top}, \btheta_{ g_{j}^{(2)}}^{(2)\top}, \alpha_{ g_{i}^{(1)}   g_{j}^{(2)}}, \gamma_{ g_{i}^{(1)}} \gamma_{ g_{j}^{(2)}})^\top \in \mR^{p_1+p_2+4}$,
	and denote $\bTheta = (\bTheta_{ij}: i \in [N_1], j \in [N_2]) \in \mR^{N_1 \times N_2 \times (p_1+p_2+4)}$. Let $\bTheta^0 = (\bTheta_{ij}^0 = (\btheta_{g_{i}^{(1)}}^{(1)0\top}, \btheta_{ g_{j}^{(2)}}^{(2)0\top}, \alpha_{ g_{i}^{(1)}   g_{j}^{(2)}}^0, \gamma^0_{ g_{i}^{(1)0}} \gamma^0_{ g_{j}^{(2)0}})^\top)$
	be the true parameter for $\bTheta$.
	Define the pseudo distance as
	\begin{align}
		d(\wh \bTheta, \bTheta)
		& =  \frac{1}{N_1}\sum_{i = 1}^{N_1}  \|\wh \btheta_{\wh g_{i}^{(1)}}^{(1)} - \btheta_{g_{i}^{(1)}}^{(1)} \|^2 +
		\frac{1}{N_2}\sum_{j = 1}^{N_2}  \|\wh \btheta_{\wh g_{j}^{(2)}}^{(2)} - \btheta_{g_{j}^{(2)}}^{(2)} \|^2 \nonumber\\
		&  + \frac{1}{N_1 N_2 }\sum_{i = 1}^{N_1}  \sum_{j = 2}^{N_2} \Big\{ |\wh \alpha_{\wh g_{i}^{(1)}  \wh g_{j}^{(2)}} - \alpha_{g_{i}^{(1)} g_{j}^{(2)}}|^2 +  |
		\wh \gamma_{\wh g_{i}^{(1)}}^{(1)} \wh \gamma_{\wh g_{j}^{(2)}}^{(2)} -  \gamma_{ g_{i}^{(1)}}^{(1)}  \gamma_{ g_{j}^{(2)}}^{(2)} |^2  \Big\}. \label{eq:pseudo_dist_int}
	\end{align}
	In the above distance, we note that the interactive effect $\wh \gamma_{\wh g_{i}^{(1)}}^{(1)} \wh \gamma_{\wh g_{j}^{(2)}}^{(2)}$ plays a similar role as $\wh \alpha_{\wh g_{i}^{(1)}  \wh g_{j}^{(2)}}$ in $d(\wh \bTheta, \bTheta)$.
	Hence, we can borrow the proof idea of Theorem \ref{thm:h_consistency2} to
	establish the estimation consistency result with $d(\wh \bTheta, \bTheta)$ for model \eqref{eq:model_int_2}.
	The required assumptions are given in Appendix \ref{subsec:assum_int}.
	We state the formal theoretical results in the following.

	\bet\label{thm:int_model_thm}
	Under Assumptions \ref{assum:tau_min}--\ref{assum:mixing} and \ref{assum:para_space_int}--\ref{assum:station_int}, the following conclusions hold.\\
	(i) Suppose $G_l \ge G_{l,0}$ for $l=1,2$, then $d(\wh\bTheta, \bTheta^0)  = O_p\{T^{-1}(\log(N_1N_2))^2 \}$.\\
	(ii) Assume $\kappa_1 = \lambda(\underline{G})/(G_1 + G_2)$ satisfies
	\begin{align}
		T^{-1}(\log (N_1 N_2))^2 \ll \kappa_1 \ll c_\gap /(G_1 G_2). \label{eq:eta_range_1}
	\end{align}
	Then $P(\wh G_1 = G_{1,0}, \wh G_2 = G_{2,0})\to 1$ as $\{N_1, N_2,T\} \to \infty$ and {$T^{-1}(\log (N_1 N_2))^2\to 0$}.\\
	(iii) Further suppose Assumption \ref{assum:group_ratio} and \ref{assum:group_diff_int} hold, and $G_l \ge G_{l,0}$ for $l=1,2$. Then for any estimated group $\wh g^{(1)} \in [G_1]$ {and $\wh g^{(2)} \in [G_2]$}, there exists true groups $g^{(1)} \in [G_{1,0}]$ {and $g^{(2)} \in [G_{2,0}]$},
	such that
	\begin{align*}
		P\Big\{\wh\cR^{(1)}_{\wh g^{(1)}} \in \cR^{(1)0}_{g^{(1)0}}\Big\} \to 1, ~~~P\Big\{\wh\cR^{(2)}_{\wh g^{(2)}} \in \cR^{(2)0}_{g^{(2)0}}\Big\} \to 1,
	\end{align*}
	where $\wh\cR^{(l)}_{\wh g^{(l)}} = \{ i_l: \wh g_{i_l}^{(l)} = \wh g^{(l)} \}$ for $l = 1,2$.
	\eet

	In Theorem \ref{thm:int_model_thm} (i), we first establish the estimation consistency based on the pseudo distance $d(\wh \bTheta, \bTheta)$ when the group numbers are possibly over-specified.
	It shows that when the interactive model is correctly specified or over-specified, the parameter $\wh\bTheta$ is a consistent estimator for $\bTheta^0$ as long as $T \gg (\log (N_1 N_2))^2$.
	Next in (ii), we show that the QIC selection method in \eqref{eq:qic_int} leads to consistent group number estimators.
	When the tuning parameter $\kappa_1$ satisfies the condition \eqref{eq:eta_range_1}, we can consistently estimate the group numbers.
	The lower bound and upper bound for $\kappa_1$ are applied for the results in over- and under-specified situations, respectively.
	Subsequently, we establish the strong group membership estimation consistency in (iii).
	The proof of Theorem \ref{thm:int_model_thm} is given in Appendix \ref{sec:proof_int}.
	Since the interactive parameter $\gamma_{g_{i}^{(1)}}^{(1)} \gamma_{g_j^{(2)}}^{(2)}$ plays a similar role as $\alpha_{g_i^{(1)} g_j^{(2)}}$ in model \eqref{eq:model_int_2}, the proofs exhibit analogous roadmap to the proofs of Theorem \ref{thm:pseudo_dist}--\ref{thm:h_consistency2} in Section \ref{sec:theory}.
	However, the new interactive term in $\cX_{i j, t}$ defined in \eqref{eq:X1} requires additional technical  Lemmas \ref{lem:Q_concent_1}--\ref{lem:tau_max_1} and Lemma \ref{lem:concenX}, which are all crucial to the proof of the theorem.

	To demonstrate the finite sample performance, we conduct several simulation experiments, which can be found in Appendix \ref{subsec:int_simu}.

	\subsection{Weighted Least Squares Estimation with Group-specific Error Variances}\label{sec:wlse_est}
	
	Next, we design a weighted least squares estimator for the case where the error terms have group-specific variances when $q=2$.
	Several simulation studies are presented in Appendix \ref{subsec:simu:WLS}.

	\subsubsection{Estimation Procedure}\label{subsec:wlse_est_procedure}
	
	In the case where the error term has group-specific variance, the GTNAR model can be modified as
	\begin{align}
		Y_{ij,t} & = \lambda^{(1)}_{g_{i}^{(1)}}\sum_{k = 1}^{N_1} w_{1ik}Y_{kj, (t-1)}
		+
		\lambda^{(2)}_{g_{j}^{(2)}}\sum_{k = 1}^{N_2} Y_{ik, (t-1)}w_{2kj}+ \alpha_{g_i^{(1)}  g_j^{(2)}}Y_{ij,(t-1)}\nonumber
		\\
		& +\bx_{it}^{(1)\top} \bzeta^{(1)}_{g_{i}^{(1)}}+\bx_{jt}^{(2)\top} \bzeta^{(2)}_{g_{j}^{(2)}}+\ve_{i j, t},	\label{eq:model0_update}
	\end{align}
	where $\ve_{i j, t}$ satisfies that $E(\varepsilon_{ij, t}) = 0$ and $\var(\varepsilon_{i j, t}) = \sigma_{g_{i}^{(1)0} g_{j}^{(2)0}}^2$, and $g_{i}^{(1)0}$ are $g_{j}^{(2)0}$ are true memberships.
	Under this situation, a weighted least squares (WLS) estimation procedure can be introduced  \citep{carroll1988asymptotic,shao1989asymptotic}.
	Specifically, similar to the calculation for \eqref{eq:Mb2}, a weighted version for the corresponding terms can be written as
	\begin{align}
		\wt\bM =  \left(
		\begin{array}{ccc}
			\wt\bM^{(1)}& \wt\bM^{(12)} & \wt\bM^{(1\alpha)} \\
			\wt\bM^{(12)\top}& \wt\bM^{(2)}& \wt\bM^{(2\alpha)}\\
			\wt\bM^{(1\alpha)\top}& \wt\bM^{(2\alpha) \top}& \wt\bM^{\alpha}
		\end{array}
		\right),~~~
		\wt\b =  \left(
		\begin{array}{c}
			\wt\b^{(1)} \\
			\wt\b^{(2)}\\
			\wt\b^{\alpha}
		\end{array}
		\right).\label{eq:w_Mb}
	\end{align}
	Take the first term $\wt\bM^{(1)}$ as an example, we set
	\begin{align}
		& \wt\bM^{(1)} = \diag\{\wt\bM_{g^{(1)}}^{(1)}:g^{(1)} \in [G_1]\} \in \mR^{G_1 (p_1+1) \times G_1 (p_1+1)}, \\
		& \wt\bM_{g^{(1)}}^{(1)} = \sum_{t, g^{(2)}} \sigma_{g^{(1)} g^{(2)}}^{-2} \mX_{g^{(1)} g^{(2)}, t}^\top\mX_{g^{(1)}g^{(2)}, t},\label{eq:M_gr}
	\end{align}
	with additional weights $\sigma_{g^{(1)} g^{(2)}}^{-2}$.
	Other terms in \eqref{eq:w_Mb} can be calculated similarly.
	Then, the WLS estimator is obtained as $\wt\btheta = \wt \bM^{-1} \wt\b$.

	However, there are still unknown parameters (e.g., $\sigma_{g^{(1)} g^{(2)}}^2$) in $\wt\btheta$. Therefore, to obtain a feasible WLS estimator, we first
	estimate $\sigma_{g^{(1)} g^{(2)}}^2$ as follows
	\begin{align}
		\wh\sigma_{g^{(1)} g^{(2)}}^2 = \frac{1}{N_{1 g^{(1)}} N_{2 g^{(2)}} T} \sum_{i \in \wh\cR_{g^{(1)}}^{(1)}} \sum_{j \in \wh\cR_{g^{(2)}}^{(2)}} \sum_t (Y_{ij,t} - \cX_{ij,t}^\top \wh \bTheta_{ij})^2\label{eq:est_sigma_gh}
	\end{align}
	and $\wh\bTheta_{ij}$ is the estimator obtained by Algorithm \ref{alg:gmnar}.
	This allows us to obtain the WLS estimator as
	\beq
	\wt\btheta^w = \widecheck{\bM}^{-1} \widecheck{\bdelta},\label{eq:WLSE}
	\eeq
	where $\widecheck{\bM}$ and $\widecheck{\bdelta}$ are obtained by substituting the estimators $\wh \sigma_{g^{(1)} g^{(2)}}^2$ and $\wh \bTheta$ in $\wt\M$ and $\wt \b$ of \eqref{eq:w_Mb}.
	In the next subsection, we establish the theoretical properties for $\wt\btheta^w$.
	
	\subsubsection{Theoretical Properties}\label{subsec:theory_wlse}

	We first show that when the group-specific error variances exist, the membership estimation consistency in Theorem \ref{thm:h_consistency2} still holds.
	Technically, we need the Assumption \ref{assum:noise_cluster} in Appendix \ref{subsec:assum_wlse} instead of the Assumption \ref{assum:sub_gaussian} in Section \ref{sec:theory}.
	We state the formal theoretical results in Theorem \ref{thm:wlse_thm}.

	\bet\label{thm:wlse_thm}
	Assume the group-specific error variances exist and the model is formed as \eqref{eq:model0_update}.
	Suppose that Assumptions \ref{assum:para_space}, \ref{assum:tau_min}, \ref{assum:mixing}--\ref{assum:group_ratio}, \ref{assum:noise_cluster} hold.
	When $G_1 = G_{1,0}, G_2 = G_{2,0}$, we have, \\
	(i) under label permutation,
	\begin{align}
		& \lim_{\min(N_1,N_2,T)\to \infty} P\Big( \wh\mG^{(1)} = \mG^{(1)0}, \wh\mG^{(2)} = \mG^{(2)0} \Big) \to 1, \label{eq:R_C_consist}\\
		& \lim_{\min(N_1,N_2,T)\to \infty} P\Big( \wt\btheta = \wt\btheta^{\text{or}} \Big) \to 1. \label{eq:theta_ora}
	\end{align}
	Further assume that there exists $n$, such that $c_1 n \le \min_l N_l \le \max_l N_l \le c_2 n$ and suppose that $\sqrt{T} \gg \log(G_1 G_2 n^2) $,\\
	(ii) it holds that
	\begin{align*}
		\sup_{g^{(1)}, g^{(2)}} \Big| \frac{1}{\wh\sigma_{g^{(1)} g^{(2)}}^2} - \frac{1}{\sigma_{g^{(1)} g^{(2)}}^2} \Big| = o_p(1).
	\end{align*}
	(iii) Assume $\lambda_{\min}(\bM^0) \ge \tau >0 $ for a positive constant $\tau$.
	Let $s = G_1(p_1 + 1) + G_2 (p_2+1) + G_1 G_2$,
	then for any $\bfeta \in \mR^{s}$ with $\|\bfeta\| = 1$, we have
	\begin{align}
		(n\sqrt{T})^{-1} \bfeta^\top (\wt\btheta^w - \btheta^0) \to_d N(\zero, \bfeta^\top {( \wt\bM^0 )^{-1}} \bfeta),
	\end{align}
	where $\wt\bM^0 = \lim_{n,T \to \infty} (n^2 T)^{-1} E(\wt\bM)$, and $\wt\bM$ is defined in \eqref{eq:w_Mb}.
	\eet
	
	The proof of Theorem \ref{thm:wlse_thm} is provided in Appendix \ref{subsec:proof_propA1}--\ref{subsec:proof_thmA2}.
	Conclusion (i) shows that in spite of the group-specific error variance, the group membership maintains strong consistency under nearly the same conditions.
	By conclusion (i), we have $ \wh\mG^{(1)} = \mG^{(1)0}$ and $\wh\mG^{(2)} = \mG^{(2)0}$ with probability tending to 1, hence we next treat the group memberships as known, denoted as $\cR_{ g^{(1)}}^{(1)}$ and $\cR_{ g^{(2)}}^{(2)}$ in the technical proof.
	Note that we substitute $\sigma_{g^{(1)} g^{(2)}}^{-2}$ in \eqref{eq:M_gr} to be its estimator $\wh \sigma_{g^{(1)} g^{(2)}}^{-2}$.
	Hence, before we show the theoretical convergence properties of
	$\wt\btheta^w$, we present the estimation consistency for $\wh \sigma_{g^{(1)} g^{(2)}}^{-2}$ in (ii), which is a critical result for later analysis.
	Conclusion (ii) shows that when $\sqrt{T} \gg \log(G_1 G_2 n^2)$, the sample weight $\wh\sigma_{g^{(1)} g^{(2)}}^{-2}$ is a consistent estimator of $\sigma_{g^{(1)} g^{(2)}}^{-2}$.
	Next, we establish the asymptotic normality for the WLS estimator $\wt\btheta^w = \widecheck\bM^{-1} \widecheck\bdelta$ in (iii).
	In practice, one can calculate the estimator for $\wt\bM^0$ as $\widecheck\bM_{nT} = (n^2 T)^{-1} \widecheck{\bM}$, where $\widecheck{\bM}$ is obtained by plugging in the error variance estimator $\wh\sigma_{g^{(1)} g^{(2)}}$ calculated by \eqref{eq:est_sigma_gh} in $\wt\bM$ defined in \eqref{eq:w_Mb}.
	In addition, we evaluate the finite sample performance of the WLS estimator $\wt\btheta^w$ in Appendix \ref{subsec:simu:WLS}.

\section{Simulation Study}\label{sec:simulation}

\subsection{Model Settings}\label{subsec:model_set}
To evaluate the finite sample performance of the GTNAR method, we conduct several simulation studies in this section for  $q=2$ (i.e., two networks).
Three different scenarios for $G_{l,0}$ are considered.
In each scenario, the node memberships $g_{i_l}^{(l)}$s are sampled from the
multinomial distribution with probability $\bpi_l= \{\pi_{g^{(l)}}^{(l)} =
G_{l,0}^{-1}: g^{(l)} = 1,\cdots, G_{l,0}\}, ~ l \in [q]$.
The dimension of exogenous covariates is set as $p_l = 3$, and
the corresponding true parameters are shown in Table
\ref{table:true_param}. For all scenarios, the covariates $\bx_{i_l t}^{(l)}, l \in [q]$ are generated from multivariate normal distribution
$N(\zero, \bI_{p_l})$.
Subsequently, we generate the noise term $\ve_{i_1 i_2, t}$ from $N(0,1)$ independently.
For each scenario, the following two network structures are considered.

\begin{table}[]
	\caption{True parameters for different scenarios.}\label{table:true_param}
	\centering
	\scalebox{0.58}{
		\begin{tabular}{cc|cc|cc}
			\hline
			\multicolumn{2}{c|}{Scenario 1}   & \multicolumn{2}{c|}{Scenario 2}   & \multicolumn{2}{c}{Scenario 3}  \\ \hline
			$G_{1,0}$ & $G_{2,0} = 2$  & $G_{1,0} = 3$      & $G_{2,0} = 2$   & $G_{1,0} = 3$ & $G_{2,0} = 3$  \\ \hline
			$\lambda_{g^{(1)}}^{(1)}$      & $\lambda_{g^{(2)}}^{(2)}$     & $\lambda_{g^{(1)}}^{(1)}$      & $\lambda_{g^{(2)}}^{(2)}$     & $\lambda_{g^{(1)}}^{(1)}$      & $\lambda_{g^{(2)}}^{(2)}$       \\
			(0.15, 0.2) & (0.25, 0.4)      & (0.15, 0.2, 0.3) & (0.25, 0.3) &    (0.15, 0.2, 0.3)   & (0.25, 0.3, 0.4) \\ \hline
			$\bzeta_{g^{(1)}}^{(1)}$   & $\bdelta_{g^{(2)}}^{(2)}$ & $\bzeta_{g^{(1)}}^{(1)}$          & $\bdelta_{g^{(2)}}^{(2)}$  & $\bzeta_{g^{(1)}}^{(1)}$   & $\bdelta_{g^{(2)}}^{(2)}$  \\
			$\left(\begin{array}{ccc}0.2 & 0.25 & -0.3 \\ 0.15 & 0.35 & -0.35 \end{array}\right)$&       $\left(\begin{array}{ccc}0.25 & -0.3 & 0.35 \\ 0.2 & -0.25 & 0.32\end{array}\right)$  & $\left(\begin{array}{ccc}0.2 & 0.25 & -0.3 \\ 0.15 & 0.35 & -0.35 \\ 0.24 & 0.3 & -0.32 \end{array}\right)$          & $\left(\begin{array}{ccc}0.25 & -0.3 & 0.35 \\ 0.2 & -0.25 & 0.32\end{array}\right)$    &    $\left(\begin{array}{ccc}0.2 & 0.25 & -0.3 \\ 0.15 & 0.35 & -0.35 \\ 0.24 & 0.30 & -0.32 \end{array}\right)$     &    $\left(\begin{array}{ccc}0.25 & -0.3 & 0.35  \\ 0.2 & -0.25 & 0.32 \\ 0.1 & -0.2 & 0.2 \end{array}\right)$                  \\ \hline
			\multicolumn{2}{c|}{$\alpha_{g^{(1)} g^{(2)}}$} & \multicolumn{2}{c|}{$\alpha_{g^{(1)} g^{(2)}}$} & \multicolumn{2}{c}{$\alpha_{g^{(1)} g^{(2)}}$} \\
			\multicolumn{2}{c|}{$\left(\begin{array}{cc}-0.2 & 0.3 \\ -0.18 & 0.35 \end{array}\right)$} & \multicolumn{2}{c|}{$\left(\begin{array}{cc} -0.2 & 0.3 \\ -0.18 & 0.35 \\ -0.15 & 0.28 \end{array}\right)$}              &       \multicolumn{2}{c}{$\left(\begin{array}{ccc}-0.2 & 0.3 & 0.4 \\ -0.18 & 0.35 & 0.4 \\ -0.15 & 0.28 & 0.2 \end{array}\right)$}  \\ \hline
	\end{tabular}}
\end{table}

{\bf Example 1.} (Stochastic Block Model, SBM) The first type of
network is the stochastic block model, in which nodes in the same block (group)
are assigned with a higher probability to be connected.
In the $l$th network, based on the group memberships $g_{i_l}^{(l)}$s and following the setting of \cite{nowicki2001estimation},
we set $P(a^{(l)}_{ij} = 1) =
20/N_l$ when the $i$th and the $j$th node are in the same group, and otherwise we set $P(a^{(l)}_{ij} = 1) = 2/N_l$.

{\bf Example 2.} (Power-Law Distribution Network) The second
type of network is generated from a power-law distribution following
\cite{clauset2009power}.
For the $i$th node in the $l$th network, its in-degree $d_i^{(l)} = \sum_{j = 1}^N a^{(l)}_{ji}$ is
assumed to be power-law distributed.
Specifically, we first generate $\wt d_i^{(l)}$ with a probability $P(\wt d_i^{(l)} = k)
\propto k^{-2.5}$,
and set $d_i^{(l)} = 4 \wt d_i^{(l)}$.
Then, $d_i^{(l)}$ followers of the $i$th node are randomly
selected to construct the adjacency matrix.
As a result, the adjacency matrix is not symmetric, which implies a directed network.

\subsection{Performance Measure and Simulation Results}\label{subsec:simu_res}

We first introduce the model performance measure and
then present the simulation results.
We set the network sizes  $(N_1, N_2) \in \{(100, 80),(200, 150),(300, 250)\}$.
The time length is set to be $T \in \{20,40\}$.
For each scenario, we repeat the experiments for $R=500$ times.
The networks are fixed throughout all replicates under one setting.
In the initialization, we use 3 trials for each clustering type, as described in Algorithm \ref{alg:init} of Appendix \ref{sec:init}.
Denote the estimated parameters in the $r$th replicate as
$\wh\lambda_{g^{(1)}}^{(1)[r]}, \wh\lambda_{g^{(2)}}^{(2)[r]}, \wh\bzeta_{g^{(1)}}^{(1)[r]}, \wh\bzeta_{g^{(2)}}^{(2)[r]},
\wh\alpha_{g^{(1)} g^{(2)}}^{[r]}$ and the
estimated group number as $\wh G_1^{[r]}$ and $\wh G_2^{[r]}$.

\subsubsection{Estimation when $G_1 = G_{1,0}$ and $G_2=G_{2,0}$}\label{subsec:simu_true_G}

We first evaluate the estimation accuracy when the group numbers are correctly specified.
Take $\blambda^{(1)} =(\lambda_{1}^{(1)}, \cdots, \lambda_{G_1}^{(1)})^\top$ for example.
Denote $\wh \blambda^{(1)[r]}$ as the estimator of $\blambda^{(1)0}=
(\lambda_1^{(1)0}, \cdots, \lambda_{G_1}^{(1)0})^\top$ in the $r$th replicate.
To evaluate the estimation accuracy, we calculate the root mean squared
error (RMSE) as $\text{RMSE}_{\blambda^{(1)}} = \{R^{-1} \sum_{r = 1}^R
(\|\wh\blambda^{(1)[r]} - \blambda^{(1)0}\|^2) \}^{1/2}$.
Next, to gauge the performance of the statistical inference, we
construct the 95\% confidence interval for each parameter.
For example, denote the estimated standard error of $\lambda_{g^{(1)}}^{(1)}$ as
$\wh{\text{SE}}_{\lambda_{g^{(1)}}^{(1)}}^{[r]}$ for the $r$th replicate, then the
95\% confidence interval for $\wh\lambda_{g^{(1)}}^{(1)[r]}$ is constructed as
$\text{CI}_{\lambda_{g^{(1)}}^{(1)}}^{[r]} = (\wh\lambda_{g^{(1)}}^{(1)[r]} - 1.96 \times
\wh{\text{SE}}_{\lambda_{g^{(1)}}^{(1)}}^{[r]}, \wh\lambda_{g^{(1)}}^{(1)[r]} +1.96 \times
\wh{\text{SE}}_{\lambda_{g^{(1)}}^{(1)}}^{[r]})$.
Here
	$\wh{\text{SE}}_{\lambda_{g^{(1)}}^{(1)}}^{[r]}$ is obtained by Theorem
	\ref{thm:normal}, plugging in the variance estimator $\wh\sigma^2$, which can be obtained by the procedure in Appendix \ref{append:normal_est}. Subsequently, the coverage probability (CP) is
formed as $\text{CP}_{\lambda_{g^{(1)}}^{(1)}} = R^{-1} \sum_{r =1}^R
I(\lambda_{g^{(1)}}^{(1)0} \in\text{CI}_{\lambda_{g^{(1)}}^{(1)}}^{[r]})$.
We calculate the CPs for other parameters similarly.
For comparison, we also calculate the RMSE and CP values for the oracle
estimators under the true group memberships (denoted as $\wh\lambda_{g^{(1)}}^{(1)\text{or}},
\wh\lambda_{g^{(2)}}^{(2)\text{or}}, \wh\bzeta_{g^{(1)}}^{(1)\text{or}}, \wh\bzeta_{g^{(2)}}^{(2)\text{or}}, \wh\alpha_{g^{(1)} g^{(2)}}^{\text{or}}$ accordingly).
Lastly, to evaluate the group memberships estimation, we calculate the
mis-clustering rates for two network groups as
$\wh\eta_l = (N_l R)^{-1}\sum_{r =  1}^R \sum_{i_l} I(\wh g_{i_l}^{(l)}\ne
g_{i_l}^{(l)0})$ for $l = 1, 2$, where $\wh g_{i_l}^{(l)}$ is the estimated group
membership of the $i_l$th node.
Here the mis-clustering rates are calculated after proper group permutations.

The simulation results of $G_{1,0} = G_{2,0} = 3$ are shown in Table \ref{table:sce3}.
The first finding across all combinations is that once the group
numbers are specified as the true values in advance, our iterative method
can estimate the true group memberships with high accuracy, especially when the
sample size is large.
As $N_1, N_2$ or $T$ increase, the mis-clustering rates for both network groups approach zero.
Furthermore, we note that the RMSEs decrease either when the network sizes
$N_1$ and $N_2$ increase or the time length $T$ increases, and they
approach the oracle RMSEs when the sample sizes  are
large.
Next, we inspect the statistical inference results. We observe that in Table
\ref{table:sce3}, the CPs are slightly smaller when the sample sizes are
not very large, but they grow up to around 0.95 as $N_1, N_2$ and $T$
increase.
This guarantees that even under the scenario with a large number
of parameters to be estimated, GTNAR can
still perform well in terms of both estimation and inference for
sufficiently large sample sizes.
The patterns are similar in the other two scenarios.

\begin{sidewaystable}[]\footnotesize
	\centering
	\caption{\small RMSEs ($\times 1000$) of estimated parameters under scenario 3 ($G_{1,0} = G_{2,0} = 3$) with 500 replications. The performances are evaluated for different sample sizes $N_1, N_2$ and the time length $T$. Results under two network structures are provided. The corresponding CPs are shown in parentheses.}\label{table:sce3}
	\scalebox{0.95}{
		\begin{tabular}{c|cc|ccc|ccccc|ccccc|cc}
			\hline
			Network                    & 			$G_1$ &
			$G_2$ &
			$N_1$ &
			$N_2$ &
			$T$ &
			$\wh\blambda^{(1)}$ &
			$\wh\blambda^{(2)}$ &
			$\wh\bzeta^{(1)}$ &
			$\wh\bzeta^{(2)}$ &
			$\wh\balpha$ &
			$\wh\blambda^{(1)\text{or}}$ &
			$\wh\blambda^{(2)\text{or}}$ &
			$\wh\bzeta^{(1)\text{or}}$ &
			$\wh\bzeta^{(2)\text{or}}$ &
			$\wh\balpha^{\text{or}}$ &
			$\wh\eta_1$ &
			$\wh\eta_2$ \\ \hline
			\multirow{11}{*}{SBM}       & \multirow{11}{*}{3} & \multirow{11}{*}{3} & \multirow{3}{*}{100} & \multirow{3}{*}{80}  & 20  & \begin{tabular}[c]{@{}c@{}}25.6\\ (0.705)\end{tabular} & \begin{tabular}[c]{@{}c@{}}12.5\\ (0.907)\end{tabular} & \begin{tabular}[c]{@{}c@{}}59.1\\ (0.595)\end{tabular} & \begin{tabular}[c]{@{}c@{}}15.9\\ (0.920)\end{tabular} & \begin{tabular}[c]{@{}c@{}}58.6\\ (0.731)\end{tabular} & \begin{tabular}[c]{@{}c@{}}11.7\\ (0.947)\end{tabular} & \begin{tabular}[c]{@{}c@{}}10.8\\ (0.938)\end{tabular} & \begin{tabular}[c]{@{}c@{}}13.6\\ (0.945)\end{tabular} & \begin{tabular}[c]{@{}c@{}}12.9\\ (0.949)\end{tabular} & \begin{tabular}[c]{@{}c@{}}19.1\\ (0.949)\end{tabular} & 0.1634   & 0.0113   \\
			&                    &                    &                      &                      & 40  & \begin{tabular}[c]{@{}c@{}}9.1\\ (0.911)\end{tabular}  & \begin{tabular}[c]{@{}c@{}}7.5\\ (0.945)\end{tabular}  & \begin{tabular}[c]{@{}c@{}}14.4\\ (0.871)\end{tabular} & \begin{tabular}[c]{@{}c@{}}9.1\\ (0.948)\end{tabular}  & \begin{tabular}[c]{@{}c@{}}15.2\\ (0.913)\end{tabular} & \begin{tabular}[c]{@{}c@{}}7.8\\ (0.951)\end{tabular}  & \begin{tabular}[c]{@{}c@{}}7.4\\ (0.945)\end{tabular}  & \begin{tabular}[c]{@{}c@{}}9.4\\ (0.938)\end{tabular}  & \begin{tabular}[c]{@{}c@{}}9.1\\ (0.949)\end{tabular}  & \begin{tabular}[c]{@{}c@{}}12.8\\ (0.947)\end{tabular} & 0.0285   & 0.0001   \\ \cline{4-18}
			&                    &                    & \multirow{3}{*}{200} & \multirow{3}{*}{150} & 20  & \begin{tabular}[c]{@{}c@{}}9.1\\ (0.825)\end{tabular}  & \begin{tabular}[c]{@{}c@{}}5.6\\ (0.925)\end{tabular}  & \begin{tabular}[c]{@{}c@{}}17.4\\ (0.805)\end{tabular} & \begin{tabular}[c]{@{}c@{}}7.0\\ (0.940)\end{tabular}  & \begin{tabular}[c]{@{}c@{}}14.5\\ (0.857)\end{tabular} & \begin{tabular}[c]{@{}c@{}}6.1\\ (0.932)\end{tabular}  & \begin{tabular}[c]{@{}c@{}}5.3\\ (0.939)\end{tabular}  & \begin{tabular}[c]{@{}c@{}}6.8\\ (0.940)\end{tabular}  & \begin{tabular}[c]{@{}c@{}}6.9\\ (0.942)\end{tabular}  & \begin{tabular}[c]{@{}c@{}}9.6\\ (0.945)\end{tabular}  & 0.0524   & 0.0006   \\
			&                    &                    &                      &                      & 40  & \begin{tabular}[c]{@{}c@{}}4.2\\ (0.942)\end{tabular}  & \begin{tabular}[c]{@{}c@{}}3.8\\ (0.939)\end{tabular}  & \begin{tabular}[c]{@{}c@{}}5.4\\ (0.940)\end{tabular}  & \begin{tabular}[c]{@{}c@{}}4.7\\ (0.949)\end{tabular}  & \begin{tabular}[c]{@{}c@{}}7.0\\ (0.941)\end{tabular}  & \begin{tabular}[c]{@{}c@{}}4.0\\ (0.948)\end{tabular}  & \begin{tabular}[c]{@{}c@{}}3.8\\ (0.941)\end{tabular}  & \begin{tabular}[c]{@{}c@{}}4.7\\ (0.950)\end{tabular}  & \begin{tabular}[c]{@{}c@{}}4.6\\ (0.950)\end{tabular}  & \begin{tabular}[c]{@{}c@{}}6.6\\ (0.948)\end{tabular}  & 0.0033   & 0.0001   \\ \cline{4-18}
			&                    &                    & \multirow{3}{*}{300} & \multirow{3}{*}{250} & 20  & \begin{tabular}[c]{@{}c@{}}4.4\\ (0.916)\end{tabular}  & \begin{tabular}[c]{@{}c@{}}3.5\\ (0.935)\end{tabular}  & \begin{tabular}[c]{@{}c@{}}6.1\\ (0.906)\end{tabular}  & \begin{tabular}[c]{@{}c@{}}4.2\\ (0.941)\end{tabular}  & \begin{tabular}[c]{@{}c@{}}6.9\\ (0.920)\end{tabular}  & \begin{tabular}[c]{@{}c@{}}3.7\\ (0.948)\end{tabular}  & \begin{tabular}[c]{@{}c@{}}3.5\\ (0.936)\end{tabular}  & \begin{tabular}[c]{@{}c@{}}4.2\\ (0.946)\end{tabular}  & \begin{tabular}[c]{@{}c@{}}4.2\\ (0.941)\end{tabular}  & \begin{tabular}[c]{@{}c@{}}5.9\\ (0.944)\end{tabular}  & 0.0092        & 0.0001        \\
			&                    &                    &                      &                      & 40  & \begin{tabular}[c]{@{}c@{}}3.0\\ (0.930)\end{tabular}  & \begin{tabular}[c]{@{}c@{}}2.3\\ (0.948)\end{tabular}  & \begin{tabular}[c]{@{}c@{}}4.2\\ (0.927)\end{tabular}  & \begin{tabular}[c]{@{}c@{}}2.9\\ (0.952)\end{tabular}  & \begin{tabular}[c]{@{}c@{}}4.8\\ (0.936)\end{tabular}  & \begin{tabular}[c]{@{}c@{}}2.5\\ (0.947)\end{tabular}  & \begin{tabular}[c]{@{}c@{}}2.3\\ (0.950)\end{tabular}  & \begin{tabular}[c]{@{}c@{}}2.9\\ (0.948)\end{tabular}  & \begin{tabular}[c]{@{}c@{}}2.9\\ (0.952)\end{tabular}  & \begin{tabular}[c]{@{}c@{}}4.1\\ (0.948)\end{tabular}  & 0.0006        & 0        \\ \hline
			\multirow{11}{*}{Power-Law} & \multirow{11}{*}{3} & \multirow{11}{*}{3} & \multirow{3}{*}{100} & \multirow{3}{*}{80}  & 20  & \begin{tabular}[c]{@{}c@{}}17.5\\ (0.769)\end{tabular} & \begin{tabular}[c]{@{}c@{}}9.2\\ (0.931)\end{tabular}  & \begin{tabular}[c]{@{}c@{}}40.2\\ (0.687)\end{tabular} & \begin{tabular}[c]{@{}c@{}}17.6\\ (0.923)\end{tabular} & \begin{tabular}[c]{@{}c@{}}34.7\\ (0.790)\end{tabular} & \begin{tabular}[c]{@{}c@{}}11.0\\ (0.938)\end{tabular} & \begin{tabular}[c]{@{}c@{}}8.6\\ (0.941)\end{tabular}  & \begin{tabular}[c]{@{}c@{}}12.9\\ (0.948)\end{tabular} & \begin{tabular}[c]{@{}c@{}}13.2\\ (0.944)\end{tabular} & \begin{tabular}[c]{@{}c@{}}17.8\\ (0.947)\end{tabular} & 0.1361   & 0.0156   \\
			&                    &                    &                      &                      & 40  & \begin{tabular}[c]{@{}c@{}}11.4\\ (0.863)\end{tabular} & \begin{tabular}[c]{@{}c@{}}5.7\\ (0.947)\end{tabular}  & \begin{tabular}[c]{@{}c@{}}20.0\\ (0.831)\end{tabular} & \begin{tabular}[c]{@{}c@{}}9.3\\ (0.937)\end{tabular}  & \begin{tabular}[c]{@{}c@{}}19.5\\ (0.886)\end{tabular} & \begin{tabular}[c]{@{}c@{}}7.2\\ (0.945)\end{tabular}  & \begin{tabular}[c]{@{}c@{}}5.6\\ (0.943)\end{tabular}  & \begin{tabular}[c]{@{}c@{}}9.5\\ (0.941)\end{tabular}  & \begin{tabular}[c]{@{}c@{}}9.3\\ (0.937)\end{tabular}  & \begin{tabular}[c]{@{}c@{}}13.2\\ (0.940)\end{tabular} & 0.0411   & 0        \\ \cline{4-18}
			&                    &                    & \multirow{3}{*}{200} & \multirow{3}{*}{150} & 20  & \begin{tabular}[c]{@{}c@{}}7.7\\ (0.843)\end{tabular}  & \begin{tabular}[c]{@{}c@{}}4.6\\ (0.939)\end{tabular}  & \begin{tabular}[c]{@{}c@{}}14.1\\ (0.840)\end{tabular} & \begin{tabular}[c]{@{}c@{}}6.8\\ (0.945)\end{tabular}  & \begin{tabular}[c]{@{}c@{}}12.1\\ (0.883)\end{tabular} & \begin{tabular}[c]{@{}c@{}}5.4\\ (0.937)\end{tabular}  & \begin{tabular}[c]{@{}c@{}}4.5\\ (0.945)\end{tabular}  & \begin{tabular}[c]{@{}c@{}}6.7\\ (0.947)\end{tabular}  & \begin{tabular}[c]{@{}c@{}}6.7\\ (0.946)\end{tabular}  & \begin{tabular}[c]{@{}c@{}}8.7\\ (0.950)\end{tabular}  & 0.0408   & 0.0001   \\
			&                    &                    &                      &                      & 40  & \begin{tabular}[c]{@{}c@{}}4.3\\ (0.933)\end{tabular}  & \begin{tabular}[c]{@{}c@{}}2.1\\ (0.948)\end{tabular}  & \begin{tabular}[c]{@{}c@{}}5.7\\ (0.927)\end{tabular}  & \begin{tabular}[c]{@{}c@{}}4.7\\ (0.948)\end{tabular}  & \begin{tabular}[c]{@{}c@{}}6.6\\ (0.940)\end{tabular}  & \begin{tabular}[c]{@{}c@{}}4.0\\ (0.946)\end{tabular}  & \begin{tabular}[c]{@{}c@{}}2.1\\ (0.949)\end{tabular}  & \begin{tabular}[c]{@{}c@{}}4.8\\ (0.940)\end{tabular}  & \begin{tabular}[c]{@{}c@{}}4.7\\ (0.948)\end{tabular}  & \begin{tabular}[c]{@{}c@{}}6.2\\ (0.948)\end{tabular}  & 0.0039   & 0        \\ \cline{4-18}
			&                    &                    & \multirow{3}{*}{300} & \multirow{3}{*}{250} & 20  & \begin{tabular}[c]{@{}c@{}}4.7\\ (0.890)\end{tabular}  & \begin{tabular}[c]{@{}c@{}}2.4\\ (0.937)\end{tabular}  & \begin{tabular}[c]{@{}c@{}}7.8\\ (0.892)\end{tabular}  & \begin{tabular}[c]{@{}c@{}}5.4\\ (0.938)\end{tabular}  & \begin{tabular}[c]{@{}c@{}}10.7\\ (0.907)\end{tabular}  & \begin{tabular}[c]{@{}c@{}}3.3\\ (0.939)\end{tabular}  & \begin{tabular}[c]{@{}c@{}}2.1\\ (0.943)\end{tabular}  & \begin{tabular}[c]{@{}c@{}}4.2\\ (0.943)\end{tabular}  & \begin{tabular}[c]{@{}c@{}}4.3\\ (0.944)\end{tabular}  & \begin{tabular}[c]{@{}c@{}}5.5\\ (0.945)\end{tabular}  & 0.0170        & 0.0001        \\
			&                    &                    &                      &                      & 40  & \begin{tabular}[c]{@{}c@{}}2.7\\ (0.942)\end{tabular}  & \begin{tabular}[c]{@{}c@{}}1.9\\ (0.936)\end{tabular}  & \begin{tabular}[c]{@{}c@{}}3.2\\ (0.950)\end{tabular}  & \begin{tabular}[c]{@{}c@{}}2.9\\ (0.948)\end{tabular}  & \begin{tabular}[c]{@{}c@{}}4.1\\ (0.947)\end{tabular}  & \begin{tabular}[c]{@{}c@{}}2.6\\ (0.944)\end{tabular}  & \begin{tabular}[c]{@{}c@{}}1.9\\ (0.935)\end{tabular}  & \begin{tabular}[c]{@{}c@{}}2.9\\ (0.952)\end{tabular}  & \begin{tabular}[c]{@{}c@{}}2.9\\ (0.948)\end{tabular}  & \begin{tabular}[c]{@{}c@{}}4.0\\ (0.949)\end{tabular}  & 0.0010        & 0        \\ \hline
		\end{tabular}
	}
\end{sidewaystable}

\subsubsection{Estimation when $G_l \ge G_{l,0}$ for $l = 1, 2$}

We next consider the case of estimation without specifying the true group numbers in advance.
Specifically, we estimate the group numbers by QIC in Section
\ref{sec:group_number}, where the tuning parameter is set to be $\kappa = 1/\{40 \log(T) T^{1/8}\}$.
Let the true group numbers be $G_{1,0} = G_{2,0} = 3$,
and the corresponding true parameters are shown in Table \ref{table:true_param}.
To evaluate the estimation accuracy, we calculate the RMSE for each
parameter as explained below.
Take $\blambda^{(1)}$ for example, define $\text{RMSE}_{\blambda^{(1)}, all} =
\{(R N_1)^{-1}\sum_{r = 1}^R \sum_{i_1 = 1}^{N_1}( \wh\lambda_{\wh
	g_{i_1}^{(1)}}^{(1)[r]} - \lambda_{g_{i_1}^{(1)0}}^{(1)0})^2\}^{1/2}$ as the RMSE for all
nodes. RMSE for other parameters is calculated similarly.
For the group memberships, the mis-clustering rates are calculated
following the idea of \cite{zhu2023simultaneous}.
Recall that we have $\wh \cR_{g^{(l)}}^{(l)} = \{i_l: \wh g_{i_l}^{(l)}  = g^{(l)}\}$, where $\wh g_{i_l}^{(l)}$ is denoted as the estimated group membership for the $i_l$th in the $l$th network.
Note that $G_l$ is not necessarily equal to $G_{l,0}$.
In this case, we define the mappings from the estimated group memberships to the true
group memberships $\chi_l: \{1,\cdots,G_l\} \to \{1, \cdots, G_{l,0}\}$ as
$\chi_l(g^{(l)}) = \argmax_{g^{(l)'} \in \{1, \cdots, G_{l,0}\}} \sum_{i_l = 1}^{N_l} I\big(i_l \in \wh\cR_{g^{(l)}}^{(l)}, g_{i_l}^{(l)0} = g^{(l)'}\big)$ for $g^{(l)} \in \{1, \cdots, G_l\}$.
Thus, the mapping $\chi_l(g^{(l)})$ maps group $g^{(l)}$ to the true membership
$g^{(l)'}$ where the majority of nodes in $\wh \cR_{g^{(l)}}^{(l)}$ belong to.
Then, for the row group memberships, the mis-clustering rate in the
$r$th replicate is  defined as
$
\wh\xi_l^{[r]} = N_l^{-1} \sum_{g^{(l)}=1}^{G_l} \sum_{i_l =1}^{N_l} I\big(i_l \in \wh\cR_{g^{(l)}}^{(l)[r]}, g_{i_l}^{(l)0} \neq \chi_l(g^{(l)}) \big),
$
where $\wh\cR_{g^{(l)}}^{(l)[r]}$ is the
estimated node set belonging to the group $g^{(l)}$ in the $r$th replicate.
Then, the overall group memberships error rate is calculated as
$\wh\xi_l = R^{-1} \sum_{r} \wh\xi_l^{[r]}$.
The results are shown in Table \ref{table:QIC_1}.

We discuss the results shown in Table \ref{table:QIC_1} from two aspects.
On one hand, when the group number is under-specified ($G_1 = 2, G_2 =
2$), the node-wise RMSEs are large and usually do not decrease when the $N_1, N_2$ and $T$ grow.
Besides, the error rates $\wh\xi_1$ and $\wh\xi_2$ are around 0.3,
indicating a low accuracy in estimating node memberships.
These results are expected since a non-ignorable estimation bias
exists in an under-fitted model.
On the other hand, when the group numbers $G_l$s are correctly ($G_1=3,
G_2=3$) or over-specified ($G_1 =4, G_2 = 4$), the RMSE values are
generally much lower.
This is consistent with our theoretical analysis in Theorem \ref{thm:pseudo_dist}.

\begin{sidewaystable}[]
	\centering
	\caption{\small Simulation results for the two network examples with pre-specified group numbers as well as the QIC selection group numbers $\wh G_l$. The true group numbers are set as $G_{1,0} = G_{2,0}= 3$. The node-wise RMSEs of different estimators are denoted as $\wh\blambda^{(1)}_{\text{all}}, \wh\blambda^{(2)}_{\text{all}}, \bzeta^{(1)}_{\text{all}},\bzeta^{(2)}_{\text{all}}, \balpha_{\text{all}}$. }\label{table:QIC_1}
	\scalebox{0.8}{
		\begin{tabular}{c|c|c|cc|ccccccc|ccccccc}
			\hline
			\multirow{2}{*}{$N_1$} & \multirow{2}{*}{$N_2$} & \multirow{2}{*}{$T$} & \multirow{2}{*}{$G_1$} & \multirow{2}{*}{$G_2$} & \multicolumn{7}{c|}{Scenario 1 (SBM)}                                                                                                                                                              & \multicolumn{7}{c}{Scenario 2 (Power-Law)}                                                                                                                                                         \\ \cline{6-19}
			&                        &                      &                      &                      & $\wh\blambda^{(1)}_{\text{all}}$ & $\wh\blambda^{(2)}_{\text{all}}$& $\wh \bzeta^{(1)}_{\text{all}}$ & $\wh \bzeta^{(2)}_{\text{all}}$ & \multicolumn{1}{c|}{$\wh\balpha_{\text{all}}$} & $\wh\xi_1$ & \multicolumn{1}{c|}{$\wh\xi_2$} & $\wh\blambda^{(1)}_{\text{all}}$ & $\wh\blambda^{(2)}_{\text{all}}$& $\wh \bzeta^{(1)}_{\text{all}}$ & $\wh \bzeta^{(2)}_{\text{all}}$ & \multicolumn{1}{c|}{$\wh\balpha_{\text{all}}$} & $\wh\xi_1$ & $\wh\xi_2$  \\ \hline
			\multirow{10}{*}{100}  & \multirow{10}{*}{80}   & \multirow{5}{*}{20}  & \multicolumn{2}{c|}{Oracle}                 & 0.0065              & 0.0059            & 0.0071           & 0.0070             & \multicolumn{1}{c|}{0.0061}             & -        & \multicolumn{1}{c|}{-}       & 0.0072              & 0.0039            & 0.0071           & 0.0072             & \multicolumn{1}{c|}{0.0060}             & -         & -            \\
			&                        &                      & 2                    & 2                    & 0.0285              & 0.0419            & 0.0427           & 0.0583             & \multicolumn{1}{c|}{0.0368}             & 0.3219   & \multicolumn{1}{c|}{0.3125}      & 0.0237              & 0.0438            & 0.0382           & 0.0647             & \multicolumn{1}{c|}{0.0367}             & 0.2705   &0.3705    \\
			&                        &                      & 3                    & 3                    & 0.0207              & 0.0096            & 0.0251           & 0.0099             & \multicolumn{1}{c|}{0.0174}             & 0.1403   & \multicolumn{1}{c|}{0.0142}     & 0.0196              & 0.0129            & 0.0263           & 0.0114             & \multicolumn{1}{c|}{0.0169}             & 0.1487   & 0.0208      \\
			&                        &                      & 4                    & 4                    & 0.0192              & 0.0099            & 0.0181           & 0.0097             & \multicolumn{1}{c|}{0.0172}             & 0.0720   & \multicolumn{1}{c|}{0.0042}        & 0.0201              & 0.0134            & 0.0223           & 0.0101             & \multicolumn{1}{c|}{0.0176}             & 0.1046   & 0.0077     \\
			&                        &                      & $\wh G$              & $\wh H$              & 0.0278              & 0.0407            & 0.0416           & 0.0565             & \multicolumn{1}{c|}{0.0358}             & -        & \multicolumn{1}{c|}{-}         & 0.0218              & 0.0374            & 0.0339           & 0.0554             & \multicolumn{1}{c|}{0.0323}             & -               & -            \\ \cline{3-19}
			&                        & \multirow{5}{*}{40}  & \multicolumn{2}{c|}{Oracle}                 & 0.0044              & 0.0043            & 0.0050           & 0.0050             & \multicolumn{1}{c|}{0.0043}             & -        & \multicolumn{1}{c|}{-}        & 0.0046              & 0.0040            & 0.0050           & 0.0049             & \multicolumn{1}{c|}{0.0040}             & -         & -            \\
			&                        &                      & 2                    & 2                    & 0.0216              & 0.0436            & 0.0430           & 0.0503             & \multicolumn{1}{c|}{0.0327}             & 0.3100   & \multicolumn{1}{c|}{0.2625}      & 0.0210              & 0.0428            & 0..0403          & 0.0607             & \multicolumn{1}{c|}{0.0361}             & 0.2800   &0.3375        \\
			&                        &                      & 3                    & 3                    & 0.0082              & 0.0070            & 0.0104           & 0.0063             & \multicolumn{1}{c|}{0.0078}             & 0.0366   & \multicolumn{1}{c|}{0.0061}      & 0.0069              & 0.0040            & 0.0085           & 0.0049             & \multicolumn{1}{c|}{0.0053}             & 0.0233   &0        \\
			&                        &                      & 4                    & 4                    & 0.0081              & 0.0061            & 0.0084           & 0.0065             & \multicolumn{1}{c|}{0.0084}             & 0.0136   & \multicolumn{1}{c|}{0.0006}         & 0.0076              & 0.0066            & 0.0078           & 0.0067             & \multicolumn{1}{c|}{0.0082}             & 0.0101   &0.0015       \\
			&                        &                      & $\wh G$              & $\wh H$              & 0.0086              & 0.0112            & 0.0121           & 0.0121             & \multicolumn{1}{c|}{0.0100}             & -        & \multicolumn{1}{c|}{-}        & 0.0069              & 0.0040            & 0.0085           & 0.0049             & \multicolumn{1}{c|}{0.0053}             & -           & -            \\ \hline
			\multirow{10}{*}{200}  & \multirow{10}{*}{150}  & \multirow{5}{*}{20}  & \multicolumn{2}{c|}{Oracle}                 & 0.0034              & 0.0031            & 0.0036           & 0.0037             & \multicolumn{1}{c|}{0.0031}             & -        & \multicolumn{1}{c|}{-}                 & 0.0036              & 0.0021            & 0.0036           & 0.0037             & \multicolumn{1}{c|}{0.0030}             & -           & -            \\
			&                        &                      & 2                    & 2                    & 0.0221              & 0.0431            & 0.0455           & 0.0592             & \multicolumn{1}{c|}{0.0351}             & 0.3505   & \multicolumn{1}{c|}{0.3133}   & 0.0217              & 0.0420            & 0.0416           & 0.0559             & \multicolumn{1}{c|}{0.0351}             & 0.3002   & 0.3200       \\
			&                        &                      & 3                    & 3                    & 0.0091              & 0.0037            & 0.0112           & 0.0040             & \multicolumn{1}{c|}{0.0068}             & 0.0504   & \multicolumn{1}{c|}{0.0015}       & 0.0076              & 0.0022            & 0.0089           & 0.0037             & \multicolumn{1}{c|}{0.0054}             & 0.0342   & 0.0001       \\
			&                        &                      & 4                    & 4                    & 0.0106              & 0.0052            & 0.0089           & 0.0054             & \multicolumn{1}{c|}{0.0096}             & 0.0296   & \multicolumn{1}{c|}{0.0003}         & 0.0086              & 0.0047            & 0.0082           & 0.0050             & \multicolumn{1}{c|}{0.0077}             & 0.0240   & 0.0004       \\
			&                        &                      & $\wh G$              & $\wh H$              & 0.0178              & 0.0334            & 0.0354           & 0.0457             & \multicolumn{1}{c|}{0.0275}             & -        & \multicolumn{1}{c|}{-}             & 0.0076              & 0.0026            & 0.0089           & 0.0042             & \multicolumn{1}{c|}{0.0056}             & -         & -            \\ \cline{3-19}
			&                         & \multirow{5}{*}{40}  & \multicolumn{2}{c|}{Oracle}                 & 0.0024              & 0.0021            & 0.0026           & 0.0026             & \multicolumn{1}{c|}{0.0022}             & -        & \multicolumn{1}{c|}{-}           & 0.0023              & 0.0011            & 0.0025           & 0.0026             & \multicolumn{1}{c|}{0.0020}             & -           & -            \\
			&                        &                      & 2                    & 2                    & 0.0199              & 0.0405            & 0.0383           & 0.0539             & \multicolumn{1}{c|}{0.0343}             & 0.2600   & \multicolumn{1}{c|}{0.3133}          & 0.0218              & 0.0409            & 0.0437           & 0.0554             & \multicolumn{1}{c|}{0.0350}             & 0.3250   & 0.3133        \\
			&                        &                      & 3                    & 3                    & 0.0029              & 0.0021            & 0.0034           & 0.0026             & \multicolumn{1}{c|}{0.0026}             & 0.0028        & \multicolumn{1}{c|}{0.0001}                & 0.0026              & 0.0011           & 0.0030           & 0.0026             & \multicolumn{1}{c|}{0.0022}             & 0.0027       & 0      \\
			&                        &                      & 4                    & 4                    & 0.0044              & 0.0038            & 0.0043           & 0.0040             & \multicolumn{1}{c|}{0.0055}             & 0.0047   & \multicolumn{1}{c|}{0.0017}       & 0.0049              & 0.0042            & 0.0045           & 0.0042             & \multicolumn{1}{c|}{0.0057}             & 0.0063   & 0.0032           \\
			&                        &                      & $\wh G$              & $\wh H$              & 0.0030              & 0.0027            & 0.0035           & 0.0034             & \multicolumn{1}{c|}{0.0029}             & -        & \multicolumn{1}{c|}{-}           & 0.0026              & 0.0011            & 0.0030           & 0.0026             & \multicolumn{1}{c|}{0.0022}             & -            & -            \\ \hline
		\end{tabular}
	}
\end{sidewaystable}

\section{Real Data Applications}\label{sec:real}

\subsection{Data Description}\label{sec:yelp}

\begin{figure}[htpb!]
	\centering
	\subfigure[Block map]{\label{fig:map}\includegraphics[width=0.4\textwidth]{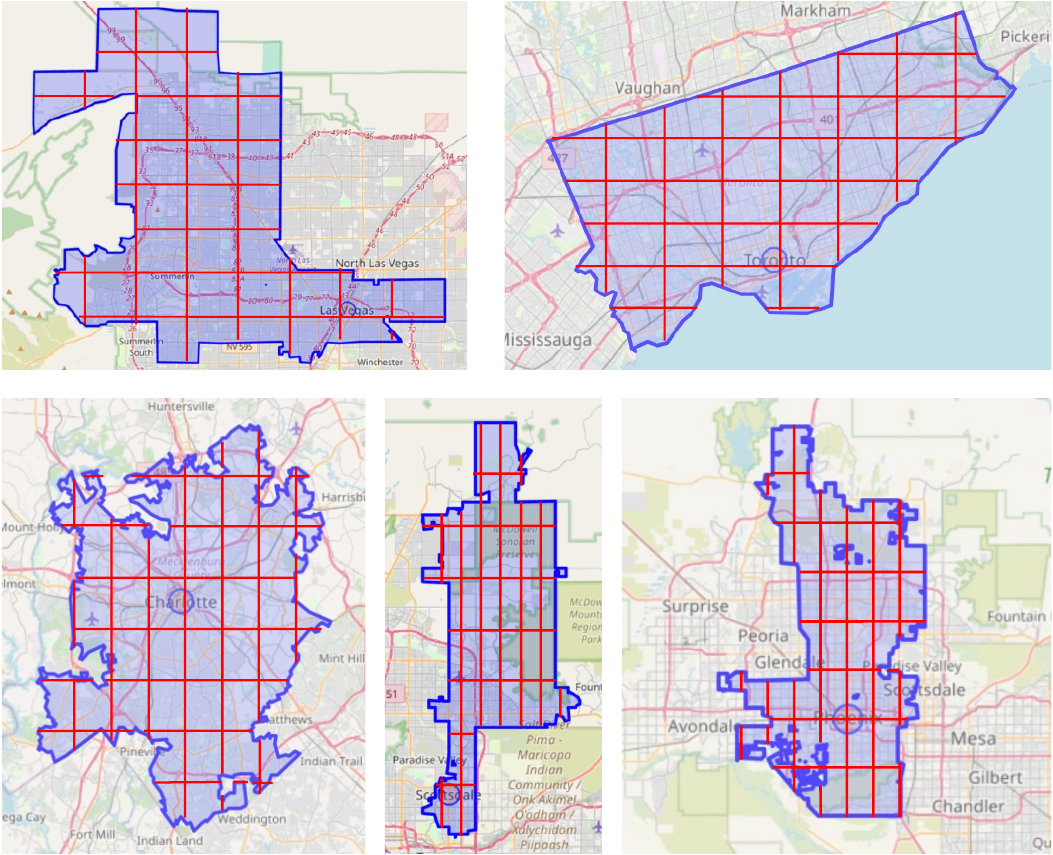}}
	\subfigure[Average review volumes]{\label{fig:des_line}\includegraphics[width=0.5\textwidth]{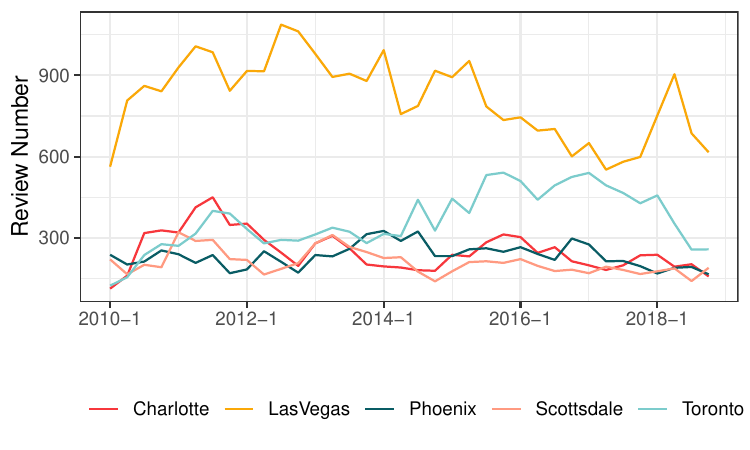}}
	\caption{\small (a) Geographical maps with split districts in each city. The city from the top left panel to the bottom right panel shows the map of Las Vegas, Toronto, Charlotte, Scottsdale, and Phoenix, respectively. (b) Average number of reviews from 2010-Q1 to 2018-Q4 in five cities.}
	\label{fig:yelp_des}
\end{figure}

The Yelp dataset spans from 2010 to 2018 and covers five North American cities: Charlotte, Las Vegas, Phoenix, Scottsdale, and Toronto. The observation period is divided into $T=36$ quarters. To ensure data quality, we filter the dataset to retain active users who have provided more than 5 reviews over this time span.
We further divide each city into districts, as illustrated in Figure
\ref{fig:map}.
Our response variable, denoted as $Y_{i j, t}$, represents the $\log(1+x)$ transformed number of
reviews by user $i$ on district $j$ during the $t$th quarter.
Here $Y_{ij, t}$ is treated as a continuous variable in our real data analysis.
To visualize temporal trends in review activity, we calculate the quarterly average responses for each city and depict them in Figure \ref{fig:des_line}.
Different patterns emerge from this analysis. For instance, Las Vegas stands out as the city with the most reviews, reflecting its bustling business environment. Charlotte, Phoenix, and Scottsdale exhibit relatively similar and stable review trends. In contrast, Toronto shows a noticeable increase in review volume after 2015, likely due to Yelp's expansion in the Toronto area during that period.

Next, we construct the adjacency matrices for users ($\bA^{(1)}$) and districts ($\bA^{(2)}$) as follows.
The user network is built based on the friend list information.
Specifically, if user $j$ is on the friend list of user $i$ on Yelp,
then we set $a^{(1)}_{ij} = 1$. Otherwise we set $a^{(1)}_{ij} = 0$.
The spatial network is built based on the geographical adjacent
relationship. Specifically,
we set $a^{(2)}_{ij} = 1$ if the district $j$ is adjacent to district $i$.

Lastly, to characterize the dynamic patterns of the responses, we
collect a number of covariates for users and districts, respectively.
For user $i$ in quarter $t$, we consider the following five
covariates: (1) the number of months after joining Yelp by
the start of the quarter $t$ ($x^{(1)}_{i t, \text{dur}}$), (2) whether the
user is VIP by the start of the quarter $t$ ($x^{(1)}_{i t, \text{vip}}$),
(3) average tags (i.e., ``useful'', ``funny'' and ``cool'') the user
$i$ obtains for his/her reviews during the last quarter ($x^{(1)}_{i t,
	\text{use}}, x^{(1)}_{i t, \text{fun}}, x^{(1)}_{i t, \text{cool}}$).
Next,
for the $j$th district in quarter $t$, we consider two covariates: (1) the average ``stars'' ($x^{(2)}_{j t, \text{star}}$), and (2)
the average review number ($x^{(2)}_{j t, \text{num}}$) obtained by the $j$th
district during the $(t-1)$th quarter.
These two covariates are indicative of the average popularity levels in the preceding time period.
We standardize all continuous covariates to the range $[0,1]$ for subsequent analysis.
In Figure \ref{fig:box} in Appendix \ref{subsec:yelp_append_des}, we visualize the relationship between these user-related covariates and the response variable. The plot reveals that users who receive more tags for their reviews tend to be motivated to contribute more reviews.
Notably, VIP users in Scottsdale and Toronto tend to write more reviews, whereas VIP users in Charlotte exhibit comparatively less activity. Subsequently, we apply the GTNAR model to each of the five cities, enabling us to analyze and understand the distinctive group patterns within each urban area.

\subsection{Estimation Results}\label{subsec:yelp_res}

There has been a lot of research finding that the activeness of users on the platform can lead to higher profit \citep{forman2008examining, pansari2017customer}, hence finding out the key factors that positively affect the review volumes can help business owners make personalized recommendations to users and develop a commercial strategy.
We employ QIC for the selection of group numbers, and show the results of Charlotte, Las Vegas, and Phoenix in Table \ref{tbl:real_res_1}.
The results of Scottsdale and Toronto are provided in Table \ref{tbl:real_res_2_yelp} in Appendix \ref{subsec:yelp_append_des}.
The numbers of user groups and district groups vary across the five cities, indicating different levels of heterogeneity among them.
For instance, consider the results for Phoenix, where there are estimated 3 user groups and 2 district groups.

\begin{figure}[htpb!]
	\begin{center}
		\includegraphics[width=0.7\textwidth]{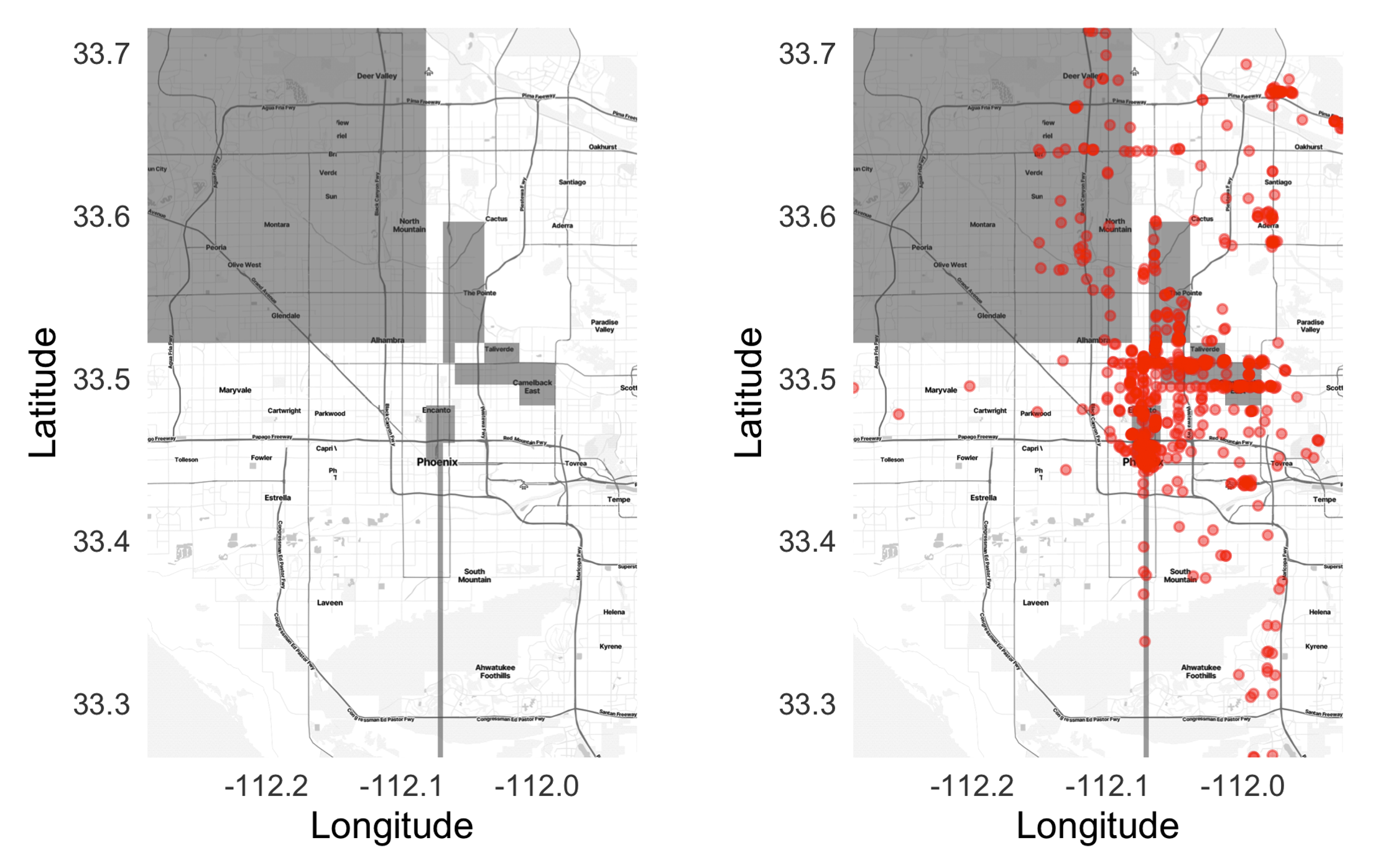}
		\caption{\small The left panel shows the two groups of districts in Phoenix (Group 2 is marked as gray). The right panel marks all shops as red points.}
		\label{fig:dis_block_map}
	\end{center}
\end{figure}

\begin{figure}[htpb!]
	\begin{center}
		\includegraphics[width=0.8\textwidth]{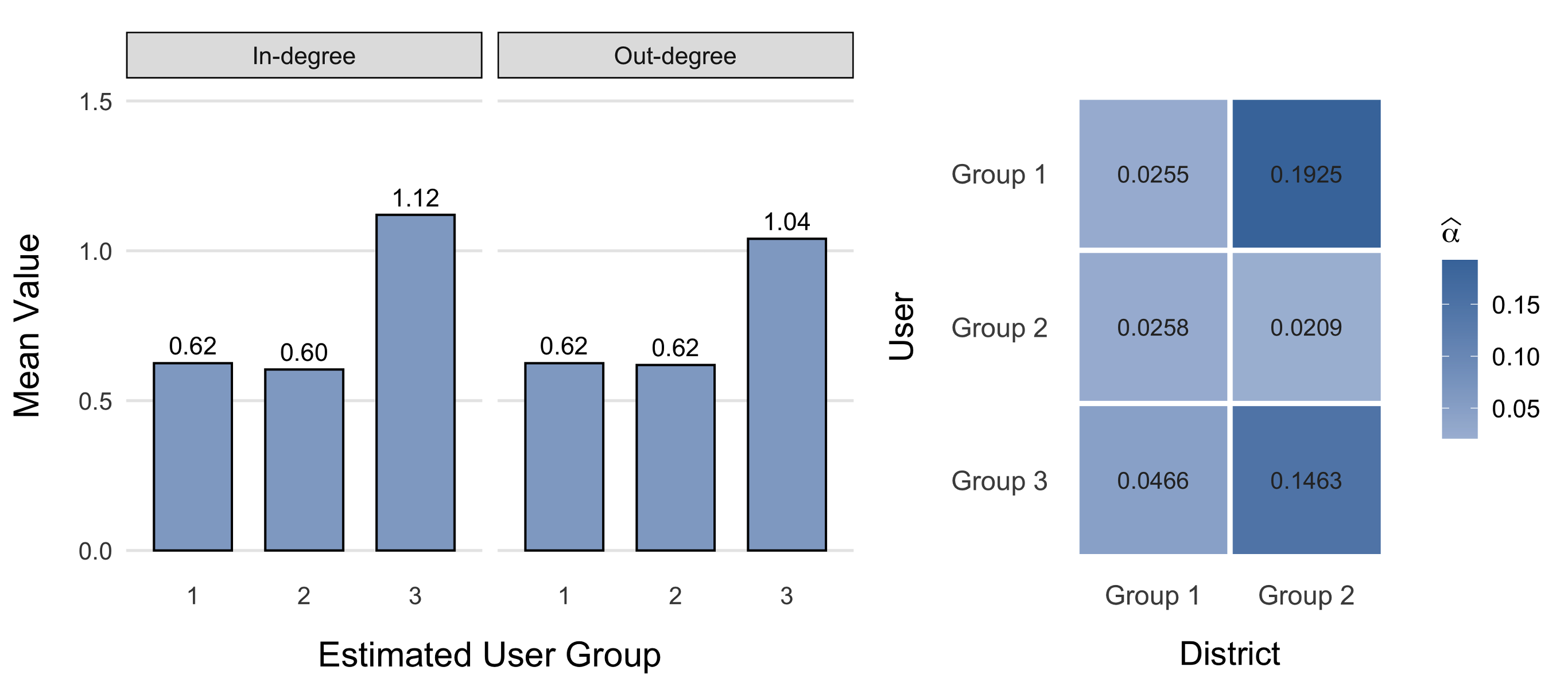}
		\caption{Left panel: average in-degree and out-degree of users in the three estimated groups in Phoenix. Right panel: heatmap of estimated momentum effect $\wh\balpha$ in Phoenix.}
		\label{fig:degree_alpha}
	\end{center}
\end{figure}

Notably, the spatial (column) network effects are consistently positive, suggesting a favorable effect from neighboring districts.
Such an observation is consistent with the findings in the literature \cite[e.g.][]{sun2017spatial}.
Furthermore, we can also observe that within the two group-wise spatial effects, $\wh\lambda_2^{(2)} = 0.270$ is larger than that of Group 1 ($\wh \lambda_1^{(2)} = 0.05$), signifying a stronger neighbor effect.
In other words, if the second group of districts' neighbors obtain more reviews in the last period, then it is likely that these districts would receive more reviews in this period.
	We further visualize the districts by estimated groups on the left
	panel of Figure \ref{fig:dis_block_map}, where the gray districts are from Group 2, and the white districts are from Group 1.
	Subsequently, we mark the shops on the map, which are shown by red points in the right panel of Figure \ref{fig:dis_block_map}. By scrutinizing the locations, we find that the shops in Group 2 are mainly located in the central business districts.
	The average number of shops per district in Group 2 is 25.16, whereas that in Group 1 is 6.57, showing a higher density of shops in the gray areas (Group 2).
	Our estimation results reveal that shops in Group 2 exhibit larger spillover effects on their neighbors.
	This finding is consistent with the existing research, which also reveals a stronger spillover effect for geometrically clustered units \citep{arzaghi2008networking, rossi2010housing, vitorino2012empirical}.
Performing a similar investigation on the user dimension, one
	can find that the social network effect in Group 1
	appears to be significantly negative ($\wh\lambda_1^{(1)} = -0.02$),
	whereas corresponding coefficients in the other two groups are
	positive. This implies that user activities in Group 1 are
	influenced oppositely by their friends' behaviors, whereas in the
	other groups, users are still positively influenced by their
	friends.
	By scrutinizing deeper into users in Group 1, we found that their average duration after joining Yelp is longer than in Groups 2 and 3. This finding aligns with existing works showing that the social effects would decrease as the users' membership duration prolongs \citep{irit2011social, aral2011creating}.
	Further, the social network effect is the largest in Group 3, i.e., $\wh \lambda_3^{(1)} = 0.024$.
	The network in-degree and out-degree of users in this group are higher than in Groups 1 and 2, which is visualized in the left panel of Figure \ref{fig:degree_alpha}.
	The phenomenon of larger network degrees associated with larger social network effects has been extensively validated in literature \citep{shawndra2006network, hinz2011seeding, susarla2012social}.
	Therefore, the business owners can implement a precision marketing strategy targeting users in Group 3, who have stronger network effects.

Besides, the estimated $\wh\balpha$ values are all positive, indicating a positive self-motivated effect overall, whose values are visualized in the right panel of Figure \ref{fig:degree_alpha}.
Specifically, the momentum effect on the review number is the largest for the Group 1 users' evaluation on the Group 2 districts, i.e., $\wh\alpha_{12} = 0.193$.
	This implies that users in Group 1 have higher self-driven persistency in the historical reviews than the other groups, while they exhibit smaller social network effects.

\begin{sidewaystable}[]
	\centering
	\caption{\small Estimation results for Charlotte, Las Vegas, and Phoenix. The $p$-values are shown in parentheses.
		Take Phoenix for example, $\blambda^{(1)}$ is clustered into three groups, with $\lambda_{g^{(1)}}^{(1)}$ representing the social network effect on the users in the $g^{(1)}$th group from their friends' behavior.
		$\blambda^{(2)}$ is clustered into two groups, with $\lambda_{g^{(2)}}^{(2)}$ meaning the spatial network effect on the districts in the $h$th group from their adjacent neighbors.
		$\bzeta^{(1)}$ and $\bzeta^{(2)}$ are the covariates effects, and we take $\zeta^{(1)}_{\text{use}}$ to illustrate. It has also been clustered into three groups, with the coefficient in the $g^{(1)}$th group meaning the effect on the users in the $g^{(1)}$th group from the average ``useful'' tag.
		$\balpha$ is clustered into three groups by row and into two groups by column, where $\alpha_{g^{(1)} g^{(2)}}$ means the self-momentum effect of the users in the $g^{(1)}$th group by row and $g^{(2)}$th group by column.
	}
	\label{tbl:real_res_1}
	\scalebox{0.68}{
		\begin{tabular}{c|cccccc|ccccc|ccccc}
			\hline
			Parameters                                    & \multicolumn{6}{c|}{Charlotte ( $N_1 = 240, ~N_2 = 60$)}                                                                                                                                                                                                                                                                                                                                                                                                                      & \multicolumn{5}{c|}{LasVegas ($N_1 = 826, ~N_2 = 64$)}                                                                                                                                                                                                                                                                                                                                                  & \multicolumn{5}{c}{Phoenix($N_1 = 323, ~ N_2 = 63$)}                                                                                                                                                                                                                                                                                                                                                    \\ \hline
			\multirow{2}{*}{}                             & \multicolumn{3}{c|}{$\lambda^{(1)}_{g^{(1)}}$}                                                                                                                                                                                                    & \multicolumn{3}{c|}{$\lambda^{(2)}_{g^{(2)}}$}                                                                                                                                                                               & \multicolumn{3}{c|}{$\lambda^{(1)}_{g^{(1)}}$}                                                                                                                                                                                                     & \multicolumn{2}{c|}{$\lambda^{(2)}_{g^{(2)}}$}                                                                                                         & \multicolumn{3}{c|}{$\lambda^{(1)}_{g^{(1)}}$}                                                                                                                                                                                                     & \multicolumn{2}{c}{$\lambda^{(2)}_{g^{(2)}}$}                                                                                                          \\
			& \begin{tabular}[c]{@{}c@{}}0.004\\ (0.017)\end{tabular}            & \begin{tabular}[c]{@{}c@{}}0.110\\ (\textless{}0.001)\end{tabular}  & \multicolumn{1}{c|}{\begin{tabular}[c]{@{}c@{}}0.047\\ (\textless{}0.001)\end{tabular}}  & \begin{tabular}[c]{@{}c@{}}0.035\\ (\textless{}0.001)\end{tabular}  & \begin{tabular}[c]{@{}c@{}}0.475\\ (\textless{}0.001)\end{tabular} & \begin{tabular}[c]{@{}c@{}}0.286\\ (\textless{}0.001)\end{tabular} & \begin{tabular}[c]{@{}c@{}}0.068\\ (\textless{}0.001)\end{tabular}  & \begin{tabular}[c]{@{}c@{}}0.008\\ (\textless{}0.001)\end{tabular}  & \multicolumn{1}{c|}{\begin{tabular}[c]{@{}c@{}}0.054\\ (\textless{}0.001)\end{tabular}}  & \begin{tabular}[c]{@{}c@{}}0.150\\ (\textless{}0.001)\end{tabular} & \begin{tabular}[c]{@{}c@{}}0.326\\ (\textless{}0.001)\end{tabular} & \begin{tabular}[c]{@{}c@{}}-0.020\\ (0.101)\end{tabular}            & \begin{tabular}[c]{@{}c@{}}0.007\\ (\textless{}0.001)\end{tabular}  & \multicolumn{1}{c|}{\begin{tabular}[c]{@{}c@{}}0.024\\ (\textless{}0.001)\end{tabular}}  & \begin{tabular}[c]{@{}c@{}}0.050\\ (\textless{}0.001)\end{tabular} & \begin{tabular}[c]{@{}c@{}}0.270\\ (\textless{}0.001)\end{tabular} \\ \hline
			& \multicolumn{3}{c|}{$\bzeta_{g^{(1)}}^{(1)}$}                                                                                                                                                                                                     & \multicolumn{3}{c|}{$\bzeta_{g^{(2)}}^{(2)}$}                                                                                                                                                                              & \multicolumn{3}{c|}{$\bzeta_{g^{(1)}}^{(1)}$}                                                                                                                                                                                                      & \multicolumn{2}{c|}{$\bzeta_{g^{(2)}}^{(2)}$}                                                                                                        & \multicolumn{3}{c|}{$\bzeta_{g^{(1)}}^{(1)}$}                                                                                                                                                                                                      & \multicolumn{2}{c}{$\bzeta_{g^{(2)}}^{(2)}$}                                                                                                         \\
			Intercept                                     & \begin{tabular}[c]{@{}c@{}}0.001\\ (0.328)\end{tabular}            & \begin{tabular}[c]{@{}c@{}}-0.004\\ (\textless{}0.001)\end{tabular}  & \multicolumn{1}{c|}{\begin{tabular}[c]{@{}c@{}}0.003\\ (0.007)\end{tabular}}             & \begin{tabular}[c]{@{}c@{}}-0.004\\ (\textless{}0.001)\end{tabular} & \begin{tabular}[c]{@{}c@{}}0.041\\ (\textless{}0.001)\end{tabular} & \begin{tabular}[c]{@{}c@{}}$-10^{-4}$ \\ (0.796)\end{tabular}           & \begin{tabular}[c]{@{}c@{}}0.011\\ (\textless{}0.001)\end{tabular}  & \begin{tabular}[c]{@{}c@{}}-0.005\\ (\textless{}0.001)\end{tabular} & \multicolumn{1}{c|}{\begin{tabular}[c]{@{}c@{}}-0.006\\ (0.586)\end{tabular}}             & \begin{tabular}[c]{@{}c@{}}0.010\\ (\textless{}0.001)\end{tabular} & \begin{tabular}[c]{@{}c@{}}0.003\\ (0.011)\end{tabular}            & \begin{tabular}[c]{@{}c@{}}0.058\\ (\textless{}0.001)\end{tabular}  & \begin{tabular}[c]{@{}c@{}}-0.052\\ (\textless{}0.001)\end{tabular} & \multicolumn{1}{c|}{\begin{tabular}[c]{@{}c@{}}-0.006\\ (\textless{}0.001)\end{tabular}} & \begin{tabular}[c]{@{}c@{}}0.050\\ (\textless{}0.001)\end{tabular} & \begin{tabular}[c]{@{}c@{}}0.021\\ (\textless{}0.001)\end{tabular} \\
			$\bzeta^{(1)}_{\text{dur}}$ / $\bzeta^{(2)}_{\text{star}}$ & \begin{tabular}[c]{@{}c@{}}-0.001\\ (0.162)\end{tabular}           & \begin{tabular}[c]{@{}c@{}}0.050\\ (\textless{}0.001)\end{tabular}  & \multicolumn{1}{c|}{\begin{tabular}[c]{@{}c@{}}0.015\\ (\textless{}0.001)\end{tabular}}  & \begin{tabular}[c]{@{}c@{}}0.002\\ (\textless{}0.001)\end{tabular}  & \begin{tabular}[c]{@{}c@{}}-0.012\\ (0.078)\end{tabular}           & \begin{tabular}[c]{@{}c@{}}0.004\\ (0.005)\end{tabular}            & \begin{tabular}[c]{@{}c@{}}0.059\\ (\textless{}0.001)\end{tabular}  & \begin{tabular}[c]{@{}c@{}}0.001\\ (0.194)\end{tabular}             & \multicolumn{1}{c|}{\begin{tabular}[c]{@{}c@{}}0.019\\ (\textless{}0.001)\end{tabular}}  & \begin{tabular}[c]{@{}c@{}}$10^{-4}$\\ (0.170)\end{tabular}            & \begin{tabular}[c]{@{}c@{}}0.001\\ (0.744)\end{tabular}            & \begin{tabular}[c]{@{}c@{}}-0.093\\ (\textless{}0.001)\end{tabular} & \begin{tabular}[c]{@{}c@{}}0.001\\ (0.150)\end{tabular}             & \multicolumn{1}{c|}{\begin{tabular}[c]{@{}c@{}}0.002\\ (0.148)\end{tabular}}             & \begin{tabular}[c]{@{}c@{}}0.003\\ (\textless{}0.001)\end{tabular} & \begin{tabular}[c]{@{}c@{}}0.004\\ (0.010)\end{tabular}            \\
			$\bzeta^{(1)}_{\text{vip}}$ / $\bzeta^{(2)}_{\text{num}}$  & \begin{tabular}[c]{@{}c@{}}0.001\\ (0.012)\end{tabular}            & \begin{tabular}[c]{@{}c@{}}-0.037\\ (\textless{}0.001)\end{tabular} & \multicolumn{1}{c|}{\begin{tabular}[c]{@{}c@{}}-0.002\\ (0.044)\end{tabular}}            & \begin{tabular}[c]{@{}c@{}}0.007\\ (\textless{}0.001)\end{tabular}  & \begin{tabular}[c]{@{}c@{}}0.008\\ (0.001)\end{tabular}            & \begin{tabular}[c]{@{}c@{}}0.003\\ (0.233)\end{tabular}            & \begin{tabular}[c]{@{}c@{}}0.028\\ (\textless{}0.001)\end{tabular}  & \begin{tabular}[c]{@{}c@{}}0.000\\ (0.093)\end{tabular}             & \multicolumn{1}{c|}{\begin{tabular}[c]{@{}c@{}}-0.010\\ (\textless{}0.001)\end{tabular}} & \begin{tabular}[c]{@{}c@{}}0.016\\ (\textless{}0.001)\end{tabular} & \begin{tabular}[c]{@{}c@{}}0.038\\ (\textless{}0.001)\end{tabular} & \begin{tabular}[c]{@{}c@{}}0.008\\ (\textless{}0.001)\end{tabular}  & \begin{tabular}[c]{@{}c@{}}$10^{-4}$\\ (0.054)\end{tabular}             & \multicolumn{1}{c|}{\begin{tabular}[c]{@{}c@{}}$10^{-4}$\\ (0.369)\end{tabular}}             & \begin{tabular}[c]{@{}c@{}}0.002\\ (\textless{}0.001)\end{tabular} & \begin{tabular}[c]{@{}c@{}}$10^{-4}$\\ (0.938)\end{tabular}            \\
			$\bzeta^{(1)}_{\text{use}}$                          & \begin{tabular}[c]{@{}c@{}}0.017\\ (\textless{}0.001)\end{tabular} & \begin{tabular}[c]{@{}c@{}}0.104\\ (\textless{}0.001)\end{tabular}  & \multicolumn{1}{c|}{\begin{tabular}[c]{@{}c@{}}0.052\\ (\textless{}0.001)\end{tabular}}  &                                                                     &                                                                    &                                                                    & \begin{tabular}[c]{@{}c@{}}1.115\\ (\textless{}0.001)\end{tabular}  & \begin{tabular}[c]{@{}c@{}}0.011\\ (\textless{}0.001)\end{tabular}  & \multicolumn{1}{c|}{\begin{tabular}[c]{@{}c@{}}0.143\\ (\textless{}0.001)\end{tabular}}  &                                                                    &                                                                    & \begin{tabular}[c]{@{}c@{}}0.385\\ (\textless{}0.001)\end{tabular}  & \begin{tabular}[c]{@{}c@{}}0.011\\ (\textless{}0.001)\end{tabular}  & \multicolumn{1}{c|}{\begin{tabular}[c]{@{}c@{}}0.071\\ (\textless{}0.001)\end{tabular}}  &                                                                    &                                                                    \\
			$\bzeta^{(1)}_{\text{fun}}$                          & \begin{tabular}[c]{@{}c@{}}-0.001\\ (0.806)\end{tabular}           & \begin{tabular}[c]{@{}c@{}}-0.020\\ (0.002)\end{tabular}            & \multicolumn{1}{c|}{\begin{tabular}[c]{@{}c@{}}$10^{-4}$\\ (0.995)\end{tabular}}             &                                                                     &                                                                    &                                                                    & \begin{tabular}[c]{@{}c@{}}-0.589\\ (\textless{}0.001)\end{tabular} & \begin{tabular}[c]{@{}c@{}}-0.001\\ (0.795)\end{tabular}            & \multicolumn{1}{c|}{\begin{tabular}[c]{@{}c@{}}-0.289\\ (\textless{}0.001)\end{tabular}} &                                                                    &                                                                    & \begin{tabular}[c]{@{}c@{}}0.064\\ (0.080)\end{tabular}             & \begin{tabular}[c]{@{}c@{}}-0.008\\ (0.002)\end{tabular}            & \multicolumn{1}{c|}{\begin{tabular}[c]{@{}c@{}}-0.033\\ (\textless{}0.001)\end{tabular}} &                                                                    &                                                                    \\
			$\bzeta^{(1)}_{\text{cool}}$                         & \begin{tabular}[c]{@{}c@{}}-0.001\\ (0.908)\end{tabular}           & \begin{tabular}[c]{@{}c@{}}0.003\\ (0.621)\end{tabular}             & \multicolumn{1}{c|}{\begin{tabular}[c]{@{}c@{}}-0.017\\ (\textless{}0.001)\end{tabular}} &                                                                     &                                                                    &                                                                    & \begin{tabular}[c]{@{}c@{}}-0.644\\ (\textless{}0.001)\end{tabular} & \begin{tabular}[c]{@{}c@{}}-0.008\\ (0.043)\end{tabular}            & \multicolumn{1}{c|}{\begin{tabular}[c]{@{}c@{}}0.217\\ (\textless{}0.001)\end{tabular}}  &                                                                    &                                                                    & \begin{tabular}[c]{@{}c@{}}0.208\\ (\textless{}0.001)\end{tabular}  & \begin{tabular}[c]{@{}c@{}}0.010\\ (0.005)\end{tabular}             & \multicolumn{1}{c|}{\begin{tabular}[c]{@{}c@{}}-0.011\\ (0.007)\end{tabular}}            &                                                                    &                                                                    \\ \hline
			\multirow{4}{*}{}                             & \multicolumn{6}{c|}{$ \balpha^\top  \in \mR^{G_1 \times G_2}$}                                                                                                                                                                                                                                                                                                                                                                                                 & \multicolumn{5}{c|}{$ \balpha^\top \in \mR^{G_1 \times G_2}$}                                                                                                                                                                                                                                                                                                                            & \multicolumn{5}{c}{$ \balpha^\top \in \mR^{G_1 \times G_2}$}                                                                                                                                                                                                                                                                                                                             \\
			& \multicolumn{2}{c}{\begin{tabular}[c]{@{}c@{}}0.019\\ (0.002)\end{tabular}}                                                              & \multicolumn{2}{c}{\begin{tabular}[c]{@{}c@{}}0.051\\ (\textless{}0.001)\end{tabular}}                                                                         & \multicolumn{2}{c|}{\begin{tabular}[c]{@{}c@{}}0.058\\ (\textless{}0.001)\end{tabular}}                                                 & \multicolumn{2}{c}{\begin{tabular}[c]{@{}c@{}}0.288\\ (\textless{}0.001)\end{tabular}}                                                    & \begin{tabular}[c]{@{}c@{}}0.021\\ (\textless{}0.001)\end{tabular}                       & \multicolumn{2}{c|}{\begin{tabular}[c]{@{}c@{}}0.174\\ (\textless{}0.001)\end{tabular}}                                                 & \multicolumn{2}{c}{\begin{tabular}[c]{@{}c@{}}0.026\\ (\textless{}0.001)\end{tabular}}                                                    & \begin{tabular}[c]{@{}c@{}}0.026\\ (\textless{}0.001)\end{tabular}                       & \multicolumn{2}{c}{\begin{tabular}[c]{@{}c@{}}0.047\\ (\textless{}0.001)\end{tabular}}                                                  \\
			& \multicolumn{2}{c}{\begin{tabular}[c]{@{}c@{}}0.005\\ (0.218)\end{tabular}}                                                              & \multicolumn{2}{c}{\begin{tabular}[c]{@{}c@{}}0.374\\ (\textless{}0.001)\end{tabular}}                                                                         & \multicolumn{2}{c|}{\begin{tabular}[c]{@{}c@{}}0.224\\ (\textless{}0.001)\end{tabular}}                                                 & \multicolumn{2}{c}{\begin{tabular}[c]{@{}c@{}}0.478\\ (\textless{}0.001)\end{tabular}}                                                    & \begin{tabular}[c]{@{}c@{}}0.036\\ (\textless{}0.001)\end{tabular}                       & \multicolumn{2}{c|}{\begin{tabular}[c]{@{}c@{}}0.330\\ (\textless{}0.001)\end{tabular}}                                                 & \multicolumn{2}{c}{\begin{tabular}[c]{@{}c@{}}0.193\\ (\textless{}0.001)\end{tabular}}                                                    & \begin{tabular}[c]{@{}c@{}}0.021\\ (\textless{}0.001)\end{tabular}                       & \multicolumn{2}{c}{\begin{tabular}[c]{@{}c@{}}0.146\\ (\textless{}0.001)\end{tabular}}                                                  \\
			& \multicolumn{2}{c}{\begin{tabular}[c]{@{}c@{}}0.017\\ (\textless{}0.001)\end{tabular}}                                                   & \multicolumn{2}{c}{\begin{tabular}[c]{@{}c@{}}0.200\\ (\textless{}0.001)\end{tabular}}                                                                         & \multicolumn{2}{c|}{\begin{tabular}[c]{@{}c@{}}0.115\\ (\textless{}0.001)\end{tabular}}                                                 & \multicolumn{2}{c}{}                                                                                                                      &                                                                                          & \multicolumn{2}{c|}{}                                                                                                                   & \multicolumn{2}{c}{}                                                                                                                      &                                                                                          & \multicolumn{2}{c}{}                                                                                                                    \\ \hline
		\end{tabular}
	}
\end{sidewaystable}

\section{Concluding Remarks}\label{sec:conclusion}

In this work, we introduce a novel Group Tensor Network Autoregression (GTNAR) model designed for time series data indexed by multi-relational networks.
By leveraging network structures on each dimension, GTNAR establishes a unique framework for
analyzing tensor-valued time series data.
This model presents a valuable tool for analyzing data collected in complex network environments, shedding light on various network effects.
From a modeling perspective, several intriguing future topics emerge.
It would be interesting to consider the high-dimensional covariates in this framework.
Furthermore, it is worth exploring models that consider multiple network effects in a multiplicative form, as investigated by \cite{chen2021autoregressive}.
Besides, introducing a hidden factor structure into the GTNAR model can potentially offer more insights into high-dimensional data, capturing more underlying information within the tensor-valued time series, and thus represents an interesting future research topic.
Lastly, it is also important to further investigate how to model categorical responses in our modeling framework.
From the theoretical perspective, although the convergence rate established in Theorem \ref{thm:pseudo_dist} is typical in existing research, how to obtain a minimax rate needs further investigation.

\acks{Yimeng Ren and Xuening Zhu's research are supported by the National Natural Science Foundation of China (nos. 72573038, 12331009), MOE Laboratory for National Development and Intelligent Governance, Fudan University. Yanyuan Ma's work is partially supported by grants from NIH.}


\newpage

\vskip 0.2in
\bibliography{GTNAR}

\newpage

\def\bet{\begin{theorem}}
	\def\eet{\end{theorem}}
\def\bep{\begin{proposition}}
	\def\eep{\end{proposition}}
\def\beq{\begin{equation}}
	\def\eeq{\end{equation}}
\def\bsq{\begin{equation*}}
	\def\esq{\end{equation*}}
\def\bse{\begin{eqnarray*}}
	\def\ese{\end{eqnarray*}}
\def\bel{\begin{lemma}}
	\def\eel{\end{lemma}}
\def\one{{\bf 1}}
\def\bA{{\mathcal A}}
\def\ve{\varepsilon}
\def\pre{\preccurlyeq}
\def\mle{\mbox{mle}}
\def\tr{\mbox{tr}}
\def\obs{\textup{obs}}
\def\lse{\mbox{lse}}
\def\red{\color{red}}
\def\blue{\color{blue}}
\def\cmle{\mbox{cmle}}
\def\n{\nonumber}
\def\ci{\perp\!\!\!\perp}
\def\e{\mathbf{e}}
\def\gap{\textup{gap}}
\def\cle{\preccurlyeq}
\def\V{\mbox{vec}}
\def\wh{\widehat}
\def\wb{\underline}
\def\lmax{\lambda_{\max}}
\def\cc{{\it correct }}
\def\svd{\mS_{\mbox{\sc svd}}}

\def\wY{\widetilde Y}
\def\ol{\overline }
\def\SE{\widehat{\mbox{SE}}}
\def\E{\mathbb{E}}
\def\bB{\mathbf{B}}
\def\mA{\mathcal A}
\def\mC{\mathcal C}
\def\mN{\mathcal{N}}
\def\mJ{\mathcal{J}}
\def\bK{\mathbf{K}}
\def\mP{\mathbb P}
\def\mD{\mathcal D}
\def\mQ{\mathbb Q}
\def\mG{\mathcal G}
\def\calR{{\cal R}}
\def\cP{\mathcal{P}}
\def\mI{\mathcal I}
\def\mL{\mathcal L}
\def\mS{\mathbb S}
\def\mH{\mathcal H}
\def\mM{\mathcal M}
\def\mm{\mathfrak m}
\def\mF{\mathcal F}
\def\F{\mathbf F}
\def\cR{\mathcal{R}}
\def\bR{\mathbf{R}}
\def\bY{\mathcal Y}
\def\mS{\mathcal S}
\def\mE{\mathcal E}
\def\mO{\mathcal O}
\def\mT{\mathcal T}
\def\bH{\mathbb H}
\def\bc{\mathbf c}
\def\mU{\mathbb{U}}
\def\mV{\mathbb{V}}
\def\mW{\mathbb{W}}
\def\cW{\mathcal{W}}
\def\mX{\mathbb{X}}
\def\cX{\mathcal{X}}
\def\wX{{\widetilde{\mathbb{X}}}}
\def\sX{{\mathbb{X}_b^*}}
\def\mY{\mathbb{Y}}
\def\mZ{\mathbb{Z}}
\def\bx{\mathbf{x}}
\def\bZ{\mathbf{Z}}
\def\bU{\mathbf{U}}
\def\bv{\mathbf{v}}
\def\var{\mbox{var}}
\def\supp{\mbox{supp}}
\def\es{E_*}
\def\vs{\mbox{var}_*}
\def\cov{\mbox{cov}}
\def\argmin{\mbox{argmin}}
\def\argmax{\mbox{argmax}}
\def\rank{\mbox{rank}}
\def\diag{\mbox{diag}}
\def\aic{\mbox{AIC}}
\def\bic{\mbox{BIC}}
\def\dbic{\mbox{DBIC}}
\def\rss{\mbox{RSS}}
\def\rmse{\mbox{RMSE}}
\def\vec{\mbox{vec}}
\def\err{\mbox{ERR}}
\def\sis{\mM_{\mbox{\sc sis}}}
\def\pis{\mM_{\mbox{\sc pis}}}
\def\pss{\mM_{\mbox{\sc pss}}}
\def\bgamma{\boldsymbol{\gamma}}
\def\bbeta{\boldsymbol{\beta}}
\def\bpsi{\boldsymbol{\psi}}
\def\bLambda{\boldsymbol{\Lambda}}
\def\bSigma{\boldsymbol{\Sigma}}
\def\bomega{\boldsymbol{\omega}}
\def\bve{\boldsymbol{\varepsilon}}
\def\bX{\mathbf{X}}
\def\bZ{\mathbf{Z}}
\def\bh{\mbox{\boldmath $h$}}
\def\by{\mathbf{y}}
\def\y{\mathbf{y}}
\def\bY{\mathbf{Y}}
\def\bA{\mathbf{A}}
\def\bG{\mathbf{G}}
\def\bD{\mathbf{D}}
\def\bW{\mathbf{W}}
\def\bS{\mathbf{S}}
\def\bH{\mathbf{H}}
\def\b{\mathbf{b}}
\def\bF{\mathbf{F}}
\def\kR{\mathfrak{R}}
\def\bg{\mbox{\boldmath $g$}}
\def\bI{\mathbf{I}}
\def\bu{\mbox{\boldmath $u$}}
\def\lag{\rm lag}
\def\cH{{\mathbb H}}
\def\cB{{\mathbb B}}
\def\cZ{{\mathcal Z}}
\def\cN{{\mathcal N}}

\def\zero{\mathbf{0}}
\def\defeq{\stackrel{\mathrm{def}}{=}}  
\def\sign{\mbox{sign}}
\def\qic{\textup{QIC}}

\def\th{^{th}}
\def\supp{\hbox{supp}}
\def\var{\hbox{var}}
\def\cov{\hbox{cov}}
\def\corr{\hbox{corr}}
\def\trace{\hbox{trace}}
\def\wh{\widehat}
\def\wc{\widecheck}
\def\eff{_{\rm eff}}
\def\sub{{\rm sub}}
\def\cat{{\rm cat}}
\def\th{^{th}}
\def\my{\mathcal Y}
\def\mL{\mathcal L}
\def\mR{\mathbb{R}}
\def\n{\nonumber}
\def\bias{\mbox{bias}}
\def\vecl{\mbox{vecl}}
\def\AIC{\mbox{AIC}}
\def\BIC{\mbox{BIC}}
\def\MSE{\mbox{MSE}}
\def\rank{\mbox{rank}}
\def\cov{\mbox{cov}}
\def\corr{\mbox{corr}}
\def\vec{\mbox{vec}}
\def\argmin{\mbox{argmin}}
\def\argmax{\mbox{argmax}}
\def\diag{\mbox{diag}}
\def\tr{\mbox{tr}}
\def\sir{{\mbox{\tiny SIR}}}
\def\save{{\mbox{\tiny SAVE}}}
\def\phd{{\mbox{\tiny PHD}}}
\def\cume{{\mbox{\tiny CUME}}}
\def\cuve{{\mbox{\tiny CUVE}}}
\def\cuhd{{\mbox{\tiny CUHD}}}
\def\cudr{{\mbox{\tiny CUDR}}}
\def\dr{{\mbox{\tiny DR}}}
\def\sumi{\sum_{i=1}^n}
\def\sumj{\sum_{j=1}^n}
\def\suml{\sum_{l=1}^n}
\def\sumk{\sum_{k=1}^n}
\def\trans{^{\top}}
\def\hDash{\bot\!\!\!\bot}
\def\mS{\mbox{ $\mathcal{S}$}}
\def\bs{\boldsymbol}
\def\bTheta{\boldsymbol\Theta}
\def\ba{\boldsymbol\alpha}
\def\bmu{\boldsymbol\mu}
\def\bnu{\boldsymbol\nu}
\def\beps{\boldsymbol\epsilon}
\def\ha{\widehat{\ba}}
\def\bb{\boldsymbol\beta}
\def\bGamma{\boldsymbol\Gamma}
\def\bOmega{\boldsymbol\Omega}
\def\bdelta{\boldsymbol\delta}
\def\bDelta{\boldsymbol\Delta}
\def\bxi{\boldsymbol\xi}
\def\bphi{\boldsymbol\phi}
\def\btau{\boldsymbol\tau}
\def\bpsi{\boldsymbol\psi}
\def\bzeta{\boldsymbol\zeta}
\def\bmu{\boldsymbol\mu}
\def\bg{\boldsymbol\gamma}
\def\hb{\widehat{\bb}}
\def\he{\widehat{\varepsilon}}
\def\defby{\stackrel{\mbox{\textrm{\tiny def}}}{=}}
\def\0{{\bf 0}}
\def\1{{\bf 1}}
\def\A{{\bf A}}
\def\U{{\bf U}}
\def\V{{\bf V}}
\def\e{{\bf e}}
\def\R{{\bf R}}
\def\G{{\bf G}}
\def\bO{{\bf O}}
\def\a{{\bf a}}
\def\B{{\bf B}}
\def\c{{\bf c}}
\def\D{{\bf D}}
\def\V{{\bf V}}
\def\K{{\bf K}}
\def\g{{\bf g}}
\def\r{{\bf r}}
\def\f{{\bf f}}
\def\h{{\bf h}}
\def\b{{\bf b}}
\def\I{{\bf I}}
\def\mU{\mathcal{U}}
\def\BB{\mbox{ $\mathcal{B}$}}
\def\N{\mbox{ $\mathcal{N}$}}
\def\M{{\bf M}}
\def\bM{{\bf M}}
\def\K{{\bf K}}
\def\t{{\bf t}}
\def\T{{\bf T}}
\def\bd{{\bf d}}
\def\bP{{\bf P}}
\def\bP{{\bf P}}
\def\bJ{{\bf J}}
\def\bV{{\bf V}}
\def\hQ{{\widehat \bQ}}
\def\U{{\bf U}}
\def\S{{\bf S}}
\def\s{{\bf s}}
\def\u{{\bf u}}
\def\m{{\bf m}}
\def\v{{\bf v}}
\def\W{{\bf W}}
\def\T{{\bf T}}
\def\bO{{\bf O}}
\def\w{{\bf w}}
\def\X{{\bf X}}
\def\L{{\bf L}}
\def\x{{\bf x}}
\def\I{{\bf I}}
\def\i{{\bf i}}
\def\tx{{\widetilde \x}}
\def\Y{{\bf Y}}
\def\H{{\bf H}}
\def\bN{{\bf N}}

\def\C{{\bf C}}
\def\tY{{\widetilde Y}}
\def\y{{\bf y}}
\def\Z{{\bf Z}}
\def\z{{\bf z}}
\def\bC{{\bf C}}
\def\Ybar{{\overline{Y}}}
\def\Xbar{{\overline{\X}}}
\def\xbar{{\overline{\x}}}
\def\wbar{{\overline{\W}}}
\def\bSig{{\bf \Sigma}}
\def\bLam{{\bf \Lambda}}
\def\diag{\hbox{diag}}
\def\dfrac#1#2{{\displaystyle{#1\over#2}}}
\def\VS{{\vskip 3mm\noindent}}
\def\refhg{\hangindent=20pt\hangafter=1}
\def\refmark{\par\vskip 2mm\noindent\refhg}
\def\naive{\hbox{naive}}
\def\itemitem{\par\indent \hangindent2\parindent \textindent}
\def\dist{\hbox{dist}}
\def\trace{\hbox{trace}}
\def\refhg{\hangindent=20pt\hangafter=1}
\def\refmark{\par\vskip 2mm\noindent\refhg}
\def\Normal{\hbox{Normal}}
\def\povr{\buildrel p\over\longrightarrow}
\def\ccdot{{\bullet}}
\def\pr{\hbox{pr}}
\def\wh{\widehat}
\def\th{^{th}}
\def\diag{\hbox{diag}}
\def\log{\hbox{log}}
\def\bias{\hbox{bias}}
\def\Siuu{\boldSigma_{i,uu}}
\def\squarebox#1{\hbox to #1{\hfill\vbox to #1{\vfill}}}
\def\btheta{{\boldsymbol \theta}}
\def\bfeta{{\boldsymbol \eta}}
\def\balpha{{\boldsymbol \alpha}}
\def\bOmega{{\boldsymbol \Omega}}
\def\bXi{{\boldsymbol \Xi}}
\def\blambda{{\boldsymbol \lambda}}
\def\bPsi{{\boldsymbol \Psi}}
\def\bzeta{{\boldsymbol \zeta}}
\def\bpi{{\boldsymbol \pi}}
\def\bx{{\bf x}}
\def\bz{{\bf z}}
\def\vec{\mathrm{vec}}
\def\mA{\mathcal{A}}
\def\cA{\mathbb{A}}
\def\bE{\mathbf{E}}
\def\mB{\mathcal{B}}
\def\mC{\mathcal{C}}
\def\mF{\mathcal{F}}
\def\cS{\mathcal{S}}
\def\mH{\mathcal{H}}
\def\mM{\mathcal{M}}
\def\my{\mathcal Y}
\def\cY{\mathcal Y}
\def\bw{\mathbf w}
\def\dfrac#1#2{{\displaystyle{#1\over#2}}}
\def\VS{{\vskip 3mm\noindent}}
\def\refhg{\hangindent=20pt\hangafter=1}
\def\refmark{\par\vskip 2mm\noindent\refhg}
\def\itemitem{\par\indent \hangindent2\parindent \textindent}
\def\var{\hbox{var}}
\def\cov{\hbox{cov}}
\def\corr{\hbox{corr}}
\def\trace{\hbox{trace}}
\def\refhg{\hangindent=20pt\hangafter=1}
\def\Normal{\hbox{Normal}}
\def\povr{\buildrel p\over\longrightarrow}
\def\dovr{\buildrel d\over\longrightarrow}
\def\ccdot{{\bullet}}
\def\pr{\hbox{pr}}
\def\br{{\bf r}}
\def\q{{\bf q}}
\def\wh{\widehat}
\def\wt{\widetilde}
\def\diag{\hbox{diag}}
\def\bias{\hbox{bias}}
\def\Siuu{\boldSigma_{i,uu}}
\def\whT{\widehat{\Theta}}
\def\diag{\hbox{diag}}
\def\th{^{th}}
\def\o{\textup{or}}
\def\logit{{\mbox{logit}}}
\def\bfa{\textbf{a}}
\def\log{\textup{log}}
\definecolor{ForestGreen}{RGB}{34,139,34}

\def\cM{\mathcal{M}}
\def\cK{\mathcal{K}}
\def\bkappa{\boldsymbol{\kappa}}
\def\bs{\mathbf{s}}

\def\boxit#1{\vbox{\hrule\hbox{\vrule\kern6pt\vbox{\kern6pt#1\kern6pt}\kern6pt\vrule}\hrule}}
\def\yimeng#1{\vskip 2mm\boxit{\vskip 2mm{\color{ForestGreen}\bf#1} {\color{blue}\bf -- Yimeng\vskip 2mm}}\vskip 2mm}
\def\xuening#1{\vskip 2mm\boxit{\vskip 2mm{\color{blue}\bf#1} {\color{blue}\bf -- Xuening\vskip 2mm}}\vskip 2mm}

\captionsetup[subfigure]{labelformat=simple}
\renewcommand\thesubfigure{(\alph{subfigure})}


%
%
%

\renewcommand{\thefigure}{A.\arabic{figure}}
\renewcommand{\thetable}{A.\arabic{table}}
\renewcommand{\thealgorithm}{A.\arabic{algorithm}}
\renewcommand{\theequation}{A.\arabic{equation}}
\renewcommand{\theassumption}{A.\arabic{assumption}}
\setcounter{equation}{0}
%



In this part, the required notations are provided in Appendix \ref{sec:notation}.
More estimation details for $q=2$ and general $q$ are provided in Appendix \ref{sec:notation_q2}--\ref{subsec:alg}.
An numerical convergence analysis for the algorithm is shown in Appendix \ref{sec:alg_converge}.
The initialization procedure and  of our main algorithm is provided in Appendix \ref{sec:init}.
We provide additional discussion about the interactive model in Appendix \ref{sec:int_est}, followed by discussion on future extensions in Appendix \ref{sec:extension}.
We provide the technical details and proofs for the main text in
Section \ref{sec:main_thm},
and several useful lemmas are given in Section
\ref{sec:tech_lemma}--\ref{sec:general_lemma}.
Besides, we provide a number of additional simulation studies in Appendix \ref{sec:add_simu}.
Lastly, some additional real data analysis are given in Appendix
\ref{sec:add_yelp} and Appendix \ref{sec:add_realdata}.

\appendix

\section{Notations}\label{sec:notation}

Define $Q^*(\bTheta) = E\{Q(\bTheta)\}$
and $Q_{i_1 \cdots i_q}^*(\bTheta_{i_1 \cdots i_q}) = E\{Q_{i_1 \cdots i_q}(\bTheta_{i_1 \cdots i_q})\}$,
where $Q(\bTheta)$ and $Q_{i_1 \cdots i_q}(\bTheta_{i_1 \cdots i_q})$ are as defined in \eqref{def:Qij} in the main text.
Denote $\mG_{-l} = \{ \mG_m: m \neq l \}$ and $\i_{-l} = (i_m: m \neq l)^\top \in \mR^{q-1}$.
Further, denote $\btheta^{-(l)} = \{(\btheta_{g^{(m)}}^{(m)}: g^{(m)} \in [G_m]) \in \mR^{G_m (p_m+1)}: m \neq
 l, 1 \le m \le q\}$, $\bxi_{g_{i_l}^{(l)}}^{(l)}=(\btheta_{ g_{i_l}^{(l)}}^{(l)\top}, \vec(\balpha_{\cdot  g_{i_l}^{(l)} \cdot})^\top)^\top$, $\bxi_{g^{-(l)}}^{-(l)} = (\btheta_{g^{(m)}}^{(m)\top}, \vec(\balpha_{\cdot  g^{(m)} \cdot})^\top:  m \neq l)^\top$, and $g_{\i_{-l}}^{-(l)} = (g_{i_m}^{(m)}: m \neq l)$.
Using these notations, we define
\begin{align}
	{Q_{i_l}(\bxi_{g_{i_l}^{(l)}}^{(l)}; \bxi_{g^{-(l)}}^{-(l)}, \mG_{-l})} &=
	\sum_{m \neq l}\sum_{i_m=1}^{N_m}\sum_{t=1}^T\Big\{Y_{i_1 \cdots i_q, t} - \sum_{l=1}^q \lambda_{g_{i_l}^{(l)}}^{(l)} \sum_{k = 1}^{N_l} w^{(l)}_{i_l k}Y_{i_1 \cdots i_{l-1} k i_{l+1} \cdots i_q,(t-1)}\nonumber\\
	&- \alpha_{g_{i_1}^{(1)} \cdots g_{i_q}^{(q)}}Y_{i_1 \cdots i_q,(t-1)}
	- \sum_{l=1}^q \bx_{i_lt}^{(l) \top} \bzeta_{g_{i_l}^{(l)}}^{(l)}\Big\}^2 \label{eq:Q_il}
\end{align}
and $Q_{i_l}^*(\bxi_{g_{i_l}^{(l)}}^{(l)}; \bxi_{g^{-(l)}}^{-(l)}, \mG_{-l})  =
E\{Q_{i_l}(\bxi_{g_{i_l}^{(l)}}^{(l)}; \bxi_{g^{-(l)}}^{-(l)}, \mG_{-l}) \}$.
By the definition, it readily follows that $Q(\bxi, \mG) = \sum_{i_l = 1}^{N_l} Q_{i_l}(\bxi_{g_{i_l}^{(l)}}^{(l)}; \bxi_{g^{-(l)}}^{-(l)}, \mG_{-l})$, where the $Q(\bxi, \mG)$ is the objective function defined in \eqref{eq:Q_obj}.
For the tensor $\balpha \in \mR^{G_1 \times \cdots \times G_q}$, we use $\balpha_{\cdot g_{i_l}^{(l)} \cdot} \in \mR^{G_1 \times  \cdots G_{l-1} \times G_{l+1} \times \cdots \times G_q}$ to denote its subset by specifying the $l$th dimension being equal to $g_{i_l}^{(l)}$.
Similarly, for the parameter tensor $\bTheta \in \mR^{N_1 \times \cdots \times N_q \times m}$, when specifying the $l$th dimension as $i_l$, we obtain the subset tensor $\bTheta_{\cdot i_l \cdot} \in \mR^{N_1 \times \cdots \times N_{l-1} \times N_{l+1} \times \cdots \times N_q \times m}$.
We define the pseudo distance
\begin{align}
	d_{i_l}(\wh \bTheta_{\cdot i_l \cdot}, \bTheta_{\cdot  i_l \cdot}) & = \frac{1}{\prod_{m \neq l}N_m} \sum_{m \neq l} \sum_{i_m = 1}^{N_m}
	\Big\|\wh \bTheta_{i_1 \cdots i_q} - \bTheta_{i_1 \cdots i_q}\Big\|^2\nonumber\\
	& = \sum_{m \neq l} \frac{1}{N_m} \sum_{i_m} \|\wh\btheta_{\wh g_{i_m}^{(m)}}^{(m)} -  \btheta_{g_{i_m}^{(m)}}^{(m)}\|^2 +
	\|\wh \btheta_{\wh g_{i_l}^{(l)}}^{(l)} - \btheta_{ g_{i_l}^{(l)}}^{(l)}\|^2 \nonumber\\
	& + \frac{1}{\prod_{m \neq l} N_m} \sum_{m \neq l} \sum_{i_m=1}^{N_m} |\wh \alpha_{\wh g_{i_1}^{(1)} \cdots \wh g_{i_q}^{(q)}} - \alpha_{ g_{i_1}^{(1)} \cdots  g_{i_q}^{(q)}}|^2.\label{eq:def_dj}
\end{align}

For the parameter $\bTheta_{i_1 \cdots i_q}$, define the element-wise pseudo distance as
	\begin{align}
		d_{i_1\cdots i_q}(\wh \bTheta_{i_1 \cdots i_q}, \bTheta_{i_1 \cdots i_q})
		& = \sum_{l=1}^q \|\wh \btheta_{\wh g_{i_l}^{(l)}}^{(l)} - \btheta_{g_{i_l}^{(l)}}^{(l)} \|^2 + |\wh \alpha_{\wh g_{i_1}^{(1)} \cdots \wh g_{i_q}^{(q)}} - \alpha_{g_{i_1}^{(1)} \cdots g_{i_q}^{(q)}}|^2.\label{eq:def_dij}
	\end{align}

Let $\mY_t = \vec(\cY_t) \in \mR^{\prod_l N_l}$ and $\E_t = \vec(\mE_t) \in \mR^{\prod_l N_l}$.
Then (\ref{eq:model}) can be rewritten as
$\mY_t =  \B_0\mY_{t-1}+\c_t + \E_t,$
where
\begin{align}
	& \B_0 = \sum_l \bI_{N_1} \otimes \cdots \otimes \bI_{N_{l-1}} \otimes (\L_{l,0} \bW^{(l)}) \otimes \bI_{N_{l+1}} \otimes \cdots \otimes \bI_{N_q} + \diag\{ \vec(\mA_0) \} \in \mR^{\prod_l N_l \times \prod_l N_l},\label{eq:def_B0}\\
	& \c_t = \sum_{l=1}^q  \vec(\one_{N_1} \circ \cdots \circ \one_{N_{l-1}} \circ \bbeta_{X_l, t}^{(l)0} \circ \one_{N_{l+1}} \circ \cdots \circ \one_{N_q}) \defeq \sum_l \c_t^{(l)} \in \mR^{\prod_l N_l},\label{eq:def_ct}
\end{align}
and $\L_{l,0}$, $\mA_0$, $\bbeta_{X_l,t}^{(l)0}$
are the true values of the corresponding terms for $l \in [q]$, where $\bbeta_{X_l, t}^{(l)} = (\bx_{i_l t}^{(l)\top} \bzeta_{g_{i_l}^{(l)}}^{(l)}:1\le i_l \le N_l)^\top  \in \mR^{N_l}$ is defined in the main text.
Recall that $E(\c_t) = \zero$ because $E(\x_{{i_l}t}^{(l)})=\0$ for all $i_l \in [N_l]$ and $l \in [q]$,
and by Assumption \ref{assum:station} that $\mY_0 = \zero$,
we have
\begin{align}
	\mY_t = \sum_{k = 0}^{t} \bB_0^k \bc_{t-k}+ \sum_{k = 0}^{t}\bB_0^k  \E_{t-k}
	\defeq \mY_t^c+ \mY_t^e.\label{eq:Yt_expan}
\end{align}
Let $\bGamma = \cov(\mY_t)$.
Denote $\tau_{\max} = \max_{i_1, \cdots, i_q} \lambda_{\max}(\bSigma_{i_1 \cdots i_q})$ and
$\bSigma_{i_1 \cdots i_q}$ is defined in Assumption \ref{assum:tau_min}.

{\sc General Notations.}
For a matrix $\bM = (m_{ij})\in \mR^{m\times n}$, let $|\bM|_e = (|m_{ij}|)_{i \in [m], j \in [n]}$.
For two matrices $\bM_1 = (m_{1ij})\in \mR^{m\times n}$ and $\bM_2 = (m_{2ij})\in \mR^{m\times n}$, $\bM_1 \cle \bM_2 $ means that
$m_{1ij}\le m_{2ij}$ for all $i\in [m]$ and $j\in [n]$.
In addition, define $\|\bM\|$ as the largest singular value of $\bM$,
$\|\bM\|_F = \tr(\bM\bM^\top)^{1/2}$ as the Frobenius norm,
$\|\bM\|_\infty = \max_i \sum_{j}|m_{ij}|$, and $\|\bM\|_{\max}$ as its largest absolute value.
For a vector $\bv$, denote the norm $\|\bv\|_1 = \sum_i |v_i|$.
Denote $\e_{i}^{(n)}$ as a vector of length $n$, whose $i$th element equals to 1 while others are 0.

\section{Additional Discussion for the Case when $q=2$}\label{sec:notation_q2}
	
	\subsection{Estimation Details}
	As shown in Section \ref{subsec:est_matrix}, the estimator when $q=2$ could be written as
	\begin{align}
		\mathbf{M}=\left(\begin{array}{ccc}
			\mathbf{M}^{(1)} & \mathbf{M}^{(12)} & \mathbf{M}^{(1 \alpha)} \\
			\mathbf{M}^{(12) \top} & \mathbf{M}^{(2)} & \mathbf{M}^{(2 \alpha)} \\
			\mathbf{M}^{(1\alpha) \top} & \mathbf{M}^{(2 \alpha) \top} & \mathbf{M}^\alpha
		\end{array}\right), \quad \mathbf{b}=\left(\begin{array}{c}
			\mathbf{b}^{(1)} \\
			\mathbf{b}^{(2)} \\
			\mathbf{b}^\alpha
		\end{array}\right),\label{eq:Mb2}
	\end{align}
	where the expressions are given as
	\begin{align*}
		& \mathbf{M}_{g^{(1)}}^{(1)}=\sum_{t, g^{(2)}} \mathbb{X}_{g^{(1)} g^{(2)} t}^{(1)\top} \mathbb{X}^{(1)}_{g^{(1)}g^{(2)} t}, \quad \mathbf{M}^{(1)}=\operatorname{diag}\left\{\mathbf{M}_{g^{(1)}}^{(1)}: g^{(1)} \in[G_1]\right\} \in \mathbb{R}^{G_1 \left(p_1+1\right) \times G_1 \left(p_1+1\right)}, \\
		& \mathbf{M}_{g^{(1)} g^{(2)}}^{(12)}=\sum_t \mathbb{X}_{g^{(1)} g^{(2)} t}^{(1)\top} \mathbb{X}^{(2)}_{g^{(1)} g^{(2)} t},\\
		&  \mathbf{M}^{(12)}=\left(\mathbf{M}_{g^{(1)} g^{(2)}}^{(12)}: g^{(1)} \in[G_1], g^{(2)} \in[G_2]\right) \in \mathbb{R}^{G_1\left(p_1+1\right) \times G_2 \left(p_2+1\right)}, \\
		& \mathbf{M}_{g^{(1)}  \mathcal{I}_{g^{(1)\prime} g^{(2)}}}^{(1\alpha)}=\sum_t \mathbb{X}_{g^{(1)} g^{(2)} t}^{(1)\top} \mathbb{Y}_{g^{(1)} g^{(2)}, (t-1)} I\left(g^{(1)}=g^{(1)\prime}\right), ~~~ \mathcal{I}_{g^{(1)\prime} g^{(2)}}=(g^{(2)}-1) G_1+g^{(1)\prime}, \\
		& \mathbf{M}^{(1 \alpha)}=\left(\mathbf{M}_{g^{(1)} \mathcal{I}_{g^{(1)\prime} g^{(2)}}}^{(1 \alpha)}: g^{(1)} \in[G_1], \mathcal{I}_{g^{(1)\prime} g^{(2)}} \in[G_1 G_2]\right) \in \mathbb{R}^{G_1 \left(p_1+1\right) \times G_1 G_2}, \\
		& \mathbf{M}_{g^{(2)}}^{(2)}=\sum_{t, g^{(1)}} \mathbb{X}_{g^{(1)} g^{(2)} t}^{(2)\top} \mathbb{X}_{g^{(1)} g^{(2)}  t}^{(2)}, \quad \mathbf{M}^{(2)}=\operatorname{diag}\left\{\mathbf{M}_{g^{(2)}}^{(2)}: g^{(2)} \in[G_2]\right\} \in \mathbb{R}^{G_2\left(p_2+1\right) \times G_2\left(p_2+1\right)},\\
		& \mathbf{M}_{g^{(2)} \mathcal{I}_{g^{(1)} g^{(2)\prime}}}^{(2) \alpha}=\sum_t \mathbb{X}_{g^{(1)} g^{(2)} t}^{(2)\top} \mathbb{Y}_{g^{(1)} g^{(2)}, (t-1)} I\left(g^{(2)}=g^{(2)\prime}\right), \\
		&  \mathbf{M}^{(2) \alpha}=\left(\mathbf{M}_{g^{(1)} \mathcal{I}_{g^{(1)} g^{(2)\prime}}}^{(2) \alpha}: g^{(2)} \in[G_2], \mathcal{I}_{g^{(1)} g^{(2)\prime}} \in[G_1 G_2]\right) \in \mathbb{R}^{G_2\left(p_2+1\right) \times G_1 G_2}, \\
		& \mathbf{M}_{\mathcal{I}_{g^{(1)} g^{(2)}} \mathcal{I}_{g^{(1)\prime} g^{(2)\prime}}}^\alpha=\sum_t\left\|\mathbb{Y}_{g^{(1)} g^{(2)},(t-1)}\right\|^2 I\left(g^{(1)}=g^{(1)\prime}, g^{(2)}=g^{(2)\prime}\right), \\
		& \mathbf{M}^\alpha=\left(\mathbf{M}_{\mathcal{I}_{g^{(1)} g^{(2)}} \mathcal{I}_{g^{(1)\prime} g^{(2)\prime}}}^\alpha: \mathcal{I}_{g^{(1)} g^{(2)}} \in[G_1 G_2], \mathcal{I}_{g^{(1)\prime} g^{(2)\prime}} \in[G_1 G_2]\right) \in \mathbb{R}^{G_1 G_2 \times G_1 G_2}, \\
		& \mathbf{b}_{g^{(1)}}^{(1)}=\sum_{t, g^{(2)}} \mathbb{X}_{g^{(1)} g^{(2)} t}^{(1)\top} \mathbb{Y}_{g^{(1)} g^{(2)}, t}, \quad \mathbf{b}^{(1)}=\left(\mathbf{b}_{g^{(1)}}^{ \top}: g^{(1)} \in[G_1]\right)^{\top} \in \mathbb{R}^{G_1 \left(p_1+1\right)}, \\
		& \mathbf{b}_g^{(2)}=\sum_{t, g^{(1)}} \mathbb{X}_{g^{(1)} g^{(2)} t}^{(2)\top} \mathbb{Y}_{g^{(1)} g^{(2)}, t}, \quad \mathbf{b}^{(2)}=\left(\mathbf{b}_{g^{(2)}}^{(2) \top}: g^{(2)} \in [G_2]\right)^{\top} \in \mathbb{R}^{G_2 \left(p_2+1\right)}, \\
		& \mathbf{b}_{\mathcal{I}_{g^{(1)} g^{(2)}}}^\alpha=\sum_t \mathbb{Y}_{g^{(1)} g^{(2)}, (t-1)}^{\top} \mathbb{Y}_{g^{(1)} g^{(2)}, t}, \quad \mathbf{b}^\alpha=\left(\mathbf{b}_{\mathcal{I}_{g^{(1)} g^{(2)}}}^\alpha: \mathcal{I}_{g^{(1)} g^{(2)}} \in[G_1 G_2]\right)^{\top} \in \mathbb{R}^{G_1 G_2} .
	\end{align*}
	In the above expressions,
	\begin{align}
		& \mX_{g^{(1)} g^{(2)}, t}^{(1)} =
		\Big(\vec(\bW_1^{(\cR^{(1)}_{g^{(1)}},\cdot)}\bY_{t-1}^{(\cdot, \cR^{(2)}_{g^{(2)}})}),
		\one_{N_{2 g^{(2)}}}\otimes (\bX_{t}^{(1)})^{(\cR^{(1)}_{g^{(1)}},\cdot)}\Big)\in \mR^{(N_{1 g^{(1)}}N_{2 g^{(2)}})\times (p_1+1)}, \label{eq:mX_1_def}\\
		& \mX_{g^{(1)} g^{(2)}, t}^{(2)} =
		\Big(\vec(\bY_{t-1}^{(\cR^{(1)}_{g^{(1)}}, \cdot)}\bW_2^{(\cdot,\cR^{(2)}_{g^{(2)}})}),
		(\bX_{t}^{(2)})^{(\cR^{(2)}_{g^{(2)}},\cdot)} \otimes \one_{N_{1 g^{(1)}}} \Big)\in \mR^{(N_{1 g^{(1)}}  N_{2 g^{(2)}})\times (p_2+1)},\label{eq:mX_2_def}
	\end{align}
	and $\bX_t^{(l)} = (\bx^{(l)}_{1t},\cdots, \bx^{(l)}_{N_1t})^\top\in \mR^{N_l\times p_l}$, $N_{l g^{(l)} }^{(l)} = |\cR_{g^{(l)}}^{(l)}|$ for $l =1, 2$, and $\mY_{g^{(1)} g^{(2)}, t} = \vec(\cY_{t}^{(\cR_{g^{(1)}}^{(1)}, \cR_{g^{(2)}}^{(2)})})\in \mR^{|\cR_{g^{(1)}}^{(1)}| |\cR_{g^{(2)}}^{(2)}|}$.
	We then provide the iterative update algorithm when $q=2$ in the following Algorithm \ref{alg:gmnar_q2}.
	
	\begin{algorithm}
		\caption{Estimation of the GTNAR Model When $q=2$}
		\label{alg:gmnar_q2}
		\begin{algorithmic}[1]
			\State {\bf Input:} $\{\cY_t, \bX_t^{(l)}, \bW^{(l)}, G_l\}$ for $l=1,2$.
			\State Obtain initial group memberships $\mG_l^{[0]}$ according to Algorithm \ref{alg:init} in Appendix \ref{sec:init}. Let $\{\bxi^{[k]}, \mG_l^{[k]}\}$ be the estimators and memberships in the $k$th iteration.
			\State Repeat {\sc Step 1} and {\sc Step 2} for {$k = 1, 2, \cdots$}
			until convergence.
			
			{\sc Step 1.} Given $\{\mG_l^{[k-1]}\}$ for all layers $l =1,2$, calculate
			$\bxi^{[k-1]} = (\bM^{[k-1]})^{-1} \b^{[k-1]}$, where $\bM^{[k-1]}$ and $\b^{[k-1]}$ are obtained from
			(\ref{eq:Mb}) with $\mG_l^{[k-1]}$s specified.
			
			{\sc Step 2.} Given $\bxi^{[k-1]}$,  update the memberships in layer $l=1,2$ sequentially by,
			\begin{align}
				g_{i}^{(1)[k]} & = \arg\min_{g_{i}^{(1)}  \in [G_1]}
				\sum_{j=1}^{N_2} \sum_{t=1}^T  \Big\{
				Y_{ij,t} \nonumber\\
				& - \Big(\lambda_{g_{i}^{(1)}}^{(1)[k]}\sum_{m= 1}^{N_1} \frac{a^{(1)}_{im}}{n_{1 i}}Y_{m j,(t-1)}
				+ \lambda_{g_{j}^{(2)[k-1]}}^{(2)[k]}\sum_{m = 1}^{N_{2}} \frac{a^{(2)}_{jm}}{n_{2 j}}Y_{i m,(t-1)} \Big) \nonumber\\
				&
				-\alpha^{[k]}_{g_{i}^{(1)} g_{j}^{(2)[k-1]} }Y_{ij,(t-1)}
				- \Big(\bx_{i t}^{(1)\top}\bzeta_{g_{i}^{(1)}}^{(1)[k]} +  \bx_{j t}^{(2)\top} \bzeta_{g_{j}^{(2)[k-1]}}^{(2)[k]} \Big)
				\Big\}^2 \label{eq:update_g1}
			\end{align}	
			and
			\begin{align}
				g_{j}^{(2)[k]} & = \arg\min_{g_{j}^{(2)}  \in [G_2]}
				\sum_{i=1}^{N_1} \sum_{t=1}^T  \Big\{
				Y_{ij,t} \nonumber\\
				& - \Big(\lambda_{g_{i}^{(1)[k]}}^{(1)[k]}\sum_{m= 1}^{N_1} \frac{a^{(1)}_{im}}{n_{1 i}}Y_{m j,(t-1)}
				+ \lambda_{g_{j}^{(2)}}^{(2)[k]}\sum_{m = 1}^{N_{2}} \frac{a^{(2)}_{jm}}{n_{2j}}Y_{i m,(t-1)} \Big) \nonumber\\
				&
				-\alpha^{[k]}_{g_{i}^{(1)[k]} g_{j}^{(2)} }Y_{ij,(t-1)}
				- \Big(\bx_{i t}^{(1)\top}\bzeta_{g_{i}^{(1)[k]}}^{(1)[k]} +  \bx_{j t}^{(2)\top} \bzeta_{g_{j}^{(2)}}^{(2)[k]} \Big)
				\Big\}^2 \label{eq:update_g2}
			\end{align}
			\State {\bf Output:} Final estimator and memberships:
			$\wh\bxi = \bxi^{[K]}$ and
			$\wh \mG = \{\mG_l^{[K]}: l =1,2\}$.
			Here $K$ is the final number of iteration rounds.
		\end{algorithmic}
	\end{algorithm}
	
	We remark here that the technical conditions in Section \ref{sec:theory} can be simplified in this matrix case for easy understanding.
	For example, the covariates Assumption \ref{assum:mixing} can be rewritten as follow.
	
	\begin{assumption}\label{assum:mixing_q2}
		{\sc(Distribution of Covariates of Matrix Case)}
		{Recall that $\bx_{i_lt}^{(l)} \in \mR^{p_l}$ is the covariate vector of the $i_l$th subject in the $l$th layer at time $t$.}
		Assume $E(\bx_{i t}^{(1)}) = \zero$ for all $i \in [N_1]$ and $t \in [T]$.
		Let $\bfeta_1 \in \mR^{p_1}$ be a constant vector satisfying $\|\bfeta_1\| \le c$, where $c$ is a positive constant.
		Define $\bx_t^{(1)\eta} = (\bx_{i t}^{(1)\top} \bfeta_{1}: i \in [N_1])^\top \in \mR^{N_1}$.
		Assume $\bx^{(1) \eta} =(\bx_t^{(1)\eta\top}: 0\le t\le T)^\top \in \mR^{N_1 (T+1)}$
		satisfies the $K$-convex concentration property for some constant $K$
		according to Definition \ref{def:convex_concen}.
	\end{assumption}

	\subsection{Model Inference}\label{append:normal_est}

	In this subsection, we provide an estimator to the asymptotic covariance when $q=2$.
	This procedure can be easily extended to $q>2$.
	With the parameter estimator $\wh\bTheta = (\wh\bTheta_{ij} = (\wh\btheta_{\wh g_{i}^{(1)}}^{(1) \top}, \wh\btheta_{\wh g_{j}^{(2)}}^{(2) \top}, \wh\alpha_{\wh g_{i}^{(1)} \wh g_{j}^{(2)}})^\top)$, we first estimate $\sigma^2$ as follows
	\begin{align}
		\wh\sigma^2 = \frac{1}{N_1 N_2 T} \sum_{i=1} \sum_{j = 1} \sum_{t = 1} (Y_{ij, t} - \cX_{ij, t}^\top \wh\bTheta_{ij})^2, \label{eq:sigma_hat}
	\end{align}
	where $\cX_{ij, t}$ is defined in Assumption \ref{assum:tau_min}.
	Next, we estimate $\M^0$ by $\wh \bM$,
	where $\wh \M$ is obtained  by plugging estimated parameters $\wh \bTheta$ into the expression in \eqref{eq:Mb}.
	%
	In the following theorem, we show that the covariance estimator is consistent.
	\bet\label{thm:cov_consistent}
	Suppose Assumption \ref{assum:para_space}--\ref{assum:station} and assume that there exists $n$, such that $c_1 n \le \min_l N_l \le \max_l N_l \le c_2 n$ for some constants $c_1, c_2 > 0$.
	When $G_1 = G_{1,0}$ $G_2 = G_{2, 0}$,
	assume $\{\log(N_1 N_2)\}^2 / {T} \to 0$, then the following holds,
	\begin{align*}
		 \wh\sigma^2 \to_p \sigma^2,~~~ (n^2 T)^{-1} \wh\sigma^2 \bfeta^\top (\wh\bM)^{-1}\bfeta  \to_p  (n^2 T)^{-1} \sigma^2 \bfeta^\top (\bM^0)^{-1} \bfeta,
	\end{align*}
	where $\bfeta$ is defined in Theorem \ref{thm:normal}.
	\eet
	The proof of Theorem \ref{thm:cov_consistent} can be found in Appendix \ref{subsec:proof_cov_consist}.
	Theorem \ref{thm:cov_consistent} indicates that
	we can obtain a consistent estimator for the
	asymptotic variance by plugging in the estimators $\wh \bTheta$ and the consistent estimator $\wh\sigma^2$.
	This assures a valid statistical inference procedure.
	We next present a number of simulation studies to examine the finite sample performances of the model estimation and inference procedures.

	\subsubsection{Proof of Theorem \ref{thm:cov_consistent}}\label{subsec:proof_cov_consist}
	
		To prove Theorem \ref{thm:cov_consistent}, we prove the following two statements separately,
		\begin{align}
			\wh\sigma^2 & \to_p \sigma^2, \label{sigma_consist}\\
			(n^2 T)^{-1} \wh\bM_{nT} & \to_p (n^2 T)^{-1} \bM^0. \label{bM_consist}
		\end{align}
		
		\noindent
		{\bf (1) Proof of (\ref{sigma_consist})}
		
		Note that $\sigma^2 = (N_1 N_2 T)^{-1} \sum_{i, j,t} E(\varepsilon_{i j, t}^2) =(N_1 N_2 T)^{-1} E \{ \sum_{i, j} Q_{ij}(\bTheta_{ij}^0)\} = (N_1 N_2 T)^{-1}\\ Q^*(\bTheta^0)$, and $\wh\sigma^2 = (N_1 N_2 T)^{-1} \sum_{i,j} Q_{ij}(\wh\bTheta_{ij}) =  (N_1 N_2 T)^{-1}  Q(\wh\bTheta)$.
		To prove \eqref{sigma_consist}, it suffices to show that $(N_1 N_2 T)^{-1}  |Q(\wh\bTheta) - Q^*(\bTheta^0)| = o_p(1)$.
		Further note that
		\begin{align*}
			\frac{1}{N_1N_2T} |Q(\wh\bTheta) - Q^*(\bTheta^0)| \le \frac{1}{N_1N_2T} \left\{|Q(\wh\bTheta) - Q^*(\wh\bTheta)| + |Q^*(\wh\bTheta) - Q^*(\bTheta^0)| \right\},
		\end{align*}
		hence we bound the two terms on the right hand side.
		
		\noindent
		{\bf (1.1) Order of $(N_1 N_2 T)^{-1}|Q(\wh\bTheta) - Q^*(\wh\bTheta)|$.}
		
		By Lemma \ref{lem:Q_diff}, we have that
		\begin{align*}
			\sup_{\| \bTheta \|_{\max} \le R} \left|\frac{1}{N_1 N_2 T} \left\{Q(\bTheta) - Q^*(\bTheta) \right\}\right| = O_p \left\{T^{-1/2} (m + \log(N_1 N_2))\right\},
		\end{align*}
		where $m = p_1 + p_2 + 3$.
		Thus when $\{m + \log(N_1 N_2)\} / \sqrt{T} \to 0$, we have $(N_1 N_2 T)^{-1}|Q(\wh\bTheta) - Q^*(\wh\bTheta)| = o_p(1)$.

		\noindent
		{\bf (1.2) Order of $(N_1 N_2 T)^{-1}|Q^*(\wh\bTheta) - Q^*(\bTheta^0)|$}
		
		By Lemma \ref{lem:Q_star_diff} we have $(N_1 N_2 T)^{-1} \{ Q^*(\wh\bTheta) - Q^*(\bTheta^0) \} \le \tau_{\max} d(\wh\bTheta, \bTheta^0)$. Since $\tau_{\max}$ is bounded due to Lemma \ref{lem:tau_max} and $d(\wh\bTheta, \bTheta^0) = O_p\{ T^{-1/2} (m + \log (N_1 N_2)) \}$ according to Theorem \ref{thm:pseudo_dist}, we have that $(N_1 N_2 T)^{-1}|Q^*(\wh\bTheta) - Q^*(\bTheta^0)| = O_p\{ T^{-1} (\log (N_1 N_2)^2) \} =o_p(1)$ when $(\log N_1 N_2)^2/T \to 0$.
		This condition can be implied by $\{m + \log(N_1 N_2)\} / \sqrt{T} \to 0$.
		
		From both (1.1) and (1.2), we have that $(N_1 N_2 T)^{-1}  |Q(\wh\bTheta) - Q^*(\bTheta^0)| = o_p(1)$ when $\{m + \log(N_1 N_2)\} / \sqrt{T} \to 0$.
		This implies that $\wh\sigma^2 \to_p \sigma^2$ under the same condition. This finishes the proof of \eqref{sigma_consist}.

		\noindent
		{\bf (2) Proof of (\ref{bM_consist})}
		
		Recall the expression of $\bM$ defined in \eqref{eq:Mb2} when $q=2$, 
		we need to show that the nine elements in $(n^2 T)^{-1} \wh\bM_{nT}$ are consistent in an analogous manner. 
		Hence, we take the first element, i.e. $ (n^2 T)^{-1} \wh\bM^{(1)} \in \mR^{G_1(p_1+1) \times G_1(p_1+1)}$, 
		as an example to show the consistency, 
		where $n$ is defined in Theorem \ref{thm:cov_consistent}.
		Other terms could be proved similarly.
		
		Recall that $\bM^{(1)} = \diag\{ \bM_{g^{(1)}}^{(1)}: g^{(1)} \in [G_1] \}$ and $\bM_{g^{(1)}}^{(1)} = \sum_{t, g^{(2)}} \mX_{g^{(1)} g^{(2)}, t}^\top \mX_{g^{(1)} g^{(2)}, t}$, we show that for any $g^{(1)}  \in [G_{1,0}]$, it holds that $\big| (n^2 T)^{-1} \wh\bM_{g^{(1)}}^{(1)} - \lim_{n, T \to \infty} (n^2 T)^{-1} E(\wh\bM_{g^{(1)}}^{(1)})\big| = o_p(1)$.
		Recall that
		\begin{align*}
			\mX_{g^{(1)} g^{(2)}, t} & =
			\Big(\vec(\bW_1^{(\cR_{g^{(1)}}^{(1)},\cdot)}\bY_{t-1}^{(\cdot, \cR_{g^{(2)}}^{(2)})}),
			\one_{N_{2 g^{(2)}}}\otimes \bX_{t}^{(\cR_{g^{(1)}}^{(1)},\cdot)}\Big) \\
			& = \Big( (\bI_{N_{2 g^{(2)}}} \otimes \bW_1^{(\cR_{g^{(1)}}^{(1)},\cdot)}) \mY_{t-1}^{\cR_{g^{(2)}}^{(2)}} ,
			\one_{N_{2 g^{(2)}}}\otimes \bX_{t}^{(\cR_{g^{(1)}}^{(1)},\cdot)}\Big) \\
			& \defeq \Big(\mW_{g^{(1)} g^{(2)}}  \mY_{t-1}^{\cR_{g^{(2)}}^{(2)}}, \one_{N_{2 g^{(2)}}}\otimes \bX_{t}^{(\cR_{g^{(1)}}^{(1)},\cdot)} \Big) \in \mR^{(N_{1 g^{(1)}}N_{2 g^{(2)}})\times (p_1+1)},
		\end{align*}
		where $\mW_{g^{(1)} g^{(2)}} \defeq \bI_{N_{2 g^{(2)}}} \otimes \bW_1^{(\cR_{g^{(1)}}^{(1)},\cdot)} \in \mR^{N_{2 g^{(2)}}N_{1 g^{(1)}} \times N_{2 g^{(2)}} N_1}$ and $ \mY_{t-1}^{\cR_{g^{(2)}}^{(2)}} \defeq \vec(\bY_{t-1}^{(\cdot, \cR_{g^{(2)}}^{(2)})}) \in \mR^{N_{2 g^{(2)}}N_1}$.
		Hence, it is equivalent to prove that
		\begin{align}
			& (n^2 T)^{-1} \sum_{t, g^{(2)}} \mY_{t-1}^{\cR_{g^{(2)}}^{(2)} \top} \mW_{g^{(1)} g^{(2)}}^\top \mW_{g^{(1)} g^{(2)}}  \mY_{t-1}^{\cR_{g^{(2)}}^{(2)}} \nonumber\\
			& \to_p \lim_{n, T \to \infty} (n^2 T)^{-1} \sum_{t, g^{(2)}} E(\mY_{t-1}^{\cR_{g^{(2)}}^{(2)} \top} \mW_{g^{(1)} g^{(2)}}^\top \mW_{g^{(1)} g^{(2)}}  \mY_{t-1}^{\cR_{g^{(2)}}^{(2)}} ),\label{Y_quad}\\
			& (n^2 T)^{-1} \sum_{t,g^{(2)}} \mY_{t-1}^{\cR_{g^{(2)}}^{(2)} \top} \mW_{g^{(1)} g^{(2)}}^\top \one_{N_{2 g^{(2)}}} \otimes \bX_{t}^{(\cR_{g^{(1)}}^{(1)},\cdot)} \nonumber\\
			& \to_p  \lim_{n, T \to \infty} (n^2 T)^{-1} \sum_{t, g^{(2)}}  E(\mY_{t-1}^{\cR_{g^{(2)}}^{(2)} \top} \mW_{g^{(1)} g^{(2)}}^\top\one_{N_{2 g^{(2)}}}\otimes \bX_{t}^{(\cR_{g^{(1)}}^{(1)},\cdot)}).\label{Y_linear}
		\end{align}
		We prove \eqref{Y_quad} for example, and \eqref{Y_linear} could be proved similarly.
		Similar as the decomposition in \eqref{eq:Yt_expan},
		write $\mY_{t-1}^{\cR_{g^{(2)}}^{(2)}} = \mY_{t-1}^{\cR_{g^{(2)}}^{(2)}, c}+ \mY_{t-1}^{\cR_{g^{(2)}}^{(2)},e} $,
		where $\mY_{t-1}^{\cR_{g^{(2)}}^{(2)}, c} = \sum_{k=0}^{t-1} (\bB_0^{\cR_{g^{(2)}}^{(2)}})^k \c_{t-k}^{\cR_{g^{(2)}}^{(2)}}$, $\bB_0^{\cR_{g^{(2)}}^{(2)}} = \bI_{N_{2 g^{(2)}}} \otimes \L_0 \bW_1 + \G_0^{(\cR_{g^{(2)}}^{(2)},\cR_{g^{(2)}}^{(2)}) \top} \bW_2^{(\cR_{g^{(2)}}^{(2)},\cR_{g^{(2)}}^{(2)})} \otimes \bI_{N_1} \in \mR^{N_1 N_{2 g^{(2)}} \times N_{2 g^{(2)}} N_1}$, and $\c_t^{\cR_{g^{(2)}}^{(2)}} \in \mR^{N_1 N_{2 g^{(2)}}}$ is a similar modification of $\c_t$. $ \mY_{t-1}^{\cR_{g^{(2)}}^{(2)},e} $ could be represented analogously.
		Then the left hand side of \eqref{Y_quad} could be separated as
		\begin{align*}
			& \mY_{t-1}^{\cR_{g^{(2)}}^{(2)} \top} \mW_{g^{(1)} g^{(2)}}^\top \mW_{g^{(1)} g^{(2)}}  \mY_{t-1}^{\cR_{g^{(2)}}^{(2)}} \\
			& = \mY_{t-1}^{\cR_{g^{(2)}}^{(2)},c  \top} \mW_{g^{(1)} g^{(2)}}^\top \mW_{g^{(1)} g^{(2)}}  \mY_{t-1}^{\cR_{g^{(2)}}^{(2)},c} +  2\mY_{t-1}^{\cR_{g^{(2)}}^{(2)},c \top} \mW_{g^{(1)} g^{(2)}}^\top \mW_{g^{(1)} g^{(2)}}  \mY_{t-1}^{\cR_{g^{(2)}}^{(2)},e} \\
			& + \mY_{t-1}^{\cR_{g^{(2)}}^{(2)},e \top} \mW_{g^{(1)} g^{(2)}}^\top \mW_{g^{(1)} g^{(2)}}  \mY_{t-1}^{\cR_{g^{(2)}}^{(2)},e}
		\end{align*}
		We take the first term as an example.
		Let $\bN_{g^{(1)} g^{(2)}} = \mW_{g^{(1)} g^{(2)}}^\top \mW_{g^{(1)} g^{(2)}} \in \mR^{N_{2 g^{(2)}} N1 \times N_{2 g^{(2)}} N_1}$, the we have
		\begin{align*}
			& \frac{1}{n^2 T} \sum_{t, g^{(2)}} \mY_{t-1}^{\cR_{g^{(2)}}^{(2)} \top} \mW_{g^{(1)} g^{(2)}}^\top \mW_{g^{(1)} g^{(2)}}  \mY_{t-1}^{\cR_{g^{(2)}}^{(2)}} \\
			&  = \frac{1}{n^2 T} \sum_{t,g^{(2)}}^{T, G_2} \sum_{l_1 = 0}^{t-1} \sum_{l_2=0}^{t-1} \c_{l_1}^{\cR_{g^{(2)}}^{(2)} \top} (\bB_0^{\cR_{g^{(2)}}^{(2)}})^{t-l_1 \top} \bN_{g^{(1)} g^{(2)}}(\bB_0^{\cR_{g^{(2)}}^{(2)}})^{t-l_2} \c_{l_2}^{\cR_{g^{(2)}}^{(2)}} \\
			& = \frac{1}{n} \sum_{g^{(2)}} \sum_{l_1 = 0}^{T-1} \sum_{l_2=0}^{T-1} \bc_{l_1}^{\cR_{g^{(2)}}^{(2)} \top} \left\{ \frac{1}{n T} \sum_{t = \max(1, l_1, l_2)}^T (\bB_0^{\cR_{g^{(2)}}^{(2)}})^{t-l_1 \top} \bN_{g^{(1)} g^{(2)}}(\bB_0^{\cR_{g^{(2)}}^{(2)}})^{t-l_2}  \right\} \c_{l_2}^{\cR_{g^{(2)}}^{(2)}}  \\
			& \defeq \frac{1}{n} \sum_{g^{(2)}} \sum_{l_1 = 0}^{T-1} \sum_{l_2=0}^{T-1} \bc_{l_1}^{\cR_{g^{(2)}}^{(2)} \top} \bE_{l_1, l_2}^{\cR_{g^{(2)}}^{(2)}}  \bc_{l_2}^{\cR_{g^{(2)}}^{(2)}}.
		\end{align*}
		Define $\bc^{\cR_{g^{(2)}}^{(2)}} = (\bc_0^{\cR_{g^{(2)}}^{(2)} \top}, \cdots, \bc_{T-1}^{\cR_{g^{(2)}}^{(2)} \top})^\top$, and $\bE^{\cR_{g^{(2)}}^{(2)}} = (\bE_{l_1, l_2}^{\cR_{g^{(2)}}^{(2)}}, 0 \le l_1, l_2 \le T-1)$, then we have
		\begin{align*}
			& \frac{1}{n^2 T} \sum_{t, g^{(2)}} \mY_{t-1}^{\cR_{g^{(2)}}^{(2)} \top} \mW_{g^{(1)} g^{(2)}}^\top \mW_{g^{(1)} g^{(2)}}  \mY_{t-1}^{\cR_{g^{(2)}}^{(2)}} = \frac{1}{n} \sum_{g^{(2)}} \bc^{\cR_{g^{(2)}}^{(2)}\top} \bE^{\cR_{g^{(2)}}^{(2)}}  \bc^{\cR_{g^{(2)}}^{(2)}} \\
			& = \frac{1}{n} \sum_{g^{(2)}} \bc^{\cR_{g^{(2)}}^{(2)}, (1)\top} \bE^{\cR_{g^{(2)}}^{(2)}}  \bc^{\cR_{g^{(2)}}^{(2)}, (1)} + \frac{2}{n} \sum_{g^{(2)}} \bc^{\cR_{g^{(2)}}^{(2)}, (1)\top} \bE^{\cR_{g^{(2)}}^{(2)}, z}  \bc^{\cR_{g^{(2)}}^{(2)}, (2)}  \\
			& + \frac{1}{n} \sum_{g^{(2)}} \bc^{\cR_{g^{(2)}}^{(2)}, (2)\top} \bE^{\cR_{g^{(2)}}^{(2)}}  \bc^{\cR_{g^{(2)}}^{(2)}, (2)}.
		\end{align*}
		Then we could follow the proof of Lemma \ref{lem:YMY} to show the result.
		Since 
		$$n^{-2} \sum_{g^{(2)}} \one_{N_{2 g^{(2)}} N_1}^\top \mW_{g^{(1)} g^{(2)}}^\top \mW_{g^{(1)} g^{(2)}} \one_{N_{2 g^{(2)}} N_1} = O(1)$$ 
		under Assumption that there exists $n$ that $c_1 n \le \min_l N_l \le max_l N_l \le c_2 n$,
		we would have
		\begin{align*}
			& P\Big\{ \Big| \frac{1}{n^2 T} \sum_{t,g^{(2)}} \big\{ \mY_{t-1}^{\cR_{g^{(2)}}^{(2)} \top} \mW_{g^{(1)} g^{(2)}}^\top \mW_{g^{(1)} g^{(2)}}  \mY_{t-1}^{\cR_{g^{(2)}}^{(2)}} - E(\mY_{t-1}^{\cR_{g^{(2)}}^{(2)} \top} \mW_{g^{(1)} g^{(2)}}^\top \mW_{g^{(1)} g^{(2)}}  \mY_{t-1}^{\cR_{g^{(2)}}^{(2)}} ) \big\} \Big| \ge u\Big\} \\
			& \le C \exp\Big\{ -c \min(T u^2, \sqrt{T} u) \Big\}.
		\end{align*}
		This would lead to the statement \eqref{Y_quad}.
		Then, \eqref{Y_linear} can be shown similarly.
		Therefore, the consistency in \eqref{bM_consist} could be obtained when $T \to \infty$.
		Overall, the consistency of the asymptotic covariance estimation in Theorem \ref{thm:cov_consistent} could be obtained by Step (1) and Step (2).

		\subsection{Existence of Asymptotic Covariance}\label{subsec:limit_exist}
		
		We illustrate the asymptotic covariance matrix $\bM^0$ in Theorem \ref{thm:normal} exists when $q=2$, which can be directly extended to the general $q$ case.
		For notation simplicity, in this subsection, we use $g$ and $h$ to represent group in the first and second dimension, respectively.
		
		\begin{assumption}\label{assum:limit_exist}
			Recall that $\mY_t = \vec(\Y_t) \in \mR^{N_1 N_2}$.
			Denote $\mY_{g h, t} = \vec(\Y_t^{(\cR_g^{(1)}, \cR_h^{(2)})})$.
			For any $g_1, g_2 \in [G_1]$ and $h_1, h_2 \in [G_2]$,
			denote $\bSigma^{\mE}_{g_1 h_1, g_2 h_2} = E(\mY^e_{g_1 h_1, t} \mY^{e \top}_{g_2 h_2, t})$, $\bSigma^{\bX}_{g_1 h_1, g_2 h_2} = E(\mY^c_{g_1 h_1, t} \mY^{c \top}_{g_2 h_2, t})$.
			Denote $\cW_{gh}^1 = (\bI_{N_{2h}} \otimes \bW_1^{(\cR_g^{(1)}, \cdot)}) \in \mR^{N_{1g} N_{2h} \times N_1 N_{2h}}$, and denote $\cW_{gh}^2 = (\bI_{N_{1g}} \otimes (\bW_2^\top)^{(\cdot, \cR_h^{(2)}) }) \in \mR^{N_{1g} N_{2h} \times N_{1g} N_2}$.
			Assume the following limits exist,
			\begin{align*}
				& \kappa^{11}_{g_1 h_1, g_2 h_2} =  \lim_{n\to \infty}  n^{-2} \tr(  \cW_{g_1 h_1}^{1} \bSigma^{\mE}_{\cdot h_1, \cdot h_2} \cW_{g_1 h_2}^{1 \top}), ~	\nu^{11}_{g_1 h_1, g_2 h_2} =  \lim_{n\to \infty}  n^{-2} \tr(  \cW_{g_1 h_1}^{1} \bSigma^{\bX}_{\cdot h_1, \cdot h_2} \cW_{g_1 h_2}^{1\top}), \\
				&  \kappa^{22}_{g_1 h_1, g_2 h_2} =  \lim_{n\to \infty}  n^{-2} \tr(  \cW_{g_1 h_1}^{2} \bSigma^{\mE}_{g_1 \cdot, g_2 \cdot } \cW_{g_1 h_1}^{2 \top}), ~	\nu^{22}_{g_1 h_1, g_2 h_2} =  \lim_{n\to \infty}  n^{-2} \tr(  \cW_{g_1 h_1}^{2} \bSigma^{\bX}_{g_1 \cdot, g_2 \cdot} \cW_{g_1 h_1}^{2\top}), \\
				&  \kappa^{12}_{g_1 h_1, g_2 h_2} =  \lim_{n\to \infty}  n^{-2} \tr(  \cW_{g_1 h_1}^{1} \bSigma^{\mE}_{\cdot h_1, g_2 \cdot} \cW_{g_1 h_1}^{2 \top}), ~\nu^{12}_{g_1 h_1, g_2 h_2} =  \lim_{n\to \infty}  n^{-2} \tr(  \cW_{g_1 h_1}^{1} \bSigma^{\bX}_{\cdot h_1, g_2 \cdot} \cW_{g_1 h_1}^{2\top}), \\
				& \kappa^1_{g_1 h_1, g_2 h_2} = \lim_{n\to \infty}  n^{-2} \tr(  \cW_{g_1 h_1}^{1} \bSigma^{\mE}_{\cdot h_1, g_2 h_2}), ~\nu^1_{g_1 h_1, g_2 h_2} =  \lim_{n\to \infty}  n^{-2} \tr( \cW_{g_1 h_1}^{1} \bSigma^{\bX}_{\cdot h_1, g_2 h_2}),\\
				& \kappa^2_{g_1 h_1, g_2 h_2} = \lim_{n\to \infty}  n^{-2} \tr(  \cW_{g_1 h_1}^{2} \bSigma^{\mE}_{g_1 \cdot, g_2 h_2}), ~\nu^2_{g_1 h_1, g_2 h_2} =  \lim_{n\to \infty}  n^{-2} \tr( \cW_{g_1 h_1}^{2} \bSigma^{\bX}_{g_1 \cdot, g_2 h_2}),\\
				& \kappa_{g_1 h_1, g_2, h_2}^{\alpha} = \lim_{n\to \infty}  n^{-2}\tr(  \bSigma^{\mE}_{g_1 h_1, g_2 h_2}), ~~~\nu_{g_1 h_1, g_2, h_2}^{\alpha} = \lim_{n\to \infty}  n^{-2}\tr( \bSigma^{\bX}_{g_1 h_1, g_2 h_2}),
			\end{align*}
			where $\bSigma^{\mE}_{\cdot h_1,\cdot h_2},\bSigma^{\bX}_{\cdot h_1,\cdot h_2} \in \mR^{N_{1} N_{2 h_1} \times N_1 N_{2h_2}}$, $\bSigma^{\mE}_{g_1 \cdot, g_2 \cdot},\bSigma^{\bX}_{g_1 \cdot, g_2 \cdot} \in \mR^{N_{1g_1} N_{2} \times N_{1g_2} N_{2}}$, and $\bSigma^{\mE}_{g_1 \cdot, g_2 h_2},\\ \bSigma^{\bX}_{g_1 \cdot, g_2 h_2} \in \mR^{N_{1g_1} N_{2} \times N_{1g_2} N_{2h_2}}$, $\bSigma^{\mE}_{g_1 h_1, g_2 \cdot},\bSigma^{\bX}_{g_1 h_1, g_2 \cdot} \in \mR^{N_{1g_1} N_{2h_2} \times N_{1g_1} N_{2}}$.
			Further assume that $\bSigma_{X_1} = E(\bx_{i t}^{(1)} \bx_{i t}^{(1)\top}) \in \mR^{p_1 \times p_1}$, $\bSigma_{X_2} = E(\bx_{jt}^{(2)} \bx_{jt}^{(2)\top}) \in \mR^{p_2 \times p_2}$ exist.
		\end{assumption}

		\begin{lemma}\label{lem:limit_exist}
			Under Assumptions \ref{assum:limit_exist},
			the asymptotic covariance matrix $\bM^0 = \lim_{n, T \to \infty} \bM_{n, T}^0$ can be written as 
			\begin{align*}
				\mathbf{M}^0=\left(\begin{array}{ccc}
					\bXi_1 & \bXi_{12} & \bXi_{1 \alpha} \\
					\bXi_{12}^\top &\bXi_2 & \bXi_{2 \alpha}\\
					\bXi_{1 \alpha}^\top & \bXi_{2 \alpha}^\top & \bXi_{\alpha}
				\end{array}\right).
			\end{align*}
			where
			\begin{align}
				&	\bXi_{1} =\diag\left( \left(\begin{array}{cc}
					s_{11, g} &  \zero^\top \\
					\zero & \bSigma_{X_1}
				\end{array}\right): g \in [G_1]\right), ~~~
				\bXi_{2, h} = \diag\left( \left(\begin{array}{cc}
					s_{22, h} &  \zero^\top \\
					\zero & \bSigma_{X_2}
				\end{array}\right): h \in [G_2] \right), \nonumber\\
				&	\bXi_{12} =\left( \left(\begin{array}{cc}
					s_{12, gh} &  \zero^\top \\
					\zero & \zero
				\end{array}\right): g \in [G_1], h \in [G_2] \right),\nonumber\\
				& \bXi_{1 \alpha} = \left( (s_{31, gh}I(g = g'), \zero^\top) : g \in [G_1], \mI_{g'h} \in [G_1 G_2] \right), \nonumber\\
				& \bXi_{2 \alpha} = \left( (s_{32, gh} I(h = h'), \zero^\top): h \in [G_2], \mI_{gh'} \in [G_1 G_2] \right), \nonumber\\
				& \bXi_{\alpha} = \left( s_{4,gh} I(g= g', h = h'): \mI_{gh} \in [G_1 G_2], \mI_{g'h'} \in [G_1 G_2] \right), \label{eq:cov_diag}
			\end{align}
			In \eqref{eq:cov_diag}, $s_{11, g} = \sum_h \nu^{11}_{gh,gh} + \kappa^{11}_{gh,gh}$, $s_{22, h} = \sum_g \nu^{22}_{gh, gh} + \kappa^{22}_{gh, gh}$, $s_{12, gh} = \nu_{gh, gh}^{12} + \kappa_{gh, gh}^{12}$, $s_{31, gh} = \nu_{gh, gh}^1 + \kappa_{gh, gh}^1$, $s_{32, gh} = \nu_{gh, gh}^2 + \kappa_{gh, gh}^2$ and $s_{4, gh} = \nu_{gh, gh}^{\alpha} + \kappa_{gh, gh}^{\alpha}$,
			with all the constants defined in Assumption \ref{assum:limit_exist}.

		\end{lemma}

		\begin{proof}
			Recall the expression of $\bM$ when $q=2$ is defined in \eqref{eq:Mb2},
			\begin{align*}
				\mathbf{M}=\left(\begin{array}{ccc}
					\mathbf{M}^{(1)} & \mathbf{M}^{(12)} & \mathbf{M}^{(1 \alpha)} \\
					\mathbf{M}^{(12) \top} & \mathbf{M}^{(2)} & \mathbf{M}^{(2 \alpha)} \\
					\mathbf{M}^{(1\alpha) \top} & \mathbf{M}^{(2 \alpha) \top} & \mathbf{M}^\alpha
				\end{array}\right).
			\end{align*}
			Due to the similar format of the terms, 
			we derive the expression of 
			\begin{align*}
				& \bs_1 \defeq \lim_{n, T \to \infty} (n^2 T)^{-1} E(\bM^{(1)}), ~~~\bs_2 \defeq \lim_{n, T \to \infty} (n^2 T)^{-1} E(\bM^{(12)}), \\
				& \bs_3 \defeq \lim_{n, T \to \infty} (n^2 T)^{-1} E(\bM^{(1 \alpha)}), ~~~ \bs_4 \defeq \lim_{n, T \to \infty} (n^2 T)^{-1} E(\bM^{\alpha})
			\end{align*}
			in the following four steps.
			
			\noindent
			{\bf (1). Expression of $\bs_1$}.
			
			We prove that $\bs_1 = \diag((s_{11, g}, \bs_{12, g}^\top; \bs_{12, g}, \bs_{13}): g \in [G_1])$ with $s_{11, g} = \sum_h s_{11, gh} =\sum_h \nu^{11}_{gh, gh} + \kappa^{11}_{gh, gh}$, $\bs_{12, g} = \sum_{h} \bs_{12, gh} = \zero$ and $\bs_{13, g} = \bSigma_{X_1}$, which is defined in Assumption \ref{assum:limit_exist}.
			Recall that $\bM^{(1)} = \diag(\bM_{g}^{(1)}: g\in [G_1]) \in \mR^{G_1 (p_1+ 1) \times G_1 (p_1+1)}$ with $\bM_{g}^{(1)} = \sum_{t, h} \mX_{gh, t}^{(1)\top} \mX_{gh, t}^{(1)}$, where $\mX_{gh, t}^{(1)}$ is defined in \eqref{eq:mX_1_def}.
			By \eqref{eq:mX_1_def}, denote $\mY_{\cdot h, (t-1)} = \vec(\Y_{t-1}^{(\cdot, \cR_h^{(2)})})$, we have
			\begin{align}
				& s_{11, gh} = \lim_{n \to \infty} (n^2 T)^{-1} \sum_{t} E\{ \mY_{\cdot h, (t-1)}^\top (\bI_{N_{2h}} \otimes \bW_1^{(\cR_{g}, \cdot) })^\top (\bI_{N_{2h}} \otimes \bW_1^{(\cR_{g}, \cdot) }) \mY_{\cdot h, (t-1)}\}, \nonumber\\
				&\bs_{12, gh} = \lim_{n, T \to \infty} (n^2 T)^{-1} \sum_{t} E\{\mY_{\cdot h,(t-1)}^\top (\bI_{N_{2h}} \otimes \bW_1^{(\cR_{g, \cdot}) })^\top (\one_{N_{2h}} \otimes  (\bX_t^{(1)})^{(\cR_g, \cdot)})\}, \nonumber\\
				& \bs_{13, gh} =\lim_{n, T \to \infty} (n^2 T)^{-1} \sum_t E\{ (\one_{N_{2h}} \otimes  (\bX_t^{(1)})^{(\cR_g, \cdot)})^\top (\one_{N_{2h}} \otimes  (\bX_t^{(1)})^{(\cR_g,  \cdot)})\}.\nonumber
			\end{align}
			Since $\mY_{\cdot h, (t-1)}$ is independent with $\bX_{t}^{(1)}$, we have $s_{12} = \zero$.
			Further note that
			\begin{align*}
				s_{11, gh}  & = \lim_{n \to \infty} (n^2T)^{-1} \sum_t  \big[ \tr( \cW_{gh}^1 \bSigma^{\bX}_{\cdot h,\cdot h} \cW_{gh}^{1\top })  + \tr(\cW_{gh}^1 \bSigma_{\cdot h,\cdot h}^{\mE} \cW_{gh}^{1\top} ) \big] \\
				& = \lim_{n \to \infty} n^{-2} \big[ \tr(\cW_{gh}^1 \bSigma^{\bX}_{\cdot h,\cdot h} \cW_{gh}^{1\top} )  + \tr( \cW_{gh}^1 \bSigma_{\cdot h,\cdot h}^{\mE} \cW_{gh}^{1\top }) \big] = \nu^{11}_{gh, gh} + \kappa^{11}_{gh, gh}.
			\end{align*}
			Recall that $\bx_{i t}^{(1)}$s are independent and identically distributed covariates across all $i \in [N_1]$ and $t \in [T]$, denote the covariance as $\bSigma_{X_1} = E(\bx_{i t}^{(1)} \bx_{i t}^{(1)\top}) \in \mR^{p_1 \times p_1}$.
			Therefore we have $\bs_{13, gh} = \bSigma_{X_1}$.

			
			\noindent
			{\bf (2). Expression of $\bs_2$}.
			
			Next, we prove that $\bs_2 = ( (s_{21, gh},\bs_{22, gh}; \bs_{22, gh}^\top, \bs_{23, gh}) : g \in [G_1], h \in [G_2])$
			with $\bs_{21, gh} = \nu_{gh, gh}^{12} + \kappa_{gh, gh}^{12}$, $\bs_{22, gh} = \zero$ and $\bs_{23, gh} = \zero$.
			Since $\mathbf{M}^{(12)}=\left(\mathbf{M}_{gh}^{(12)}: g\in[G_1], h \in[G_2]\right)$, we derive the expression of $\lim_{n,T \to \infty} (n^2T)^{-1} E(\mathbf{M}_{gh}^{(12)}) = \lim_{n,T \to \infty}  (n^2T)^{-1} \sum_t E(\mathbb{X}_{gh, t}^{(1)\top} \mathbb{X}^{(2)}_{gh, t})$.
			By the definition of $\mX_{gh, t}^{(1)}$ and $\mX_{gh, t}^{(2)}$ in \eqref{eq:mX_1_def} and \eqref{eq:mX_2_def}, we have
			\begin{align*}
				& s_{21, gh} = \lim_{n, T\to \infty} (n^2 T)^{-1} \sum_{t} E\{\mY_{\cdot h, (t-1)}^\top (\bI_{N_{2h}} \otimes \bW_1^{(\cR_{g}^{(1)}, \cdot) })^\top (\bI_{N_{1g}} \otimes \bW_2^{(\cdot, \cR_h^{(2)}) \top} ) \mY_{g \cdot,( t-1)}\} , \\
				& \bs_{22, gh} = \lim_{n, T\to \infty} (n^2 T)^{-1} \sum_{t}E\{ \mY_{\cdot h, (t-1)}^\top (\bI_{N_{2h}} \otimes \bW_1^{(\cR_{g}^{(1)}, \cdot) })^\top  ((\bX_t^{(2)})^{(\cR_h^{(2)}, \cdot)} \otimes \one_{N_{1g}})\}, \\
				& \bs_{23, gh} = \lim_{n, T\to \infty} (n^2 T)^{-1} \sum_{t} E\{((\bX_t^{(2)})^{(\cR_h^{(2)}, \cdot)} \otimes \one_{N_{1g}})^\top  ( \one_{N_{2h}} \otimes  (\bX_t^{(1)})^{(\cR_g^{(1)}, \cdot)})\}.
			\end{align*}
			Since $\mY_{\cdot h, (t-1)}$ is independent with $\bX_t^{(2)}$, we have $\bs_{22} = \zero$, and 
			due to the independence between $\bX_t^{(1)}$ and $\bX_t^{(2)}$, we have $\bs_{23} = \zero$.
			Note that 
			\begin{align*}
				s_{21, gh} =  \lim_{n, T\to \infty} (n^2 T)^{-1} \sum_{t} \big[\tr(\cW_{gh}^{1} \bSigma^{\bX}_{\cdot h, g\cdot} \cW_{gh}^{2\top}) + \tr(\cW_{gh}^{1} \bSigma^{\mE}_{\cdot h, g\cdot} \cW_{gh}^{2\top}) \big] = \nu_{gh, gh}^{12} + \kappa_{gh, gh}^{12},
			\end{align*}
			we obtain the final expression.

			\noindent
			{\bf (3). Expression of $\bs_3$}.

			We show that $\bs_3 = ( (s_{31, gh} I(g = g'), \bs_{32, gh}) : g \in [G_1], \mI_{g'h} \in [G_1 G_2])$ with $\mI_{g'h} = (h-1) G_1 + g'$, where $s_{31, gh} = \nu^1_{gh, gh} + \kappa^1_{gh, gh}$ and $\bs_{32, gh} = \zero$.
			Since $\bM^{(1\alpha)} = ( \bM_{g \mI_{g'h}}^{(1 \alpha)} : g \in [G_1], \mI_{g'h} \in [G_1 G_2])$ and $\bM_{g \mI_{g'h}}^{(1 \alpha)} = \sum_t \mX_{gh, t}^{(1)\top} \mY_{gh, (t-1)} I(g= g')$ with $\mI_{g'h} = (h-1) G_1 + g'$,
			we derive the expression of $\lim_{n,T \to \infty} (n^2 T)^{-1} \sum_t \mX_{gh, t}^{(1)\top} \mY_{gh, (t-1)}$, and we have
			\begin{align*}
				& s_{31, gh} = \lim_{n, T\to \infty} (n^2 T)^{-1} \sum_t E\{ \mY_{\cdot h, t}^\top (\bI_{N_{2h}} \otimes \bW_1^{(\cR_g^{(1)}, \cdot)} )^\top \mY_{g h, (t-1)}\}, \\
				& \bs_{32, gh} = \lim_{n, T\to \infty} (n^2 T)^{-1} \sum_t E\{ ( \one_{N_{2h}} \otimes  (\bX_t^{(1)})^{(\cR_g^{(1)}, \cdot)} )^\top \mY_{g h, (t-1)}\}.
			\end{align*}
			Since $\mY_{(t-1)}$ is independent with $\bX_{t}^{(1)}$, we have $\bs_{32, gh} = \zero$.
			Note that 
			\begin{align*}
				s_{31, gh} & = \lim_{n, T\to \infty} (n^2 T)^{-1} \sum_t E\{ \mY_{\cdot h, (t-1)}^\top (\bI_{N_{2h}} \otimes \bW_1^{(\cR_g^{(1)}, \cdot)} )^\top \mY_{g h, (t-1)}\} \\
				&  = \lim_{n, T\to \infty} (n^2 T)^{-1} \sum_t \big[  \tr(\cW_{gh}^{(1)} \bSigma_{\cdot h,g h}^{\bX} ) + \tr(\cW_{gh}^{(1)} \bSigma_{\cdot h,g h}^{\mE})\big] = \nu^1_{gh, gh} + \kappa^1_{gh, gh}.
			\end{align*}
			Therefore, we obtain the final expression.
			
			\noindent
			{\bf (4). Expression of $\bs_4$}.
			
			We next show that $\bs_4 = ( s_{4,gh} I(g= g', h = h'): \mI_{gh} \in [G_1 G_2], \mI_{g'h'} \in [G_1 G_2])$ with $s_{4, gh} = \nu_{gh, gh}^{\alpha} + \kappa_{gh, gh}^{\alpha}$.
			Since $\bM^{\alpha} = (\bM_{\mI_{gh} \mI_{g'h'}}^{\alpha}: \mI_{gh} \in [G_1 G_2], \mI_{g'h'} \in [G_1 G_2])$, where $\bM_{\mI_{gh} \mI_{g'h'}}^{\alpha} = \sum_t \mY_{g h, (t-1)}^\top  \mY_{g h, (t-1)} I(g = g', h = h')$,
			we derive $s_{4, gh} = \lim_{n,T \to \infty} (n^2 T)^{-1} \sum_t \\ E (\sum_t \mY_{gh, (t-1)}^\top \mY_{gh, (t-1)})$.
			\begin{align*}
				s_{4, gh}& = \lim_{n,T \to \infty} (n^2 T)^{-1} \sum_ t E(\mY_{gh, (t-1)}^\top \mY_{gh, (t-1)}) \\
				& = \lim_{n,T \to \infty} (n^2 T)^{-1} \sum_ t \big[\tr(\bSigma_{g h,g h}^{\bX}) + \tr(\bSigma_{g h,g h}^{\mE})\big] = \nu_{gh, gh}^{\alpha} + \kappa_{gh, gh}^{\alpha}.
			\end{align*}
			Therefore, we obtain the final expression.
		\end{proof}

	\subsection{Additional Discussion of Weighted Least Squares Estimation with Group-specific Error Variances}\label{sec:wlse}

	\subsubsection{Technical Assumption}\label{subsec:assum_wlse}

	We need the Assumption \ref{assum:noise_cluster} in Appendix \ref{subsec:assum_wlse} instead of the Assumption \ref{assum:sub_gaussian} in Section \ref{sec:theory}.
	
		\begin{assumption}\label{assum:noise_cluster}
		{\sc (Distribution of Noise Term with Group-specific Variance)} Assume $\varepsilon_{ij,t}$ is a zero-mean sub-Gaussian variable, and independently distributed across $i \in [N_1], j \in [N_2]$ and $t \in [T]$ and independent with $\{\bY_s: s \le t-1\}$, $\{\bX_s^{(1)} = (\bx^{(1)}_{is}: i \in [N_1])^\top: s \le t\}$ and $\{\bX^{(2)}_s= (\bx^{(2)}_{js}: j \in [N_2])^\top: s \le t\}$.
		Assume that $\var(\varepsilon_{ij,t}) = \sigma_{ g_{i}^{(1)0} g_{j}^{(2)0}}^2$, where $g_{i}^{(1)0}$ and $g_{j}^{(2)0}$ are true group memberships. It holds $c_1 \le \min_{g^{(1)}, g^{(2)}}\sigma_{g^{(1)} g^{(2)}}^2\le \max_{g^{(1)}, g^{(2)}}\sigma_{g^{(1)} g^{(2)}}^2 \le c_2$, where $c_1 $ and $c_2$ are two finite positive constants.
	\end{assumption}

		\subsubsection{Proof of Theorem \ref{thm:wlse_thm} (i)}\label{subsec:proof_propA1}
		
		\begin{proof}
			We prove the proposition in three steps.
			In step (1), we first show that $d(\bTheta, \bTheta^0) = O_p\{T^{-1} (\log(N_1 N_2)^2\}$ holds under the clustered error case. This is a required condition for the Theorem \ref{thm:h_consistency2}.
			In step (2), we show that when $G_1 \ge G_{1,0}$, $G_2 \ge G_{2,0}$ and $c_{\text{gap}} \gg d(\bTheta, \bTheta^0)$, the three conclusions in Theorem \ref{thm:h_consistency2} still hold.
			In the final step (3), we come to the proposition.
			
			\noindent
			{\bf Step (1). Order of $d(\bTheta, \bTheta^0)$.}
			
			We first show that Lemma \ref{lem:Q_diff} and Lemma \ref{lem:Q_star_diff} holds in step (A), and then show that Theorem \ref{thm:pseudo_dist} holds in step (B).
			
			\noindent
			{\sc Step (A). Lemma \ref{lem:Q_diff} and Lemma \ref{lem:Q_star_diff}.}
			
			First, we show that \eqref{eq:Q_ij_diff} and \eqref{eq:sup_Q_ij_diff} hold in Lemma \ref{lem:Q_diff}.
			To validate \eqref{eq:Q_ij_diff}, it suffices to show \eqref{eq:eps2_conv}--\eqref{eq:X2_conv} hold.
			By scrutinizing the three inequality, we only need to take care of \eqref{eq:eps2_conv} and \eqref{eq:epsX_conv}, which are related to the error term.
			Rewrite the two inequality by
			\begin{align}
				& P\Big(\Big|\frac{1}{T}\sum_t \ve_{ij, t}^2 - \sigma_{g_{i}^{(1)}}^2\nu_{g_{j}^{(2)}}^2\Big|>x/3\Big)\le
				2\exp\{-c_1T\min(x^2, x)\}, \label{eq:eps2_conv_hetero}\\
				& P\Big(\sup_{\|\bxi\|_{\max} < 2R}\Big|\frac{2}{T}\sum_t
				\ve_{ij, t}\cX_{ij,t}^{\top}\bxi\Big|>x/3\Big)\le \exp\Big\{-c_1 \min(Tx^2,T^{1/2}x)+c_3m\Big\} \label{eq:epsX_conv_hetero},
			\end{align}
			where \eqref{eq:eps2_conv_hetero} could be obtained by Lemma \ref{lem:xAx_convex}. Since $E(\varepsilon_{ij,t}) = 0$, \eqref{eq:epsX_conv_hetero} could be proved following the similar procedure to prove \eqref{eq:epsX_conv}.
			Hence, we have
			\begin{align*}
				P\Big(\sup_{\|\bTheta_{ij}\|_{\max} <R} \Big| \frac{1}{T} Q_{ij}(\bTheta_{ij}) - \frac{1}{T} Q^*_{ij}(\bTheta_{ij}) > x \Big| \Big) \le \exp\Big\{ -c_1 \min(T x^2, T^{1/2} x) + c_2 m \Big\}
			\end{align*}
			still holds for clustered error variance.
			Likewise, \eqref{eq:sup_Q_ij_diff} could be validated similarly.
			
			Next, we show that Lemma \ref{lem:Q_star_diff} holds.
			First note that Lemma \ref{lem:tau_max} still holds under Assumption \ref{assum:mixing} and Assumption \ref{assum:station}, and the two assumptions could still be satisfied if the error terms have group-specific variance.
			Hence, \eqref{eq:Qij_star_diff}--\eqref{eq:Qj_star_diff} in Lemma \ref{lem:Q_star_diff} are obtained when Assumption \ref{assum:tau_min} holds.
			
			\noindent
			{\sc Step (B). Theorem \ref{thm:pseudo_dist} holds.}
			
			By Step (A) we know that Lemma \ref{lem:Q_diff} and Lemma \ref{lem:Q_star_diff} hold, then we could borrow the idea of proof for Theorem \ref{thm:pseudo_dist}, and show that $d(\bTheta, \bTheta^0) = O_p\{T^{-1} (\log(N_1 N_2)^2 \}$.
			
			\noindent
			{\bf Step (2). Proof of Theorem \ref{thm:h_consistency2}.}
			
			We prove the two statements in Lemma \ref{lem:h_consistency_prepare} first in Step (A)--(B), and then we prove the statement in 
			Theorem \ref{thm:h_consistency2} in Step (C).
			
			\noindent
			{\sc Step (A). Proof of statement (i) of Lemma \ref{lem:h_consistency_prepare}}. 
			
			Since the definition of $\mN_\eta^{(1)}$ and $\mA_{\eta}(\bxi, g^{(2)0}, \mG_{-1}^0)$ are the same as the case under homogeneity, hence by the similar techniques in the proof of (i) of Theorem \ref{thm:h_consistency2}, one could keep the conclusion under heterogeneity case.
			
			\noindent
			{\sc Step (B). Proof of statement (ii) of Lemma \ref{lem:h_consistency_prepare}}.
			
			First, we show some required lemmas still hold.
			Note that in the clustered error case, the Assumption \ref{assum:tau_min} for convexity and Assumption \ref{assum:group_diff} still maintain.
			Since we have shown that Lemma \ref{lem:Q_diff} and Lemma \ref{lem:Q_star_diff} holds, we next show that Lemma \ref{lem:tau_max} also holds.
			Because in the clustered error case, Assumption \ref{assum:mixing} and Assumption \ref{assum:station} are not affected, Lemma \ref{lem:tau_max} could be proved similarly.
			
			Then it comes to the proof for statement (ii).
			Similar as the proof procedure in Theorem \ref{thm:h_consistency2}, by using Lemma \ref{lem:Q_star_diff} and Assumption \ref{assum:tau_min}, one could have for any $g^{(2)} \notin \mA_\eta(\btheta, g_{j}^{(2)0})$,
			\begin{align}
				\frac{1}{N_1T}Q_{j}^*(\bxi_{g^{(2)}}^{(2)};\bxi_{g^{(1)}}^{(1)},\wh\mG_1(\bxi))-
				\frac{1}{N_1T}Q_{j}^*(\bxi_{g_j^{(2)0}}^{(2)};\bxi_{g^{(1)0}}^{(1)0},\mG_1^0 )\ge \tau_{\min}(c_\gap c_\pi/2 - \eta), \label{eq:Q_star_j_diff_1_hetero}
			\end{align}
			when $N_1 \to \infty$.
			On the other hand, for $\wt g_{j}^{(2)} \in \mA_\eta^{(2)}(\bxi, g_{j}^{(2)0}, \mG_1^0)$, the following still holds,
			\begin{align}
				&\frac{1}{N_1T}Q_{ j}^*(\bxi_{\wt g_j^{(2)}}^{(2)};\bxi_{g^{(1)}}^{(1)},\wh\mG_1(\bxi))-
				\frac{1}{N_1T}Q_{ j}^*(\bxi_{g_j^{(2)0}}^{(2)};\bxi_{g^{(1)0}}^{(1)0},\mG_1^0 ) \le \tau_{\max}\big\{d(\bTheta, \bTheta^0)+\eta\big\}.\label{eq:Q_star_j_diff2_hetero}
			\end{align}
			Combining \eqref{eq:Q_star_j_diff_1_hetero} and \eqref{eq:Q_star_j_diff2_hetero}, and by the order of $d(\bTheta, \bTheta^0)$ derived in Step (1), similar conclusion as in the proof of Theorem \ref{thm:h_consistency2} could be obtained,
			\begin{align*}
				W_{jg^{(2)}}(\bxi) \le 2 I\left(\sup_{i,j}\sup_{\|\bTheta_{ij}\|_{\max} <R}
				\Big|\frac{1}{T}Q_{ij}(\bTheta_{ij}) - \frac{1}{T}Q_{ij}^*(\bTheta_{ij})\Big| \ge \epsilon_\eta/2\right).
			\end{align*}
			Further by Step (1), the results could be obtained.
			
			\noindent
			{\sc Step (C). Proof of statement (iii) of Lemma \ref{lem:h_consistency_prepare}.} 
			
			We first show that Lemma \ref{lem:sig_h_bound} and Proposition \ref{pro:gh_consistency} hold, and then (iii) could be proved.	
			Since Theorem \ref{thm:pseudo_dist} holds, one could use the same procedure to prove Lemma \ref{lem:sig_h_bound}.
			To prove Proposition \ref{pro:gh_consistency}, we need a new assumption for the distribution of noise term, namely Assumption \ref{assum:noise_cluster}.
			Together with the previously stated Lemma \ref{lem:Q_diff}, Lemma \ref{lem:Q_star_diff}, Lemma \ref{lem:sig_h_bound} and Theorem \ref{thm:pseudo_dist}, Proposition \ref{pro:gh_consistency} holds under the Assumption \ref{assum:para_space}--\ref{assum:tau_min}, \ref{assum:mixing}--\ref{assum:station} and \ref{assum:noise_cluster}.
			
			Then it comes to the proof of (iii) of Lemma \ref{lem:h_consistency_prepare}.
			Similar as the proof of (iii) for Theorem \ref{thm:h_consistency2}, first it could be obtained that
			\begin{align*}
				& \max_{\wt g^{(2)}\in [G_2]}\min_{g^{(2)}\in [G_{2,0}]}
				\Big(\|\wh\btheta_{\wt g^{(2)}}^c - \btheta_{g^{(2)}}^{(2)0} \|^2+
				\frac{1}{N_1}\sum_i|\wh\alpha_{\wh g_{i}^{(1)}\wt g^{(2)}} -
				\alpha_{g_{i}^{(1)0} g^{(2)}}^0|^2\Big) \\
				& = O_p(c_\pi^{-1}
				T^{-1}(\log(N_1N_2))^2)
			\end{align*}
			by Proposition \ref{pro:gh_consistency}, and due to Lemma \ref{lem:sig_h_bound}, it also holds that
			\begin{align*}
				& \max_{g^{(2)}\in [G_2,0]}\min_{\wt g^{(2)}\in [G_2]}
				\Big(\|\wh\btheta_{\wt g^{(2)}}^{(2)} - \btheta_{g^{(2)}}^{(2)0} \|^2+
				\frac{1}{N_1}\sum_i|\wh\alpha_{\wh g_{I}^{(1)}\wt g^{(2)}} -
				\alpha_{g_{I}^{(1)0} g^{(2)}}^0|^2\Big) \\
				& = O_p(c_\pi^{-1}
				T^{-1}(\log(N_1N_2))^2).
			\end{align*}
			Then we have $\wh\bxi \in \mN_\eta^{(2)}$ with $\eta =
			\tau_{\min}c_\gap c_\pi/\{8(\tau_{\min}+\tau_{\max})\}$ under the condition that $c_{\gap} \gg T^{-1}\log(N_1N_2)^2$. Then the statements (i) and (ii) in Lemma \ref{lem:h_consistency_prepare} hold, and then one could demonstrate Theorem \ref{thm:h_consistency2} by using the conclusion Lemma \ref{lem:h_consistency_prepare}.
			Other proof details are consistent with those in the i.i.d. error case.
			
			\noindent
			{\bf Step (3). Membership Consistency.}
			
			Similarly as the proof for Corollary \ref{coro:group_consistency}, using the statement in Theorem \ref{thm:h_consistency2} proved in Step (2), the arguments \eqref{eq:R_C_consist} and \eqref{eq:theta_ora} hold.
		\end{proof}

		\subsubsection{Proof of Theorem \ref{thm:wlse_thm} (ii)}\label{subsec:proof_thmA1}

		\begin{proof}
			We prove the result by two steps.
			In step (1), we first show that $|\wh\sigma_{g^{(1)} g^{(2)}}^{-2} - \sigma_{g^{(1)} g^{(2)}}^{-2} | = o_p(1)$ when $\log (N_{1 g^{(1)}} N_{2 g^{(2)}})^2 /T \to 0$.
			Then in step (2), we show that $\sup_{g^{(1)}, g^{(2)}} | \wh\sigma_{g^{(1)} g^{(2)}}^{-2}-\sigma_{g^{(1)} g^{(2)}}^{-2} | = o_p(1)$ under the condition $ \sqrt{T} \gg \log(G_1 G_2 n^2) + m$.

			\noindent
			{\bf Step (1). Order of $|\wh\sigma_{g^{(1)} g^{(2)}}^{-2} - \sigma_{g^{(1)} g^{(2)}}^{-2}|$.}
			
			By Assumption \ref{assum:noise_cluster}, and define the event $\mO_{g^{(1)} g^{(2)}} = \{\wh\sigma_{g^{(1)} g^{(2)}}^2 \ge  \sigma_{g^{(1)} g^{(2)}}^{2}/2\} = \{\wh\sigma_{g^{(1)} g^{(2)}}^{-2} \le 2 \sigma_{g^{(1)} g^{(2)}}^{-2} \le 2c \}$, where $c$ is a positive constant.
			Then, we have
			\begin{align*}
				P\Big\{ |\wh\sigma_{g^{(1)} g^{(2)}}^{-2} - \sigma_{g^{(1)} g^{(2)}}^{-2} | \ge t \Big\} & = P\Big\{ \Big| \sigma_{g^{(1)} g^{(2)}}^{-2}  (\sigma_{g^{(1)} g^{(2)}}^{2} - \wh\sigma_{g^{(1)} g^{(2)}}^{2}) \wh\sigma_{g^{(1)} g^{(2)}}^{-2}  \Big| \ge t \Big\}\\
				& \le P\Big\{ \Big| \sigma_{g^{(1)} g^{(2)}}^{-4}  (\sigma_{g^{(1)} g^{(2)}}^{2} - \wh\sigma_{g^{(1)} g^{(2)}}^{2})\Big| \ge t/2 \Big\} + P(\mO_{g^{(1)} g^{(2)}}^c) \\
				& \le P\Big\{ \Big| (\sigma_{g^{(1)} g^{(2)}}^{2} - \wh\sigma_{g^{(1)} g^{(2)}}^{2})\Big| \ge \frac{t}{2 c^2 }\Big\} + P(\mO_{g^{(1)} g^{(2)}}^c).
			\end{align*}
			Note that for the event $\mO_{g^{(1)} g^{(2)}}^c$, we could calculate the probability as
			\begin{align*}
				P(\mO_{g^{(1)} g^{(2)}}^c) = P \Big\{ \wh\sigma_{g^{(1)} g^{(2)}}^2 < \frac{1}{2} \sigma_{g^{(1)} g^{(2)}}^2 \Big\} & = P\Big\{ \wh\sigma_{g^{(1)} g^{(2)}}^2 - \sigma_{g^{(1)} g^{(2)}}^2 \le -\frac{1}{2} \sigma_{g^{(1)} g^{(2)}}^2 \Big\} \\
				& \le P\Big\{ | \wh\sigma_{g^{(1)} g^{(2)}}^2 - \sigma_{g^{(1)} g^{(2)}}^2 | \ge \frac{1}{2} \sigma_{g^{(1)} g^{(2)}}^2 \Big\} \\
				& \le  P\Big\{ | \wh\sigma_{g^{(1)} g^{(2)}}^2 - \sigma_{g^{(1)} g^{(2)}}^2 | \ge \frac{1}{2c} \Big\}.
			\end{align*}
			Hence, we have
			\begin{align*}
				& P\Big\{ |\wh\sigma_{g^{(1)} g^{(2)}}^{-2} - \sigma_{g^{(1)} g^{(2)}}^{-2} | \ge t \Big\} \\
				& \le P\Big\{ \Big| \sigma_{g^{(1)} g^{(2)}}^{2} - \wh\sigma_{g^{(1)} g^{(2)}}^{2} \Big| \ge \frac{t}{2 c^2}\Big\}  + P\Big\{ | \wh\sigma_{g^{(1)} g^{(2)}}^2 - \sigma_{g^{(1)} g^{(2)}}^2 | \ge \frac{1}{2c} \Big\}\\
				& \le P\Big\{ \Big| \sigma_{g^{(1)} g^{(2)}}^{2} - \wh\sigma_{g^{(1)} g^{(2)}}^{2} \Big| \ge \frac{t}{2 c^2}\Big\}  + P\Big\{ | \wh\sigma_{g^{(1)} g^{(2)}}^2 - \sigma_{g^{(1)} g^{(2)}}^2 | \ge \frac{t}{2c} \Big\} \\
				& \le 2 P\Big\{ \Big| \sigma_{g^{(1)} g^{(2)}}^{2} - \wh\sigma_{g^{(1)} g^{(2)}}^{2} \Big| \ge \frac{t}{C}\Big\}.
			\end{align*}
			when $0<t<1$ and $C = \max(2 c^2, 2c)$.
			Note that
			\begin{align*}
				|\wh\sigma_{g^{(1)} g^{(2)}}^2 - \sigma_{g^{(1)} g^{(2)}}^2| & = (N_{1 g^{(1)}} N_{2 g^{(2)}} T)^{-1} \big| \sum_{i \in \cR_{g^{(1)}}^{(1)}} \sum_{j \in \cR_{g^{(2)}}^{(2)}} \big\{Q_{ij}(\wh\bTheta_{ij})  - Q_{ij}^*(\bTheta_{ij}^0)\big\} \big| \\
				& \le  (N_{1 g^{(1)}} N_{2 g^{(2)}} T)^{-1} \Big[\big| \sum_{i \in \cR_{g^{(1)}}^{(1)}} \sum_{j \in \cR_{g^{(2)}}^{(2)}} \big\{Q_{ij}(\wh\bTheta_{ij})  - Q_{ij}^*(\wh\bTheta_{ij})\big\} \big| \\
				& + \big| \sum_{i \in \cR_{g^{(1)}}^{(1)}} \sum_{j \in \cR_{g^{(2)}}^{(2)}} \big\{Q_{ij}^*(\wh\bTheta_{ij})  - Q_{ij}^*(\bTheta_{ij}^0)\big\} \big| \Big]
			\end{align*}
			Then we derive the order of the two terms on the right hands separately.
			First, by Lemma \ref{lem:Q_diff}, we could directly obtain that
			\begin{align*}
				& (N_{1 g^{(1)}} N_{2 g^{(2)}} T)^{-1} \big| \sum_{i \in \cR_{g^{(1)}}^{(1)}} \sum_{j \in \cR_{g^{(2)}}^{(2)}} \big\{Q_{ij}(\wh\bTheta_{ij})  - Q_{ij}^*(\wh\bTheta_{ij})\big\} \big| \\
				& \le (N_{1 g^{(1)}} N_{2 g^{(2)}} T)^{-1} \big| \sum_{i \in \cR_{g^{(1)}}^{(1)}} \sum_{j \in \cR_{g^{(2)}}^{(2)}} \sup_{\| \bTheta \|_{\max} \le R} \big\{Q_{ij}(\bTheta)  - Q_{ij}^*(\bTheta)\big\} \big| \\
				&\le \sup_{i\in \cR_{g^{(1)}}^{(1)},j \in \cR_{g^{(2)}}^{(2)}} \sup_{\| \bTheta \|_{\max} \le R} \big| T^{-1} \big\{Q_{ij}(\bTheta)  - Q_{ij}^*(\bTheta)\big\} \big|,
			\end{align*}
			and
			\begin{align*}
				& P\Big\{  \sup_{i\in \cR_{g^{(1)}}^{(1)},j \in \cR_{g^{(2)}}^{(2)}} \sup_{\| \bTheta \|_{\max} \le R} \big| T^{-1} \big\{Q_{ij}(\bTheta)  - Q_{ij}^*(\bTheta)\big\} \big| \ge t \Big\} \\
				&  \le N_{1 g^{(1)}} N_{2 g^{(2)}}  \exp\big\{ -c_1 \min(T t^2, \sqrt{T} t) + c_2 m\big\}.
			\end{align*}
			Then, for the second term, one could borrow the idea of the proof of Lemma \ref{lem:Q_star_diff} to obtain,
			\begin{align*}
				& P \Big\{\frac{1}{N_{1 g^{(1)}} N_{2 g^{(2)}} T} \big| \sum_{i \in \cR_{g^{(1)}}^{(1)}} \sum_{j \in \cR_{g^{(2)}}^{(2)}} \big\{Q_{ij}^*(\wh\bTheta_{ij})  - Q_{ij}^*(\bTheta_{ij}^0)\big\} \big| \ge t \Big\} \\
				& \le N_{1 g^{(1)}} N_{2 g^{(2)}}  \exp\big\{ -c_1 \min(T t^2, \sqrt{T} t) + c_2 m\big\}.
			\end{align*}
			Hence, we have
			\begin{align*}
				P\Big\{ |\wh\sigma_{g^{(1)} g^{(2)}}^{-2} - \sigma_{g^{(1)} g^{(2)}}^{-2} | \ge t \Big\} & \le2 P\Big\{ \Big| \sigma_{g^{(1)} g^{(2)}}^{2} - \wh\sigma_{g^{(1)} g^{(2)}}^{2} \Big| \ge \frac{t}{C}\Big\} \\
				& \le N_{1 g^{(1)}} N_{2 g^{(2)}}  \exp\big\{ -c_1 \min(T t^2, \sqrt{T} t) + c_2 m\big\}.
			\end{align*}
			
			\noindent
			{\bf Step (2). Order of $\sup_{g^{(1)} g^{(2)}} |\wh\sigma_{g^{(1)} g^{(2)}}^{-2} - \sigma_{g^{(1)} g^{(2)}}^{-2} |$.}
			
			By step (1), we have
			\begin{align*}
				P\Big\{ \sup_{g^{(1)} g^{(2)}} |\wh\sigma_{g^{(1)} g^{(2)}}^{-2} - \sigma_{g^{(1)} g^{(2)}}^{-2} | \ge t \Big\} \le  G_1 G_2 n^2 \exp \{ -c_1 \min(T t^2, \sqrt{T} t) + c_2 m \},
			\end{align*}
			where $n$ is defined in the statement of Theorem \ref{thm:wlse_thm}.
			This implies that $\sup_{g^{(1)} g^{(2)}} |\wh\sigma_{g^{(1)} g^{(2)}}^{-2} - \sigma_{g^{(1)} g^{(2)}}^{-2} | = O_p\{ (\log (G_1 G_2 n^2) + m)/\sqrt{T} \}$, and thus the proof is completed.
		\end{proof}

		\subsubsection{Proof of Theorem \ref{thm:wlse_thm} (iii)}\label{subsec:proof_thmA2}
		
		\begin{proof}
			Note that
			\begin{align*}
				n T^{1/2} (\wt\btheta^w - \btheta^0) = n T^{1/2}   \widecheck\bM^{-1} \widecheck\bdelta =((n^2 T)^{-1}\widecheck{\bM})^{-1} (n T^{1/2})^{-1} \widecheck\bdelta \defeq \widecheck\bM_{nT}^{-1} \bLambda (n T^{1/2})^{-1} \widecheck\bdelta,
			\end{align*}
			where $\widecheck\bdelta =  (\widecheck\bdelta^{(1)\top}, \widecheck\bdelta^{(2)\top}, \widecheck\bdelta^{\alpha\top})^\top$ and
			\begin{align*}
				&\widecheck\bdelta_{g^{(1)}}^{(1)} = \sum_{t, g^{(2)}} (\wh\sigma_{g^{(1)} g^{(2)}})^{-2}  \mX_{g^{(1)}g^{(2)},t}^{(1)\top} \E_{g^{(1)}g^{(2)},t},~~ \widecheck\bdelta^{(1)} = (\widecheck\bdelta_g^{(1)\top}: g^{(1)}\in [G_1])^\top,\\
				& \widecheck\bdelta_{g^{(2)}}^{(2)} = \sum_{t, g^{(1)}} (\wh\sigma_{g^{(1)} g^{(2)}})^{-2}  \mX_{g^{(1)}g^{(2)},t}^{(2)\top } \E_{g^{(1)}g^{(2)},t},~~\widecheck\bdelta^{(2)} = (\widecheck\bdelta_{g^{(2)}}^{(2)\top}: g^{(2)}\in [G_2])^\top,\\
				& \widecheck\bdelta_{\mI_{g^{(1)} g^{(2)}}}^\alpha = \sum_t (\wh\sigma_{g^{(1)} g^{(2)}})^{-2} \mY_{g^{(1)} g^{(2)}(t-1)}^\top
				\E_{g^{(1)}g^{(2)},t}, ~~\widecheck\bdelta^\alpha = (\widecheck\bdelta_{\mI_{g^{(1)} g^{(2)}}}^{\alpha\top}:\mI_{g^{(1)} g^{(2)}}\in [g^{(1)} g^{(2)}])^\top.
			\end{align*}
			Then, we have
			\begin{align*}
				n T^{1/2}\bfeta^\top (\wt\btheta^w - \btheta^0)& = \bfeta^\top \widecheck\bM_{nT}^{-1} (n T^{1/2})^{-1} \widecheck\bdelta \\
				& = \bfeta^\top \{ \widecheck\bM_{nT}^{-1} - (\wt\bM_{nT}^0)^{-1} \} (n T^{1/2})^{-1} \widecheck\bdelta + \bfeta^\top  (\wt\bM_{nT}^0)^{-1} (n T^{1/2})^{-1} \widecheck\bdelta \\
				& = \bfeta^\top  (\wt\bM_{nT}^0)^{-1} (n T^{1/2})^{-1} \widecheck\bdelta + \bfeta^\top  \widecheck\bM_{nT}^{-1} \{  \wt\bM_{nT}^0 - \widecheck\bM_{nT} \}(\wt\bM_{nT}^0)^{-1}  (n T^{1/2})^{-1} \widecheck\bdelta \\
				& = \bfeta^\top  (\wt\bM_{nT}^0)^{-1} (n T^{1/2})^{-1} \wt\bdelta + \bfeta^\top  (\wt\bM_{nT}^0)^{-1} (n T^{1/2})^{-1} (\widecheck\bdelta - \wt\bdelta ) \\
				& + \bfeta^\top  \widecheck\bM_{nT}^{-1} \{  \wt\bM_{nT}^0 - \widecheck\bM_{nT} \}(\wt\bM_{nT}^0)^{-1}  (n T^{1/2})^{-1} \widecheck\bdelta.
			\end{align*}
			Hence, it suffices to show that
			\begin{align}
				&\bfeta^\top  (\wt\bM_{nT}^0)^{-1} (n T^{1/2})^{-1} \wt\bdelta  \to_d
				N(0, \bfeta^\top (\wt\bM^0)^{-1}\bfeta),\label{eq:wt_normal}\\
				&  \bfeta^\top  (\wt\bM_{nT}^0)^{-1} (n T^{1/2})^{-1} (\widecheck\bdelta - \wt\bdelta ) = o_p(1),\label{eq:check_wt_op1}\\
				& \bfeta^\top  \widecheck\bM_{nT}^{-1} \{  \wt\bM_{nT}^0 - \widecheck\bM_{nT} \}(\wt\bM_{nT}^0)^{-1}  (n T^{1/2})^{-1} \widecheck\bdelta = o_p(1).\label{eq:check_op1}
			\end{align}

While it is obvious that \eqref{eq:check_op1} holds when $n, T \to \infty$, we prove the \eqref{eq:wt_normal}--\eqref{eq:check_wt_op1} in the following two steps.

			\noindent
			{\bf 1. Proof of (\ref{eq:wt_normal})}
			
			By noting that \eqref{eq:wt_normal} and \eqref{eq:normal1} have similar forms, one could borrow the idea from the proof for \eqref{eq:normal1} with the new Assumption \ref{assum:noise_cluster}. Then we could obtain the similar results by using the central limit theorem of the martingale difference array.

			\noindent
			{\bf 2. Proof of (\ref{eq:check_wt_op1})}
			
			Note that $|\bfeta^\top (\wt\bM_{nT}^0)^{-1} (n T^{1/2})^{-1} (\widecheck{\bdelta} - \wt\bdelta)| \le \|(\wt\bM_{nT}^0)^{-1} \bfeta\| \|(n T^{1/2})^{-1} (\widecheck{\bdelta} - \wt\bdelta)\|$, and we have assumed that $\lambda_{\min}(\bM^0) >0 $ and $\max_{g^{(1)} g^{(2)}}\sigma_{g^{(1)} g^{(2)}}^2 < \infty$, hence it holds that $\|(\wt\bM_{nT}^0)^{-1} \bfeta\| < c < \infty$.
			Then to derive the order of $\|(n T^{1/2})^{-1} (\widecheck{\bdelta} - \wt\bdelta)\|$, we focus on the order of $(n^2 T)^{-1} (\widecheck\bdelta - \wt\bdelta)^\top (\widecheck\bdelta - \wt\bdelta)$.
			We take the term $\widecheck\bdelta^{(1)}\in \mR^{G  \times (p_1 + 1)}$ and $\wt\bdelta^{(1)} \in \mR^{G \times(p_1 + 1)}$ for example to illustrate.
			Note that
			\begin{align*}
				& (n^2 T)^{-1} (\widecheck\bdelta^{(1)} - \wt\bdelta^{(1)})^\top (\widecheck\bdelta^{(1)} - \wt\bdelta^{(1)}) \\
				& = (n^2 T)^{-1}  \sum_{g^{(1)}, g^{(2)}, t} (\wh\sigma_{g^{(1)} g^{(2)}}^{-2} - \sigma_{g^{(1)} g^{(2)}}^{-2})^2 \E_{g^{(1)}g^{(2)},t}^\top  \mX^{(1)}_{g^{(1)}g^{(2)},t} \mX_{g^{(1)}g^{(2)},t}^{(1)\top} \E_{g^{(1)}g^{(2)},t}\\
				& \le (n^2 T)^{-1} \sum_{g^{(1)}, g^{(2)}, t} \big\{\sup_{g^{(1)} g^{(2)}}(\wh\sigma_{g^{(1)} g^{(2)}}^{-2} - \sigma_{g^{(1)} g^{(2)}}^{-2})^2 \big\} \E_{g^{(1)} g^{(2)}, t}^\top  \mX^{(1)}_{g^{(1)} g^{(2)}, t} \mX_{g^{(1)} g^{(2)}, t}^{(1)\top} \E_{g^{(1)} g^{(2)}, t}.
			\end{align*}
			By Theorem \ref{thm:wlse_thm} (ii), one has $\sup_{g^{(1)} g^{(2)}}(\wh\sigma_{g^{(1)} g^{(2)}}^{-2} - \sigma_{g^{(1)} g^{(2)}}^{-2})^2  = O_p\{ T^{-1} (m + \log (G_1 G_2 n^2) \}$,
			hence we next derive the order of $(n^2 T)^{-1} \sum_{g^{(1)}, g^{(2)}, t} E (\E_{g^{(1)} g^{(2)}, t}^\top  \mX_{g^{(1)} g^{(2)}, t}^{(1)} \mX_{g^{(1)} g^{(2)}, t}^{(1)\top} \E_{g^{(1)} g^{(2)}, t})$.

			By Assumption \ref{assum:noise_cluster} and $ \max_{g^{(1)} g^{(2)}}\sigma_{g^{(1)} g^{(2)}} \le c$, we know that with $\bfeta_1 \in \mR^{p+1}$ and $\|\bfeta_1\| = 1$,
			\begin{align*}
				& (n^2 T)^{-1} \sum_{g^{(1)}, g^{(2)}, t} E (\E_{g^{(1)} g^{(2)}, t}^\top  \mX_{g^{(1)} g^{(2)}, t}^{(1)} \mX_{g^{(1)} g^{(2)}, t}^{(1)\top} \E_{g^{(1)} g^{(2)}, t} )\\
				& = (n^2 T)^{-1}  \sum_{g^{(1)}, g^{(2)}, t} E \{\sigma_{g^{(1)} g^{(2)}}^2  \tr( \mX_{g^{(1)} g^{(2)}, t}^{(1)} \mX_{g^{(1)} g^{(2)}, t}^{(1)\top}) \}\nonumber \\
				& \le c^2 (n^2 T)^{-1} \sum_{g^{(1)}, g^{(2)}, t} E \{ \tr \big(\bW_1^{(\cR_{g^{(1)}}^{(1)}, \cdot)} \bY_t^{(\cdot, \cR_{g^{(2)}}^{(2)})}  \bY_t^{(\cdot, \cR_{g^{(2)}}^{(2)})\top} \bW_1^{(\cR_{g^{(1)}}^{(1)}, \cdot)\top} \big)\}  \nonumber \\
				& + c^2 (n^2 T)^{-1} \sum_{g^{(1)}, g^{(2)}, t} E \{ \vec(\one_{N_{2 g^{(2)}}} \otimes (\bX_t^{(1)})^{(\cR_{g^{(1)}}^{(1)}, \cdot)})^\top \vec(\one_{N_{2 g^{(2)}}} \otimes (\bX_t^{(1)})^{(\cR_{g^{(1)}}^{(1)}, \cdot)}) \} \nonumber \\
				& =  c^2 (n^2 T)^{-1} \sum_{t} E \{ \tr \big(\bW_1 \bY_t  \bY_t^{\top} \bW_1^{\top} \big)\} +  c^2N_2   (n^2 T)^{-1} \sum_{t} \sum_{i \in[N_1]} E \|\bx_{it}^{(1)}\|^2.
			\end{align*}
			Note that
			\begin{align*}
				\sum_{t} E \{ \tr \big(\bW_1\bY_t  \bY_t^{\top} \bW_1^\top \big) \}&\le c_1  \sum_{t}  \tr \big(\bW_1 \one_{N_1} \one_{N_2}^\top \one_{N_2} \one_{N_1}^\top \bW_1^{\top} \big) = O(n^2 T),
			\end{align*}
			where $c_1 = E(Y_{ij,t}^2) <\infty$ by Lemma \ref{lem:10}, and also note that
			\begin{align*}
				\sum_{t} \sum_{i \in [N_1]} E \|\bx_{it}^{(1)}\|^2 \le c_K p_1 T N_{1} = O(nTs)
			\end{align*}
			due to the $K$-convexity property of $\bx_{it}^{(1)}$ and the definition of $n$,
			we have that 
			$$(n^2 T)^{-1}  \sum_{g^{(1)}, g^{(2)}, t} (\E_{g^{(1)} g^{(2)}, t}^\top  \mX_{g^{(1)} g^{(2)}, t}^{(1)} \mX_{g^{(1)} g^{(2)}, t}^{(1)\top} \E_{g^{(1)} g^{(2)}, t})  = O_p(1).$$
			Together with $\sup_{g^{(1)} g^{(2)}}(\wh\sigma_{g^{(1)} g^{(2)}}^{-2} - \sigma_{g^{(1)} g^{(2)}}^{-2})^2  = O_p\{ T^{-1} (m + \log (GHn^2) \}$, we know that
			\begin{align*}
				(n^2 T)^{-1} (\widecheck\bdelta^{(1)} - \wt\bdelta^{(1)})^\top (\widecheck\bdelta^{(1)} - \wt\bdelta^{(1)}) = O_p \{(m+ \log(G_1 G_2 n^2)) /T \} = o_p(1)
			\end{align*}
			when $T \gg (\log(G_1 G_2 n^2))$ due to that $m$ is a fixed constant.

			By step 1--2, we could arrive at the final conclusion.
			
		\end{proof}

\subsubsection{Simulation Results of Weighted Least Squares Estimation}\label{subsec:simu:WLS}

In this section, we conduct the simulation experiment when $G_{1,0} = G_{2,0} = 3$. We set the group-specific error variances as $\wt\bSigma = (\sigma_{g^{(1)} g^{(2)}}: g^{(1)} \in [G_{1,0}], g^{(2)} \in [G_{2,0}]) \in\mR^{3 \times 3}$ as
\begin{align*}
	\wt\bSigma=\left(\begin{array}{ccc}
		1 & 2 & 3 \\
		4 & 5 & 6 \\
		7 & 8 & 9
	\end{array}\right).
\end{align*}
Then we set $(N_1, N_2) \in \{(100, 80), (200,150)\}$ and set $T \in\{20, 40, 80\}$.
Next, we repeat the experiment for $R = 300$ times, and show the RMSEs results of $\wt\btheta^w$ in Table \ref{tbl:wls}.
From the table, we first focus on the estimation results under weighted least squares method (marked as ``WLS'' in the table).
We could see that when the error has group-specific variances, our proposed estimation could estimated the true group memberships with high accuracy as the sample size grows.
Furthermore, we note that RMSEs of all parameters decrease toward the corresponding oracle results when the sample sizes are large.
For the inference results of WLS, when both $N_1, N_2, T$ are large, the CPs are all around 0.95.
Then, we compare the results of WLS and the ordinary method in the main text (marked as ``OLS'').
One could see that the RMSEs of WLS estimator are always smaller than those of OLS estimator, and the RMSEs of the corresponding oracle estimators have the same pattern.
This shows that when the error has group-specific variances, our proposed weighted least squares estimator outperforms the ordinary least squares estimator.

\begin{sidewaystable}
	\centering
	\caption{RMSEs ($\times 1000$) of estimated parameters under scenario 3 ($G_{1,0} =3, G_{2,0} = 3$) with 300 replications. The performances are evaluated for different sample sizes $N_1, N_2$ and the time length $T$. ``OLS'' means the results estimated using the ordinary least squares method proposed in Section \ref{sec:estimate}, ``WLS'' means the results using weighted least squares proposed in Appendix \ref{sec:wlse}.
		The corresponding CPs are shown in the parenthesis.}
	\label{tbl:wls}
	\scalebox{0.75}{
		\begin{tabular}{cc|c|c|ccccc|ccccc|cc}
			\hline
			$N_1$                & $N_2$                & $T$                 & Estimation & $\wh\blambda^{(1)}$                                                & $\wh\blambda^{(2)}$                                                 & $\wh\bzeta^{(1)}$                                                   & $\wh\bzeta^{(2)}$                                                  & $\wh\balpha$                                                  & $\wh\blambda^{(1)\text{or}}$                                    & $\wh\blambda^{(2)\text{or}}$                                     & $\wh\bzeta^{(1)\text{or}}$                                      & $\wh\bzeta^{(2)\text{or}}$                                     & $\wh\balpha^{\text{or}}$                                     & $\eta_1$                & $\eta_2$                \\ \hline
			\multirow{11}{*}{100} & \multirow{11}{*}{80}  & \multirow{3}{*}{20} & OLS        & \begin{tabular}[c]{@{}c@{}}37.34\\      (0.671)\end{tabular} & \begin{tabular}[c]{@{}c@{}}38.54\\      (0.637)\end{tabular} & \begin{tabular}[c]{@{}c@{}}128.98\\      (0.746)\end{tabular} & \begin{tabular}[c]{@{}c@{}}107.75\\      (0.822)\end{tabular} & \begin{tabular}[c]{@{}c@{}}104.58\\      (0.515)\end{tabular} & \begin{tabular}[c]{@{}c@{}}22.8\\      (0.864)\end{tabular}  & \begin{tabular}[c]{@{}c@{}}16.12\\      (0.919)\end{tabular} & \begin{tabular}[c]{@{}c@{}}70.85\\      (0.944)\end{tabular} & \begin{tabular}[c]{@{}c@{}}72.94\\      (0.924)\end{tabular} & \begin{tabular}[c]{@{}c@{}}22.17\\      (0.914)\end{tabular} & \multirow{2}{*}{0.3025} & \multirow{2}{*}{0.0930} \\
			&                      &                     & WLS        & \begin{tabular}[c]{@{}c@{}}36.64\\      (0.657)\end{tabular} & \begin{tabular}[c]{@{}c@{}}38.57\\      (0.676)\end{tabular} & \begin{tabular}[c]{@{}c@{}}99.58\\      (0.541)\end{tabular}  & \begin{tabular}[c]{@{}c@{}}106.13\\      (0.87)\end{tabular}  & \begin{tabular}[c]{@{}c@{}}103.84\\      (0.462)\end{tabular} & \begin{tabular}[c]{@{}c@{}}15.61\\      (0.966)\end{tabular} & \begin{tabular}[c]{@{}c@{}}14.92\\      (0.949)\end{tabular} & \begin{tabular}[c]{@{}c@{}}40.94\\      (0.948)\end{tabular} & \begin{tabular}[c]{@{}c@{}}69.28\\      (0.952)\end{tabular} & \begin{tabular}[c]{@{}c@{}}21.87\\      (0.945)\end{tabular} &                         &                         \\ \cline{3-16}
			&                      & \multirow{3}{*}{40} & OLS        & \begin{tabular}[c]{@{}c@{}}24.79\\      (0.71)\end{tabular}  & \begin{tabular}[c]{@{}c@{}}12.47\\      (0.919)\end{tabular} & \begin{tabular}[c]{@{}c@{}}87.59\\      (0.774)\end{tabular}  & \begin{tabular}[c]{@{}c@{}}52.22\\      (0.93)\end{tabular}   & \begin{tabular}[c]{@{}c@{}}46.10\\      (0.696)\end{tabular}  & \begin{tabular}[c]{@{}c@{}}15.10\\      (0.883)\end{tabular} & \begin{tabular}[c]{@{}c@{}}11.18\\      (0.93)\end{tabular}  & \begin{tabular}[c]{@{}c@{}}51.11\\      (0.945)\end{tabular} & \begin{tabular}[c]{@{}c@{}}50.26\\      (0.936)\end{tabular} & \begin{tabular}[c]{@{}c@{}}14.68\\      (0.936)\end{tabular} & \multirow{2}{*}{0.1960} & \multirow{2}{*}{0.0077} \\
			&                      &                     & WLS        & \begin{tabular}[c]{@{}c@{}}24.45\\ (0.673)\end{tabular}      & \begin{tabular}[c]{@{}c@{}}12.02\\      (0.928)\end{tabular} & \begin{tabular}[c]{@{}c@{}}68.28\\      (0.576)\end{tabular}  & \begin{tabular}[c]{@{}c@{}}49.96\\      (0.951)\end{tabular}  & \begin{tabular}[c]{@{}c@{}}45.45\\      (0.635)\end{tabular}  & \begin{tabular}[c]{@{}c@{}}11.20\\      (0.952)\end{tabular} & \begin{tabular}[c]{@{}c@{}}10.23\\      (0.947)\end{tabular} & \begin{tabular}[c]{@{}c@{}}28.77\\      (0.947)\end{tabular} & \begin{tabular}[c]{@{}c@{}}47.34\\      (0.954)\end{tabular} & \begin{tabular}[c]{@{}c@{}}14.53\\      (0.955)\end{tabular} &                         &                         \\ \cline{3-16}
			&                      & \multirow{3}{*}{80} & OLS        & \begin{tabular}[c]{@{}c@{}}15.86\\      (0.774)\end{tabular} & \begin{tabular}[c]{@{}c@{}}8.31\\      (0.949)\end{tabular}  & \begin{tabular}[c]{@{}c@{}}53.54\\      (0.87)\end{tabular}   & \begin{tabular}[c]{@{}c@{}}33.93\\      (0.952)\end{tabular}  & \begin{tabular}[c]{@{}c@{}}22.80\\      (0.861)\end{tabular}  & \begin{tabular}[c]{@{}c@{}}9.39\\      (0.909)\end{tabular}  & \begin{tabular}[c]{@{}c@{}}8.30\\      (0.952)\end{tabular}  & \begin{tabular}[c]{@{}c@{}}39.10\\      (0.952)\end{tabular} & \begin{tabular}[c]{@{}c@{}}33.94\\      (0.952)\end{tabular} & \begin{tabular}[c]{@{}c@{}}10.71\\      (0.95)\end{tabular}  & \multirow{2}{*}{0.0877} & \multirow{2}{*}{0.0000} \\
			&                      &                     & WLS        & \begin{tabular}[c]{@{}c@{}}14.50\\      (0.783)\end{tabular} & \begin{tabular}[c]{@{}c@{}}7.61\\      (0.949)\end{tabular}  & \begin{tabular}[c]{@{}c@{}}39.13\\      (0.765)\end{tabular}  & \begin{tabular}[c]{@{}c@{}}32.16\\      (0.947)\end{tabular}  & \begin{tabular}[c]{@{}c@{}}22.56\\      (0.81)\end{tabular}   & \begin{tabular}[c]{@{}c@{}}7.24\\      (0.948)\end{tabular}  & \begin{tabular}[c]{@{}c@{}}7.38\\      (0.959)\end{tabular}  & \begin{tabular}[c]{@{}c@{}}22.81\\      (0.946)\end{tabular} & \begin{tabular}[c]{@{}c@{}}31.87\\      (0.946)\end{tabular} & \begin{tabular}[c]{@{}c@{}}10.63\\      (0.949)\end{tabular} &                         &                         \\ \hline
			\multirow{11}{*}{200} & \multirow{11}{*}{150} & \multirow{3}{*}{20} & OLS        & \begin{tabular}[c]{@{}c@{}}23.50\\      (0.584)\end{tabular} & \begin{tabular}[c]{@{}c@{}}12.61\\      (0.818)\end{tabular} & \begin{tabular}[c]{@{}c@{}}87.96\\      (0.616)\end{tabular}  & \begin{tabular}[c]{@{}c@{}}41.33\\      (0.904)\end{tabular}  & \begin{tabular}[c]{@{}c@{}}52.51\\      (0.562)\end{tabular}  & \begin{tabular}[c]{@{}c@{}}11.6\\      (0.876)\end{tabular}  & \begin{tabular}[c]{@{}c@{}}8.55\\      (0.912)\end{tabular}  & \begin{tabular}[c]{@{}c@{}}36.19\\      (0.947)\end{tabular} & \begin{tabular}[c]{@{}c@{}}38.21\\      (0.929)\end{tabular} & \begin{tabular}[c]{@{}c@{}}10.66\\      (0.924)\end{tabular} & \multirow{2}{*}{0.2564} & \multirow{2}{*}{0.0189} \\
			&                      &                     & WLS        & \begin{tabular}[c]{@{}c@{}}25.15\\      (0.524)\end{tabular} & \begin{tabular}[c]{@{}c@{}}12.5\\      (0.827)\end{tabular}  & \begin{tabular}[c]{@{}c@{}}79.29\\      (0.409)\end{tabular}  & \begin{tabular}[c]{@{}c@{}}40.03\\      (0.94)\end{tabular}   & \begin{tabular}[c]{@{}c@{}}51.98\\      (0.47)\end{tabular}   & \begin{tabular}[c]{@{}c@{}}8.47\\      (0.95)\end{tabular}   & \begin{tabular}[c]{@{}c@{}}8.09\\      (0.933)\end{tabular}  & \begin{tabular}[c]{@{}c@{}}20.16\\      (0.957)\end{tabular} & \begin{tabular}[c]{@{}c@{}}36.66\\      (0.95)\end{tabular}  & \begin{tabular}[c]{@{}c@{}}10.63\\      (0.954)\end{tabular} &                         &                         \\ \cline{3-16}
			&                      & \multirow{3}{*}{40} & OLS        & \begin{tabular}[c]{@{}c@{}}12.26\\      (0.706)\end{tabular} & \begin{tabular}[c]{@{}c@{}}5.48\\      (0.928)\end{tabular}  & \begin{tabular}[c]{@{}c@{}}40.14\\      (0.84)\end{tabular}   & \begin{tabular}[c]{@{}c@{}}26.10\\      (0.929)\end{tabular}  & \begin{tabular}[c]{@{}c@{}}16.32\\      (0.79)\end{tabular}   & \begin{tabular}[c]{@{}c@{}}7.85\\      (0.867)\end{tabular}  & \begin{tabular}[c]{@{}c@{}}5.48\\      (0.93)\end{tabular}   & \begin{tabular}[c]{@{}c@{}}26.85\\      (0.945)\end{tabular} & \begin{tabular}[c]{@{}c@{}}26.10\\      (0.928)\end{tabular} & \begin{tabular}[c]{@{}c@{}}7.59\\      (0.933)\end{tabular}  & \multirow{2}{*}{0.1033} & \multirow{2}{*}{0.0000} \\
			&                      &                     & WLS        & \begin{tabular}[c]{@{}c@{}}11.06\\      (0.946)\end{tabular} & \begin{tabular}[c]{@{}c@{}}5.09\\      (0.942)\end{tabular}  & \begin{tabular}[c]{@{}c@{}}32.77\\      (0.673)\end{tabular}  & \begin{tabular}[c]{@{}c@{}}25.12\\      (0.949)\end{tabular}  & \begin{tabular}[c]{@{}c@{}}16.07\\      (0.74)\end{tabular}   & \begin{tabular}[c]{@{}c@{}}5.66\\      (0.95)\end{tabular}   & \begin{tabular}[c]{@{}c@{}}4.98\\      (0.94)\end{tabular}   & \begin{tabular}[c]{@{}c@{}}15.58\\      (0.952)\end{tabular} & \begin{tabular}[c]{@{}c@{}}24.89\\      (0.949)\end{tabular} & \begin{tabular}[c]{@{}c@{}}7.54\\      (0.95)\end{tabular}   &                         &                         \\ \cline{3-16}
			&                      & \multirow{3}{*}{80} & OLS        & \begin{tabular}[c]{@{}c@{}}5.89\\      (0.863)\end{tabular}  & \begin{tabular}[c]{@{}c@{}}4.21\\      (0.910)\end{tabular}  & \begin{tabular}[c]{@{}c@{}}17.95\\      (0.944)\end{tabular}  & \begin{tabular}[c]{@{}c@{}}19.14\\      (0.920)\end{tabular}  & \begin{tabular}[c]{@{}c@{}}5.49\\      (0.919)\end{tabular}   & \begin{tabular}[c]{@{}c@{}}5.80\\      (0.864)\end{tabular}  & \begin{tabular}[c]{@{}c@{}}4.21\\      (0.910)\end{tabular}  & \begin{tabular}[c]{@{}c@{}}17.71\\      (0.949)\end{tabular} & \begin{tabular}[c]{@{}c@{}}19.14\\      (0.920)\end{tabular} & \begin{tabular}[c]{@{}c@{}}5.29\\      (0.922)\end{tabular}  & \multirow{2}{*}{0.0078} & \multirow{2}{*}{0.0000} \\
			&                      &                     & WLS        & \begin{tabular}[c]{@{}c@{}}4.12\\      (0.950)\end{tabular}  & \begin{tabular}[c]{@{}c@{}}3.90\\      (0.949)\end{tabular}  & \begin{tabular}[c]{@{}c@{}}10.83\\      (0.939)\end{tabular}  & \begin{tabular}[c]{@{}c@{}}18.19\\      (0.948)\end{tabular}  & \begin{tabular}[c]{@{}c@{}}5.42\\      (0.944)\end{tabular}   & \begin{tabular}[c]{@{}c@{}}3.99\\      (0.952)\end{tabular}  & \begin{tabular}[c]{@{}c@{}}3.89\\      (0.944)\end{tabular}  & \begin{tabular}[c]{@{}c@{}}10.46\\      (0.954)\end{tabular} & \begin{tabular}[c]{@{}c@{}}18.17\\      (0.948)\end{tabular} & \begin{tabular}[c]{@{}c@{}}5.23\\      (0.951)\end{tabular}  &                         &                         \\ \hline
	\end{tabular}}
\end{sidewaystable}

\section{Specific Expressions of $\bM$ in Equation \eqref{eq:Mb}}\label{sec:M_express}

{\allowdisplaybreaks
	\begin{align*}
		& \bM^{(l)}_{g^{(l)}} = \sum_{t, g^{-(l)}} \mX_{g^{(1)} \cdots g^{(q)},t}^{ (l) \top}  \mX_{g^{(1)} \cdots g^{(q)},t}^{(l)}, ~~~ \bM^{(l)} = \diag\big(\bM^{(l)}_{g^{(l)}}: g^{(l)} \in [G_l] \big) \in \mR^{G_l(p_l+1) \times G_l (p_l + 1)}, \\
		& \bM^{(lm)}_{g^{(l)} g^{(m)}} = \sum_{t, g^{-(l, m)}} \mX_{g^{(1)} \cdots g^{(q)},t}^{(l)\top} \mX_{g^{(1)} \cdots g^{(q)},t}^{(m)}, \\
		&  \bM^{(lm)} = \big(\bM^{(lm)}_{g^{(l)} g^{(m)}}: g^{(l)} \in [G_l], g^{(m)} \in [G_m]\big) \in \mR^{G_l(p_l+1) \times G_m(p_m+1)},\\
		& \bM^{(l\alpha)}_{g^{(l)} \mI_{g^{(l)'} g^{-(l)}}} = \sum_t \mX_{g^{(1)} \cdots g^{(q)},t}^{ (l) \top} \mY_{g^{(1)} \cdots g^{(q)},(t-1)} I(g^{(l)} = g^{(l)'}), \\
		& \text{with}~ \mI_{g^{(l)'} g^{-(l)}} = g^{(l)'} + \sum_{m \neq l}((g^{(m)}-1) G_l),\\
		& \bM^{(l\alpha)} = \Big( \bM^{(l\alpha)}_{g^{(l)} \mI_{g^{(l)'} g^{-(l)}}} : g^{(l)} \in [G_l], \mI_{g^{(l)'} g^{-(l)}} \in \Big[\prod_l G_l \Big] \Big) \in \mR^{G_l(p_l+1) \times (\prod_l G_l)},\\
		& \bM^{\alpha}_{\mI_{g^{(1)}, \cdots, g^{(q)}}, \mI_{g^{(1)'}, \cdots  g^{(q)'}}} = \sum_t \|\mY_{g^{(1)} \cdots g^{(q)},(t-1)} \|^2  I\{(g^{(1)}, \cdots, g^{(q)}) =  (g^{(1)'}, \cdots, g^{(q)'})\}, \\
		& \bM^{\alpha} = \Big( \bM^{\alpha}_{\mI_{g^{(1)}, \cdots, g^{(q)}}, \mI_{g^{(1)'}, \cdots  g^{(q)'}}} : \mI_{g^{(1)}, \cdots, g^{(q)}} \in \Big[\prod_l G_l \Big], \mI_{g^{(1)'}, \cdots  g^{(q)'}} \in \Big[\prod_l G_l \Big]\Big) \\
		&  \in \mR^{(\prod_l G_l) \times (\prod_l G_l)},\\
		& \b_{g^{(l)}}^{(l)} = \sum_{t, g^{-(l)}} \mX_{g^{(1)} \cdots g^{(q)},t}^{(l)\top} \mY_{g^{(1)} \cdots g^{(q)},t}, ~~~ \b^{(l)} = \big(\b_{g^{(l)}}^{(l)}: g^{(l)} \in [G_l]\big) \in \mR^{G_l(p_l+1)},\\
		& \b_{\mI_{g^{(1)}, \cdots, g^{(q)}}}^{\alpha} = \sum_t \mY_{g^{(1)} \cdots g^{(q)},(t-1)}^\top \mY_{g^{(1)} \cdots g^{(q)},t}, \\
		&  \b^\alpha = \Big( \b_{\mI_{g^{(1)}, \cdots, g^{(q)}}}^{\alpha} : \mI_{g^{(1)}, \cdots, g^{(q)}} \in \Big[\prod_l G_l\Big]\Big) \in \mR^{\prod_l G_l},
	\end{align*}
}
where $\sum_{t, g^{-(l, m)}}$ is the simplified notation for $\sum_{\{ g^{(1)}, \cdots, g^{(q)} \} \setminus \{g^{(l)}, g^{(m)}\} }$.

\section{GTNAR Algorithm}\label{subsec:alg}

We summarize the GTNAR algorithm in the following Algorithm \ref{alg:gmnar}.

\begin{algorithm}
	\caption{Estimation of the GTNAR Model}
	\label{alg:gmnar}
	\begin{algorithmic}[1]
		\State {\bf Input:} $\{\cY_t, \bX_t^{(l)}, \bW^{(l)}, G_l\}$ for $1 \le l \le q$.
		\State Obtain initial group memberships $\mG_l^{[0]}$ according to Algorithm \ref{alg:init} in Appendix \ref{sec:init}. Let $\{\bxi^{[k]}, \mG_l^{[k]}\}$ be the estimators and memberships in the $k$th iteration.
		\State Repeat {\sc Step 1} and {\sc Step 2} for {$k = 1, 2, \cdots$}
		until convergence.
		
		{\sc Step 1.} Given $\{\mG_l^{[k-1]}\}$ for all layers $l\in[q]$, calculate
		$\bxi^{[k-1]} = (\bM^{[k-1]})^{-1} \b^{[k-1]}$, where $\bM^{[k-1]}$ and $\b^{[k-1]}$ are obtained from
		(\ref{eq:Mb}) with $\mG_l^{[k-1]}$s specified.
		
		{\sc Step 2.} Given $\bxi^{[k-1]}$, sequentially update memberships $\mG_l^{[k]} = ( g_{i_l}^{(l)[k]}: 1 \le i_l \le N_l)^\top$ for $1\le l\le q$ as follows,
			\begin{align}
				& g_{i_l}^{(l)[k]}  = \arg\min_{g_{i_l}^{(l)}  \in [G_l]}
				\sum_{i_{-l}} \sum_{t=1}^T  \Big\{
				Y_{i_1i_2...i_q,t} - \Big(\sum_{l'=1}^{l-1} \lambda_{g_{i_{l'}}^{(l')[k]}}^{(l')[k]}\sum_{m= 1}^{N_{l'}} \frac{a^{(l')}_{i_{l'}m}}{n_{l'i_{l'}}}Y_{i_1...i_{l'-1}m i_{l'+1}...i_q,(t-1)} \nonumber\\
				& + \lambda_{g_{i_l}^{(l)}}^{(l)[k]}\sum_{m = 1}^{N_l} \frac{a^{(l)}_{i_{l}m }}{n_{li_l}}Y_{i_1...i_{l-1} m i_{l+1}...i_q,(t-1)} +  \sum_{l''=l+1}^q \lambda_{g_{i_{l''}}^{(l'')[k-1]}}^{(l'')[k]}\sum_{m = 1}^{N_{l''}} \frac{a^{(l'')}_{i_{l''}m}}{n_{l'' i_{l''}}}Y_{i_1...i_{l''-1}m i_{l''+1}...i_q,(t-1)} \Big) \nonumber\\
				&
				-\alpha^{[k]}_{g_{i_1}^{(1)[k]} \cdots g_{i_{l-1}}^{(l-1)[k]} g_{i_l}^{(l)} g_{i_{l+1}}^{(l+1)[k-1]} \cdots g_{i_q}^{(q)[k-1]}}Y_{i_1i_2...i_q,(t-1)}
				\nonumber\\
				& - \Big(\sum_{l' = 1}^{l-1} \bx_{i_l' t}^{(l')\top} \bzeta_{g_{i_l'}^{(l')[k]}}^{(l')[k]}  + \bx_{i_lt}^{(l)\top}\bzeta_{g_{i_l}^{(l)}}^{(l)[k]} + \sum_{l'' = l+1}^{q} \bx_{i_l'' t}^{(l'')\top} \bzeta_{g_{i_l''}^{(l'')[k-1]}}^{(l'')[k]} \Big)
				\Big\}^2 \label{eq:update_g}
		\end{align}
		\State {\bf Output:} Final estimator and memberships:
		$\wh\bxi = \bxi^{[K]}$ and
		$\wh \mG = \{\mG_l^{[K]}: l \in [q]\}$.
		Here $K$ is the final number of iteration rounds.
	\end{algorithmic}
\end{algorithm}

\section{Local Convergence of the Numerical Algorithm}\label{sec:alg_converge}

	In this section, we would show that our estimation Algorithm \ref{alg:gmnar} is convergent, and its solution is indeed a local minimizer of \eqref{eq:Q_obj}.
	We first show the convergence of the algorithm in (1), followed by the local minimization proof in (2).
	
	\noindent
	{\bf (1) Convergence of the algorithm.}
	
	We show that the objective function defined in \eqref{eq:Q_obj} is monotonically decreasing and hence could obtain the convergence.
	Denote the estimators in the $k$th iteration as $(\bxi^{(k)}, \mG^{(k)})$.
	Then we have that
	\begin{align*}
		Q(\bxi^{(k+1)}, \mG^{(k+1)}) & = \min_{\bxi} Q(\bxi, \mG^{(k+1)}) \le Q(\bxi^{(k)}, \mG^{(k+1)}) \\
		& \stackrel{\circled{1}}{\le} Q(\bxi^{(k)}, \{\mG_1^{(k)},  \mG_2^{(k+1)}, \cdots, \mG_q^{(k+1)}\}) \\
		&  \stackrel{\circled{2}}{\le} \cdots \cdots \\
		&  \stackrel{\circled{3}}{\le} Q(\bxi^{(k)}, \{\mG_1^{(k)},  \mG_2^{(k)}, \cdots, \mG_q^{(k)}\}) = Q(\bxi^{(k)}, \mG^{(k)}),
	\end{align*}
	where the inequalities $\circled{1}$--$\circled{3}$ are derived by
	\begin{align*}
		\mG_l^{(k+1)} = \argmin_{\mG_l} Q(\bxi^{(k)},  \{ \mG_1^{(k+1)}, \cdots, \mG_{q-1}^{(k+1)}, \mG_l, \mG_{l+1}^{(k)}, \cdots, \mG_q^{(k)} \} )
	\end{align*}
	in our updating mechanism \eqref{eq:g_i}.
	Then, we have that
	\begin{align*}
		Q(\bxi^{(0)}, \mG^{(0)}) \ge Q(\bxi^{(1)}, \mG^{(1)}) \ge \cdots  \ge Q(\bxi^{(k^*)}, \mG^{(k^*)})  = Q(\bxi^{(k^*+1)}, \mG^{(k^*+1)})
	\end{align*}
	Consequently, the estimators could be obtained by $Q(\bxi^{(k^*)}, \mG^{(k^*)})$.
	
	\noindent
	{\bf (2) Local Minimality.}
	
	Next, we prove that the solution $(\bxi^{(k^*)}, \mG^{(k^*)})$ is the local minimum of the function $\bxi \defeq \min_{\mG} Q(\bxi, \mG)$.
	Define $\wt\mG \defeq \argmin_{\mG} Q(\bxi^{(k^*)} + \bdelta, \mG)$, where $\bdelta$ is a small perturbation.
	When $\|\bdelta\|$ is small enough, we have that $\wt\mG = \mG^{(k^*)}$ due to the group memberships' discreteness.
	Then, we have
	\begin{align*}
		& \min_{\mG} Q(\bxi^{(k^*)} + \bdelta, \mG) = Q(\bxi^{(k^*)} + \bdelta, \wt\mG) \\
		& = Q(\bxi^{(k^*)} + \bdelta, \mG^{(k^*)}) \ge Q(\bxi^{(k^*)}, \mG^{(k^*)}) = \min_{\mG} Q(\bxi^{(k^*)}, \mG).
	\end{align*}
	This completes the proof.

\section{Initialization}\label{sec:init}

We provide the initialization procedure for the group memberships $\mG_l$ in the Algorithm \ref{alg:init}.

\begin{algorithm}
	\caption{Initialization of the GTNAR Model}
	\label{alg:init}
	\begin{algorithmic}[1]
		\State {\bf Input:} $\{\cY_t, \bX_t^{(l)}, \bW^{(l)}, G_l\}$ for $1 \le l \le q$.
		\State Treat each node as a group for all dimensions $l \in [q]$, estimate $\wh\btheta^{(l)}$ and $\wh\balpha$ by \eqref{eq:Mb}. Here $\wh\btheta^{(l)} \in \mR^{N_l(p_l+1)}$ and $\wh\balpha \in \mR^{N_1 \times \cdots \times N_q}$.
		\State Run the $k$-means clustering for the above estimators for $T_{\text{init}}=3$ trials. For each trial $t = 1, \cdots, T_{\text{init}}$, try the following two clustering types:
		
		{\bf Type 1. (Clustering by time effect)}
		
		{\sc Step 1.} Clustering the nodes in the $l$th dimension by the mode-$l$ matricization of
		{\it self-driven time effect} $\wh\balpha_{(l)} \in \mR^{N_l \times (\prod_{m \neq l} N_m)}$.
		Then one obtain the $\mG_l^{1[t]}$ for $1 \le l \le q$.
		
		{\sc Step 2.} Calculate the loss in the $t$th trial for type 1 by $Q^{[t]}(\wh\bxi, \mG^{1[t]})$, where $\mG^{1[t]} = \{ \mG_l^{1[t]}: l \in [q]\}$ is the first type initial group memberships.
		
		{\bf Type 2. (Clustering by network and covariate effect)}
		
		{\sc Step 1.} Clustering by {\it network
			effects} and {\it covariate effect} on $\wh\btheta^{(l)} = (\wh\btheta_{g^{(l)}}^{(l)}: g^{(l)} \in [G_l])$.
		Then one obtain the $\mG_l^{2[t]}$ for $1 \le l \le q$.
		
		{\sc Step 2.} Calculate the loss in the $t$th trial for type 2 by $Q^{[t]}(\wh\bxi, \mG^{2[t]})$, where $\mG^{2[t]} = \{ \mG_l^{2[t]}: l \in [q]\}$ is the second type initial group memberships.
		
		\State Select the best initial trial $t^* = \argmin_{t} [\min\{ Q^{[t]}(\wh\bxi, \mG^{1[t]}), Q^{[t]}(\wh\bxi, \mG^{2[t]})\}$], and the corresponding initial memberships are denoted as $\mG^{[0]}$.
		\State {\bf Output:} Best initial memberships: $\mG^{[0]}$.
	\end{algorithmic}
\end{algorithm}

\section{Additional Discussion of Mixed GTNAR Model \eqref{eq:model_int}}\label{sec:int_est}

		\subsection{Estimation Procedure}\label{subsec:int_est_alg}

	The element-wise model form can be given as
	\begin{align*}
		Y_{ij, t} & = \gamma_{g_{i}^{(1)}}^{(1)} \gamma_{g_{j}^{(2)}}^{(2)} \sum_k \sum_m   \frac{a_{i m}^{(1)}}{n_{1 i}} Y_{mk, t-1} \frac{a_{k j}^{(2)}}{n_{2 j}} \\
		& +  \Big( \lambda_{g_{i}^{(1)}}^{(1)} \sum_k \frac{a_{i k}^{(l)}}{n_{1 i}} Y_{kj, t-1} + \lambda_{g_j^{(2)}}^{(2)} \sum_k \frac{a_{kj}^{(2)}}{n_{2j}} Y_{ik, t-1} \Big) \nonumber \\
		& + \alpha_{g_{i}^{(1)} g_{j}^{(2)} }Y_{ij,(t-1)}
		+ \Big(\bx_{i t}^{(1)\top}\bzeta_{g_{i}^{(1)}}^{(1)} +  \bx_{j t}^{(2)\top} \bzeta_{g_{j}^{(2)}}^{(2)}  \Big) + \ve_{ij, t}.
	\end{align*}
	
	For simplicity, denote the interactive parameters as $\bpsi = (\bpsi^{(1)\top}, \bpsi^{(2)\top})^\top \in \mR^{G_1 + G_2}$ with $\bpsi^{(1)} = (\gamma_1^{(1)}, \cdots, \gamma_{G_1}^{(1)})^\top \in \mR^{G_1}$ and $\bpsi^{(2)} = (\gamma_1^{(2)}, \cdots, \gamma_{G_2}^{(2)})^\top \in \mR^{G_2}$.
	Recall that $\bxi = (  \btheta^{(1)\top}, \btheta^{(2)\top}, \vec(\balpha)^\top)^\top \in \mR^{G_1(p_1+1)+G_2(p_2+1) + G_1G_2}$ includes the other group parameters
	except for the interactive parameters $\bpsi$.
	To estimate the unknown parameters,
	we minimize the following objective function,
		\begin{align}
			Q(\bxi, \bpsi, \mG) = \sum_{i=1}^{N_1}  \sum_{j = 1}^{N_2} \sum_{t=1}^T \Big\{Y_{ij, t} & -\gamma_{g_{i}^{(1)}}^{(1)} \gamma_{g_{j}^{(2)}}^{(2)} \sum_k \sum_m   \frac{a_{i m}^{(1)}}{n_{1 i}} Y_{mk, t-1} \frac{a_{k j}^{(2)}}{n_{2 j}} \nonumber\\
			& - \Big( \lambda_{g_{i}^{(1)}}^{(1)} \sum_k \frac{a_{i k}^{(l)}}{n_{1 i}} Y_{kj, t-1} + \lambda_{g_j^{(2)}}^{(2)} \sum_k \frac{a_{kj}^{(2)}}{n_{2j}} Y_{ik, t-1} \Big) \nonumber\\
			& - \alpha_{g_{i}^{(1)} g_{j}^{(2)} }Y_{ij,(t-1)}
			- \Big(\bx_{i t}^{(1)\top}\bzeta_{g_{i}^{(1)}}^{(1)} + \bx_{j t}^{(2)\top} \bzeta_{g_{j}^{(2)}}^{(2)}  \Big)   \Big\}^2.\label{eq:Q_obj_int}
		\end{align}
	
	We utilize the following iterative algorithm.

	\noindent
	{\bf (1) Update the group parameters $\bxi$ and $\bpsi$.}
	
	First, fix the group memberships $g_i^{(1)}$ and  $g_j^{(2)}$ and then we update the group parameters.
	In this parameter updating step, we iteratively update $\bpsi$ and $\bxi$ as follows, which can be conducted in a fast speed.
	{To be more specific,}
	given interactive network effects $\bpsi$, we could calculate
	\begin{align*}
		\wt \Y_t = \Y_t -(\bGamma^{(1)} \W^{(1)})  \Y_{t-1} (\W^{(2)} \bGamma^{(2)}).
	\end{align*}
	Then the parameter $\bxi$ can be obtained by $\wh \bxi  = (\bM )^{-1} \b$, whose specific expressions can be found in \eqref{eq:Mb2},
	with $\mY_{g^{(1)} g^{(2)}, t}$ replaced by $\vec \big(\wt \Y_t^{( \cR_{g^{(1)}}^{(1)}, \cR_{g^{(2)}}^{(2)} )} \big)$.
	
	Subsequently, given $\bxi$, 
	denote $ \breve \Y_t = \Y_t - \L^{(1)} \W^{(1)} \Y_{t-1} - \Y_{t-1} \W^{(2)} \L^{(2)} - \A \circ \Y_{t-1} - \bbeta_{X_1, t}^{(1)} \one_{N_2}^\top - \one_{N_1} \bbeta_{X_2, t}^{(2)\top}$. We need to minimize the following least squares problem,
	\begin{align*}
		\min_{\bGamma^{(1)}, \bGamma^{(2)}} \sum_t \| \breve{\Y}_t  - \bGamma^{(1)} \W^{(1)} \Y_{t-1} \W^{(2)} \bGamma^{(2)} \|_F^2.
	\end{align*}
	Denote that $\wt\mX_t = \W^{(1)} \bY_{t-1} \W^{(2)} \bGamma^{(2)} \in \mR^{N_1 \times N_2}$.
	By the first order condition, we have that
	\begin{align}
		& \wh\gamma_{g^{(1)}}^{(1)} = \Big(\sum_t \sum_{i \in \cR_{g^{(1)}}^{(1)}} \sum_j \wt\mX_{ij, t} \breve{\Y}_{ij, t} \Big) \Big/ \sum_t  \Big\|\wt\mX_t^{(\cR_{g^{(1)}}^{(1)}, \cdot)} \Big\|_F^2, ~~~g^{(1)} \in [G_1] \label{eq:update_L1}\\
		& \wh\gamma_{g^{(2)}}^{(2)} = \Big(\sum_t  \sum_i  \sum_{j \in \cR_{g^{(2)}}^{(2)}}\wt\mX_{ij, t} \breve{\Y}_{ij, t} \Big) \Big/ \sum_t  \Big\|\wt\mX_t^{(\cdot, \cR_{g^{(2)}}^{(2)})} \Big\|_F^2, ~~~g^{(2)} \in [G_2],\label{eq:update_L2}
	\end{align}
	where $\cR_{g^{(1)}}^{(1)} = \{i \in [N_1]: g_i^{(1)} = g^{(1)}\}$ and $\cR_{g^{(2)}}^{(2)} = \{j \in [N_2]: g_j^{(2)} = g^{(2)}\}$.
	Hence, we can iteratively update $\bGamma^{(1)}$ and $\bGamma^{(2)}$ using the above equations \eqref{eq:update_L1} and \eqref{eq:update_L2}.

	\noindent
	{\bf (2) Update the group memberships $\mG_1$ and $\mG_2$.}

	Second, {we consider given $\bxi$ and $\bpsi$}, updating the group memberships $\mG_1 = (g_{i}^{(1)}: i \in [N_1])^\top $ and $\mG_2 = (g_{j}^{(2)}: j \in [N_2])^\top$ iteratively. Given $\mG_2$, update
	\begin{align}
		\wh g_{i}^{(1)} & = \argmin_{g_{i}^{(1)} \in [G_1]}  \sum_{j = 1}^{N_2} \sum_t \Big\{ Y_{ij, t} -  \gamma_{g_{i}^{(1)}}^{(1)} \gamma_{g_{j}^{(2)}}^{(2)} \sum_k \sum_m    \frac{a_{i m}^{(1)}}{n_{1 i}} Y_{mk, t-1} \frac{a_{k j}^{(2)}}{n_{2 j}}  \nonumber \\
		&  -  \Big( \lambda_{g_{i}^{(1)}}^{(1)} \sum_k \frac{a_{i k}^{(l)}}{n_{1 i}} Y_{kj, t-1} + \lambda_{g_j^{(2)}}^{(2)} \sum_k \frac{a_{kj}^{(2)}}{n_{2j}} Y_{ik, t-1} \Big) \nonumber \\
		& - \alpha_{g_{i}^{(1)} g_{j}^{(2)} }Y_{ij,(t-1)}
		- \Big(\bx_{i t}^{(1)\top}\bzeta_{g_{i}^{(1)}}^{(1)} +  \bx_{j t}^{(2)\top} \bzeta_{g_{j}^{(2)}}^{(2)}  \Big)\Big\}^2.
	\end{align}
	The parallel computation is implement to update $\mG_2$ given $\mG_1$.
	To summarize, the iterative algorithm is shown in Algorithm \ref{alg:gmnar_int}.

	To estimate the group numbers for model \eqref{eq:model_int_2}, denoted as $\underline{G} = (G_1, G_2)$, we minimize the information criterion.
	\begin{align}
		\qic(\underline{G})  = \log \{ Q(\wh \bxi(\underline{G}), \wh \bpsi(\underline{G}), \wh \mG(\underline{G})) \} + \lambda(\underline{G}),\label{eq:qic_int}
	\end{align}
	where $\wh\bxi_1(\underline{G})$, $\wh\bpsi(\underline{G})$ and $\wh \mG(\underline{G})$ are the estimated parameters when specifying the group numbers as $G_1$ and $G_2$, $Q(\cdot, \cdot, \cdot)$ is the objective function defined in \eqref{eq:Q_obj_int}, and $\lambda(\underline{G})$ is the penalty function.
	Specifically, we set $\lambda(\underline{G}) = \kappa_1 (G_1 + G_2)$ with $\kappa_1$ being a tuning parameter.

	We summarize the estimation algorithm in the following Algorithm \ref{alg:gmnar_int}.

		\begin{algorithm}
		\caption{Estimation of the GTNAR Model with Interactive Term When $q=2$}
		\label{alg:gmnar_int}
		\begin{algorithmic}[1]
			\State {\bf Input:} $\{\Y_t, \bX_t^{(l)}, \bW^{(l)}, G_l\}$ for $1 \le l \le q$.
			\State Obtain initial group memberships $\mG_l^{[0]}$ according to Algorithm \ref{alg:init}. Let $\{\bxi^{[k]}, \mG_l^{[k]}\}$ be the estimators and memberships in the $k$th iteration.
			\State Repeat {\sc Step 1} and {\sc Step 2} for {$k = 1, 2, \cdots$}
			until convergence.
			
			{\sc Step 1.} Given $\{\mG_l^{[k-1]}\}$ for all layers $l\in[q]$
			
			{\sc Step (1a).} Fix the interactive network effects $\bGamma^{(1)[k-1]} $ and $\bGamma^{(2)[k-1]}$, calculate $\wt \Y_t = \Y_t -(\bGamma^{(1)[k-1]} \W^{(1)})  \Y_{t-1} (\W^{(2)} \bGamma^{(2)[k-1]})$.
			Update $\bxi^{[k]} = (\bM^{[k-1]})^{-1} \b^{[k-1]}$..
			
			{\sc Step (1b).} Fix the estimated parameters $\bxi^{[k]}$. Given $\bGamma^{(2)[k-1]}$, obtain $\bGamma^{(1)[k]}$ using \eqref{eq:update_L1}; then given $\bGamma^{(1)[k]}$, obtain $\bGamma^{(2)[k]}$ using \eqref{eq:update_L2}.
			
			{\sc Step 2.} Given $\gamma_{g^{(1)[k-1]}}^{(1)[k]}$, $\gamma_{g^{(2)[k-1]}}^{(2)[k]}$, $\lambda_{g^{(1)[k-1]}}^{(1)[k]}$, $\lambda_{g^{(2)[k-1]}}^{(2)[k]}$, $\alpha^{[k]}_{g^{(1)[k-1]} g^{(2)[k-1]}}$ and $\bzeta_{g^{(1)[k-1]}}^{(1)[k]}$, $\bzeta_{g^{(2)[k-1]}}^{(2)[k]}$, update the memberships $\mG_1^[(k)]$ and $\mG_2^[(k)]$ sequentially by,
			\begin{align*}
				g_{i}^{(1)[k]} & = \argmin_{g_{i}^{(1)} \in [G_1]} \sum_{j} \sum_{t} \Big\{ Y_{ij, t} - \Big( \gamma_{g_{i}^{(1)}}^{(1)[k]}  \gamma_{g_{j}^{(2)[k-1]}}^{(2)[k]} \sum_k \sum_m \frac{a_{i m}^{(1)}}{n_{1 i}} Y_{mk, t-1} \frac{a_{k j}^{(2)}}{n_{2 j}}  \Big) \\
				&  -  \Big( \lambda_{g_{i}^{(1)[k]}}^{(1)[k]} \sum_k \frac{a_{i k}^{(l)}}{n_{1 i}} Y_{kj, t-1} + \lambda_{g_j^{(2)[k-1]}}^{(2)[k]} \sum_k \frac{a_{kj}^{(2)}}{n_{2j}} Y_{ik, t-1} \Big)  \\
				& - \alpha^{[k]}_{g_{i}^{(1)} g_{j}^{(2)[k-1]} }Y_{ij,(t-1)}
				- \Big(\bx_{i t}^{(1)\top}\bzeta_{g_{i}^{(1)}}^{(1)[k]} +  \bx_{j t}^{(2)\top} \bzeta_{g_{j}^{(2)[k-1]}}^{(2)[k]} \Big)
				\Big\}^2
			\end{align*}
			and
			\begin{align*}
				g_{j}^{(2)[k]} & = \argmin_{g_{j}^{(2)} \in [G_2]} \sum_{i} \sum_{t} \Big\{ Y_{ij, t} - \Big( \gamma_{g_{i}^{(1)[k]}}^{(1)[k]}  \gamma_{g_{j}^{(2)}}^{(2)[k]} \sum_k \sum_m \frac{a_{i m}^{(1)}}{n_{1 i}} Y_{mk, t-1} \frac{a_{k j}^{(2)}}{n_{2 j}}  \Big) \\
				& -  \Big( \lambda_{g_{i}^{(1)[k]}}^{(1)[k]} \sum_k \frac{a_{i k}^{(l)}}{n_{1 i}} Y_{kj, t-1} + \lambda_{g_j^{(2)}}^{(2)[k]} \sum_k \frac{a_{kj}^{(2)}}{n_{2j}} Y_{ik, t-1} \Big) \\
				& - \alpha^{[k]}_{g_{i}^{(1)[k]} g_{j}^{(2)} }Y_{ij,(t-1)}
				- \Big(\bx_{i t}^{(1)\top}\bzeta_{g_{i}^{(1)[k]}}^{(1)[k]} +  \bx_{j t}^{(2)\top} \bzeta_{g_{j}^{(2)}}^{(2)[k]} \Big)
				\Big\}^2
			\end{align*}
			\State {\bf Output:} Final estimator and memberships:
			$\wh \bxi = \bxi^{[K]}$,
			$\wh \bpsi = \bpsi^{[K]}$, and
			$\wh \mG = \{\mG_l^{[K]}: l = 1,2 \}$.
			Here $K$ is the final number of iteration rounds.
		\end{algorithmic}
	\end{algorithm}

	\subsection{Technical Assumptions}\label{subsec:assum_int}

	Denote
	\begin{align}
	\cX_{ij, t} = (\sum_{k=1}^{N_1} w_{ik}^{(1)} Y_{kj, t}, \bx_{1t}^{(1)\top}, \sum_{k=1}^{N_2} w_{kj}^{(2)} Y_{ik, t}, \bx_{jt}^{(2)\top}, Y_{ij, t-1}, \sum_k^{N_1} \sum_m^{N_2} w_{ik}^{(1)} Y_{km, t-1} w_{mj}^{(2)}) \in \mR^{p_1 + p_2 + 4}. \label{eq:X1}
	\end{align}
	By using $\cX_{i j, t}$, we rewrite the objective function as $Q(\bTheta)  = \sum_{i, j, t} (Y_{ij, t} - \cX_{i j, t}^\top \bTheta_{ij})^2$.
	Denote $\wh \bTheta$ as the estimator of $\bTheta$ by minimizing the loss function $Q(\bTheta)$.
	We require the following assumptions.

	\begin{assumption}\label{assum:para_space_int}
		{\sc (Parameter Space)} The parameter satisfies that $\|\bTheta\|_{\max} < \infty$.
	\end{assumption}

	\begin{assumption}\label{assum:convex_int}
		{\sc (Convexity)} Denote $\bSigma_{ij} = E(\cX_{ij, t} \cX_{ij, t}^\top)$, where $\cX_{ij, t}$ is defined in \eqref{eq:X1}. Assume that $\tau_{\min}^1 \defeq \min_{i,j} \bSigma_{ij}>0$ is a constant.
	\end{assumption}
	
	\begin{assumption}\label{assum:station_int}
		{\sc(Stability)}
		{Suppose} $\mY_0 \defeq \vec(\cY_0) = \zero$, and assume that
		\begin{align*}
			\max_{g^{(1)} \in [G_{1,0}], g^{(2)} \in [G_{2,0}]}  \Big| \lambda_{g^{(1)}}^{(1)0} +  \lambda_{g^{(2)}}^{(2)0}  + \alpha_{g^{(1)} g^{(2)}}^0  +\gamma^{(1)0}_{g^{(1)}} \gamma^{(2)0}_{g^{(2)}}  \Big| \le \kappa_{\max} <1,
		\end{align*}
		where $G_{l, 0}$ is the true number of groups in the $l$th dimension, and $\kappa_{\max}$ is a positive constant.
	\end{assumption}

	\begin{assumption}\label{assum:group_diff_int}
		{\sc(Group Difference)}
		For the first dimension, assume
		$$\min_{g_1^{(1)} \ne g_2^{(1)}}  \Big[ \Big\|\btheta_{g_1^{(1)}}^{(1)0}  - \btheta_{g_2^{(1)}}^{(1)0}\Big\|^2 +
		\max_{g^{(2)} \in [G_2^0]}  \Big\{ \Big| \alpha_{g_1^{(1)} g^{(2)}}^0 - \alpha_{g_2^{(1)} g^{(2)}}^0 \Big|^2 +  \Big| \gamma^{(1)0}_{g_1^{(1)}} \gamma^{(2)0}_{g^{(2)}} -  \gamma^{(1)0}_{g_2^{(1)}} \gamma^{(2)0}_{g^{(2)}} \Big|^2 \Big\} \Big] \ge  c_\gap,$$
		where $c_\gap \gg T^{-1}(\log (N_1  N_2))^2 (G_1 G_2)$ as $T \to \infty$.
		{Here} $G_1, G_2$ are finite.
		Assume the same holds for the group $g_1^{(2)} \neq g_2^{(2)}$ in the second dimension.
	\end{assumption}

\subsection{Numerical Studies}\label{subsec:int_simu}

The estimation results in Table \ref{tbl:simu_int} demonstrate excellent finite sample performance: all RMSEs decrease with increasing time length $T$ or increasing sizes $N_1, N_2$ (e.g., RMSE of $\widehat{\boldsymbol{\lambda}}^{(1)}$ declines from 0.0131 at $T=40$ to 0.0028 at $T=80$), while the coverage probabilities in parentheses converge closely to the nominal 95\% level at larger $T$.
Furthermore, the group membership errors $\widehat{\eta}_1$ and $\widehat{\eta}_2$ remain near-zero across all configuration with consistently diminishing magnitudes at larger sample sizes, indicating exceptionally accurate group classification even for moderate sample sizes.

\begin{table}[]
	\centering
	\caption{RMSEs of estimated parameters under when $G_{1,0} = G_{2,0} = 3$ with 100 replications. The performances are evaluated for different sample sizes $N_1, N_2$ and the time length $T$. The corresponding CPs are shown in the parenthesis.}
	\label{tbl:simu_int}
	\scalebox{0.9}{
	\begin{tabular}{cc|cc|c|ccccc|c|cc}
		\hline
		$G_1$              & $G_2$              & $N_1$                & $N_2$                & $T$ & $\wh \blambda^{(1)}$                                    & $\wh \blambda^{(2)}$                                    & $\wh \bzeta^{(1)}$                                      & $\wh \bzeta^{(2)}$                                      & $\wh \balpha$                                            & $\wh\bGamma_\textup{norm}$ & $\wh\eta_1$ & $\wh\eta_2$ \\ \hline
		\multirow{11}{*}{3} & \multirow{11}{*}{3} & \multirow{5}{*}{100} & \multirow{5}{*}{80}  & 20  & \begin{tabular}[c]{@{}c@{}}0.0131\\ (0.90)\end{tabular} & \begin{tabular}[c]{@{}c@{}}0.0119\\ (0.93)\end{tabular} & \begin{tabular}[c]{@{}c@{}}0.0137\\ (0.93)\end{tabular} & \begin{tabular}[c]{@{}c@{}}0.0150\\ (0.92)\end{tabular} & \begin{tabular}[c]{@{}c@{}}0.0369\\ (0.92)\end{tabular}  & 3.0282                     & 0.0122      & 0.0086      \\
		&                    &                      &                      & 40  & \begin{tabular}[c]{@{}c@{}}0.0079\\ (0.93)\end{tabular} & \begin{tabular}[c]{@{}c@{}}0.0081\\ (0.95)\end{tabular} & \begin{tabular}[c]{@{}c@{}}0.0089\\ (0.95)\end{tabular} & \begin{tabular}[c]{@{}c@{}}0.0094\\ (0.95)\end{tabular} & \begin{tabular}[c]{@{}c@{}}0.0131\\ (0.95)\end{tabular}  & 1.2326                     & 0.0005      & 0.0001      \\
		&                    &                      &                      & 80  & \begin{tabular}[c]{@{}c@{}}0.0055\\ (0.93)\end{tabular} & \begin{tabular}[c]{@{}c@{}}0.0052\\ (0.95)\end{tabular} & \begin{tabular}[c]{@{}c@{}}0.0064\\ (0.95)\end{tabular} & \begin{tabular}[c]{@{}c@{}}0.0067\\ (0.95)\end{tabular} & \begin{tabular}[c]{@{}c@{}}0.0091\\ (0.95)\end{tabular}  & 0.5685                     & 0.0000      & 0.0001      \\ \cline{3-13}
		&                    & \multirow{5}{*}{200} & \multirow{5}{*}{150} & 20  & \begin{tabular}[c]{@{}c@{}}0.0061\\ (0.94)\end{tabular} & \begin{tabular}[c]{@{}c@{}}0.0053\\ (0.93)\end{tabular} & \begin{tabular}[c]{@{}c@{}}0.0070\\ (0.94)\end{tabular} & \begin{tabular}[c]{@{}c@{}}0.0069\\ (0.94)\end{tabular} & \begin{tabular}[c]{@{}c@{}}0.0095\\ (0.94)\end{tabular}  & 2.5897                     & 0.0004      & 0.0002      \\
		&                    &                      &                      & 40  & \begin{tabular}[c]{@{}c@{}}0.0040\\ (0.95)\end{tabular} & \begin{tabular}[c]{@{}c@{}}0.0039\\ (0.93)\end{tabular} & \begin{tabular}[c]{@{}c@{}}0.0047\\ (0.95)\end{tabular} & \begin{tabular}[c]{@{}c@{}}0.0048\\ (0.96)\end{tabular} & \begin{tabular}[c]{@{}c@{}}0.0068\\ (0..95)\end{tabular} & 1.0622                     & 0.0000      & 0.0002      \\
		&                    &                      &                      & 80  & \begin{tabular}[c]{@{}c@{}}0.0027\\ (0.95)\end{tabular} & \begin{tabular}[c]{@{}c@{}}0.0028\\ (0.93)\end{tabular} & \begin{tabular}[c]{@{}c@{}}0.0032\\ (0.96)\end{tabular} & \begin{tabular}[c]{@{}c@{}}0.0035\\ (0.96)\end{tabular} & \begin{tabular}[c]{@{}c@{}}0.0047\\ (0.94)\end{tabular}  & 0.6653                     & 0.0000      & 0.0001      \\ \hline
	\end{tabular}
}
\end{table}

\subsection{Technical Proofs}\label{sec:proof_int}

\subsubsection{Proof of Theorem \ref{thm:int_model_thm} (i)}

	By using the $\cX_{i j, t}$ defined in \eqref{eq:X1}, we define the objective function,
	\begin{align*}
		Q({\bTheta}) = \sum_{i=1}^{N_1} \sum_{j = 2}^{N_2} \sum_{t = 1}^T (Y_{i j, t} - \cX_{i j, t}^\top \bTheta_{i j})^2 = \sum_{i = 1}^{N_1} \sum_{j = 1}^{N_2} \sum_{t=1}^T  Q_{i j}(\bTheta_{ i j}).
	\end{align*}
	Given this objective function, we can follow the proof of Theorem \ref{thm:pseudo_dist} using the partition of the new parameter space.
	To obtain the conclusion, we use the Lemma \ref{lem:Q_star_diff_1} and Lemma \ref{lem:Q_concent_1} instead of the usage of Lemma \ref{lem:Q_star_diff} and Lemma \ref{lem:Sij_concent} in the original proof of Theorem \ref{thm:pseudo_dist}.
	Since the rest of the proof remains the same, we omit the details here.

\subsubsection{Proof of Theorem \ref{thm:int_model_thm} (iii)}

We first declare some notations in the proof.
Denote $\wt\bgamma^{(1)}_{g^{(1)}} = (\gamma^{(1)}_{g^{(1)}} \gamma^{(2)}_{g^{(2)}} : g^{(2)} \in [G_2])^\top \in \mR^{G_2}$ and $\wt\bgamma^{(2)}_{g^{(2)}} = (\gamma^{(2)}_{g^{(1)}} \gamma^{(2)}_{g^{(2)}} : g^{(1)} \in [G_1])^\top \in \mR^{G_1}$.
Denote $ {\wt\bgamma }= (\wt\bgamma^{(1)\top}_{g^{(1)}} : g^{(1)}\in [G_1])^\top \in \mR^{G_1 G_2}$.
Further denote that ${ \wt\balpha} = (\vec(\balpha_{g^{(1)} \cdot})^\top: g^{(1)} \in [G_1] )^\top \in \mR^{G_1 G_2}$.
Recall that $\btheta^{(1)} = (\btheta^{(1)\top}_{g^{(1)}}: g^{(1)}\in [G_1])^\top$.
Denote $\bxi_{g^{(1)}}^{(1)}=(\btheta_{g^{(1)}}^{(1)\top}, \vec(\balpha_{g^{(1)} \cdot})^\top, \wt \bgamma_{g^{(1)}}^{(1)\top})^\top$ as the parameters of the $g^{(1)}$th group $(g^{(1)}\in [G_1])$.
Correspondingly, denote $\bxi_{g^{(1)0}}^{(1)0}$ as the true parameters of the true group $g^{(1)0} \in [G_{1,0}]$.
	Let $\bxi^{(1)} = (\btheta^{(1)\top}, { \wt\balpha^{\top}, \wt\bgamma^{\top} } )^\top$.
	In parallel, we can define $\bxi_{g^{(2)}}^{(2)} = (\btheta_{g^{(2)}}^{(2)\top}, \wt\balpha^\top, \wt\bgamma^\top)^\top$ and $\bxi^{(2)} = (\btheta^{(2)\top}, \wt\balpha^\top, \wt\bgamma^\top)^\top$.
	At last, denote $\bxi = (\btheta^{(1)\top}, \btheta^{(2)\top}, \wt\balpha^\top, \wt\bgamma^\top)^\top$.
	In the following, we prove the theorem for the first dimension, and the second dimension group membership consistency can be proved similarly.
	For $g^{(1)}\in[G_1]$ and $g^{(1)0}\in [G_{1,0}]$,
	define
	\begin{align}
	& {\mL^{(1)}(\bxi_{g^{(1)}}^{(1)}, \bxi_{g^{(1)0}}^{(1)0}; \mG_2, \mG_2^0)} = \big \| \btheta_{g^{(1)}}^{(1)} - \btheta^{(1)0}_{g^{(1)0}} \big \|^2 \nonumber\\
	& + \frac{1}{N_2} \sum_{j} \Big\{\big|\alpha_{g^{(1)}g_{j}^{(2)}} - \alpha^0_{g^{(1)0} g_j^{(2)0}}\big|^2 +  \big|  \gamma_{g^{(1)}}^{(1)} \gamma_{g_j^{(2)}}^{(2)} - \gamma_{g^{(1)0}}^{(1)0} \gamma_{g_j^{(2)0}}^{(2)0} \big|^2 \Big\}.\label{eq:mL}
	\end{align}
	For notation simplicity, denote $\mL_{g^{(1)}, g^{(1)0}}^{(1)} \defeq \mL^{(1)}(\bxi_{g^{(1)}}^{(1)}, \bxi_{g^{(1)0}}^{(1)0}; { \mG_2, \mG_2^0})$.
	Define the distance
	\begin{align*}
		d_L^{(1)}(\bxi^{(1)}, \bxi^{(1)0}; \mG_2, \mG_2^0) = \max\Big\{ \max_{g^{(1)0}\in[G_{1,0}]} \min_{g^{(1)}\in [G_1]} \mL_{g^{(1)}, g^{(1)0}}^{(1)}, \max_{g^{(1)}\in[G_1]} \min_{g^{(1)0}\in [G_{1,0}]}   \mL_{g^{(1)}, g^{(1)0}}^{(1)}\Big\}.
	\end{align*}
	In addition, given parameters $\bxi$, denote the estimated group memberships in the second dimension as $\wh\mG_2(\bxi) = (\wh g_j^{(2)}(\bxi): j \in[N_2])^\top$, define
	\begin{align*}
		\cN_{\eta}^{(1)} = \{ \bxi: d_L^{(1)}(\bxi^{(1)}, \bxi^{(1)0}; \wh\mG_2(\bxi), \mG_2^0)< \eta \}
	\end{align*}
	and denote
	\begin{align*}
		& \mA_\eta^{(1)}({ \bxi}, g^{(1)0}, { \mG_2^0}) = \Big\{g^{(1)}\in[G_1]: \\
		& \big \| \btheta_{g^{(1)}}^{(1)} - \btheta^{(1)0}_{g^{(1)0}} \big \|^2 + \frac{1}{N_2} \sum_{j} \Big\{\big|\alpha_{g^{(1)}\wh g_j^{(2)}(\bxi)} - \alpha^0_{g^{(1)0} g_j^{(2)0}}\big|^2 +  \big|  \gamma_{g^{(1)}}^{(1)} \gamma_{\wh g_j^{(2)}(\bxi)}^{(2)} - \gamma_{g^{(1)0}}^{(1)0} \gamma_{g_j^{(2)0}}^{(2)0} \big|^2 \Big\} \le \eta \Big\}.
	\end{align*}
	
	Based on the above notations, we first give the following lemma, whose proof can be found in Appendix \ref{append.proof_lem_int}.
	\bel\label{lem:g_consistency_1_prepare}
	Under Assumptions \ref{assum:sub_gaussian}, \ref{assum:mixing}, \ref{assum:group_ratio}, \ref{assum:para_space_int}--\ref{assum:group_diff_int}, and assume that $G_1 \ge G_{1,0}$ and $G_2 \ge G_{2,0}$.
	{For $\bTheta$ satisfying the condition
	\begin{align}
		d(\bTheta, \bTheta^0) = O_p\{ T^{-1} (\log (N_1 N_2))^2 \} = o_p(c_\gap)\label{eq:d_condition}
	\end{align}
	as $N_l \to \infty, ~l=1,2$, the following conclusions hold,}\\
	(i) For all $\bxi \in \cN^{(1)}_{\eta}$ with $\eta < c_\gap  c_\pi / 4$, then we have that $\{\mA_\eta^{(1)}(\bxi, g^{(1)0}, {\mG_2^0}): g^{(1)0}\in[G_{1,0}]\}$ is a partition of $[G_1]$. \\
	(ii) For all $\bxi \in \cN^{(1)}_{\eta}$ with $\eta \le \tau_{\min}^1 c_\gap c_\pi / (8 (\tau_{\min}^1 + \tau_{\max}^1))$, define the event $\mO = \{ \wh g_{i}^{(1)}(\bxi) \in \mA_\eta^{(1)}(\bxi, g_i^{(1)0}, { \mG_2^0}): i \in [N_1] \}$.
	Then we have $P(\mO^c) \le C \exp(-c_1 \sqrt{T} c_\gap + c_2 m + \log N_1 + \log N_2)$, where $C, c_1, c_2$ are positive constants, and $m = p_1 + p_2 + 4$.
	\eel

	Next, by using Lemma \ref{lem:g_consistency_1_prepare}, we first show that ${\wh\bxi } \in \cN^{(1)}_{\eta}$ when $\eta = \tau_{\min}^1 c_\gap c_\pi / (8 (\tau_{\min}^1 + \tau_{\max}^1))$ with probability tending to 1, then we use (i) and (ii) in Lemma \ref{lem:g_consistency_1_prepare} to obtain the final results.

	\noindent
	{\bf Step 1. $P\big(\wh\bxi \in \cN_{\eta}^{(1)}\big) \to 1$.}
	

	Using the similar procedure in the Step 1 of proof of Theorem \ref{thm:h_consistency2},
	we can show that $d_L^{(1)}(\wh\bxi^{(1)}, \bxi^{(1)0}; \wh\mG_2(\wh\bxi), \mG_2^0) = O_p\{ T^{-1} (\log (N_1 N_2))^2 \}$.
	By the condition that 
	$$c_{\gap} \gg T^{-1} (\log (N_1 N_2))^2,$$
	we have $\wh\bxi \in \cN_{\eta}^{(1)}$ with probability tending to 1.

	\noindent
	{\bf Step 2. Conclusion of Theorem \ref{thm:int_model_thm} (iii).}
	
	For $\wt g^{(1)} \in [G_1]$, suppose there are $i_1, i_2 \in \wh \cR_{\wt g^{(1)}}^{(1)}$, then $\wh g_{i_1}^{(1)} = \wh g_{i_2}^{(2)} = \wt g^{(1)}$.
	By Lemma \ref{lem:g_consistency_1_prepare}(ii), we have $\wt g^{(1)} \in \mA_{\eta}^{(1)}(\wh\bxi, g_{i_1}^{(1)0}, \mG_2^0)$ and $\wt g^{(1)} \in \mA_{\eta}^{(1)}(\wh\bxi, g_{i_2}^{(1)0}, \mG_2^0)$ hold with probability tending to 1.
	By Lemma \ref{lem:g_consistency_1_prepare}(i), we know that $\mA^{(1)}_{\eta}(\wh\bxi, \cdot, \mG_2^0)$ is a partition of $[G_1]$, hence $g_{i_1}^{(1)0} = g_{i_2}^{(1)0}$.
	Then, by setting $g^{(1)} = g_{i_1}^{(1)0} = g_{i_2}^{(1)0}$, we have
	$i_1, i_2 \in \cR_{ g^{(1)}}^{(1)0}$.

	\subsubsection{Proof of Lemma \ref{lem:g_consistency_1_prepare}}\label{append.proof_lem_int}

		
		\noindent
		{\bf (1) Proof of (i).}
		
		For all $\bxi \in \cN_{\eta}^{(1)}$, by the definition of $\cN_{\eta}^{(1)}$ and the group set definition $ \mA_\eta^{(1)}(\bxi, g^{(1)0}, \mG_2^0)$, we have $\cup_{g^{(1)0} = 1}^{G_{1,0}}  \mA_\eta^{(1)}(\bxi, g^{(1)0}, \mG_2^0) = [G_1]$.
		Then we prove $\mA_\eta^{(1)}(\bxi, g_1^{(1)0}, \mG_2^0) \cap \mA_\eta^{(1)}(\bxi, g_2^{(1)0}, \mG_2^0)  = \emptyset$ for $g_1^{(1)0}, g_2^{(1)0} \in [G_{1,0}]$ and $g_1^{(1)0} \neq g_2^{(1)0}$.
		We use the contradiction.
		Assume there exists $g_{3}^{(1)} \in [G_1]$ s.t. $g_3^{(1)} \in \mA_\eta^{(1)}(\bxi, g_1^{(1)0}, \mG_2^0) \cap \mA_\eta^{(1)}(\bxi, g_2^{(1)0}, \mG_2^0)$. Then under Assumption \ref{assum:group_diff_int}, we have
		\begin{align*}
			c_{\gap} &  \le \Big\|\btheta_{g_1^{(1)0}}^{(1)0}  - \btheta_{g_2^{(1)0}}^{(1)0}\Big\|^2 +
			\max_{g^{(2)0} \in [G_2^0]} \Big\{ \Big| \alpha_{g_1^{(1)0} g^{(2)0}}^0 - \alpha_{g_2^{(1)0} g^{(2)0}}^0 \Big|^2 + \Big| \gamma^{(1)0}_{g_1^{(1)0}} \gamma^{(2)0}_{g^{(2)0}} -  \gamma^{(1)0}_{g_2^{(1)0}} \gamma^{(2)0}_{g^{(2)0}} \Big|^2 \Big\} \\
			& \le 2 \Big\|\btheta_{g_1^{(1)0}}^{(1)0}  - \btheta_{g_3^{(1)}}\Big\|^2 + 2  \Big\|\btheta_{g_1^{(1)0}}^{(1)0}  - \btheta_{g_3^{(1)}}\Big\|^2 \\
			& + \max_{g^{(2)0} \in [G_{2,0}]} \Big\{ \frac{1}{\pi_{g^{(2)0}, N_2}^{(2)} N_2} \sum_j  \Big| \alpha_{g_1^{(1)0} g_j^{(2)0}}^0 - \alpha_{g_2^{(1)0} g_j^{(2)0}}^0 \Big|^2 I(g_j^{(2)0} = g^{(2)0})  \Big\} \\
			& + \max_{g^{(2)0} \in [G_{2,0}]} \Big\{ \frac{1}{\pi_{g^{(2)0}, N_2}^{(2)} N_2} \sum_j  \Big| \gamma^{(1)0}_{g_1^{(1)0}} \gamma^{(2)0}_{g_j^{(2)0}} - \gamma_{g_2^{(1)0}}^{(1)0} \gamma_{g_j^{(2)0}}^{(2)0} \Big|^2 I(g_j^{(2)0} = g^{(2)0})  \Big\} \\
			& \le 2 \Big\|\btheta_{g_1^{(1)0}}^{(1)0}  - \btheta_{g_3^{(1)}}\Big\|^2 + 2  \Big\|\btheta_{g_1^{(1)0}}^{(1)0}  - \btheta_{g_3^{(1)}}\Big\|^2 \\
			& +  \max_{g^{(2)0} \in [G_{2,0}]} \Big( \frac{1}{\pi_{g^{(2)0}, N_2}^{(2)} N_2} \Big) \sum_j  \Big\{  \Big|  \alpha_{g_1^{(1)0} g_j^{(2)0}}^0 - \alpha_{g_2^{(1)0} g_j^{(2)0}}^0 \Big|^2 +  \Big| \gamma^{(1)0}_{g_1^{(1)0}} \gamma^{(2)0}_{g_j^{(2)0}} - \gamma_{g_2^{(1)0}}^{(1)0} \gamma_{g_j^{(2)0}}^{(2)0} \Big|^2 \Big\}  \\
			& \le 2 \Big\|\btheta_{g_1^{(1)0}}^{(1)0}  - \btheta_{g_3^{(1)}}\Big\|^2 + 2  \Big\|\btheta_{g_1^{(1)0}}^{(1)0}  - \btheta_{g_3^{(1)}}\Big\|^2 \\
			& + \Big( \frac{1}{ \min_{g^{(2)0} \in [G_{2,0}]}  \pi_{g^{(2)0}, N_2}^{(2)} N_2 } \Big) \sum_j  \Big\{  \Big|  \alpha_{g_1^{(1)0} g_j^{(2)0}}^0 - \alpha_{g_2^{(1)0} g_j^{(2)0}}^0 \Big|^2 \\
			& \hspace{14em} +  \Big| \gamma^{(1)0}_{g_1^{(1)0}} \gamma^{(2)0}_{g_j^{(2)0}} - \gamma_{g_2^{(1)0}}^{(1)0} \gamma_{g_j^{(2)0}}^{(2)0} \Big|^2 \Big\} \\
			& \le \Big( \frac{2}{ \min_{g^{(2)0} \in [G_{2,0}]}  \pi_{g^{(2)0}, N_2}^{(2)}} \Big) \Big\{ \Big\|\btheta_{g_1^{(1)0}}^{(1)0}  - \btheta_{g_3^{(1)}}\Big\|^2 + 2  \Big\|\btheta_{g_1^{(1)0}}^{(1)0}  - \btheta_{g_3^{(1)}}\Big\|^2 \\
			& + \frac{1}{N_2} \sum_j \Big| \alpha_{g_1^{(1)0} g_j^{(2)0}}^0 - \alpha_{g_3^{(1)} \wh g_j^{(2)}(\bxi)}  \Big|^2 +  \frac{1}{N_2} \sum_j \Big| \alpha_{g_2^{(1)0} g_j^{(2)0}}^0 - \alpha_{g_3^{(1)} \wh g_j^{(2)}(\bxi)}  \Big|^2\\
			& + \frac{1}{N_2} \sum_j \Big| \gamma^{(1)0}_{g_1^{(1)0}} \gamma^{(2)0}_{g_j^{(2)0}} - \gamma_{g_3^{(1)}}^{(1)} \gamma_{\wh g_j^{(2)}(\bxi)}^{(2)} \Big|^2 +  \frac{1}{N_2} \sum_j \Big| \gamma^{(1)0}_{g_2^{(1)0}} \gamma^{(2)0}_{g_j^{(2)0}} - \gamma_{g_3^{(1)}}^{(1)} \gamma_{\wh g_j^{(2)}(\bxi)}^{(2)} \Big|^2  \Big\} \\
			& \le \frac{4 \eta}{\min_{g^{(2)0} \in [G_{2,0}]}  \pi_{g^{(2)0}, N_2}^{(2)}},
		\end{align*}
		where the last line holds due to $g_3^{(1)} \in \mA_\eta^{(1)}(\bxi, g_1^{(1)0}, \mG_2^0) \cap \mA_\eta^{(1)}(\bxi, g_2^{(1)0}, \mG_2^0)$.
		However, this conclusion contradicts $\eta < c_\gap c_\pi /4$ as $N_2 \to \infty$.
		Therefore, we prove that $\mA_\eta^{(1)}(\bxi; g_1^{(1)0}, \mG_2^0) \cap \mA_\eta^{(1)}(\bxi; g_2^{(1)0}, \mG_2^0)  = \emptyset$, which completes the proof of (i).
		
		\noindent
		{\bf (2) Proof of (ii).}
		
		Denote $\bxi_{g_i^{(1)}}^{(1)} = (\btheta_{g_i^{(1)}}^{(1)\top}, \vec(\balpha_{g_i^{(1)} \cdot})^\top, \wt\bgamma_{g_i^{(1)}}^{(1)\top})$.
		Define the loss function
		\begin{align}
			Q_{i}(\bxi_{g_i^{(1)}}^{(1)}; \bxi_{g^{(2)}}^{(2)}, \mG_2) & = \sum_j \sum_t \Big\{ Y_{ij,t} - \gamma_{g_i^{(1)}}^{(1)} \gamma_{g_j^{(2)}}^{(2)} \sum_k \sum_m w_{im}^{(1)} Y_{mk, t-1} w_{kj}^{(2)} \nonumber\\
			& - \lambda_{g_i^{(1)}}^{(1)} \sum_k w_{ik}^{(1)} Y_{kj, t-1} - \lambda_{g_j^{(2)}}^{(2)} \sum_k w_{kj} Y_{ik, t-1}  \nonumber\\
			& - \alpha_{g_i^{(1)} g_j^{(2)}} Y_{ij,t-1} - \Big(\bx_{it}^{(1)\top} \bzeta_{g_i^{(1)}}^{(1)} + \bx_{jt}^{(2)\top} \bzeta_{g_j^{(2)}}^{(2)}  \Big)  \Big\}^2.\label{eq:loss_Qi1}
		\end{align}
		For simplicity, for any given $\bxi$ satisfying $\bxi \in \cN_{\eta}^{(1)}$, denote $\wh g_i^{(1)}(\bxi)$ as $\wh g_i^{(1)}$.
		Hence we have that for any $\wt g^{(1)} \neq g^{(1)}$,
		\begin{align*}
			\big\{\wh g_i^{(1)} = g^{(1)}\big\} \subseteq \big\{ Q_{i}(\bxi_{g^{(1)}}^{(1)}; \bxi_{g^{(2)}}^{(2)}, \wh \mG_2(\bxi)) \le Q_{i}(\bxi_{\wt g^{(1)}}^{(1)}; \bxi_{g^{(2)}}^{(2)}, \wh \mG_2(\bxi))\big\}.
		\end{align*}
		Hence, for $\wt g_i^{(1)} \in \mA_{\eta}^{(1)} (\bxi, g_i^{(1)0}, \mG_2^0)$,
		\begin{align*}
			& I \big( \wh g_i^{(1)} \notin \mA_{\eta}^{(1)} (\bxi, g_{i}^{(1)0},  \mG_2^0) \big) = \sum_{g^{(1)} = 1}^{G_1}
			I \big(g^{(1)} \notin \mA^{(1)}_{\eta}( \bxi, g_i^{(1)0}, \mG_2^0) \big)
			I \big(\wh g_{i}^{(1)} = g^{(1)} \big) \\
			& \le \sum_{g^{(1)} =1}^{G_1}
			I \big(g^{(1)} \notin \mA_{\eta}^{(1)} ( \bxi, g_i^{(1)0}, \mG_2^0)\big) I \big\{Q_{i}(\bxi_{g^{(1)}}^{(1)}; \bxi_{g^{(2)}}^{(2)} , \wh \mG_2(\bxi) )\big\} \defeq \sum_{g^{(1)} =1}^{G_1}  W_{i g^{(1)}}(\bxi).
		\end{align*}
		On one hand, for all $g^{(1)} \notin \mA_{\eta}^{(1)}(\bxi, g_i^{(1)0}, \mG_2^0)$ and $g^{(1)} \in [G_1]$, there exists a $g_{i_2}^{(1)0} \neq g_{i}^{(1)0}$, such that $g^{(1)} \in \mA_{\eta}^{(1)}(\bxi, g_{i_2}^{(1)0}, \mG_2^0)$ due to the conclusion of (i).
		Then we have
		\begin{align*}
			& \big\| \btheta_{g^{(1)}}^{(1)} -\btheta^{(1)0}_{g_i^{(1)0}}\big\|^2 + \frac{1}{N_2} \sum_j \Big\{ \big| \alpha_{g^{(1)} \wh g_j^{(2)}(\bxi)} -  \alpha^0_{g_i^{(0)}  g_j^{(2)0}} \big|^2 + \big| \gamma^{(1)}_{g^{(1)}} \gamma^{(2)}_{\wh g_j^{(2)}(\bxi)} -  \gamma^{(1)0}_{g_i^{(0)}} \gamma^{(2)0}_{g_j^{(2)0}} \big|^2 \Big\} \\
			& \ge \frac{1}{2} \big\| \btheta_{g_{i_2}^{(1)0}}^{(1)0} -\btheta^{(1)0}_{g_i^{(1)0}}\big\|^2 + \frac{1}{2 N_2} \sum_j \Big\{ \big| \alpha^0_{g_{i_2}^{(1)0} g_j^{(2)0}} -  \alpha^0_{g_i^{(1)0}  g_j^{(2)0}} \big|^2 + \big| \gamma^{(1)0}_{g_{i_2}^{(1)0}} \gamma^{(2)0}_{g_j^{(2)0}} -  \gamma^{(1)0}_{g_i^{(1)0}} \gamma^{(2)0}_{g_j^{(2)0}} \big|^2 \Big\} \\
			& - \Big[ \big\| \btheta_{g^{(1)}}^{(1)} -\btheta^{(1)0}_{g_{i_2}^{(1)0}}\big\|^2 + \frac{1}{N_2} \sum_j \Big\{ \big| \alpha_{g^{(1)} \wh g_j^{(2)}(\bxi)} -  \alpha^0_{g_{i_2}^{(1)0}  g_j^{(2)0}} \big|^2 + \big| \gamma^{(1)}_{g^{(1)}} \gamma^{(2)}_{\wh g_j^{(2)}(\bxi)} -  \gamma^{(1)0}_{g_{i_2}^{(1)0}} \gamma^{(2)0}_{g_j^{(2)0}} \big|^2 \Big\} \Big] \\
			& \ge c_\gap c_\pi/2 - \eta,
		\end{align*}
		when $N_2 \to \infty$,
		where $c_\gap$ and $c_\pi$ are defined in Assumption \ref{assum:group_diff_int} and Assumption \ref{assum:group_ratio}, respectively.
		By Lemma \ref{lem:Q_star_diff_1}, it holds for any $g^{(1)} \notin \mA_{\eta}^{(1)}(\bxi, g_i^{(1)0}, \mG_2^0)$,
		\begin{align}
			\frac{1}{N_2 T} \Big\{ Q^*_{i}(\bxi_{g^{(1)}}^{(1)}; \bxi_{g^{(2)}}^{(2)}, \wh\mG_2(\bxi)) - Q^*_{i}(\bxi_{g_i^{(1)0}}^{(1)}; \bxi_{g^{(2)}}^{(2)}, \mG_{2}^0)  \Big\} \ge \tau_{\min}^1 (c_\gap c_\pi/2 - \eta).\label{eq:Qi_star_diff_1}
		\end{align}
		On the other hand, for $\wt g_i^{(1)} \in \mA_{\eta}^{(1)} (\bxi, g_i^{(1)0}, \mG_2^0)$, it holds that
		\begin{align}
			& \frac{1}{N_2 T} \Big\{ Q^*_{i}(\bxi_{\wt g_i^{(1)}}^{(1)}; \bxi_{g^{(2)}}^{(2)}, \wh\mG_2(\bxi)) - Q^*_{i}(\bxi_{g_i^{(1)0}}^{(1)}; \bxi_{g^{(2)}}^{(2)}, \mG_{2}^0)  \Big\} \nonumber\\
			&  \le \tau_{\max}^1 \Big\{ \big\|  \btheta_{\wt g_i^{(1)}}^{(1)} -\btheta^{(1)0}_{g_{i}^{(1)0}} \big\|^2 + \frac{1}{N_2} \sum_j  \big\|  \btheta_{\wh g_j^{(2)}(\bxi)}^{(1)} -\btheta^{(1)0}_{g_{j}^{(2)0}} \big\|^2 \nonumber\\
			&  + \frac{1}{N_2} \sum_j  \big(  \big| \alpha_{\wt g_i^{(1)} \wh g_j^{(2)}(\bxi)} -  \alpha^0_{g_{i}^{(1)0} g_j^{(2)0}} \big|^2 + \big| \gamma^{(1)}_{\wt g_i^{(1)} } \gamma^{(2)}_{\wh g_j^{(2)}(\bxi)} -  \gamma^{(1)0}_{g_{i}^{(1)0}} \gamma^{(2)0}_{g_j^{(2)0}} \big|^2  \big)  \Big\} \nonumber\\
			& \le \tau_{\max}^1 (d(\bTheta, \bTheta^0) + \eta) \le \tau_{\max}^1 \{ C T^{-1} (\log N_1 + \log N_2)^2 + \eta \}, \label{eq:Qi_star_diff_2}
		\end{align}
		with probability tending to 1,
		where the first inequality holds due to Lemma \ref{lem:Q_star_diff_1}, and
		the last inequality holds due to \eqref{eq:d_condition}.
		Together with \eqref{eq:Qi_star_diff_1}, we have that with probability tending to 1,
		\begin{align*}
			& \frac{1}{N_2 T}  Q^*_{i}(\bxi_{g^{(1)}}^{(1)}; \bxi_{g^{(2)}}^{(2)}, \wh\mG_2(\bxi)) - Q^*_{i}(\bxi_{\wt g_i^{(1)}}^{(1)}; \bxi_{g^{(2)}}^{(2)}, \wh\mG_2(\bxi)) \\
			& \ge \tau_{\min}^1 (c_\gap c_\pi/2 - \eta) - \tau_{\max}^1 \{ C T^{-1} (\log N_1 + \log N_2)^2 + \eta \} \defeq  \epsilon_{\eta}.
		\end{align*}
		Then we apply the similar techniques as in \eqref{eq:W_il} and \eqref{eq:g_il_indicator}, to obtain that
		\begin{align*}
			& P \Big\{ \sup_i I(\wh g_i^{(1)} \notin \mA_{\eta}^{(1)}(\bxi, g_i^{(1)0}, \mG_2^0)) \Big\}\\
			& \le \sum_{g^{(1)}=1}^{G_1} P\Big\{ \sup_i W_{ig^{(1)}}(\bxi) > 0 \Big\} \\
			& \le G_1 \exp \Big\{ -C_2 \min(T \epsilon_{\eta}^2, \sqrt{T} \epsilon_{\eta}) + C_3 m + \log N_1  + \log N_2\Big\}.
		\end{align*}
		by Lemma \ref{lem:Q_concent_1}.
		Since $c_{\gap} \gg T^{-1} (\log (N_1 N_2)^2$, by the condition on $\eta$, we have $\eta < \{ \tau_{\min}^1 c_{\gap} c_{\pi} /4 - \tau_{\max}^1 C T^{-1} (\log (N_1 N_2))^2 \}/(\tau_{\min}^1 + \tau_{\max}^1)$.
		Hence, by the definition of $\epsilon_{\eta}$, we have
		$\epsilon_{\eta} > \tau_{\min}^1 c_{\gap} c_{\pi} /4$ as $N_2, T \to \infty$.
		This leads to that
		\begin{align*}
			P \Big\{ \sup_i I(\wh g_i^{(1)} \notin \mA_{\eta}^{(1)}(\bxi, g_i^{(1)0}, \mG_2^0)) \Big\} \le C \exp\big\{ -c_1 \sqrt{T} c_\gap + c_2 m + \log N_1 + \log N_2 \big\},
		\end{align*}
		where $m = p_1 + p_2 + 4$.

	\subsubsection{Technical Lemmas for Interactive Model}\label{append.lemma_int}

	\bel\label{lem:Q_concent_1}
	Under Assumptions \ref{assum:sub_gaussian}--\ref{assum:mixing} and \ref{assum:para_space_int}--\ref{assum:station_int}, we have
	\begin{align}
	& 	P\Big\{\sup_{\|\bTheta_{ij}\|_{\max}\le R} \Big| T^{-1} Q_{ij}(\bTheta_{ij})  - Q^*_{ij} (\bTheta_{ij})  \Big| \ge x  \Big\} \nonumber\\
		& \le C_1 \exp\Big\{ -C_2 \min(T x^2, \sqrt{T} x) + C_3 m \Big\}\label{eq:Qij_concent_int}
	\end{align}
	Further recall that $S_{ij}(\bTheta_{ij}) = Q_{ij}(\bTheta_{ij}) - Q_{ij}(\bTheta_{ij}^0)$, we have that
	\begin{align}
		P \Big\{ \sup_{\bTheta_{ij}} T^{-1} \frac{|S_{ij}(\bTheta_{ij}) - S_{ij}^*(\bTheta_{ ij})|}{d_{ij} (\bTheta_{ij}, \bTheta_{ ij}^0)} > x \Big\} \le C_1 \exp\Big\{ -C_2 \min(T x^2, \sqrt{T}x) + C_3m \Big\}.\label{eq:Sij_concent_int}
	\end{align}
	with $m = p_1 + p_2 + 4$, and $C_1, C_2, C_3$ are positive constants, and
	\begin{align*}
		d_{ij} (\bTheta_{ij}, \bTheta_{ij}^0)& = \|\wh \btheta_{\wh g_{i}^{(1)}}^{(1)} - \btheta_{g_{i}^{(1)}}^{(1)} \|^2 + \|\wh \btheta_{\wh g_{j}^{(2)}}^{(2)} - \btheta_{g_{j}^{(2)}}^{(2)} \|^2 \\
		&  +|\wh \alpha_{\wh g_{i}^{(1)}  \wh g_{j}^{(2)}} - \alpha_{g_{i}^{(1)} g_{j}^{(2)}}|^2 +  |\wh \gamma_{\wh g_{i}^{(1)}}^{(1)} \wh \gamma_{\wh g_{j}^{(2)}}^{(2)} -  \gamma_{ g_{i}^{(1)}}^{(1)}  \gamma_{ g_{j}^{(2)}}^{(2)} |^2.
	\end{align*}
	\eel

	\begin{proof}
		
		\noindent
		{\bf 1. Proof of \eqref{eq:Qij_concent_int}.}
		
		Note that
		\begin{align*}
			T^{-1} Q_{ij}(\bTheta_{ ij}) = T^{-1} \sum_t \big(\ve_{ij, t} + \cX_{ij, t}^\top \bTheta_{ ij}^0 - \cX_{ij, t}^\top \bTheta_{ij}\big)^2,
		\end{align*}
		where $\cX_{i j, t}$ is defined in \eqref{eq:X1}.
		We use the similar procedure in the proof of Lemma \ref{lem:Q_diff}.
		The three core steps in the proof of Lemma \ref{lem:Q_diff} are \eqref{eq:eps2_conv}--\eqref{eq:X2_conv}.
		While the implementation of \eqref{eq:eps2_conv} remains the same, the interactive network terms in $\cX_{i j, t}$ and $\bTheta_{ij}$ cause changes in using \eqref{eq:epsX_conv} and \eqref{eq:X2_conv}.
		We take the change in \eqref{eq:X2_conv} for example.
		
		Specifically, the diagonal elements of $T^{-1} \sum_t \cX_{i j, t} \cX_{i j, t}^\top$ take the forms $T^{-1} \sum_t \bw^\top \mY_t \mY_t^\top \bw$, $T^{-1} \sum_t \|\bx_{i t}^{(1)} \|^2$, and $T^{-1} \sum_t \|\bx_{j t}^{(2)} \|^2$.
		For the additive network term, $\bw$ is replaced by $\bW^{(1)\top}_{i\cdot} \otimes \one_{N_2} \in \mR^{N_1 \times N_2}$, while for the interactive network term, $\bw$ is replaced by $\bW_{i \cdot}^{(1) \top} \otimes \bW_{\cdot j}^{(2)} \in \mR^{N_1 \times N_2}$.
		Both types of network terms satisfy that $\|\bw\|_1 = 1$.
		Hence, for $T^{-1} \sum_t \bw^\top \mY_t \mY_t^\top \bw$, we apply Lemma \ref{lem:concenX} to prove the concentration inequality.
		For $T^{-1} \sum_t \|\bx_{i t}^{(1)} \|^2$, and $T^{-1} \sum_t \|\bx_{j t}^{(2)} \|^2$, we use Lemma \ref{lem:concenX} and Assumptions \ref{assum:mixing}.
		
		Then, similar with the Step 2 in the proof of Lemma \ref{lem:Q_diff}, by taking union upper bound for all $\|\bTheta_{ij}\|_{\max}\le R$, we obtain the final conclusion.
		
		\noindent
		{\bf 2. Proof of \eqref{eq:Sij_concent_int}.}
		
		Note that
		\begin{align}
			S_{ij}(\bTheta_{ij}) - S_{ij}^*(\bTheta_{ij}) & = \big[(\bTheta_{ij} - \bTheta_{ ij}^0)^\top \cX_{i j, t} \cX_{i j, t}^\top(\bTheta_{ij} - \bTheta_{ij}^0) \big] \nonumber\\
			&  - E  \big[(\bTheta_{ij} - \bTheta_{ ij}^0)^\top \cX_{i j, t} \cX_{i j, t}^\top(\bTheta_{ij} - \bTheta_{ ij}^0) \big] \label{eq:Sij_concent_term1}\\
			& + 2 \cX_{i j, t}^\top (\bTheta_{ij} - \bTheta_{ ij}^0) \ve_{ij, t}.\label{eq:Sij_concent_term2}
		\end{align}
		We borrow the idea to prove Lemma \ref{lem:Q_diff} to derive the results.
		For instance, we can apply \eqref{eq:Q_ij_diff} on \eqref{eq:Sij_concent_term1} by using the fact that $\|\bTheta_{ij} - \bTheta_{ij}^0\|/\sqrt{d_{ij} (\bTheta_{ij}, \bTheta_{ij}^0)} \le cR$ for some constant $c$.
		Similar techniques can be applied on proof of \eqref{eq:Sij_concent_term2}.	
	\end{proof}

\bel\label{lem:S_Xtheta_tail_int}
Under Assumptions \ref{assum:sub_gaussian}--\ref{assum:mixing} and \ref{assum:para_space_int}--\ref{assum:station_int},
for vector $\bDelta \in \mR^{p_1+ p_2+4}$ in the parameter space of the interactive model \eqref{eq:model_int_2},
we have
\begin{align}
	& P \Big\{\sup_{i}	\sup_{\| \bDelta\|^2 \le \omega^2}  (N_2 T)^{-1}
	\Big| \sum_{j}  \sum_t  \Big\{\bDelta^\top \cX_{ij,t} \cX_{ij,t}^\top \bDelta  -  \big[\bDelta^\top E(\cX_{ij,t} \cX_{ij,t}^\top) \bDelta \big] \Big\}  \Big| \ge x \Big\} \nonumber \\
	& \le C_1 \exp\Big\{ -C_2 \min\Big( T x^2/\omega^2, \sqrt{T}x/\omega \Big) + C_3 m  + \log (N_1 N_2 )\Big\}, \nonumber\\
	& P \Big\{ \sup_{i} \sup_{\| \bDelta\|^2 \le \omega^2}  (N_2 T)^{-1} \Big|\sum_{j}  \sum_t \cX_{ij,t}^\top \bDelta \ve_{ij, t} \Big| \ge x \Big\} \nonumber\\
	& \le  C_1 \exp\Big\{ -C_2 \min\Big( T x^2/\omega^2, \sqrt{T}x/\omega \Big) + C_3 m  + \log (N_1 N_2) \Big\},\nonumber
\end{align}
where $\cX_{ij, t}$ is defined in equation \eqref{eq:X1}, 
$m = p_1+ p_2 + 4$, $C_1, C_2, C_3$ are positive constants.
\eel

\begin{proof}
	The proof is similar with the proof of Lemma \ref{lem:S_Xtheta_tail}. The key difference lies in the expression of $\cX_{i j, t}$. The technical details can refer to the proof of Lemma \ref{lem:Q_concent_1}, which is omitted here.
\end{proof}

	\bel\label{lem:Q_star_diff_1}
			Under Assumption \ref{assum:convex_int}, it holds that
			\begin{align*}
				& \tau_{\min}^1 d_{ i}(\bTheta_{i \cdot, 1}, \bTheta_{i \cdot}^0) \le \frac{1}{N_2 T} \Big\{ Q_{i}^*(\bxi_{g_i^{(1)}}^{(1)}; \bxi_{g^{(2)}}^{(2)}, \mG_2) - Q_{i}^*(\bxi_{g_i^{(1)0}}^{(1)0}; \bxi_{g^{(2)0}}^{(2)0}, \mG_2^0)  \Big\} \\
				& \le \tau_{\max}^1  d_{i}(\bTheta_{i \cdot}, \bTheta_{i \cdot}^0), \\
				& \tau_{\min}^1 d(\bTheta, \bTheta^0)  \le \frac{1}{N_2 T} \Big\{ Q^*(\bTheta) - Q^*(\bTheta^0)  \Big\} \le \tau_{\max}^1  d(\bTheta, \bTheta^0)
			\end{align*}
			where $\tau_{\min}^1$ is defined in Assumption \ref{assum:convex_int}, and $\tau_{\max}^1$ is defined in Lemma \ref{lem:tau_max_1}.
			Besides, $ \bTheta_{i \cdot} = (\bTheta_{ij}^\top: j \in [N_2])\in \mR^{N_2 \times m}$, $\bTheta_{ij} = (\btheta_{ g_{i}^{(1)}}^{(1)\top}, \btheta_{ g_{j}^{(2)}}^{(2)\top}, \alpha_{ g_{i}^{(1)}   g_{j}^{(2)}}, \gamma_{ g_{i}^{(1)}} \gamma_{ g_{j}^{(2)}})^\top \in \mR^{m}$, and
			\begin{align*}
				d_{ i}(\bTheta_{i \cdot}, \bTheta_{i \cdot}^0) & = N_2^{-1} \sum_j \| \bTheta_{ij} - \bTheta_{ij}^0 \|^2 \\
				& =  \| \btheta_{g_i^{(1)}}^{(1)} - \btheta_{g_i^{(1)0}}^{(1)0} \|^2 + N_2^{-1} \sum_j \|  \btheta_{g_j^{(2)}}^{(2)} - \btheta_{g_j^{(2)0}}^{(2)0}\|^2 \\
				& + N_2^{-1} \sum_j \Big\{ \big| \alpha_{ g_{i}^{(1)}   g_{j}^{(2)}} - \alpha_{ g_{i}^{(1)0}   g_{j}^{(2)0}}^0 \big|^2 + \big| \gamma_{ g_{i}^{(1)}}^{(1)} \gamma_{ g_{j}^{(2)}}^{(2)} - \gamma_{ g_{i}^{(1)0}}^{(1)0} \gamma_{ g_{j}^{(2)0}}^{(2)0} \big|^2 \Big\}.
			\end{align*}
	\eel
	\begin{proof}
		By the definition of loss function $Q_{i}(\bxi_{g_i^{(1)}}^{(1)}; \bxi_{g^{(2)}}^{(2)}, \mG_2)$ in \eqref{eq:loss_Qi1},
		we know $Q_{i}^*(\bxi_{g_i^{(1)}}^{(1)}; \bxi_{g^{(2)}}^{(2)}, \\ \mG_2) = \sum_j Q_{ij}^*(\bTheta_{ij})$ and $Q_{i}^*(\bxi_{g_i^{(1)0}}^{(1)0}; \bxi_{g^{(2)0}}^{(2)0}, \mG_2^0) = \sum_j Q_{ij}^*(\bTheta_{ij}^0)$ {by following the similar derivation in \eqref{eq:Q_il}}.
		Since $Q_{ij}(\bTheta_{ ij}) = \sum_t (\ve_{ij, t} + \cX_{ij, t}^\top \bTheta_{ij}^0 - \cX_{ij, t}^\top \bTheta_{ij})^2$, where $\cX_{i j, t}$ is defined in \eqref{eq:X1},
		we can use similar techniques in the proof of Lemma \ref{lem:Q_star_diff} to obtain the conclusion.
	\end{proof}

	\bel\label{lem:tau_max_1}
	Under Assumption \ref{assum:mixing} and \ref{assum:station_int}, we have $\tau_{\max}^1 = \max_{i, j} \lambda_{\max}(E(\cX_{i j, t} \cX_{i j, t}^\top)) < \infty$ with $\cX_{i j, t}$ defined in \eqref{eq:X1}.
	\eel
	\begin{proof}
		We use the similar techniques in the proof of Lemma \ref{lem:tau_max}.
		Due to the interactive network term, the different term in $\cX_{i j, t}$ is $ (\bW_{\cdot j}^{(2)} \otimes \bW_{i \cdot}^{(1)\top})^\top \mY_t$, with $\|\bW_{\cdot j}^{(2)} \otimes \bW_{i \cdot}^{(1)\top}\|_1 = 1$.
		{Denote $\bw = \bW_{\cdot j}^{(2)} \otimes \bW_{i \cdot}^{(1)\top}$,}
		this typical term has the property that $\var(\bw^\top \mY_t) = \bw^\top \bGamma \bw \le \| \bGamma \|_{\max}  |\bw^\top \one \one^\top \bw|= \| \bGamma \|_{\max} < c_\Gamma$, where $c_\Gamma$ is the upper bound conclusion in Lemma \ref{lem:10}.
		Other terms are the same as in the Lemma \ref{lem:tau_max}.
	\end{proof}

\section{Discussion about Future Extension}\label{sec:extension}

We remark that the GTNAR model can be easily extended to a general-purpose machine learning model by considering nonlinear terms.
Take $q=2$ for example, model \eqref{eq:model00} can be extended to 
	\begin{align}
	Y_{ij, t} & = \lambda_{g_i^{(1)}}^{(1)} \sum_{k = 1}^{N_1} \frac{a^{(1)}_{ik}}{n_{1i}} Y_{kj, (t-1)}
	+ \lambda_{g_j^{(2)}}^{(2)} \sum_{k = 1}^{N_2} \frac{a^{(2)}_{kj}}{n_{2j}}Y_{ik, (t-1)}\\
	&  + \alpha_{g_i^{(1)} g_j^{(2)}}Y_{ij, (t-1)} + f_1(\bx_{it}^{(1)})+
	f_2(\bx_{jt}^{(2)}) +\ve_{ij, t},\label{eq:model_nonlinear}
\end{align}
where $f_1(\cdot)$ and $f_2(\cdot)$ are non-linear functions.
To estimate model \eqref{eq:model_nonlinear}, we could conduct the following (1) and (2) iteratively.

{\bf (1) Estimate the linear part}. First, given the non-linear part $f(\bx_{it}^{(1)})$ and $f(\bx_{jt}^{(2)})$, calculate $\wt Y_{ij, t} = Y_{ij, t} - f_1(\bx_{it}^{(1)}) - f_2(\bx_{jt}^{(2)})$.
Then we implement Algorithm \ref{alg:gmnar} to estimate the other parameters and group memberships iteratively.

{\bf (2) Estimate the non-linear part}. Given the parameters $\lambda_{g_i^{(1)}}^{(1)}$, $ \lambda_{g_j^{(2)}}^{(2)}$, and $\alpha_{g_i^{(1)} g_j^{(2)}}$, calculate $\widecheck{Y}_{ij, t} = Y_{ij, t} - \lambda_{g_i^{(1)}}^{(1)} \sum_{k = 1}^{N_1} \frac{a^{(1)}_{ik}}{n_{1i}} Y_{kj, (t-1)}
-\lambda_{g_j^{(2)}}^{(2)} \sum_{k = 1}^{N_2} \frac{a^{(2)}_{kj}}{n_{2j}}Y_{ik, (t-1)} - \alpha_{g_i^{(1)} g_j^{(2)}}Y_{ij, (t-1)}$. Subsequently, one can use deep learning model to fit the function $f_1(\cdot)$ and $f_2(\cdot)$.

Since the discussion of theoretical guarantees is beyond our scope, we leave it as an interesting future topic.

\section{Proof of Main Theorems}\label{sec:main_thm}

	\subsection{Proof of Theorem \ref{thm:pseudo_dist}}\label{append:proof_dist}
	
	Recall that $d(\wh\bTheta, \bTheta)$ is defined in \eqref{eq:pseudo_dist}.
	To prove $d(\wh\bTheta, \bTheta^0) = O_p(a_{NT}^2)$, it suffices to show that
	\begin{align*}
		P\Big\{a_{NT}^{-1} \sqrt{d(\wh\bTheta, \bTheta^0)} >K \Big\} \to 0 ~\textup{as}~ K \to \infty.
	\end{align*}
	
	Denote $\bOmega$ as the parameter space, and partition the space into shells $\bOmega_{j} = \{ \bTheta \in \bOmega: j-1 < a_{NT}^{-1} \sqrt{d(\bTheta, \bTheta^0)} \le j \}$, where $j \ge 1$ takes over all positive integers.
	For a positive integer $K$, if $a_{NT}^{-1} \sqrt{d(\wh\bTheta, \bTheta^0)} \ge K$, then $\wh\bTheta \in \bOmega_{j}$ with $j \ge K$.
	Since $\wh\bTheta$ minimizes objective function $Q(\bTheta)$, we have $(\prod_l N_l T)^{-1} S(\wh\bTheta) \defeq (\prod_l N_l T)^{-1}Q(\wh\bTheta) -(\prod_l N_l T)^{-1} Q(\bTheta^0) \le 0$.
	Denote $S^*(\bTheta) = E\{ S(\bTheta) \}$.
	Then we have
	\begin{align*}
		P\Big\{ a_{NT}^{-1} \sqrt{d(\bTheta, \wh\bTheta)} > K  \Big\}&  \le P \Big\{ \inf_{\bTheta \in \bigcup_{  j \ge K} \bOmega_j}  (\prod_l N_l T)^{-1} S(\bTheta)  \le 0 \Big\}\\
		&\le \sum_{j \ge K} P \Big\{ \inf_{\bTheta \in \bOmega_j} (\prod_l N_l T)^{-1} S(\bTheta) \le 0 \Big\} \\
		&\le \sum_{j \ge K} P \Big\{ \inf_{\bTheta \in \bOmega_j}  (\prod_l N_l T)^{-1} \{ S(\bTheta) -  S^*(\bTheta)\} \\
		& +  \inf_{\bTheta \in \bOmega_j}  (\prod_l N_l T)^{-1}S^*(\bTheta) \le 0\Big\}
	\end{align*}
	By Lemma \ref{lem:Q_star_diff}, we have that
	$(\prod_l N_l T)^{-1} S^*(\bTheta) = (\prod_l N_l T)^{-1} \{Q^*(\bTheta) -  Q^*(\bTheta^0)\} \ge \tau_{\min} d(\bTheta, \bTheta^0)$ with $\tau_{\min}$ defined in Assumption \ref{assum:tau_min}.
	Since $d(\bTheta, \bTheta^0) > (j-1)^2 a_{NT}^2$ for all $\bTheta \in \bOmega_j$, we have
	\begin{align}
		P\Big\{ a_{NT}^{-1} \sqrt{d(\bTheta, \wh\bTheta)} > K  \Big\}  & \le  \sum_{j \ge K} P \Big\{ \inf_{\bTheta \in \bOmega_j}  (\prod_l N_l T)^{-1} \{ S(\bTheta) -  S^*(\bTheta)\} \le - \tau_{\min} (j-1)^2 a_{NT}^2 \Big\} \nonumber\\
		&\le  \sum_{j \ge K} P \Big\{ \sup_{\bTheta \in \bOmega_j}  (\prod_l N_l T)^{-1} \big| S(\bTheta) -  S^*(\bTheta)\big|  \ge \tau_{\min} (j-1)^2 a_{NT}^2 \Big\}.\label{eq:distance_prepare}
	\end{align}

	Denote $S_{i_1 \cdots i_q}(\bTheta_{i_1 \cdots i_q}) = Q_{i_1 \cdots i_q}(\bTheta_{i_1 \cdots i_q}) - Q_{i_1 \cdots i_q}(\bTheta_{i_1 \cdots i_q}^0)$, where $Q_{i_1 \cdots i_q}(\bTheta_{i_1 \cdots i_q})$ is defined in \eqref{def:Qij} in the main text.
	For simplicity we denote $d_{i_1 \cdots i_q} = d_{i_1 \cdots i_q} (\bTheta_{i_1 \cdots  i_q}, \bTheta_{i_1 \cdots  i_q}^0)$.
	We next show that
	\begin{align*}
		&  P \Big\{  \sup_{d(\bTheta, \bTheta^0)\le \omega^2} (\prod_l N_l T)^{-1} \big| S(\bTheta) -  S^*(\bTheta)\big|  \ge x \Big\} \\
		 & \le C_1 \exp\Big\{-C_2 \min(T x^2/\omega^2, \sqrt{T} x /\omega) + C_3 m + \log \big(\prod_l N_l \big)  \Big\}.
	\end{align*}
	We have
	\begin{align*}
		& P \Big\{  \sup_{d(\bTheta, \bTheta^0)\le \omega^2} (\prod_l N_l T)^{-1} \big| S(\bTheta) -  S^*(\bTheta)\big|  \ge x \Big\} \\
		& \le P \Big\{  \sup_{d(\bTheta, \bTheta^0)\le \omega^2} (\prod_l N_l)^{-1} \sum_l \sum_{i_l} T^{-1} \big| S_{i_1 \cdots i_q}(\bTheta_{i_1 \cdots i_q}) -  S_{i_1 \cdots i_q}^*(\bTheta_{i_1 \cdots  i_q})\big|  \ge x \Big\}  \\
		& = P \Big\{  \sup_{d(\bTheta, \bTheta^0)\le \omega^2} \frac{1}{\prod_l N_l} \sum_l \sum_{i_l} \frac{ T^{-1} \big| S_{i_1 \cdots i_q}(\bTheta_{i_1 \cdots i_q}) -  S_{i_1 \cdots i_q}^*(\bTheta_{i_1 \cdots  i_q})\big| }{\sqrt{d_{i_1 \cdots i_q} }} \sqrt{d_{i_1 \cdots i_q} } \ge x \Big\} \\
		& \le P \Big\{  \sup_{d(\bTheta, \bTheta^0)\le \omega^2} \sqrt{ \frac{1}{\prod_l N_l} \sum_l \sum_{i_l}  \frac{T^{-2}\big| S_{i_1 \cdots i_q}(\bTheta_{i_1 \cdots i_q}) -  S_{i_1 \cdots i_q}^*(\bTheta_{i_1 \cdots  i_q})\big| ^2}{d_{i_1 \cdots i_q} }  }  \sqrt{d_{i_1 \cdots i_q} } \ge x \Big\} \\
		& = P \Big\{  \sup_{d(\bTheta, \bTheta^0)\le \omega^2} { \frac{1}{\prod_l N_l} \sum_l \sum_{i_l}  \frac{T^{-2}\big| S_{i_1 \cdots i_q}(\bTheta_{i_1 \cdots i_q}) -  S_{i_1 \cdots i_q}^*(\bTheta_{i_1 \cdots  i_q})\big| ^2}{d_{i_1 \cdots i_q} }  }    {d_{i_1 \cdots i_q} }\ge x^2\Big\} \\
		& \le P \Big\{  \sup_{d(\bTheta, \bTheta^0)\le \omega^2} { \frac{1}{\prod_l N_l} \sum_l \sum_{i_l}  \frac{T^{-2}\big| S_{i_1 \cdots i_q}(\bTheta_{i_1 \cdots i_q}) -  S_{i_1 \cdots i_q}^*(\bTheta_{i_1 \cdots  i_q})\big| ^2}{d_{i_1 \cdots i_q} }  }  \omega^2 \ge x^2\Big\} \\
		& \le \sum_l \sum_{i_l}  P \Big\{   \sup_{d(\bTheta, \bTheta^0)\le \omega^2}{ \frac{T^{-2}\big| S_{i_1 \cdots i_q}(\bTheta_{i_1 \cdots i_q}) -  S_{i_1 \cdots i_q}^*(\bTheta_{i_1 \cdots  i_q})\big| ^2}{d_{i_1 \cdots i_q} }  }  \omega^2 \ge x^2\Big\} \\
		& \le \sum_l \sum_{i_l}  P \Big\{   \sup_{\bTheta_{i_1 \cdots  i_q}}{ \frac{T^{-1}\big| S_{i_1 \cdots i_q}(\bTheta_{i_1 \cdots i_q}) -  S_{i_1 \cdots i_q}^*(\bTheta_{i_1 \cdots  i_q})\big|}{\sqrt{d_{i_1 \cdots i_q} }  }}  {\omega} \ge x\Big\} \\
		& \le C_1 \exp\Big\{-C_2 \min(T x^2/\omega^2, \sqrt{T} x /\omega) + C_3 m + \log \big(\prod_l N_l \big)  \Big\},
	\end{align*}
	where the third inequality holds due to $\sum_i (a_i^2 b_i^2) \le (\sum_i a_i^2) (\sum_i b_i^2)$ and the last inequality holds due to Lemma \ref{lem:Sij_concent}.
	Here $m = \sum_l (p_l+1) + 1$ is a fixed constant.
	Then plugging the last line into \eqref{eq:distance_prepare}, we have
	\begin{align}
		& P\Big\{ a_{NT}^{-1} \sqrt{d(\bTheta, \wh\bTheta)} > K  \Big\}  \le  \sum_{j \ge K} P \Big\{ \sup_{\bTheta \in \bOmega_j}  (\prod_l N_l T)^{-1} \big| S(\bTheta) -  S^*(\bTheta)\big|  \ge \tau_{\min} (j-1)^2 a_{NT}^2 \Big\} \nonumber\\
		& \le \sum_{j \ge K} C_1 \exp\Big\{ -C_2 \min \Big( \frac{T (j-1)^4 a_{NT}^2}{j^2}, \frac{\sqrt{T}(j-1)^2 a_{NT}}{j} \Big) + C_3 m + \log \big(\prod_l N_l \big)  \Big\} \nonumber\\
		& \le  \sum_{j \ge K}  C_1 \exp\Big\{ -C_2 \min \Big( \frac{(j-1)^4 \log^2(\prod_l N_l)  }{j^2}, \frac{ (j-1)^2 \log(\prod_l N_l) }{j} \Big) + C_3 m + \log \big(\prod_l N_l \big)\Big\}  \nonumber\\
		& = \sum_{j \ge K}  C_1 \exp\Big\{ -C_2  \frac{ (j-1)^2  \log(\prod_l N_l) }{j} + C_3 m + \log \big(\prod_l N_l \big)\Big\} \nonumber\\
		& \le \sum_{j \ge K}  C_1 \exp \Big\{-C_2 (j-1) (1-K^{-1}) \log(\prod_l N_l) + C_3 m + \log \big(\prod_l N_l \big) \Big\} \nonumber\\
		&  =   \frac{\exp \big\{-C_2 (K-1) (1-K^{-1})  \log(\prod_l N_l) + C_3 m + \log \big(\prod_l N_l \big) \big\}}{1 -  \exp \big\{ -C_2(1-K^{-1}) \log(\prod_l N_l)\big\}} \nonumber \\
		& \le  \frac{\exp(-C_2 (K-1) (1-K^{-1}) - 1)  \log(\prod_l N_l)+ C_3 m }{1 - \exp(-C_2 \log(\prod_l N_l))} \to 0 \label{eq:tail_K}
	\end{align}
	when $K \to \infty$.
	The third inequality holds by plugging in the expression of $a_{NT}^2 = \log^2 (\prod_l N_l) T^{-1}$.
	The equality holds due to the sum of a geometric sequence.
	Then we finish the proof.

\subsection{Proof of Theorem \ref{thm:select_GH}}

In the following proof we write $Q(\wh \bxi(\underline{G}), \wh \mG(\underline{G}))$
as $Q(\wh \bTheta(\underline{G}))$ as we defined in \eqref{def:Qij}.
We use the notations here to denote $\wh \bTheta(\underline{G})$ as the estimate when group numbers $G_1, \cdots, G_q$ are specified.
Define $G_{-l} = (G_k:k\ne l)^\top \in \mR^{q-1}$ and we use $G_{-l}\ge G_{-l,0}$ to denote that  $G_{k}\ge G_{k,0}$
for $k\ne l$.
To establish the consistency property, we consider both the underfitted model
(there exists an $l$ with $G_l < G_{l,0}$) and the overfitted model ($\{G_l>G_{l,0}, G_{-l} \ge G_{-l,0}\}$ for any $l$).
In the following we show $\qic(\underline{G}_0) < \qic(\underline{G})$ with probability tending to 1 under both circumstances.
We write the difference of the two criteria as
\begin{align*}
	\qic(\underline{G}) - \qic(\underline{G}_0) &= \log\{Q(\wh \bTheta(\underline{G}))\}
	-\log\{Q(\wh \bTheta(\underline{G}_0))\} + \lambda(\underline{G})-\lambda(\underline{G}_0)\\
	&  = \log \Big\{1+\frac{Q(\wh \bTheta(\underline{G})) -
		Q(\wh \bTheta(\underline{G}_0))}{Q(\wh \bTheta(\underline{G}_0))}\Big\}+
	\lambda(\underline{G})-\lambda(\underline{G}_0).
\end{align*}

\noindent
{\sc 1. Overfitted Model.}
In the following we show that $ (\prod_l N_l T)^{-1} \{ Q(\wh\bTheta(\underline{G}_0)) -  Q(\wh\bTheta(\underline{G}))\} = (\prod_l N_l T)^{-1} \{ [Q(\wh\bTheta(\underline{G}_0)) - Q(\bTheta^0)]  - [ Q(\wh\bTheta(\underline{G})) -Q(\bTheta^0) ]\} = (\prod_l N_l T)^{-1} \{ S(\wh\bTheta(G_0))  - S(\wh\bTheta(\underline{G}))\}  =  O_p(a_{NT})$,
where $a_{NT} = T^{-1} (\sum_l \log N_l)^2$.
Note that we have
\begin{align*}
	&\frac{1}{(\prod_l N_l) T}\big| S(\wh \bTheta(\underline{G})) -
	S(\wh \bTheta(\underline{G}_0))\big|
	\le
	\frac{1}{(\prod_l N_l)T}\big|S(\wh \bTheta(\underline{G}_0)) - S^*(\wh \bTheta(\underline{G}_0))\big|\\
	&+
	\frac{1}{(\prod_l N_l)T}\big|S^*(\wh \bTheta(\underline{G}_0)) - S^*(\wh \bTheta(\underline{G}))\big|+
	\frac{1}{(\prod_l N_l)T}\big|S(\wh \bTheta(\underline{G})) - S^*(\wh \bTheta(\underline{G}))\big|\\
	& = O_p(T^{-1}(\sum_l \log N_l)^2) + \frac{1}{(\prod_l N_l)T}\big|S^*(\wh \bTheta(\underline{G}_0)) - S^*(\wh \bTheta(\underline{G}))\big|
\end{align*}
by Lemma \ref{lem:Sij_concent}.

Further by Lemma \ref{lem:Q_star_diff} and Theorem \ref{thm:pseudo_dist}, it holds that
\begin{align*}
	& \frac{1}{(\prod_l N_l)T}\big|S^*(\wh \bTheta(\underline{G}_0)) - S^*(\wh \bTheta(\underline{G}))\big|  \\
	&\le \frac{1}{(\prod_l N_l)T}\big|Q^*(\wh \bTheta(\underline{G}_0)) - Q^*(
	\bTheta^0)\big|+\frac{1}{(\prod_l N_l)T}\big|Q^*(\bTheta^0)- Q^*(\wh
	\bTheta(\underline{G}))\big|\\
	&\le \tau_{\max} \Big\{d(\wh \bTheta(\underline{G}_0), \bTheta^{0}) +
	d(\wh \bTheta(\underline{G}), \bTheta^{0})\Big\}=  O_p (T^{-1}( \sum_l \log  N_l)^2),
\end{align*}
where $\tau_{\max}$ is defined in Assumption \ref{assum:tau_min}.
Since $m$ is a fixed constant,
$((\prod_l N_l)T)^{-1}\big|Q(\wh \bTheta(\underline{G}_0)) - Q(\wh
\bTheta(\underline{G}))\big|
= O_p \{T^{-1}(\sum_l \log N_l)^2\}$,
i.e.,
\bse
\frac{1}{(\prod_l N_l)T}Q(\wh \bTheta(\underline{G})) =
\frac{1}{(\prod_l N_l)T}Q(\wh
\bTheta(\underline{G}_0))+ O_p \{T^{-1}(\sum_l \log N_l)^2\}.
\ese

Next we show $((\prod_l N_l)T)^{-1}Q(\wh \bTheta(\underline{G}_0))=\sigma^2 +o_p(1)$.
Similar to above while replacing $\wh\bTheta(\underline{G})$
by $\bTheta^0$, we have
\begin{align}
	((\prod_l N_l)T)^{-1}Q(\wh \bTheta(\underline{G}_0))
	= ((\prod_l N_l)T)^{-1}Q^*(\bTheta^{0}) +  O_p \{T^{-1}(\sum_l \log N_l)^2\}.\label{eq:diff_Q_hat_Qstar_Theta0}
\end{align}
Here we have $((\prod_l N_l)T)^{-1}Q^*(\bTheta^{0}) = \sigma^2 >0$.
This shows that $((\prod_l N_l)T)^{-1}Q(\wh \bTheta(\underline{G}_0))=\sigma^2 +o_p(1)$.
It implies
$\qic(\underline{G}) - \qic(\underline{G}_0) =  O_p \{T^{-1}(\sum_l \log N_l)^2\} + \lambda(\underline{G})-\lambda(\underline{G}_0).$
We can verify that $ \lambda(\underline{G}) - \lambda(\underline{G}_0) \ge \kappa$
under the overfitting case, where $\kappa = \lambda(\underline{G})/(\sum_l G_l)$.
Further note that $ \kappa \gg T^{-1}(\sum_l \log N_l)^2$ as we assume,
this implies $\qic(\underline{G}_0)<\qic(\underline{G})$ with probability tending to 1 under the overfitting case.

\noindent
{\sc 2. Underfitted Model.}
From the specification of $\lambda(\underline{G})$, we have
$ |\lambda(\underline{G})-\lambda(\underline{G}_0) |= o(c_\gap  /(\prod_l G_l))$.
It suffices to show that
$((\prod_l N_l)T)^{-1}\{Q(\wh \bTheta(\underline{G})) -
Q(\wh \bTheta(\underline{G}_0))\}\ge C_1c_\gap (c_\pi)^q / \\ (\prod_l G_l)$
for a positive constant $C_1$ {when $N_l \to \infty$.}
Note that
\begin{align*}
& ((\prod_l N_l)T)^{-1}\{Q(\wh \bTheta(\underline{G})) -
Q(\wh \bTheta(\underline{G}_0))\}\\
& =((\prod_l N_l)T)^{-1}\{
Q(\wh \bTheta(\underline{G})) -Q^*(\bTheta^{0})-
Q(\wh \bTheta(\underline{G}_0))+Q^*(\bTheta^{0})\},
\end{align*}
and we have already proved that
$((\prod_l N_l)T)^{-1}\{Q(\wh \bTheta(\underline{G}_0)) -Q^*(\bTheta^{0})\} =
 O_p(T^{-1}(\sum_l \log N_l)^2) =  o_p\{c_\gap  /(\prod_l G_l)\}$  in \eqref{eq:diff_Q_hat_Qstar_Theta0}. Also note that $c_\pi$ is a constant according to Assumption \ref{assum:group_ratio}.
Hence it suffices to show $((\prod_l N_l)T)^{-1}\{Q(\wh \bTheta(\underline{G}))-Q^*(\bTheta^{0})\}
\ge Cc_\gap (c_\pi)^q /(\prod_l G_l)$ for some positive constant $C$ when $N_l \to \infty$.

Without loss of generality, we consider $G_l<G_{l,0}$.
To prove the result, we use two steps.
First, we show that by Assumption \ref{assum:group_diff}, we have
\begin{align}
	\max_{g^{(l)} \in [G_{l,0}]}
	\max_{g^{-(l)} \in [G_{-l}^0]}
	\min_{g^{(l)'} \in [G_{l}]}\Big\{
	\|\wh \btheta_{g^{(l)'}}^{(l)} - \btheta_{g^{(l)}}^{(l)0}\|^2
	& + |\wh \alpha_{\varphi_1(g^{(1)}) \cdots \varphi_{l-1}(g^{(l-1)}) g^{(l)'} \varphi_{l+1}(g^{(l+1)}) \cdots \varphi_q(g^{(q)})} \nonumber\\
	&- \alpha_{g^{(1)} \cdots g^{(l-1)} g^{(l)} g^{(l+1)} \cdots g^{(q)}}^0|^2\Big\}\ge  c_\gap/4,\label{eq:max_min_lower}
\end{align}
where $\varphi_l(g^{(l)}) = \argmax_{g^{(l)'} \in [G_l]}\sum_{i_l} I(g_{i_l}^{(l)0} = g^{(l)}, \wh g_{i_l}^{(l)} = g^{(l)'})$,
and $\{g^{-(l)} \in [G_{-l}^0]\}$ denotes $\{g^{(1)} \in [G_{1,0}], \cdots, g^{(l-1)} \in [G_{l-1,0}], g^{(l+1)} \in [G_{l+1, 0}], \cdots, g^{(q)} \in [G_{q,0}]\}$.
We prove \eqref{eq:max_min_lower} by contradiction.
Assume \eqref{eq:max_min_lower} does not hold.
Define
$\sigma_l: [G_{l,0}]\to [G_l]$, such that
\begin{align*}
	\sigma_l(g^{(l)}) & = \argmin_{g^{(l)'} \in [G_l]} \Big\{
	\|\wh \btheta_{g^{(l)'}}^{(l)} - \btheta_{g^{(l)}}^{(l)0}\|^2\\
	& + \max_{g^{-(l)} \in [G_{-l}^0]}  |\wh \alpha_{\varphi_1(g^{(1)}) \cdots \varphi_{l-1}(g^{(l-1)}) g^{(l)'} \varphi_{l+1}(g^{(l+1)}) \cdots \varphi_q(g^{(q)})} \\
	& - \alpha_{\varphi_1(g^{(1)}) \cdots \varphi_{l-1}(g^{(l-1)}) g^{(l)} \varphi_{l+1}(g^{(l+1)}) \cdots \varphi_q(g^{(q)})}^0|^2\Big\}.
\end{align*}
Since $G_{l,0}>G_l$, there exists
at least two $g_1^{(l)},g_2^{(l)}\in [G_{l,0}]$, such that
$\sigma_l(g_1^{(l)}) = \sigma_l(g_2^{(l)})$.
Then we have for all $g^{-(l)}\in [G_{-l}^0]$,
\begin{align}
	& \|\btheta_{g_1^{(l)}}^{(l)0} -\btheta_{g_2^{(l)}}^{(l)0}\|^2 + |\alpha_{g^{(1)} \cdots g^{(l-1)} g_1^{(l)}g^{(l+1)} \cdots g^{(q)}}^0 - \alpha_{g^{(1)} \cdots g^{(l-1)} g_2^{(l)}g^{(l+1)} \cdots g^{(q)}}^0|^2 \nonumber\\
	&\le 2\|\btheta_{g_1^{(l)}}^{(l)0}  -
	\wh\btheta^{(l)}_{\sigma_l(g_1^{(l)})}\|^2+2
	\|\btheta_{g_2^{(l)}}^{(l)0} -  \wh\btheta^{(l)}_{\sigma_l(g_2^{(l)})}\|^2 \nonumber\\
	&+ 2\Big\{|\wh \alpha_{\varphi_1(g^{(1)}) \cdots \varphi_{l-1}(g^{(l-1)}) \sigma_l(g_1^{(l)}) \varphi_{l+1}(g^{(l+1)}) \cdots \varphi_q(g^{(q)})} - \alpha_{g^{(1)} \cdots g^{(l-1)} g_1^{(l)}g^{(l+1)} \cdots g^{(q)}}^0 |^2 \nonumber\\
	&+
	|\wh \alpha_{\varphi_1(g^{(1)}) \cdots \varphi_{l-1}(g^{(l-1)}) \sigma_l(g_2^{(l)}) \varphi_{l+1}(g^{(l+1)}) \cdots \varphi_q(g^{(q)})}  - \alpha_{g^{(1)} \cdots g^{(l-1)} g_2^{(l)}g^{(l+1)} \cdots g^{(q)}}^0|^2\Big\} \nonumber\\
	& < c_\gap/2 + c_\gap/2 = c_\gap, \label{eq:contradict}
\end{align}
which contradicts Assumption \ref{assum:group_diff}.
We explain the last inequality as follows.
For any $g^{(l)}\in[G_{l,0}],
g^{-(l)}\in[G_{-l}^0], g^{(l)'}\in[G_l]$, we have
\begin{align*}
	&\|\btheta_{g^{(l)}}^{(l)0} -
	\wh \btheta_{\sigma_l(g^{(l)})}^{(l)}\|^2+
	|\wh \alpha_{\varphi_1(g^{(1)}) \cdots \varphi_{l-1}(g^{(l-1)}) \sigma_l(g^{(l)}) \varphi_{l+1}(g^{(l+1)}) \cdots \varphi_q(g^{(q)})} - \alpha_{g^{(1)} \cdots g^{(l-1)} g^{(l)}g^{(l+1)} \cdots g^{(q)}}^0|^2 \\
	\le & \max_{g^{-(l)}\in[G_{-l}^0]}\Big\{\|\btheta_{g^{(l)}}^{(l)0} -
	\wh \btheta_{\sigma_l(g^{(l)})}^{(l)}\|^2 \\
	& +
	|\wh \alpha_{\varphi_1(g^{(1)}) \cdots \varphi_{l-1}(g^{(l-1)}) \sigma_l(g^{(l)}) \varphi_{l+1}(g^{(l+1)}) \cdots \varphi_q(g^{(q)})} - \alpha_{g^{(1)} \cdots g^{(l-1)} g^{(l)}g^{(l+1)} \cdots g^{(q)}}^0|^2 \Big\}\\
	& \le \max_{g^{-(l)}\in[G_{-l}^0]}\Big\{\|\btheta_{g^{(l)}}^{(l)0} -
	\wh \btheta_{g^{(l)'}}^{(l)}\|^2 \\
	& +
	|\wh \alpha_{\varphi_1(g^{(1)}) \cdots \varphi_{l-1}(g^{(l-1)}) g^{(l)'} \varphi_{l+1}(g^{(l+1)}) \cdots \varphi_q(g^{(q)})} - \alpha_{g^{(1)} \cdots g^{(l-1)} g^{(l)}g^{(l+1)} \cdots g^{(q)}}^0|^2 \Big\} \\
	& \le \max_{g^{(l)}\in[G_{l,0}]} \max_{g^{-(l)}\in[G_{-l}^0]} \Big\{\|\btheta_{g^{(l)}}^{(l)0} -
	\wh \btheta_{g^{(l)'}}^{(l)}\|^2 \\
	& +
	|\wh \alpha_{\varphi_1(g^{(1)}) \cdots \varphi_{l-1}(g^{(l-1)}) g^{(l)'} \varphi_{l+1}(g^{(l+1)}) \cdots \varphi_q(g^{(q)})} - \alpha_{g^{(1)} \cdots g^{(l-1)} g^{(l)}g^{(l+1)} \cdots g^{(q)}}^0|^2 \Big\},
\end{align*}
where the second inequality holds by the definition of $\sigma_l(g^{(l)})$.
Since the above inequality holds for any $g^{(l)'}$, hence by setting
\begin{align*}
	g^{(l)'}=\argmin_{g^{(l)'}\in[G_l]} \left\{\|\btheta_{g^{(l)}}^{(l)0} - \wh \btheta_{g^{(l)'}}^{(l)}\|^2+
	|\wh \alpha_{\varphi_1(g^{(1)}) \cdots \varphi_{l-1}(g^{(l-1)}) g^{(l)'} \varphi_{l+1}(g^{(l+1)}) \cdots \varphi_q(g^{(q)})} \right.\\
	-\left. \alpha_{g^{(1)} \cdots g^{(l-1)} g^{(l)}g^{(l+1)} \cdots g^{(q)}}^0|^2 \right\},
\end{align*}
we get
\begin{align*}
	\|\btheta_{g^{(l)}}^{(l)0} -
	\wh \btheta_{\sigma_l(g^{(l)})}^{(l)}\|^2+
	|\wh \alpha_{\varphi_1(g^{(1)}) \cdots \varphi_{l-1}(g^{(l-1)}) \sigma_l(g^{(l)}) \varphi_{l+1}(g^{(l+1)}) \cdots \varphi_q(g^{(q)})} - \alpha_{g^{(1)} \cdots g^{(l-1)} g^{(l)}g^{(l+1)} \cdots g^{(q)}}^0|^2\\
	\le\max_{g^{(l)}\in[G_{l,0}]} \max_{g^{-(l)}\in[G_{-l}^0]} \min_{g^{(l)'}\in[G_l]} \left\{\|\btheta_{g^{(l)}}^{(l)0} - \wh \btheta_{g^{(l)'}}^{(l)}\|^2+
	|\wh \alpha_{\varphi_1(g^{(1)}) \cdots \varphi_{l-1}(g^{(l-1)}) g^{(l)'} \varphi_{l+1}(g^{(l+1)}) \cdots \varphi_q(g^{(q)})} \right.\\
	-\left. \alpha_{g^{(1)} \cdots g^{(l-1)} g^{(l)}g^{(l+1)} \cdots g^{(q)}}^0|^2 \right\}< c_\gap/4.
\end{align*}
Hence, by the contradiction \eqref{eq:contradict}, we could obtain that \eqref{eq:max_min_lower} holds.

Subsequently, define the mapping $\varphi_l(g^{(l)}) = \argmax_{g^{(l)'} \in [G_l]}\sum_{i_l} I(g_{i_l}^{(l)0} = g^{(l)}, \wh g_{i_l}^{(l)} = g^{(l)'})$.
In addition, let
\begin{align}
	& (g^{(1)*}, \cdots, g^{(q)*}) = \argmax_{g^{(1)} \in [G_{1,0}], \cdots, g^{(q)} \in [G_{q,0}]}
	\Big[\min_{g^{(l)'}\in [G_l]}\Big\{
	\|\wh \btheta_{g^{(l)'}}^{(l)} - \btheta_{g^{(l)}}^{(l)0}\|^2
	\nonumber\\
	&+ |\wh \alpha_{\varphi_1(g^{(1)}) \cdots \varphi_{l-1}(g^{(l-1)}) g^{(l)'} \varphi_{l+1}(g^{(l+1)}) \cdots \varphi_q(g^{(q)})}
	- \alpha_{g^{(1)} \cdots g^{(l-1)} g^{(l)}g^{(l+1)} \cdots g^{(q)}}^0|^2\Big\}\Big].\label{eq:def_gh_star}
\end{align}
In this case, we have
\begin{align*}
	&d(\wh \bTheta(\underline{G}), \bTheta^0) = \frac{1}{\prod_l N_l} \sum_{i_1 =1}^{N_1} \cdots \sum_{i_q = 1}^{N_q}  \Big\|\wh \bTheta_{i_1 \cdots i_q}  - \bTheta_{i_1 \cdots i_q}  \Big\|^2\\
	& = \sum_l \frac{1}{N_l}  \sum_{i_l =1}^{N_l}  \sum_{g^{(l)'} = 1}^{G_l} \sum_{g^{(l)} = 1}^{G_{l,0}} I(\wh g_{i_l}^{(l)} = g^{(l)'}, g_{i_l}^{(l)0} = g^{(l)} ) \| \wh\btheta_{ g^{(l)'}}^{(l)} - \btheta_{g^{(l)}}^{(l)0} \|^2 \\
	& + \frac{1}{\prod_l N_l }
	\sum_{i_1,\cdots,i_q} \sum_{g^{(1)'} =1 }^{G_1} \cdots \sum_{g^{(q)'} =1 }^{G_q} \sum_{g^{(1)} = 1}^{G_{1,0}} \cdots \sum_{g^{(q)} = 1}^{G_{q,0}} \\
	& I(\wh g_{i_1}^{(1)} = g^{(1)'}, \cdots, \wh g_{i_q}^{(q)} = g^{(q)'}, g_{i_1}^{(1)0} = g^{(1)}, \cdots, g_{i_q}^{(q)0} = g^{(q)})
	|\wh \alpha_{g^{(1)'} \cdots g^{(q)'}} - \alpha_{g^{(1)} \cdots g^{(q)}}^0|^2\\
	&\ge \sum_l \frac{1}{N_l}\sum_{i_l = 1}^{N_l}
	I(\wh g_{i_l}^{(l)} = \varphi_l(g^{(l)*}), g_{i_l}^{(l)0} = g^{(l)*} )\|\wh \btheta_{\varphi_l(g^{(l)*})}^{(l)} - \btheta_{g^{(l)*}}^{(l)0}\|^2\\
	&  + \frac{1}{\prod_l N_l}
	\sum_{i_1,\cdots,i_q}   I(\wh g_{i_1}^{(1)} = \varphi_1(g^{(1)*}), \cdots, \wh g_{i_q}^{(q)} = \varphi_q(g^{(q)*}), g_{i_1}^{(1)0} = g^{(1)*}, \cdots, g_{i_q}^{(q)0} = g^{(q)*})\\
	& |\wh \alpha_{\varphi_1(g^{(1)*}) \cdots \varphi(g^{(q)*})} - \alpha_{g^{(1)*} \cdots g^{(q)*}}^0|^2.
\end{align*}
By (\ref{eq:max_min_lower}), we further have
\begin{align*}
	& d(\wh \bTheta(\underline{G}), \bTheta^0)\\
	&\ge \frac{1}{\prod_l N_l} \sum_{i_1, \cdots, i_q}
	I(\wh g_{i_1}^{(1)} = \varphi_1(g^{(1)*}), \cdots, \wh g_{i_q}^{(q)} = \varphi_q(g^{(q)*}), g_{i_1}^{(1)0} = g^{(1)*}, \cdots, g_{i_q}^{(q)0} = g^{(q)*}) \\
	& \hspace{1em} \|\wh \btheta_{\varphi_l(g^{(l)*})}^{(l)} - \btheta_{g^{(l)*}}^{(l)0}\|^2\\
	&+ \frac{1}{\prod_l N_l}
	\sum_{i_1, \cdots, i_q} I(\wh g_{i_1}^{(1)} = \varphi_1(g^{(1)*}), \cdots, \wh g_{i_q}^{(q)} = \varphi_q(g^{(q)*}), g_{i_1}^{(1)0} = g^{(1)*}, \cdots, g_{i_q}^{(q)0} = g^{(q)*})\\
	&|\wh \alpha_{\varphi_1(g^{(1)*}) \cdots \varphi(g^{(q)*})} - \alpha_{g^{(1)*} \cdots g^{(q)*}}^0|^2\\
	&\ge \frac{c_\gap}{4\prod_l N_l}
	\sum_{i_1, \cdots, i_q}
	I(\wh g_{i_1}^{(1)} = \varphi_1(g^{(1)*}), \cdots, \wh g_{i_q}^{(q)} = \varphi_q(g^{(q)*}), g_{i_1}^{(1)0} = g^{(1)*}, \cdots, g_{i_q}^{(q)0} = g^{(q)*}) \\
	&= \frac{c_\gap}{4\prod_l N_l}
	\sum_{l=1}^{q}\sum_{i_l=1}^{N_l}
	I\{\wh g_{i_l}^{(l)} = \varphi_l(g^{(l)*}), g_{i_l}^{(l)0} = g^{(l)*}\}.
\end{align*}
where the last inequality is due to that
\begin{align*}
	& \|\wh \btheta_{\varphi_l(g^{(l)*})}^{(l)} - \btheta_{g^{(l)*}}^{(l)0}\|^2 +
	|\wh \alpha_{\varphi_1(g^{(1)*}) \cdots \varphi(g^{(q)*})} - \alpha_{g^{(1)*} \cdots g^{(q)*}}^0|^2\\
	&\ge
	\min_{g^{(l)'}\in [G_l]}
	\Big\{\|\wh \btheta_{g^{(l)'}}^{(l)} - \btheta_{g^{(l)*}}^{(l)0}\|^2 +
	|\wh \alpha_{\varphi_1(g^{(1)*}) \cdots \varphi_{l-1}(g^{(l-1)*}) g^{(l)'} \varphi_{l+1}(g^{(l+1)*}) \cdots \varphi_q(g^{(q)*})} - \alpha_{g^{(1)*} \cdots g^{(q)*}}^0|^2\Big\}\\
	& = \max_{g^{(l)} \in [G_{l,0}]}
	\max_{g^{-(l)}\in [G_{-l}^0]}\min_{g^{(l)'}\in [G_l]}\Big\{
	\|\wh \btheta_{g^{(l)'}}^{(l)} - \btheta_{g^{(l)}}^{(l)0}\|^2 \\
	&  +
	|\wh \alpha_{\varphi_1(g^{(1)}) \cdots \varphi_{l-1}(g^{(l-1)}) g^{(l)'} \varphi_{l+1}(g^{(l+1)}) \cdots \varphi_q(g^{(q)})} - \alpha_{g^{(1)} \cdots g^{(q)}}^0|^2\Big\} \ge \frac{c_\gap}{4}
\end{align*}
by the definition (\ref{eq:def_gh_star}) and (\ref{eq:max_min_lower}).

By the definition of $\varphi_l(\cdot)$,
we have
\begin{align*}
	N_l^{-1}\sum_{i_l} I(\wh g_{i_l}^{(l)} = \varphi_l(g^{(l)*}), g_{i_l}^{(l)0} = g^{(l)*})\ge
	N_l^{-1} G_l^{-1}\sum_{i_l} I(g_{i_l}^{(l)0} = g^{(l)*})  = G_l^{-1} \pi_{g^{(l)*}, N_l}^{(l)}
\end{align*}
From Assumption \ref{assum:group_ratio},
we know that there exists $M>0$, such that for $ N_l>M$ it holds
$G_l^{-1} \pi_{g^{(l)*}, N_l}^{(l)}  > c_{\pi}/2G_l $.
This yields
$d(\wh \bTheta(\underline{G}), \bTheta^0) \ge \frac{c_\gap (c_\pi)^q}{ 2^{q+2} \prod_l G_l}$
as $N_l \to\infty$ for all $l \in [q]$ by Assumption \ref{assum:group_ratio}.
Then by Lemmas \ref{lem:Sij_concent} and
\ref{lem:Q_star_diff},
we have
\begin{align*}
	&((\prod_l N_l)T)^{-1}\{Q(\wh \bTheta(\underline{G})) -
	Q(\wh \bTheta(\underline{G}_0))\}\\
	& = O_p(T^{-1}(\sum_l \log N_l)^2) +
	((\prod_l N_l)T)^{-1} \Big\{Q^*(\wh \bTheta(\underline{G}))-Q^*(\bTheta^{0})\Big\}\\
	& \ge \tau_{\min} d(\wh \bTheta(\underline{G}), \bTheta^0) + O_p(T^{-1}(\sum_l \log N_l)^2))\\
	&\ge
	\frac{\tau_{\min}c_\gap (c_\pi)^q}{ 2^{q+2} \prod_l G_l }+ O_p(T^{-1/2}(m+\sum_l \log N_l))
	\ge  \frac{C c_\gap (c_\pi)^q}{\prod_l G_l}.
\end{align*}
with probability tending to 1, where $C = \tau_{\min}/2^{q+2}$
is a positive constant, and the last inequality holds since the condition \eqref{eq:eta_range} that $ T^{-1}(\sum_l \log N_l)^2 \ll c_\gap/ (\prod_l G_l)$.

\subsection{Proof of Proposition \ref{pro:gh_consistency}}

Recall that
\begin{align*}
	{Q_{i_l}(\bxi_{g_{i_l}^{(l)}}^{(l)}; \bxi_{g^{-(l)}}^{-(l)}, \mG_{-l})} &=
	\sum_{m \neq l}\sum_{i_m=1}^{N_m}\sum_{t=1}^T\Big\{Y_{i_1 \cdots i_q, t} - \sum_{l=1}^q \lambda_{g_{i_l}^{(l)}}^{(l)} \sum_{k = 1}^{N_l} w^{(l)}_{i_l k}Y_{i_1 \cdots i_{l-1} k i_{l+1} \cdots i_q,(t-1)}\nonumber\\
	&- \alpha_{g_{i_1}^{(1)} \cdots g_{i_q}^{(q)}}Y_{i_1 \cdots i_q,(t-1)}
	- \sum_{l=1}^q \bx_{i_lt}^{(l) \top} \bzeta_{g_{i_l}^{(l)}}^{(l)}\Big\}^2,
\end{align*}
which is defined in \eqref{eq:Q_il}.
Define
\begin{align*}
	\sigma_l(g^{(l)}) & = \argmin_{\wt g^{(l)} \in [G_l]}\Big\{
	\|\wh \btheta_{\wt g^{(l)}}^{(l)} - \btheta_{g^{(l)}}^{(l)0}\|^2
	\\
	&+  \frac{1}{\prod_{m \neq l} N_m} \sum_{i_{-l}} |\wh \alpha_{\wh g_{i_1}^{(1)} \cdots \wh g_{i_{l-1}}^{(l-1)} \wt g^{(l)} \wh g_{i_{l+1}}^{(l+1)} \cdots \wh g_{i_q}^{(q)} }
	- \alpha_{ g_{i_1}^{(1)0} \cdots g_{i_{l-1}}^{(l-1)0} g^{(l)} g_{i_{l+1}}^{(l+1)0} \cdots g_{i_q}^{(q)0} }^0|^2\Big\}.
\end{align*}
	

Let $R_1 = \sup_l \sup_{i_l} \sup_{\|\bTheta_{i_1 \cdots i_q}\|_{\max} <R} |
\frac{1}{T}\{Q_{i_1 \cdots i_q}(\bTheta_{i_1 \cdots i_q}) - Q_{i_1 \cdots i_q}^*(\bTheta_{i_1 \cdots i_q})\}|$
and we have
$R_1 = O_p(T^{-1/2}(m + \sum_l \log N_l))$
due to Lemma \ref{lem:Q_diff}.
Therefore again by Lemma \ref{lem:Q_diff}
and the fact that $\wh \bxi_{\wh g_{i_l}^{(l)}}^{(l)} = (\wh\btheta_{\wh g_{i_l}^{(l)}}^{(l)\top}, \vec(\wh\balpha_{\cdot, \wh g_{i_l}^{(l)} \cdot})^\top)^\top$
minimize $ Q_{i_l}(\bxi_{g_{i_l}^{(l)}}^{(l)};   \bxi_{g^{-(l)}}^{-(l)}, \mG_{-l})$,
we have
\begin{align*}
	&\frac{1}{(\prod_{m \neq l} N_m) T}
	Q_{i_l}^*( \wh \bxi_{\wh g_{i_l}^{(l)}}^{(l)}; \wh \bxi_{g^{-(l)}}^{-(l)}, \wh\mG_{-l})
	-R_1\le
	\frac{1}{(\prod_{m \neq l} N_m) T}
	Q_{i_l}( \wh \bxi_{\wh g_{i_l}^{(l)}}^{(l)} ;\wh \bxi_{g^{-(l)}}^{-(l)} , \wh\mG_{-l})\\
	&\le
	\frac{1}{(\prod_{m \neq l} N_m) T}
	Q_{i_l}( \wh \bxi_{\sigma_l(g_{i_l}^{(l)0})}^{(l)};\wh \bxi_{g^{-(l)}}^{-(l)}, \wh\mG_{-l}) \le
	\frac{1}{(\prod_{m \neq l} N_m)  T}
	Q_{i_l}^*(\wh \bxi_{\sigma_l(g_{i_l}^{(l)0})}^{(l)}; \wh  \bxi_{g^{-(l)}}^{-(l)}, \wh\mG_{-l}) + R_1
\end{align*}
for all $i_l\in [N_l]$.
Therefore it holds
\begin{align}
	\sup_{i_l} \left\{
	\frac{1}{(\prod_{m \neq l} N_m)  T}Q_{i_l}^*( \wh \bxi_{\wh g_{i_l}^{(l)}}^{(l)}; \wh  \bxi_{g^{-(l)}}^{-(l)} , \wh\mG_{-l}) -
	\frac{1}{(\prod_{m \neq l} N_m) T} Q_{i_l}^*(\wh \bxi_{\sigma_l(g_{i_l}^{(l)0})}^{(l)}; \wh \bxi_{g^{-(l)}}^{-(l)}, \wh\mG_{-l})
	\right\} \le 2R_1.\label{eq:Q_1j_diff1}
\end{align}
On the other hand by Lemma \ref{lem:Q_star_diff} we have
\begin{align}
	0\le & \sup_{i_l} \Big\{ \frac{1}{(\prod_{m \neq l} N_m) T}Q_{i_l}^*(\wh \bxi_{\sigma_l(g_{i_l}^{(l)0})}^{(l)}; \wh \bxi_{g^{-(l)}}^{-(l)}, \wh\mG_{-l})
	-\frac{1}{(\prod_{m \neq l} N_m)T}
	Q_{i_l}^*(\bxi_{g_{i_l}^{(l)0}}^{(l)0};   \bxi_{g^{-(l)}}^{-(l)0}, \mG_{-l}^0)\Big\} \nonumber\\
	&\le  \tau_{\max} \sup_{i_l}   \Big\{  \frac{1}{\prod_{m \neq l} N_m}\sum_{m \neq l} \sum_{i_m} \|\wh\btheta_{\wh g_{i_m}^{(m)}}^{(m)} -  \btheta_{g_{i_m}^{(m)0}}^{(m)0}\|^2 +
	\|\wh \btheta_{\sigma_l( g_{i_l}^{(l)0})}^{(l)} - \btheta_{ g_{i_l}^{(l)0}}^{(l)0}\|^2 \nonumber\\
	& + \frac{1}{\prod_{m \neq l} N_m} \sum_{m \neq l} \sum_{i_m=1}^{N_m} |\wh \alpha_{\wh g_{i_1}^{(1)} \cdots \wh g_{i_{l-1}}^{(l-1)} \sigma_l(g_{i_l}^{(l)0}) \wh g_{i_{l+1}}^{(l+1)} \cdots \wh g_{i_q}^{(q)}} - \alpha^0_{g_{i_1}^{(1)0} \cdots g_{i_q}^{(q)0}}|^2 \Big\} \nonumber\\
	& =O_p(T^{-1}(\sum_l \log  N_l)^2  ) + O_p(T^{-1}(\sum_l \log  N_l)^2 )= O_p(T^{-1}(\sum_l \log N_l)^2)\label{eq:Q_1j_diff2}
\end{align}
where the second last equation is obtained by
Theorem \ref{thm:pseudo_dist}
and
Lemma \ref{lem:sig_h_bound}.

	Next, we use the order in \eqref{eq:Q_1j_diff2} to improve the order of $R_1$ in \eqref{eq:Q_1j_diff1}.
	Denote $\wh\bTheta_{\cdot i_l \cdot}^{\sigma}$ as the parameter by replacing the $\wh g_{i_l}^{(l)}$ by $\sigma(g_{i_l}^{(l)0})$ in $\wh\bTheta_{\cdot i_l \cdot}$ while keep the other terms the same.
	By \eqref{eq:Q_1j_diff2}, there exists a positive constant $C$, the event $\mO_1 = \{  \sup_{i_l}d_{i_l}(\wh\bTheta_{\cdot i_l \cdot}^{\sigma}, \bTheta_{\cdot i_l \cdot}^0) \le C a_{NT}^2 \}$ has the probability $P(\mO_1) > 1- \epsilon$ for $\forall \epsilon>0$, where $a_{NT} = (\sum_l \log N_l)/\sqrt{T}$.
	To derive the order of $R_1$, we need to derive the order of distance $d_{i_l}(\wh\bTheta_{\cdot i_l \cdot}, \wh\bTheta_{\cdot i_l \cdot}^{\sigma})$.
	We next show that $P( a_{NT}^{-1} \sup_{i_l} \sqrt{d_{i_l}(\wh\bTheta_{\cdot i_l \cdot}, \wh\bTheta_{\cdot i_l \cdot}^\sigma)} > K) \to 0$ as $K \to \infty$, which implies that $d_{i_l}(\wh\bTheta_{\cdot i_l \cdot}, \wh\bTheta_{\cdot i_l \cdot}^\sigma) = O_p(a_{NT}^2)$.
	Partition the parameter $\bTheta_{\cdot i_l \cdot}$ into space $\bOmega_j = \{ \bTheta_{\cdot i_l \cdot} \in \bOmega: \gamma(j-1) < a_{NT}^{-1} \sqrt{d_{i_l}(\bTheta_{\cdot i_l \cdot}, \wh\bTheta_{\cdot i_l \cdot}^\sigma)} \le \gamma j \}$, with $j \ge 1$ taking over all positive integers and $\gamma = \sqrt{1 + C (\tau_{\max} + \tau_{\min})/\tau_{\min}}$ is a positive constant.
	For a positive integer $K$, if $a_{NT}^{-1} \sqrt{d_{i_l}(\wh\bTheta_{\cdot i_l \cdot}, \wh\bTheta_{\cdot i_l \cdot}^\sigma)} > K$, then $\wh \bTheta_{\cdot  i_l \cdot} \in \bigcup_{j \ge K} \bOmega_j$.
	Define
	\begin{align*}
		S_{i_l}(\bTheta_{\cdot i_l \cdot}) = Q_{i_l}( \bxi_{\wh g_{i_l}^{(l)}}^{(l)}; \wh  \bxi_{g^{-(l)}}^{-(l)} , \wh\mG_{-l}) - Q_{i_l}(\wh \bxi_{\sigma_l(g_{i_l}^{(l)0})}^{(l)}; \wh \bxi_{g^{-(l)}}^{-(l)}, \wh\mG_{-l}).
	\end{align*}
	By the definition of $\wh \bTheta_{\cdot  i_l \cdot}$, we know that $S_{i_l}(\wh\bTheta_{\cdot i_l \cdot})\le 0$.
	Hence, we have
	\begin{align*}
		& P \Big\{ a_{NT}^{-1} \sup_{i_l} \sqrt{d_{i_l}(\wh\bTheta_{\cdot i_l \cdot}, \wh\bTheta_{\cdot i_l \cdot}^\sigma)} \ge K \Big\} \le P \Big\{  \inf_{i_l} \inf_{\bTheta_{\cdot  i_l \cdot} \in \bigcup_{j \ge K} \bOmega_j} (\prod_{m \ne l} N_m T)^{-1} S_{i_l}(\wh\bTheta_{\cdot i_l \cdot}) \le 0 \Big\} \\
		& \le \sum_{j \ge K} P \Big\{   \inf_{i_l} \Big[ \inf_{\wh\bTheta_{\cdot  i_l \cdot} \in  \bOmega_j}  (\prod_{m \ne l} N_m T)^{-1}\{S_{i_l}(\wh\bTheta_{\cdot i_l \cdot}) - S_{i_l}^*(\wh\bTheta_{\cdot i_l \cdot})\}  \\
		& +\inf_{\wh\bTheta_{\cdot  i_l \cdot} \in  \bOmega_j}  (\prod_{m \ne l} N_m T)^{-1} S_{i_l}^*(\wh\bTheta_{\cdot i_l \cdot})\Big] \le 0 \Big\}
	\end{align*}
	Under event $\mO_1$, we know that if $\wh \bTheta_{\cdot  i_l \cdot} \in \bOmega_j$,
	\begin{align*}
		(\prod_{m \ne l} N_m T)^{-1}  \inf_{i_l} S_{i_l}^*(\wh\bTheta_{\cdot i_l \cdot}) & = (\prod_{m \ne l} N_m T)^{-1} \inf_{i_l}  \big\{ Q_{i_l}^*(\wh \bxi_{\wh g_{i_l}^{(l)}}^{(l)}; \wh  \bxi_{g^{-(l)}}^{-(l)} , \wh\mG_{-l}) - Q_{i_l}^*(\wh \bxi_{\sigma_l(g_{i_l}^{(l)0})}^{(l)}; \wh \bxi_{g^{-(l)}}^{-(l)}, \wh\mG_{-l})\big\} \\
		& = (\prod_{m \ne l} N_m T)^{-1}  \inf_{i_l} \Big[ \big\{ Q_{i_l}^*(\wh \bxi_{\wh g_{i_l}^{(l)}}^{(l)}; \wh  \bxi_{g^{-(l)}}^{-(l)} , \wh\mG_{-l}) - Q_{i_l}^*(\bxi_{g_{i_l}^{(l)0}}^{(l)0};   \bxi_{g^{-(l)}}^{-(l)0}, \mG_{-l}^0) \big\} \\
		& + \big\{Q_{i_l}^*(\bxi_{g_{i_l}^{(l)0}}^{(l)0};   \bxi_{g^{-(l)}}^{-(l)0}, \mG_{-l}^0)- Q_{i_l}^*(\wh \bxi_{\sigma_l(g_{i_l}^{(l)0})}^{(l)}; \wh \bxi_{g^{-(l)}}^{-(l)}, \wh\mG_{-l})\big\}\Big] \\
		&\stackrel{\circled{1}}{\ge} \tau_{\min}  \inf_{i_l}  d_{i_l}(\wh\bTheta_{\cdot i_l \cdot}, \bTheta^0_{\cdot i_l \cdot}) - \tau_{\max} \sup_{i_l}d_{i_l}(\wh\bTheta_{\cdot i_l \cdot}^\sigma, \bTheta^0_{\cdot i_l \cdot})\\
		& \stackrel{\circled{2}}{\ge} \tau_{\min}  \inf_{i_l} d_{i_l}(\wh\bTheta_{\cdot i_l \cdot}, \wh\bTheta_{\cdot i_l \cdot}^\sigma)  - (\tau_{\min} + \tau_{\max}) \sup_{i_l}d_{i_l}(\wh\bTheta_{\cdot i_l \cdot}^\sigma, \bTheta^0_{\cdot i_l \cdot})  \\
		& \stackrel{\circled{3}}{\ge} \tau_{\min}  (j-1)^2 (1 +C  \frac{\tau_{\max} + \tau_{\min}}{\tau_{\min}} )a_{NT}^2- C (\tau_{\max} + \tau_{\min}) a_{NT}^2 \\
		& =  \tau_{\min}a_{NT}^2  (j-1)^2  + C \{(j-1)^2-1\} (\tau_{\min} + \tau_{\max}) a_{NT}^2 \\
		& \stackrel{\circled{4}}{\ge}  \tau_{\min} a_{NT}^2 (j-1)^2 ,
	\end{align*}
	where the inequality $\circled{1}$ holds due to Lemma \ref{lem:Q_star_diff}, inequality $\circled{2}$ holds due to the triangle inequality, inequality $\circled{3}$ holds under $\mO_1$ and the definition of the partition, and inequality $\circled{4}$ holds when $j >1$.
	Therefore, we have
	\begin{align}
		&  P \Big\{  a_{NT}^{-1}\sup_{i_l}  \sqrt{d_{i_l}(\wh\bTheta_{\cdot i_l \cdot}, \wh\bTheta_{\cdot i_l \cdot}^\sigma)} \ge K \Big\} \nonumber\\
		 & \le \sum_{j \ge K} P \Big\{ \inf_{i_l}\Big[ \inf_{\wh\bTheta_{\cdot  i_l \cdot} \in  \bOmega_j}  (\prod_{m \ne l} N_m T)^{-1}\{S_{i_l}(\wh\bTheta_{\cdot i_l \cdot}) - S_{i_l}^*(\wh\bTheta_{\cdot i_l \cdot})\}  \nonumber\\
		 & +\inf_{\wh\bTheta_{\cdot  i_l \cdot} \in  \bOmega_j}  (\prod_{m \ne l} N_m T)^{-1} S_{i_l}^*(\wh\bTheta_{\cdot i_l \cdot})\Big] \le 0 \Big\} \nonumber\\
		 & \le  \sum_{j \ge K} P \Big\{ \sup_{i_l} \sup_{\wh\bTheta_{\cdot  i_l \cdot} \in  \bOmega_j}  (\prod_{m \ne l} N_m T)^{-1} | S_{i_l}(\wh\bTheta_{\cdot i_l \cdot}) - S_{i_l}^*(\wh\bTheta_{\cdot i_l \cdot})| \ge \tau_{\min} a_{NT}^2 (j-1)^2  \Big\} + P(\mO_1^c).\label{eq:d_Theta_il_bound}
	\end{align}
	Next, we derive the upper bound for the first term in \eqref{eq:d_Theta_il_bound} under event $\mO_1$.
	Note that
	\begin{align*}
		S_{i_l}(\wh\bTheta_{\cdot  i_l \cdot}) & = Q_{i_l}(\wh\bTheta_{\cdot  i_l \cdot}) - Q_{i_l} (\wh \bTheta_{\cdot  i_l \cdot}^\sigma) \\
		& = \sum_{m \neq l} \sum_{i_m} \sum_t \Big\{ \Big(Y_{i_1 \cdots i_q, t}  - \cX_{i_1 \cdots i_q, t}^\top \wh\bTheta_{i_1 \cdots i_q}  \Big)^2 - \Big(Y_{i_1 \cdots i_q, t}  - \cX_{i_1 \cdots i_q, t}^\top \wh\bTheta_{i_1 \cdots i_q}^\sigma  \Big)^2  \Big\} \\
		& = \sum_{m \neq l} \sum_{i_m}  \sum_t \Big\{ (\bTheta_{i_1 \cdots i_q}^0 - \wh\bTheta_{i_1 \cdots i_q})^\top \cX_{i_1 \cdots i_q, t} \cX_{i_1 \cdots i_q, t}^\top (\bTheta_{i_1 \cdots i_q}^0 - \wh\bTheta_{i_1 \cdots i_q}) \\
		&  - (\bTheta_{i_1 \cdots i_q}^0 - \wh\bTheta_{i_1 \cdots i_q}^\sigma)^{ \top} \cX_{i_1 \cdots i_q, t} \cX_{i_1 \cdots i_q, t}^\top (\bTheta_{i_1 \cdots i_q}^0 - \wh\bTheta_{i_1 \cdots i_q}^\sigma) \\
		& + 2 \cX_{i_1 \cdots i_q, t}^\top (\wh\bTheta_{i_1 \cdots i_q}^\sigma - \wh\bTheta_{i_1 \cdots i_q}) \ve_{i_1 \cdots i_q, t} \Big\}
	\end{align*}
	Hence we have
	\begin{align}
		& |S_{i_l}(\wh\bTheta_{\cdot i_l \cdot} ) - S^*_{i_l}(\wh\bTheta_{\cdot i_l \cdot} )| \le \nonumber\\
		& \le 2 \Big| \sum_{m \neq l} \sum_{i_m}  \sum_t \Big\{(\bTheta_{i_1 \cdots i_q}^0 - \wh\bTheta_{i_1 \cdots i_q}^\sigma)^\top \cX_{i_1 \cdots i_q,t} \cX_{i_1 \cdots i_q,t}^\top (\bTheta_{i_1 \cdots i_q}^0 - \wh\bTheta_{i_1 \cdots i_q}^\sigma) \nonumber\\
		&   - \big[(\bTheta_{i_1 \cdots i_q}^0 - \wh\bTheta_{i_1 \cdots i_q}^\sigma)^\top E( \cX_{i_1 \cdots i_q,t} \cX_{i_1 \cdots i_q,t}^\top) (\bTheta_{i_1 \cdots i_q}^0 - \wh\bTheta_{i_1 \cdots i_q}^\sigma)\big] \Big\} \Big| \label{eq:S_il_concent_1}\\
		& + \Big|  \sum_{m \neq l} \sum_{i_m}  \sum_t \Big\{ (\wh \bTheta_{i_1 \cdots  i_q}^\sigma - \wh\bTheta_{i_1 \cdots  i_q})^\top \cX_{i_1 \cdots i_q,t} \cX_{i_1 \cdots i_q,t}^\top (\wh \bTheta_{i_1 \cdots  i_q}^\sigma - \wh\bTheta_{i_1 \cdots  i_q}) \nonumber\\
		& -  \big[  (\wh \bTheta_{i_1 \cdots  i_q}^\sigma - \wh\bTheta_{i_1 \cdots  i_q})^\top E(\cX_{i_1 \cdots i_q,t} \cX_{i_1 \cdots i_q,t}^\top) (\wh \bTheta_{i_1 \cdots  i_q}^\sigma - \wh\bTheta_{i_1 \cdots  i_q}) \big] \Big\} \Big|  \label{eq:S_il_concent_2}\\
		& +  2 \Big|  \sum_{m \neq l} \sum_{i_m}  \sum_t \cX_{i_1 \cdots i_q, t}^\top (\wh\bTheta_{i_1 \cdots i_q}^\sigma - \wh\bTheta_{i_1 \cdots i_q}) \ve_{i_1 \cdots i_q, t} \Big| \label{eq:S_il_concent_3}
	\end{align}
	Due to the definition of parameter $\bTheta_{\cdot i_l \cdot}$'s partition $\bOmega_j$ and under event $\mO_1$,
	{the concentration inequalities for \eqref{eq:S_il_concent_1}--\eqref{eq:S_il_concent_3} can be obtained by Lemma \ref{lem:S_Xtheta_tail} that,}
	\begin{align*}
		& \sum_{j \ge K} P \Big\{ \sup_{i_l} \sup_{\wh\bTheta_{\cdot i_l \cdot} \in \bOmega_j} (\prod_{m \neq l} N_m T)^{-1}
	\\
	& \Big| \sum_{m \neq l} \sum_{i_m}  \sum_t
		\Big\{(\bTheta_{i_1 \cdots i_q}^0 - \wh\bTheta_{i_1 \cdots i_q}^\sigma)^\top \cX_{i_1 \cdots i_q,t} \cX_{i_1 \cdots i_q,t}^\top (\bTheta_{i_1 \cdots i_q}^0 - \wh\bTheta_{i_1 \cdots i_q}^\sigma)  \\
		& -  E \big[(\bTheta_{i_1 \cdots i_q}^0 - \wh\bTheta_{i_1 \cdots i_q}^\sigma)^\top \cX_{i_1 \cdots i_q,t} \cX_{i_1 \cdots i_q,t}^\top (\bTheta_{i_1 \cdots i_q}^0 - \wh\bTheta_{i_1 \cdots i_q}^\sigma)\big] \Big\}  \Big| \ge \tau_{\min} a_{NT}^2 (j-1)^2 \Big\} \\
		& \le \sum_{j \ge K} C_1 \exp\Big\{ -C_2 \min \Big( \frac{T(j-1)^4 a_{NT}^2}{ C }, \frac{\sqrt{T} (j-1)^2 a_{NT} }{\sqrt{C}} \Big) + C_3 m + \log(\prod_l N_l) \Big\} \\
		&  = \sum_{j \ge K}  C_1 \exp \Big\{- \wt C_2 ( j-1)^2 \log(\prod_l N_l) + C_3 m + \log(\prod_l N_l )  \Big\} \\
		& \le \sum_{j \ge K}  C_1 \exp \Big\{- \frac{\wt C_2 ( j-1)^2 \log(\prod_l N_l) } {j}+ C_3 m + \log(\prod_l N_l)  \Big\} \to 0
	\end{align*}
	as $K \to \infty$, where the equality holds as $a_{NT} = \log (\prod_l N_l)/\sqrt{T}$, the last inequality holds when $j \ge 1$, and the limit holds by the same calculation in \eqref{eq:tail_K}.
	Therefore, by substituting the tail bounds for the first term in \eqref{eq:d_Theta_il_bound}, we obtain that $P \Big\{ a_{NT}^{-1} \sup_{i_l} \sqrt{d_{i_l}(\wh\bTheta_{\cdot i_l \cdot}, \wh\bTheta_{\cdot i_l \cdot}^\sigma)} \ge K \Big\} \to 0$ as $K \to \infty$.
	
	This implies that $ \sup_{i_l}d_{i_l}( \wh\bTheta_{\cdot i_l \cdot}, \wh\bTheta_{\cdot i_l \cdot}^\sigma) = O_p(a_{NT}^2)$. By using Lemma \ref{lem:Q_star_diff}, we have that
	\begin{align*}
		& \sup_{i_l} \left\{
		\frac{1}{(\prod_{m \neq l} N_m)  T}Q_{i_l}^*( \wh \bxi_{\wh g_{i_l}^{(l)}}^{(l)}; \wh  \bxi_{g^{-(l)}}^{-(l)} , \wh\mG_{-l}) -
		\frac{1}{(\prod_{m \neq l} N_m) T} Q_{i_l}^*(\wh \bxi_{\sigma_l(g_{i_l}^{(l)0})}^{(l)}; \wh \bxi_{g^{-(l)}}^{-(l)}, \wh\mG_{-l})
		\right\}  \\
		& = O_p(T^{-1} (\sum_l \log N_l)^2).
	\end{align*}

Therefore, together with \eqref{eq:Q_1j_diff2}, we have
\begin{align}
	& \sup_{i_l} \left\{
	\frac{1}{(\prod_{m \neq l} N_m)  T}Q_{i_l}^*(\wh \bxi_{\wh g_{i_l}^{(l)}}^{(l)};\wh \bxi_{g^{-(l)}}^{-(l)}, \wh\mG_{-l})-
	\frac{1}{(\prod_{m \neq l} N_m) T} Q_{i_l}^*(\bxi_{g_{i_l}^{(l)0}}^{(l)0};  \bxi_{g^{-(l)}}^{-(l)0}, \mG_{-l}^0)
	\right\} \nonumber \\
	& = O_p(T^{-1}(\sum_l \log N_l)^2).\label{eq:Q_il_dist}
\end{align}
By Lemma \ref{lem:Q_star_diff}, we get
\begin{align*}
	& \sup_{i_l} \left\{ \| \wh\btheta_{\wh g_{i_l}^{(l)}}^{(l)} - \btheta_{g_{i_l}^{(l)0}}^{(l)0} \|^2 + \frac{1}{\prod_{m \neq l} N_m} \sum_{i_{-l}} |\wh\alpha_{\wh g_{i_1}^{(1)} \cdots \wh g_{i_q}^{(q)}} - \alpha_{g_{i_1}^{(1)0} \cdots g_{i_q}^{(q)0}}^0 |^2 \right\} \\
	& = O_p \big\{\tau_{\min}^{-1} T^{-1}(\sum_l \log N_l)^2\big\}
	= O_p\big\{T^{-1}(\sum_l \log N_l)^2\big\}.
\end{align*}

\subsection{Proof of Theorem \ref{thm:h_consistency2}}\label{subsec:proof_thm3}

\begin{figure}
	\centering
	\includegraphics[width=0.5\textwidth]{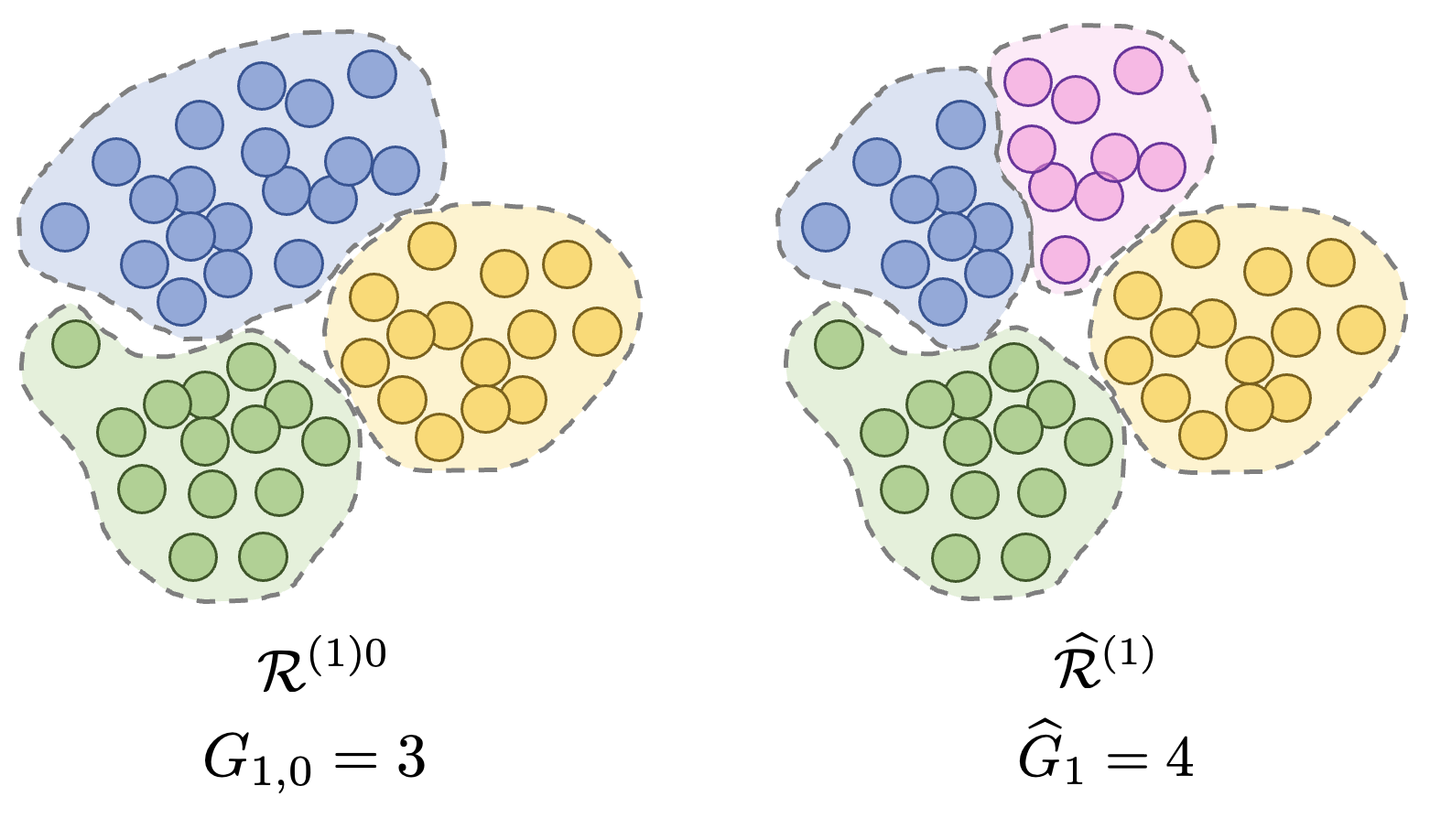}
	\caption{The group memberships when $G_{1,0}=3$ and $\wh G_1 = 4$. The units belonging to the blue group are split into two sub groups (marked as blue and pink).}
	\label{fig:memb_consist}
\end{figure}

The Figure \ref{fig:memb_consist} visualizes the strong group memberships consistency when true group number $G_{1,0}=3$ and estimated group number $\wh G_{1}  =4$.

To prove the result, we use Lemma \ref{lem:h_consistency_prepare} by setting $\bxi = \wh \bxi$.
First we note that $\wh \bTheta$ satisfies \eqref{eq:Theta_cond}
by Theorem \ref{thm:pseudo_dist} and the condition for $c_{\gap}$.
In the following we first show that
$\wh \bxi \in \mN_\eta^{(l)}$ with $\eta =  \tau_{\min}c_\gap (c_\pi)^{q-1}/\{8(\tau_{\min}+\tau_{\max})\}$ with probability tending to one. Then the conclusions of Lemma \ref{lem:h_consistency_prepare} hold.
Second, we use the conclusion to prove the result.
We remark that the notations in the following proof are borrowed in the statement in Lemma \ref{lem:h_consistency_prepare}.

\noindent
{\bf Step 1. Proof of $\wh \bxi \in \mN^{(l)}_\eta$ with probability tending to one.}

First by
Lemma \ref{lem:sig_h_bound}, we have
\begin{align*}
	& \max_{g^{(l)}\in [G_{l,0}]}\min_{\wt g^{(l)}\in [G_l]}
	\Big(\|\wh \btheta_{\wt g^{(l)}}^{(l)} - \btheta_{g^{(l)}}^{(l)0}\|^2+\frac{1}{\prod_{m \neq l} N_m} \sum_{m \neq l} \sum_{i_m} |\wh \alpha_{\wt g^{(l)} \wh g_{\i_{-l}}^{-(l)}(\bxi)} - \alpha_{g^{(l)} g_{\i_{-l}}^{-(l)0}}^0|^2\Big)\\
	&  = O_p(T^{-1}(\sum_l \log N_l)^2).
\end{align*}
Next we show
\begin{align*}
	&\max_{\wt g^{(l)}\in [G_l]}\min_{g^{(l)}\in [G_{l,0}]}
	\Big(\|\wh \btheta_{\wt g^{(l)}}^{(l)} - \btheta_{g^{(l)}}^{(l)0}\|^2+\frac{1}{\prod_{m \neq l} N_m} \sum_{m \neq l} \sum_{i_m} |\wh \alpha_{\wt g^{(l)} \wh g_{\i_{-l}}^{-(l)}(\bxi)} - \alpha_{g^{(l)} g_{\i_{-l}}^{-(l)0}}^0|^2\Big) \\
	& = O_p(T^{-1}(\sum_l \log N_l)^2).
\end{align*}
Define the set $\mO_{\wt g^{(l)}} = \{i_l \in [N_l]: \wh g_{i_l}^{(l)} = \wt g^{(l)}\}$, where $\wh g_{i_l}^{(l)}$ is obtained by \eqref{eq:g_i} by using $\wh\bxi$.
Therefore for all $i_l \in \mO_{\wt g^{(l)}}$, it holds
\begin{align*}
	&\min_{g^{(l)}\in [G_{l,0}]}
	\Big(\|\wh \btheta_{\wt g^{(l)}}^{(l)} - \btheta_{g^{(l)}}^{(l)0}\|^2+\frac{1}{\prod_{m \neq l} N_m} \sum_{m \neq l}\sum_{i_m} |\wh \alpha_{\wt g^{(l)} \wh g_{\i_{-l}}^{-(l)}(\bxi)} - \alpha_{g^{(l)} g_{\i_{-l}}^{-(l)0}}^0|^2\Big)\\
	&\le \|\wh \btheta_{\wt g^{(l)}}^{(l)} - \btheta_{g_{i_l}^{(l)0}}^{(l)0}\|^2+\frac{1}{\prod_{m \neq l} N_m} \sum_{m \neq l}\sum_{i_m} |\wh \alpha_{\wt g^{(l)} \wh g_{\i_{-l}}^{-(l)}(\bxi)} - \alpha_{g_{i_l}^{(l)0} g_{\i_{-l}}^{-(l)0}}^0|^2\\
	& = \|\wh \btheta_{\wh g_{i_l}^{(l)}}^{(l)} - \btheta_{g_{i_l}^{(l)0}}^{(l)0}\|^2+\frac{1}{\prod_{m \neq l} N_m}\sum_{m \neq l}\sum_{i_m} |\wh \alpha_{\wh g_{i_l}^{(l)} \wh g_{\i_{-l}}^{-(l)}(\bxi)} - \alpha_{g_{i_l}^{(l)0} g_{\i_{-l}}^{-(l)0}}^0|^2.
\end{align*}
Then it yields
\begin{align*}
	&\max_{\wt g^{(l)}\in [G_l]}\min_{g^{(l)}\in [G_{l,0}]}
	\Big(\|\wh \btheta_{\wt g^{(l)}}^{(l)} - \btheta_{g^{(l)}}^{(l)0}\|^2+\frac{1}{\prod_{m \neq l} N_m} \sum_{m \neq l}\sum_{i_m} |\wh \alpha_{\wt g^{(l)} \wh g_{\i_{-l}}^{-(l)}(\bxi)} - \alpha_{g^{(l)} g_{\i_{-l}}^{-(l)0}}^0|^2\Big) \\
	&\le\max_{\wt g^{(l)}\in [G_l]} \sup_{i_l \in \mO_{\wt g^{(l)}}}
	\left\{\|\wh \btheta_{\wh g_{i_l}^{(l)}}^{(l)} - \btheta_{g_{i_l}^{(l)0}}^{(l)0}\|^2+\frac{1}{\prod_{m \neq l} N_m} \sum_{m \neq l}\sum_{i_m} |\wh \alpha_{\wh g_{i_l}^{(l)} \wh g_{\i_{-l}}^{-(l)}(\bxi)} - \alpha_{g_{i_l}^{(l)0} g_{\i_{-l}}^{-(l)0}}^0|^2\right\}\\
	&=\max_{i_l \in [N_l]} \left\{\|\wh \btheta_{\wh g_{i_l}^{(l)}}^{(l)} - \btheta_{g_{i_l}^{(l)0}}^{(l)0}\|^2+\frac{1}{\prod_{m \neq l} N_m} \sum_{m \neq l}\sum_{i_m} |\wh \alpha_{\wh g_{i_l}^{(l)} \wh g_{\i_{-l}}^{-(l)}(\bxi)} - \alpha_{g_{i_l}^{(l)0} g_{\i_{-l}}^{-(l)0}}^0|^2\right\} \\
	& = O_p\big\{
	T^{-1}(\sum_l \log N_l)^2\big\},
\end{align*}
where the last equation is obtained by Proposition \ref{pro:gh_consistency}.
By the condition that $ c_{\gap} \gg T^{-1} (\sum_l \log N_l)^2$, and the definition of $\mN^{(l)}_\eta$,
we can conclude $\wh \bxi \in \mN^{(l)}_\eta$ with probability tending to one.

\noindent
{\bf Step 2. Proof of conclusion of Theorem \ref{thm:h_consistency2}.}

Since $\wh \bxi \in \mN^{(l)}_\eta$, we have that the two conclusions in Lemma \ref{lem:h_consistency_prepare} holds for $\wh\bxi$.
Then it should be equivalent to prove that for $i_{l1}, i_{l2} \in \wh \cR^{(l)}_{\wt g^{(l)}}$ for some $\wt g^{(l)} \in [G_l]$, it holds $i_{l1}, i_{l2} \in \cR_{g^{(l)}}^{(l)0}$ for some $g^{(l)0} \in [G_{l,0}]$.
Suppose $i_{l1}, i_{l2} \in \wh \cR^{(l)}_{\wt g^{(l)}}$ for some $\wt g^{(l)} \in [G_l]$, then $\wh g_{i_{l1}}^{(l)} = \wh g_{i_{l2}}^{(l)} = \wt g^{(l)}$ in this case.
By conclusion (ii) in Lemma \ref{lem:h_consistency_prepare}, we have
$\wt g^{(l)} \in \mA^{(l)}_\eta(\wh \bxi, g_{i_{l1}}^{(l)0}, \mG_2^0)$ and $\wt g^{(l)} \in \mA^{(l)}_\eta(\wh\bxi, g_{i_{l2}}^{(l)0}, \mG_2^0)$ with probability tending to 1.
Next, by conclusion (i) in Lemma \ref{lem:h_consistency_prepare}, it holds $g_{i_{l1}}^{(l)0} = g_{i_{l2}}^{(l)0}$ since $\mA^{(l)}_\eta(\wh \bxi, \cdot, \mG_2^0)$ is a partition of $[G_l]$.
By defining $g^{(l)} \defeq g_{i_{l1}}^{(l)0} = g_{i_{l2}}^{(l)0} \in [G_{l,0}]$,
we have $i_{l1}, i_{l2} \in \cR_{g^{(l)}}^{(l)0}$.
Then the conclusion we finish the proof.

\subsection{Proof of Corollary \ref{coro:group_consistency}}

By the Step 1 in the proof of Theorem \ref{thm:h_consistency2}, we have $\wh \bxi \in \mN^{(l)}_\eta$ with probability tending to one.
For $G_l = G_{l,0}$, since $\{\mA^{(l)}_\eta(\wh \bxi, g^{(l)}, \mG_2^0), g^{(l)}\in [G_{l,0}]\}$ is a partition of $[G_l]$,
then we can conclude that $\mA^{(l)}_\eta(\wh \bxi, g^{(l)}, \mG_2^0)$ contains only one element in $[G_l]$.
Consequently, we can define a permutation by $\mA^{(l)}_\eta(\wh \bxi, \cdot, \mG_2^0): [G_{l,0}]\to [G_l]$. By Theorem \ref{thm:h_consistency2},
this implies that
$\lim_{\min\{N_1, \cdots, N_q,T\}\to \infty} P\left(\wh \mG_l = \mG_l^0
\right)\to 1$.
Next, given that the probability of $ \{\wh \mG_1 = \mG_1^0, \cdots, \wh \mG_q = \mG_q^0\}$ goes to 1,
(\ref{eq:oracle}) can be directly obtained.

\subsection{Proof of Theorem \ref{thm:normal}}\label{subsec:proof_normal}

By Corollary \ref{coro:group_consistency}, we have $\wh \bxi = \wh \bxi^\o$ with probability tending to 1.
Then it suffices to treat $\wh \bxi$ as $\wh \bxi^\o$ equivalently.
{Recall that $\mE_{t}^{(\cR_{g^{(1)}}^{(1)}, \cdots, \cR_{g^{(q)}}^{(q)})} \in \mR^{N_{1g^{(1)}} \times \cdots \times N_{q g^{(q)}}}$ is defined as the subset of the tensor $\mE_t$ with each dimension selected by the index sets $\cR_{g^{(l)}}^{(l)}$.
	By the vectorization of tensor defined in the main text,}
denote $\mY_{g^{(1)} \cdots g^{(q)}, t} = \vec\big(\cY_{t}^{(\cR_{g^{(1)}}^{(1)}, \cdots, \cR_{g^{(q)}}^{(q)})}\big)$ and $\E_{g^{(1)} \cdots g^{(q)}, t} = \vec\big(\mE_{t}^{(\cR_{g^{(1)}}^{(1)}, \cdots, \cR_{g^{(q)}}^{(q)})}\big)$, then we have
\begin{align*}
	n^{q/2} T^{1/2}(\wh \bxi - \bxi^0)  = n^{q/2} T^{1/2}\bM^{-1}\bdelta = (n^{-q} T^{-1}\bM)^{-1}(n^{-q/2} T^{-1/2}\bdelta),
\end{align*}
where $\bdelta =  (\bdelta^{(1)\top}, \cdots, \bdelta^{(q)\top}, \bdelta^{\alpha\top})^\top$ and
\begin{align*}
	&\bdelta_{g^{(l)}}^{(l)} = \sum_{t, g^{-(l)}} \mX_{g^{(1)} \cdots g^{(q)}, t}^{(l)\top} \E_{g^{(1)} \cdots g^{(q)}, t},~~\bdelta^{(l)} = (\bdelta_{g^{(l)}}^{(l)\top}: g^{(l)} \in [G_l])^\top \in \mR^{G_l (p_l+1)},\\
	& \bdelta_{\mI_{g^{(1)}, \cdots, g^{(q)}}}^\alpha = \sum_t \mY_{g^{(1)} \cdots g^{(q)} , (t-1)}^\top
	\E_{g^{(1)} \cdots g^{(q)} ,t}, ~~\bdelta^\alpha = (\bdelta_{\mI_{g^{(1)}, \cdots, g^{(q)}}}^{\alpha\top}:\mI_{g^{(1)}, \cdots, g^{(q)}}\in \Big[\prod_l G_l\Big])^\top.
\end{align*}
Here
\begin{align*}
	\mX_{g^{(1)} \cdots g^{(q)},t}^{(l)} &= \Big(\vec\big\{(\cY_{t-1} \times_l \bW^{(l)})_{g^{(1)} \cdots g^{(q)}}\big\},\nonumber\\
	& \one_{N_{1 g^{(1)}}} \otimes \cdots \otimes (\bX_t^{(l)})^{(\cR_{g^{(l)}}^{(l)}, \cdot)} \otimes \cdots \otimes \one_{N_{q g^{(q)}}}   \Big) \in \mR^{(N_{1 g^{(1)}} \cdots N_{q g^{(q)}} ) \times (p_l + 1)}
\end{align*}
is defined in \eqref{eq:X_gt}.
Recall that $ \bM_{nT} = n^{-q} T^{-1}\bM$ and then
we have
\begin{align*}
	& n^{q/2} T^{1/2}\bfeta^\top(\wh \bxi - \bxi^0)  = \bfeta^\top \bM_{nT}^{-1}(n^{-q/2} T^{-1/2}\bdelta) \\
	& =
	n^{-q/2} T^{-1/2}\bfeta^\top (\bM_{nT}^0)^{-1} \bdelta + n^{-q/2} T^{-1/2} \bfeta^\top \left\{
	\bM_{nT}^{-1} - (\bM_{nT}^0)^{-1}\right\}\bdelta\\
	& = n^{-q/2} T^{-1/2} \bfeta^\top (\bM_{nT}^0)^{-1}\bdelta
	+n^{-q/2} T^{-1/2} \bfeta^\top \bM_{nT}^{-1}\left(
	\bM_{nT}^0-\bM_{nT}\right)(\bM_{nT}^0)^{-1}\bdelta.
\end{align*}
{Since we have $n^{-q/2} T^{-1/2} \bfeta^\top \bM_{nT}^{-1}\left(
	\bM_{nT}^0-\bM_{nT}\right)(\bM_{nT}^0)^{-1}\bdelta = o_p(1)$,}
then it suffices to show
\begin{align}
	&n^{-q/2} T^{-1/2} \bfeta^\top (\bM_{nT}^0)^{-1}\bdelta  \to_d
	N\{0,\sigma^2\bfeta^\top (\bM^0)^{-1}\bfeta\}.\label{eq:normal1}
\end{align}
Next, we provide the proof of \eqref{eq:normal1}.

Define $\wt \bfeta \defeq  (\bM_{nT}^0)^{-1}\bfeta =
(\wt\bfeta_1^{(1)\top}, \cdots, \wt\bfeta_{G_1}^{(1)\top}, \cdots,
\wt \bfeta_1^{(q)\top}, \cdots, \wt \bfeta_{G_q}^{(q)\top},
\wt \bfeta_{1 \cdots 1}^{\alpha},\cdots , \wt \bfeta_{G_1 \cdots G_q}^{\alpha})^\top$, where
$\wt \bfeta_{g^{(l)}}^{(l)}\in \mR^{p_l+1}$ and
$\wt \bfeta_{g^{(1)} \cdots g^{(q)}}^\alpha \in \mR$.
Then we have
\begin{align*}
	n^{-q/2} T^{-1/2} \bfeta^\top (\bM_{nT}^0)^{-1}\bdelta
	& = \frac{1}{\sqrt{n^q T}} \sum_l \sum_{g^{(l)}} \wt \bfeta_{g^{(l)}}^{(l)\top}\sum_{t, g^{-(l)}}\mX_{g^{(1)} \cdots g^{(q)}, t}^{(l)\top} \E_{g^{(1)} \cdots g^{(q)}, t} \\
	&+ \frac{1}{\sqrt{n^q T}} \sum_l  \sum_{g^{(l)} = 1}^{G_l} \wt \bfeta_{g^{(1)} \cdots g^{(q)}}^{\alpha} \sum_{t}\mY_{g^{(1)} \cdots g^{(q)}, (t-1)}^\top \E_{g^{(1)} \cdots g^{(q)}, t}\\
	& \defeq \sum_t  \sum_{g^{(l)}}\cZ_{g^{(1)} \cdots g^{(q)}, t}^{\top} \E_{g^{(1)} \cdots g^{(q)}, t}\defeq
	\sum_t\cZ_{t}^{\top} \E_{t}.
\end{align*}
Define $\mF_{t}$ as the sigma field generated by $\{\ve_{i_1 \cdots i_q,s}: i_l\in
[N_l], l \in [q], s\le t\}$.
We have $\{\cZ_{t}^\top \E_{t}, \mF_t\}$ as a martingale difference
array.
By the central limit theorem of \cite{hall2014martingale}, it suffices to verify for any $u>0$
\begin{align}
	&\sum_{t = 1}^T E\left\{(\cZ_{t}^\top \E_{t})^2I(|\cZ_{t}^\top \E_{t}|>u)|\mF_{t-1}\right\}\to_p 0\label{eq:martingale1}\\
	&\sum_{t = 1}^T E\left\{(\cZ_{t}^\top \E_{t})^2|\mF_{t-1}\right\}\to_p  \bfeta^\top  (\bM^0)^{-1}\bM_{b}^0 (\bM^0)^{-1}\bfeta.\label{eq:martingale2}
\end{align}
where $\bM_{b}^0 \defeq \lim_{\min(n,T)\to \infty} n^{-q} T^{-1}\var(\bdelta) =
\sigma^2 \lim_{\min(n,T)\to \infty}  \bM_{nT}^{0} = \sigma^2 \bM^{0}$, and $ (\bM^0)^{-1} =  (\bM^0)^{-1}\bM_{b}^0 (\bM^0)^{-1}$ by the condition that $\ve_{i_1 \cdots i_q,t}$ is i.i.d over $i_1,\cdots, i_q,t$ with $\var(\ve_{i_1 \cdots i_q,t}) = \sigma^2$ and it is also independent of $
\{\cY_{s-1}, \bX_s^{(1)}, \cdots, \bX_s^{(q)}: s \le t\}$,
where $\bX_s^{(l)} = (\bx_{i_l, s}^{(l)}: i_l \in [N_l])^\top$.

\noindent
{\bf (1) Proof of (\ref{eq:martingale1})}

We have
\begin{align*}
	&\sum_{t = 1}^T E\left\{(\cZ_{t}^\top \E_{t})^2I(|\cZ_{t}^\top \E_{t}|>u)|\mF_{t-1}\right\}
	\le u^{-2}\sum_t E\left\{(\cZ_{t}^\top \E_{t})^4|\mF_{t-1}\right\}
	\le cu^{-2} \sum_t (\cZ_t^\top\cZ_t)^2,
\end{align*}
where $c$ is the fourth moment of $\varepsilon_{i_1 \cdots i_q, t}$, which is
a finite constant by Assumption \ref{assum:sub_gaussian}.
Then it suffices to show $E\{\sum_t (\cZ_t^\top\cZ_t)^2\}\to 0$.
Note that we have
\begin{align*}
	\cZ_{g^{(1)} \cdots g^{(q)}, t} & =
	\sum_{l} \frac{1}{\sqrt{n^qT}} \mX^{(l)}_{g^{(1)} \cdots g^{(q)}, t} \wt \bfeta_{g^{(l)}}^{(l)} + \frac{1}{\sqrt{n^qT}} \mY_{g^{(1)} \cdots g^{(q)} , (t-1)} \wt \bfeta_{g^{(1)} \cdots g^{(q)}}^\alpha \\
	&   \defeq \sum_l \cZ_{g^{(1)} \cdots g^{(q)}, t}^{(l)} + \cZ_{g^{(1)} \cdots g^{(q)}, t}^\alpha.
\end{align*}

Then it suffices to show
\begin{align}
	&E\{\sum_t(\sum_l \sum_{g^{(l)}}\cZ_{g^{(1)} \cdots g^{(q)}, t}^{(l)\top}\cZ_{g^{(1)} \cdots g^{(q)}, t}^{(l)})^2\}\to 0, \label{eq:Z_top_Z_l}\\
	&E\{\sum_t
	(\sum_l \sum_{g^{(l)}}\cZ_{g^{(1)} \cdots g^{(q)}, t}^{\alpha\top}\cZ_{g^{(1)} \cdots g^{(q)}, t}^\alpha)^2\}\to 0 \label{eq:Z_top_Z_alpha}
\end{align}
due to Cauchy-Schwarz inequality.
We first show \eqref{eq:Z_top_Z_l} in Step (1.1), and then show \eqref{eq:Z_top_Z_alpha} in Step (1.2).

\noindent
{\bf (1.1) Proof of $E\{\sum_t(\sum_l \sum_{g^{(l)}}\cZ_{g^{(1)} \cdots g^{(q)}, t}^{(l)\top}\cZ_{g^{(1)} \cdots g^{(q)}, t}^{(l)})^2\}\to 0$}

Write $\wt\bfeta_{g^{(l)}}^{(l)} = (\wt\eta_{g^{(l)},1}^{(l)}, \wt\bfeta_{g^{(l)},(-1)}^{{(l)}\top})^\top$.
By the definition of $\cY_{t-1} \times_l \bW^{(l)}$ in the main text,
we have
\begin{align*}
	& \cY_{t-1} \times_l \bW^{(l)(\cR_{g^{(l)}}, \cdot)} \\
	& = (\sum_{i_l = 1}^{N_l} Y_{i_1, \cdots, i_l, \cdots, i_q, (t-1)} (a_{s, i_l}^{(l)}/ n_{ls}))_{i_1 \in [N_1], \cdots, s \in \cR_{g^{(l)}}^{(l)}, \cdots, i_q \in [N_q]}  \in \mR^{N_1 \times \cdots \times N_{l g^{(l)}} \times N_q}.
\end{align*}
We have
\begin{align*}
	&E\left\{\sum_t (\sum_l \sum_{g^{(l)}}\cZ_{g^{(1)} \cdots g^{(q)}, t}^{(l)\top}\cZ_{g^{(1)} \cdots g^{(q)}, t}^{(l)})^2\right\} \le c\frac{1}{n^{2q} T}  E\left\{
	\left(\sum_{g^{(1)}, \cdots, g^{(q)}}(\wt\eta_{g^{(l)},1}^{(l)})^2
	\|\mX_{g^{(1)} \cdots g^{(q)},t}^{(l)\dag}\|^2
	\right)^2\right\}\\
	&+c\frac{1}{n^{2q} T} E\left\{\left(\sum_{g^{(1)}, \cdots, g^{(q)}} (\prod_{m \ne l} N_{mg^{(m)}}) \wt\bfeta_{g^{(l)},(-1)}^{(l)\top}
	\mX_{g^{(1)} \cdots g^{(q)},t}^{(l)\dag\dag\top}
	\mX_{g^{(1)} \cdots g^{(q)},t}^{(l)\dag\dag}   \wt\bfeta_{g^{(l)},(-1)}^{(l)}\right)^2\right\}\\
	&= c\frac{1}{n^{2q} T} E\left\{\left(\sum_{g^{(l)}}(\wt\eta_{g^{(l)},1}^{(l)})^2
	\vec\big\{ \cY_{t-1} \times_l \bW^{(l)(\cR_{g^{(l)}}, \cdot)}  \big\}^\top \vec \big\{ \cY_{t-1} \times_l \bW^{(l)(\cR_{g^{(l)}}, \cdot)} \big\}
	\right)^2\right\} \\
	&+
	c\frac{1}{n^{2q} T} E\left\{\left(\sum_{g^{(l)}} (\prod_{m \neq l} N_m) \wt\bfeta_{g^{(l)},(-1)}^{(l)\top} (\bX_t^{(l)})^{(\cR_{g^{(l)}}^{(l)},\cdot)\top}
	(\bX_t^{(l)})^{(\cR_{g^{(l)}}^{(l)},\cdot)} \wt\bfeta_{g^{(l)},(-1)}^{(l)}\right)^2\right\},
\end{align*}
where $c$ is a constant.
By Lemma \ref{lem:Y4_bound} and Cauchy-Schwarz inequality, it holds
\begin{align*}
	&E\left\{\left(\sum_{g}(\wt\eta_{g^{(l)},1}^{(l)})^2
	\vec\big\{\cY_{t-1} \times_l \bW^{(l)(\cR_{g^{(l)}}, \cdot)} \big\}^\top \vec\big\{\cY_{t-1} \times_l \bW^{(l)(\cR_{g^{(l)}}, \cdot)} \big\}
	\right)^2\right\}\\
	&\le c_4\sum_{g^{(l)}}(\wt\eta_{g^{(l)},1}^{(l)})^2
	\tr\Big(\bW^{(l) (\cR_{g^{(l)}}^{(l)},\cdot)}
	(\one_{N_1}^\top \cdots \one_{N_{l-1}}^\top \one_{N_l} \one_{N_{l+1}}^\top \cdots \one_{N_q}^\top) \\
	& (\one_{N_q} \cdots \one_{N_{l+1}} \one_{N_l}^\top \one_{N_{l-1}} \cdots \one_{N_1})
	\bW^{(l) (\cR_{g^{(l)}}^{(l)},\cdot) \top}\Big)^2 \le c_4\left(\sum_{g^{(l)}} (\wt\eta_{g^{(l)},1}^{(l)})^2 c_2 n^q
	\right)^2\le c n^{2q}
\end{align*}
where $c_4 = E(Y_{i_1 \cdots i_q, t-1}^4)<\infty$ and $c$ is a constant.
Here the last second inequality holds because $N_{l} \le c_2 n$ and $c_2$ is a positive constant.
Next, we have
\begin{align*}
	& \frac{1}{n^{2q} T}E\left\{\left(\sum_{g^{(l)}}(\prod_{m \ne l} N_m)\wt\bfeta_{g^{(l)},(-1)}^{(l)\top}(\bX_t^{(l)})^{(\cR_{g^{(l)}}^{(l)},\cdot)\top}
	(\bX_t^{(l)})^{(\cR_{g^{(l)}}^{(l)},\cdot)} \wt\bfeta_{g^{(l)},(-1)}^{(l)}\right)^2\right\}\\
	&=  \frac{1}{n^{2q}T}(\prod_{m \ne l} N_m)^2 E\left(\sum_{g^{(l)}} \sum_{i_l \in\cR_{g^{(l)}}^{(l)}} (\wt\bfeta_{g^{(l)},(-1)}^{(l)\top}\bx_{i_l t}^{(l)})^2\right)^2\le \frac{c}{n^2 T} E\left(\sum_{g^{(l)}} \sum_{i_l \in\cR_{g^{(l)}}^{(l)}} (\wt\bfeta_{g^{(l)},(-1)}^{(l)\top}\bx_{i_l t}^{(l)})^2\right)^2.
\end{align*}
By the Cauchy's inequality we have
\begin{align*}
	E\left\{(\wt\bfeta_{g_1^{(l)},(-1)}^{(l)\top} \bx_{i_{l1}t}^{(l)})^2(\wt\bfeta_{g_2^{(l)},(-1)}^{(l)\top} \bx_{i_{l2}t}^{(l)})^2\right\} \le \left[E\left\{\big(\wt\bfeta_{g_1^{(l)},(-1)}^{(l)\top} \bx_{i_{l1}t}^{(l)}\big)^4\right\}E\left\{\big(\wt\bfeta_{g_2^{(l)},(-1)}^{(l)\top} \bx_{i_{l2}t}^{(l)}\big)^4\right\}\right]^{1/2}.
\end{align*}
We also have
\begin{align*}
	E\left\{\big(\wt\bfeta_{g^{(l)},(-1)}^{(l)\top} \bx_{i_{l}t}^{(l)}\big)^4\right\}& =
	\Big\|\wt\bfeta_{g^{(l)},(-1)}^{(l)}\Big\|^4E\left\{\Big(\wt\bfeta_{g^{(l)},(-1)}^{(l)\dag\top} \bx_{i_{l}t}^{(l)}\Big)^4\right\}\\
	&=4\Big\|\wt\bfeta_{g^{(l)},(-1)}^{(l)}\Big\|^4\int_0^\infty t^3 P\left\{\big|\langle \wt\bfeta_{g^{(l)},(-1)}^{(l)\dag},\bx_{i_{l}t}^{(l)} \rangle\big|>t\right\}dt\le c_K
	\|\wt\bfeta_{g^{(l)},(-1)}^{(l)}\Big\|^4
\end{align*}
where $\wt\bfeta_{g^{(l)},(-1)}^{(l)\dag} = \wt\bfeta_{g^{(l)},(-1)}^{(l)}/\|\wt\bfeta_{g^{(l)},(-1)}^{(l)}\|$,
and $c_K$ is a constant related to $K$.
Here the last inequality is obtained by noting that $\langle \wt\bfeta_{g^{(l)},(-1)}^{(l)\dag},
\bx_{i_lt}^{(l)}\rangle$ is a $1$-Lipschitz convex function of $\x_{i_lt}^{(l)}$ with
mean zero and $\x_{i_lt}^{(l)}$ is $K$-convex defined in Definition \ref{def:convex_concen}.
Consequently we have
\begin{align*}
	&\frac{c}{n^2T} E\left(\sum_{g_1^{(l)},g_2^{(l)}}\sum_{i_{l1} \in \cR_{g_1^{(l)}}^{(l)}}\sum_{i_{l2} \in \cR_{g_2^{(l)}}^{(l)}} \big(\wt\bfeta_{g_1^{(l)},(-1)}^{(l)\top} \bx_{i_{l1}t}^{(l)}\big)^2
	\big(\wt\bfeta_{g_2^{(l)},(-1)}^{(l)\top} \bx_{i_{l2}t}^{(l)}\big)^2\right)\\
	& \le
	\frac{C}{n^2T}\left (\sum_{g_1^{(l)},g_2^{(l)}} c_K\|\wt\bfeta_{g_1^{(l)},(-1)}^{(l)}\|^2
	\|\wt\bfeta_{g_2^{(l)},(-1)}^{(l)}\|^2n^2\right) = O(T^{-1}) = o(1).
\end{align*}
Consequently, it yields
$E\left\{\sum_t(\sum_l \sum_{g^{(l)}}\cZ_{g^{(1)} \cdots g^{(q)}, t}^{(l)\top}\cZ_{g^{(1)} \cdots g^{(q)}, t}^{(l)})^2\right\} = O(T^{-1}) = o(1)$.

\noindent
{\bf (1.2) Proof of $E\{\sum_t
	(\sum_l \sum_{g^{(l)}}\cZ_{g^{(1)} \cdots g^{(q)}, t}^{\alpha\top}\cZ_{g^{(1)} \cdots g^{(q)}, t}^\alpha)^2\}\to 0$}

Next, we have
\begin{align*}
	&E\left\{\sum_t
	(\sum_l \sum_{g^{(l)}}\cZ_{g^{(1)} \cdots g^{(q)}, t}^{\alpha\top}\cZ_{g^{(1)} \cdots g^{(q)}, t}^\alpha)^2\right\} \\
	& = \frac{1}{n^{2q}T}E\left\{\sum_l \sum_{g^{(l)}}
	(\wt\bfeta_{g^{(1)} \cdots g^{(q)}}^{\alpha})^2\mY_{g^{(1)} \cdots g^{(q)},(t-1)}^\top \mY_{g^{(1)} \cdots g^{(q)}, (t-1)}
	\right\}^2\\
	&\le  \frac{1}{n^{2q}T}c \left\{\sum_l \sum_{g^{(l)}}
	(\wt\bfeta_{g^{(1)} \cdots g^{(q)}}^{\alpha})^2 n^q
	\right\}^2 = O(T^{-1}) = o(1),
\end{align*}
where the inequality is obtained by Lemma \ref{lem:Y4_bound}.

\noindent
{\bf(2) Proof of (\ref{eq:martingale2})}

We have
\begin{align*}
	&\sum_{t = 1}^T E\left\{(\cZ_{t}^\top \E_{t})^2|\mF_{t-1}\right\}
	= \sigma^2 \sum_t \cZ_{t}^\top\cZ_{t}
	=\sigma^2 \sum_t \sum_l \sum_{g^{(l)}} \cZ_{g^{(1)} \cdots g^{(q)}, t}^\top \cZ_{g^{(1)} \cdots g^{(q)},t}\\
	& = \frac{\sigma^2}{n^q T} \sum_t \sum_l   \sum_{g^{(1)}, \cdots, g^{(q)}} \wt\bfeta_{g^{(l)}}^{(l)\top} \mX_{g^{(1)} \cdots g^{(q)}, t}^{(l)\top} \mX^{(l)}_{g^{(1)} \cdots g^{(q)}, t}\wt\bfeta_{g^{(l)}}^{(l)}\\
	& + \frac{\sigma^2}{n^qT}\sum_t \sum_{g^{(1)}, \cdots, g^{(q)}} (\wt\bfeta_{g^{(1)}, \cdots, g^{(q)}}^{\alpha})^2 \mY_{g^{(1)} \cdots g^{(q)},(t-1)}^\top\mY_{g^{(1)} \cdots g^{(q)}, (t-1)}.
\end{align*}
In the following we prove that
\begin{align}
	& \Big| \frac{1}{n^q T} \sum_t  \sum_{g^{(1)}, \cdots, g^{(q)}} \wt\bfeta_{g^{(l)}}^{(l)\top} \mX_{g^{(1)} \cdots g^{(q)}, t}^{(l)\top} \mX^{(l)}_{g^{(1)} \cdots g^{(q)}, t}\wt\bfeta_{g^{(l)}}^{(l)} \nonumber\\
	& -
	\lim_{n\to \infty} \frac{1}{n^q} \sum_{g^{(1)}, \cdots, g^{(q)}} \wt\bfeta_{g^{(l)}}^{(l)\top} E(\mX_{g^{(1)} \cdots g^{(q)}, t}^{(l)\top} \mX^{(l)}_{g^{(1)} \cdots g^{(q)}, t}) \wt\bfeta_{g^{(l)}}^{(l)}
	\Big| \to_p 0,\label{eq:XX_conv}\\
	&  \Big| \frac{1}{n^q T}\sum_t \sum_{g^{(1)}, \cdots, g^{(q)}} (\wt\bfeta_{g^{(1)}, \cdots, g^{(q)}}^{\alpha})^2 \mY_{g^{(1)} \cdots g^{(q)},(t-1)}^\top\mY_{g^{(1)} \cdots g^{(q)}, (t-1)} \nonumber\\
	& - \lim_{n \to \infty}
	\frac{1}{n^q}  \sum_{g^{(1)}, \cdots, g^{(q)}} (\wt\bfeta_{g^{(1)}, \cdots, g^{(q)}}^{\alpha})^2 E(\mY_{g^{(1)} \cdots g^{(q)},(t-1)}^\top\mY_{g^{(1)} \cdots g^{(q)}, (t-1)}) \Big| \to_p 0.\label{eq:YY_conv}
\end{align}

We first show (\ref{eq:XX_conv}) in the following.
Note that by the step (1.1), we have
\begin{align}
	& \frac{1}{n^q T} \sum_t  \sum_{g^{(1)}, \cdots, g^{(q)}} \wt\bfeta_{g^{(l)}}^{(l)\top} \mX_{g^{(1)} \cdots g^{(q)}, t}^{(l)\top} \mX^{(l)}_{g^{(1)} \cdots g^{(q)}, t}\wt\bfeta_{g^{(l)}}^{(l)} \nonumber\\
	& = \frac{1}{n^q T}\sum_t\sum_{g^{(l)}}
	(\wt\eta_{g^{(l)},1}^{(l)})^2
	\vec\big\{ \cY_{t-1} \times_l \bW^{(l)(\cR_{g^{(l)}}, \cdot)}  \big\}^\top \vec \big\{ \cY_{t-1} \times_l \bW^{(l)(\cR_{g^{(l)}}, \cdot)} \big\}\nonumber\\
	&+\frac{1}{n^q T}\sum_t
	\sum_{g^{(l)}}
	(\prod_{m \neq l} N_m) \wt\bfeta_{g^{(l)},(-1)}^{(l)\top} (\bX_t^{(l)})^{(\cR_{g^{(l)}}^{(l)},\cdot)\top}
	(\bX_t^{(l)})^{(\cR_{g^{(l)}}^{(l)},\cdot)} \wt\bfeta_{g^{(l)},(-1)}^{(l)}.\label{eq:etaXXeta}
\end{align}
Note that
\begin{align*}
	&\frac{1}{n^q T}\sum_t\sum_{g^{(l)}}
	(\wt\eta_{g^{(l)},1}^{(l)})^2 \vec\big\{ \cY_{t-1} \times_l \bW^{(l)(\cR_{g^{(l)}}, \cdot)}  \big\}^\top \vec \big\{ \cY_{t-1} \times_l \bW^{(l)(\cR_{g^{(l)}}, \cdot)} \big\}\\
	& = \frac{1}{n^q T}\sum_t\sum_{g^{(l)}}(\wt\eta_{g^{(l)},1}^{(l)})^2
	\left((\I_{N_{1}}\otimes \cdots \otimes \I_{N_{l-1}} \otimes \bW^{(l)(\cR_{g^{(l)}}^{(l)},\cdot)} \otimes \I_{N_{l+1}} \cdots \otimes \I_{N_q}) \mY_t\right)^\top \\
	&\left((\I_{N_{1}}\otimes \cdots \otimes \I_{N_{l-1}} \otimes \bW^{(l)(\cR_{g^{(l)}}^{(l)},\cdot)} \otimes \I_{N_{l+1}} \cdots \otimes \I_{N_q}) \mY_t \right).
\end{align*}
Further note that
\begin{align}
	& \Big| \frac{1}{n^q T}\sum_t\sum_{g^{(l)}} 	(\wt\eta_{g^{(l)},1}^{(l)})^2
	\left((\I_{N_{1}}\otimes \cdots \otimes \I_{N_{l-1}} \otimes \bW^{(l)(\cR_{g^{(l)}}^{(l)},\cdot)} \otimes \I_{N_{l+1}} \cdots \otimes \I_{N_q}) \mY_t\right)^\top \nonumber\\
	&\left((\I_{N_{1}}\otimes \cdots \otimes \I_{N_{l-1}} \otimes \bW^{(l)(\cR_{g^{(l)}}^{(l)},\cdot)} \otimes \I_{N_{l+1}} \cdots \otimes \I_{N_q}) \mY_t \right)\nonumber\\
	& -\frac{1}{n^q}\sum_{g^{(l)}}  (\wt\eta_{g^{(l)},1}^{(l)})^2
	\tr\left\{ \big(\I_{N_{q}}\otimes \cdots \otimes \I_{N_{l-1}} \otimes  \bW^{(l)(\cR_{g^{(l)}}^{(l)},\cdot) \top}\bW^{(l)(\cR_{g^{(l)}}^{(l)},\cdot)} \otimes \I_{N_{l+1}} \otimes \cdots \otimes \I_{N_{q}} \big) \bGamma\right\}  \Big| \nonumber\\
	&= o_p(1) \label{eq:WYYW}
\end{align}
by Lemma \ref{lem:YMY},
where  $\wt
\bW$ in Lemma \ref{lem:YMY} is set to be $ (\prod_{m \neq l} N_m)^{-1/2} \big( \I_{N_{1}} \otimes \cdots \I_{N_{l-1}} \otimes (\wt\eta_{g^{(l)},1}^{(l)}
\bW^{(l)(\cR_{g^{(l)}}^{(l)},\cdot)}, 1\le g^{(l)}\le G_l
)^\top \otimes \I_{N_{l+1}} \otimes \cdots \otimes \I_{N_{q}} \big) \in \mR^{(\prod_l N_l)\times (\prod_l N_l)}$ and  we can verify that
$n^{-1}\1_{\prod_l N_l}^\top\wt \bW^\top \wt
\bW\1_{\prod_l N_l}$ is bounded under when $N_l = N_{lg^{(l)}} = O(n)$ for all $l \in [q]$.

Next define $\bx_t^{(l)\dag} \defeq((\bX_t^{(l)(\cR_1,\cdot)} \wt\bfeta_{1,(-1)}^{(l)\dag})^\top, (\bX_t^{(l)(\cR_2,\cdot)} \wt\bfeta_{2,(-1)}^{(l)\dag})^\top,\cdots,
(\bX_t^{(l)(\cR_{G_l},\cdot)} \wt\bfeta_{G_l,(-1)}^{(l)\dag})^\top)^\top
$, where $ \wt\bfeta_{g^{(l)},(-1)}^{(l)\dag} =
\wt\bfeta_{g^{(l)},(-1)}^{(l)}/\|\wt\bfeta_{g^{(l)},(-1)}^{(l)}\|$.
Then we have
\begin{align}
	\frac{1}{nT}\sum_t \left(
	\sum_{g^{(l)}}\wt\bfeta_{g^{(l)},(-1)}^{(l) \top}\bX_t^{(l)(\cR_{g^{(l)}}^{(l)},\cdot)\top}
	\bX_t^{(l)(\cR_{g^{(l)}},\cdot)} \wt\bfeta_{g^{(l)},(-1)}^{(l)}\right)
	= \frac{1}{nT}\sum_t \bx_t^{(l)\dag \top}\bA^{(l)} \bx_t^{(l)\dag}\label{eq:xAx_conv}
\end{align}
where $\bA^{(l)} = \diag\{\|\wt\bfeta_{g^{(l)},(-1)}^{(l)}\|^2 \one_{N_{lg^{(l)}}}:g^{(l)}\in
[G_l]\}$.
Note that $\bx_t^\dag$
satisfies
$K$-convex concentration property defined in Definition \ref{def:convex_concen}.
Then by using Lemma \ref{lem:xAx_convex},  we have
\begin{align*}
	\big| (nT)^{-1} \sum_t \bx_t^{(l)\dag \top}\bA^{(l)} \bx_t^{(l)\dag} - n^{-1}
	E(\bx_t^{(l)\dag \top}\bA^{(l)} \bx_t^{(l)\dag})\big| = o_p(1).
\end{align*}
Together with (\ref{eq:WYYW}) and (\ref{eq:etaXXeta}), and by taking the limit, we have proved
(\ref{eq:XX_conv}).

Next, (\ref{eq:YY_conv}) can be similarly proved by using the same technique as
(\ref{eq:WYYW}) and Lemma \ref{lem:YMY}.
Together with the fact that $n^{-q/2} T^{-1/2} \bfeta^\top \bM_{nT}^{-1}\left(
\bM_{nT}^0-\bM_{nT}\right)(\bM_{nT}^0)^{-1}\bdelta = o_p(1)$,
we reach the conclusion.

\section{Technical Lemmas}\label{sec:tech_lemma}

\bel\label{lem:Q_diff}
Suppose Assumptions \ref{assum:para_space}--\ref{assum:station} hold.
Then we have
\begin{align}
	&P\Big(\sup_{\|\bTheta_{i_1 \cdots i_q}\|_{\max} <R}\Big|\frac{1}{T}Q_{i_1\cdots i_q}(\bTheta_{i_1 \cdots i_q}) - \frac{1}{T}Q^*_{i_1\cdots i_q}(\bTheta_{i_1 \cdots i_q})\Big|>x\Big)\nonumber\\
	&\le
	\exp\Big\{
	-c_1 \min(Tx^2,T^{1/2}x)+c_2m\Big\},\label{eq:Q_ij_diff}
\end{align}
where $m=\sum_l (p_l+1)+1$,
and $c_1,c_2$ are positive constants,
In addition, we have
\begin{align}
	&\sup_{\|\bTheta\|_{\max} < R}\Big|
	\frac{1}{(\prod_l N_l) T}\Big\{Q(\bTheta) - Q^*(\bTheta)\Big\}\Big|\nonumber\\
	&\le \sup_{i_1, \cdots, i_q}\sup_{\|\bTheta_{i_1 \cdots i_q}\|_{\max} <R} \Big|
	\frac{1}{T}\Big\{ Q_{i_1\cdots i_q}(\bTheta_{i_1 \cdots i_q}) - \frac{1}{T}Q^*_{i_1\cdots i_q}(\bTheta_{i_1 \cdots i_q}) \Big\}\Big| \nonumber\\
	& = O_p\Big(T^{-1/2}(m+ \sum_l \log N_l)\Big).\label{eq:sup_Q_ij_diff}
\end{align}
\eel

\begin{proof}
	Note that we have
	\begin{align*}
		\frac{1}{T}Q_{i_1\cdots i_q}(\bTheta_{i_1 \cdots i_q}) &=
		\frac{1}{T}\sum_{t = 1}^T
		\big(\ve_{i_1 \cdots i_q, t} + \cX_{i_1 \cdots i_q, t}^\top \bTheta^0_{i_1 \cdots i_q} - \cX_{i_1 \cdots i_q, t}^\top \bTheta_{i_1 \cdots i_q} \big)^2\\
		& = \frac{1}{T}\sum_{t = 1}^T\Big\{\ve_{i_1 \cdots i_q, t}^2 +
		2\ve_{i_1 \cdots i_q, t}\cX_{i_1 \cdots i_q, t}^\top(\bTheta^0_{i_1 \cdots i_q}- \bTheta_{i_1 \cdots i_q} )\\
		&+
		(\bTheta_{i_1 \cdots i_q}- \bTheta^0_{i_1 \cdots i_q} )^\top\cX_{i_1 \cdots i_q, t}\cX_{i_1 \cdots i_q, t}^\top (\bTheta_{i_1 \cdots i_q}- \bTheta^0_{i_1 \cdots i_q})
		\Big\}.
	\end{align*}
	Recall that $\bSigma_{i_1 \cdots i_q} = E(\cX_{i_1 \cdots i_q, t}\cX_{i_1 \cdots i_q, t}^\top)$.
	It is sufficient to show
	\begin{align}
		&P\Big(\Big|\frac{1}{T}\sum_t \ve_{i_1 \cdots i_q, t}^2 - \sigma^2\Big|>x/3\Big)\le
		2\exp\{-c_1T\min(x^2, x)\} \label{eq:eps2_conv}\\
		& P\Big(\sup_{\|\bfa\|_{\max} < 2R}\Big|\frac{2}{T}\sum_t
		\ve_{i_1 \cdots i_q, t}\cX_{i_1 \cdots i_q, t}^{\top}\bfa\Big|>x/3\Big)\le \exp\Big\{-c_1 \min(Tx^2,T^{1/2}x)+c_3m\Big\},\label{eq:epsX_conv}\\
		&P\Big(\sup_{\|\bfa\|_{\max} < 2R}\Big|\frac{1}{T}\sum_t \bfa^\top\cX_{i_1 \cdots i_q, t} \cX_{i_1 \cdots i_q, t}^{\top}\bfa -
		\bfa^\top \bSigma_{i_1 \cdots i_q}\bfa\Big|>x/3\Big)\nonumber\\
		& \le \exp\Big\{
		- c_1 \min(Tx^2,T^{1/2}x)+c_3m\Big\}
		\label{eq:X2_conv}
	\end{align}
	where $m = \sum_l (p_l+1) + 1$ as defined in Theorem \ref{thm:pseudo_dist}, and
	$c_1$, $c_3$ are positive constants.
	First by Lemma
	\ref{lem:xAx_convex}, (\ref{eq:eps2_conv}) holds directly.
	We next prove (\ref{eq:X2_conv}) in two steps.
	First we establish the upper bound for a fixed $\bfa$,
	then we establish the upper bound for all $\bfa$ satisfying $\|\bfa\|_{\max} < 2R$.
	This yields (\ref{eq:Q_ij_diff}) and (\ref{eq:sup_Q_ij_diff})
	holds subsequently.

	\noindent
	{\bf Step 1.}
	We first show
	\begin{align*}
		P\Big(\Big|\frac{1}{T}\sum_t \bfa^\top \cX_{i_1 \cdots i_q, t} \cX_{i_1 \cdots i_q, t}^{\top}\bfa -
		\bfa^\top \bSigma_{i_1 \cdots i_q}\bfa\Big|>x/3\Big)\le
		c_0\exp\Big(-c_3\min(Tx^2,T^{1/2}x)\Big)
	\end{align*}
	for any $\bfa$,
	where $c_3$ and $c_0$ are positive constants.
	For the sake of similarity, we only verify the concentration inequality for diagonal elements of $T^{-1}\sum_t \cX_{i_1 \cdots i_q, t}  \cX_{i_1 \cdots i_q, t}^{\top}$.
	Note that the diagonal elements of $T^{-1}\sum_t \cX_{i_1 \cdots i_q, t}  \cX_{i_1 \cdots i_q, t}^{\top}$ takes
	the form $ T^{-1}\sum_t\bw^\top\mY_t\mY_t^\top\bw$ (with $\bw \in \mR^{\prod_l N_l}$ and $\|\bw\|_1 = 1$), or
	$T^{-1} \sum_t (\bx_{i_l j, t}^{(l)})^2$, where $ \bx_{i_l j, t}^{(l)}$ is the $j$th element of the $p_l$-length vector $ \bx_{i_l, t}^{(l)}$.
	For $T^{-1}\sum_t \bx_{i_l, t}^{(l)\top} \bx_{i_l, t}^{(l)}$.
	We use Lemma
	\ref{lem:xAx_convex} and Assumption \ref{assum:mixing},
	and  note that
	$c_0\exp\{-c_3 T\min(x^2,x)\}
	\le c_0\exp\{-c_3 \min(Tx^2,T^{1/2}x)\}$. This yields
	the
	concentration inequality.
	For the $T^{-1}\sum_t\bw^\top\mY_t\mY_t^\top\bw$, we use
	Lemma \ref{lem:concenX} to obtain the result.

	\noindent
	{\bf Step 2.}
	Here we follow Lemma F.2 of \cite{basu2015regularized} to prove the result.
	First we define
	$\mM = \{\bv\in \mR^{m}:
	\|\bv\|_{\max} \le 2R\}$, where $m = \sum_l (p_l+1) + 1$.
	Let $\mA = \{\bu_{1},\cdots, \bu_{|\mA|}\}$
	be a $R/5$-net of $\mM$, where $|\mA|\le 10^m$.
	Hence for any $\bv\in \mM$, there exists some $\bu_{i}\in \mA$ such that $\|\Delta \bv\|_{\max} \le R/5$, where
	$\Delta\bv = \bv - \bu_{i}$.
	Define $\bM = |T^{-1}\sum_t\cX_{i_1 \cdots i_q,t}\cX_{i_1 \cdots i_q, t}^\top - \bSigma_{i_1 \cdots i_q}|_e$.
	\begin{align*}
		\tau&\defeq \sup_{\bv\in\mM}|\bv^\top \bM \bv|\le
		\max_{1\leq k\leq |\mA|} |\bu_k^\top\bM\bu_k|+ 2\sup_{\bv\in \mM}
		|\max_{1\le k\le |\mA|} \Delta\bv^\top  \bM \bu_k| +
		\sup_{\bv\in \mM}
		|\Delta\bv^\top \bM \Delta\bv|\\
		&\le2\max_{1\leq k\leq |\mA|} |\bu_k^\top\bM\bu_k|+
		2\sup_{\bv\in \mM}
		|\Delta\bv^\top \bM \Delta\bv| \le 2\max_{1\leq k\leq |\mA|} |\bu_k^\top\bM\bu_k|+
		\tau/50,
	\end{align*}
	where the last step is because  $10\Delta\bv\in \mM$.
	Thus $\tau\le (100/49)\max_{1\leq k\leq |\mA|} |\bu_k^\top\bM\bu_k|$.
	This leads to
	\begin{align*}
		&\sup_{\|\bfa\|_{\max}\le 2R}P\Big(\Big|\frac{1}{T}\sum_t \bfa^\top \cX_{i_1 \cdots i_q, t}\cX_{i_1 \cdots i_q, t}^{\top}\bfa -
		\bfa^\top \bSigma_{i_1 \cdots i_q}\bfa\Big|>x/3\Big)\\
		&=
		P\Big\{\sup_{\bv\in\mM}|\bv^\top \bM
		\bv|>\frac{x}{3}\Big\}\le P\Big\{(100/49)\max_{1\le k\le |\mA|}|\bu_k^\top \bM \bu_k|>x/3\Big\}\\
		&\le
		c_0|\mA|\exp\Big(-c_4 \min(T(49x/100)^2,T^{1/2}49x/100)\Big) \le \exp\Big\{
		-c_1 \min(Tx^2,T^{1/2}x)+c_3m\Big\},
	\end{align*}
	where $c_3, c_4$ are positive constants.
	(\ref{eq:epsX_conv}) can be proved similarly hence we skip the details.
	
\end{proof}


\bel\label{lem:concenX}
Under Assumption \ref{assum:mixing}, then we have
\begin{align}
	& P\Big(\Big|\frac{1}{T}\sum_t\bw^\top\mY_t\mY_t^\top\bw - \bw^\top\bGamma\bw\Big|>u\Big)\le 2(q^2 + q +1)\exp\Big(-C_1\min(Tu^2,T^{1/2}u)\Big)\label{eq:wYYw},
\end{align}
where $\bw \in \mR^{\prod_l N_l}$ is element-wisely non-negative and $\|\bw\|_1 = 1$ and
$\bGamma = \cov(\mY_t)$,
and $C_1, C_2, C_3$ are positive constants.
\eel

\begin{proof}	
	By (\ref{eq:Yt_expan}), we have
	\begin{align*}
		\frac{1}{T}\sum_t\bw^\top\mY_t\mY_t^\top\bw =
		\frac{1}{T}\sum_t\bw^\top\mY_t^c\mY_t^{c\top}\bw +
		\frac{2}{T}\sum_t\bw^\top\mY_t^c\mY_t^{e\top}\bw +
		\frac{1}{T}\sum_t\bw^\top\mY_t^e\mY_t^{e\top}\bw.
	\end{align*}
	Specifically, let $\bGamma^c = \cov(\mY_t^c)$
	and $\bGamma^e = \cov(\mY_t^e)$.
	It suffices to show
	\begin{align}
		&P\Big(\Big|\frac{1}{T}\sum_t\bw^\top\mY_t^c\mY_t^{c\top}\bw
		- \bw^\top\bGamma^c\bw\Big|>u/3\Big)\le
		2q^2\exp\{-c_1 \min(
		Tu^2,T^{1/2}u)\}
		\label{eq:Y_t_c2}\\
		&P\Big(\Big|\frac{2}{T}\sum_t\bw^\top\mY_t^c\mY_t^{e\top}\bw
		\Big|>u/3\Big)\le 2q \exp\{-c_2\min(
		Tu^2,T^{1/2}u
		)\}\label{eq:Y_t_ce}\\
		&P\Big(\Big|\frac{1}{T}\sum_t\bw^\top\mY_t^e\mY_t^{e\top}\bw
		- \bw^\top\bGamma^e\bw\Big|>u/3\Big)\le 2\exp\Big(-c_3\min\Big(Tu^2,
		T^{1/2}u\Big)\Big).\label{eq:Y_t_e2}
	\end{align}
	where $c_1, c_2, c_3$ are positive constants.
	In summary we obtain the final conclusion.

	\noindent
	{\bf 1. Proof of (\ref{eq:Y_t_c2})}

	Let $\bA = \bw\bw^\top \in \mR^{(\prod_l N_l) \times (\prod_l N_l)}$. By (\ref{eq:Yt_expan}),
	\begin{align}
		& \frac{1}{T}\sum_t\bw^\top\mY_t^c\mY_t^{c\top}\bw
		=
		\frac{1}{T}\sum_{t = 1}^T\mY_t^{c\top}
		\A\mY_t^c \nonumber=
		\frac{1}{T}\sum_{t = 1}^T \sum_{l_1 = 0}^t\sum_{s_2 = 0}^t \bc_{l_1}^\top (\bB_0^{t-l_1})^\top \A\bB_0^{t-s_2}\bc_{s_2}\nonumber\\
		&  = \sum_{l_1 = 0}^T\sum_{s_2=
			0}^T \bc_{l_1}^\top \Big\{\frac{1}{T}\sum_{t = \max\{1, l_1,s_2\}}^T
		(\bB_0^{t-l_1})^\top \A\bB_0^{t-s_2}\Big\} \bc_{s_2}\defeq
		\sum_{l_1 = 0}^T\sum_{s_2= 0}^T \bc_{l_1}^\top\bD_{l_1s_2}\bc_{s_2}.\label{eq:YcAYc}
	\end{align}
	For any $l\ge 0$, we define
	$\bc = (\bc_{0}^\top, \bc_{1}^\top,\cdots, \bc_{T}^\top)^\top$,
	and
	\begin{align}
		\bD = \left(
		\begin{array}{cccc}
			\bD_{00}&\bD_{01}&\cdots&\bD_{0T}\\
			\bD_{10}&\bD_{11}&\cdots & \bD_{1T}\\
			\vdots &\vdots &\ddots&\vdots\\
			\bD_{T0}&\bD_{T1}&\cdots & \bD_{TT}
		\end{array}
		\right).\label{def_D}
	\end{align}
	Then we have
		$T^{-1}\sum_{t = 1}^T\mY_t^{c\top} \A\mY_t^c = \c^\top \bD\c =
		\sum_{l=1}^q \bc^{(l)\top} \bD\bc^{(l)} +
		2 \sum_{l \neq m} \bc^{(l)\top} \bD\bc^{(m)}$,
	where $\bc^{(l)} = (\bc_0^{(l)\top}, \cdots, \bc_T^{(l)\top})$.
	Obviously,
	\begin{align*}
		&P\Big(\Big|\frac{1}{T}\sum_t\bw^\top\mY_t^c\mY_t^{c\top}\bw
		-\bw^\top\bGamma^c\bw|>u/3\Big)\\
		&=P\Big(\Big|\sum_{l=1}^q \bc^{(l)\top} \bD\bc^{(l)} +
		2 \sum_{l \neq m} \bc^{(l)\top} \bD\bc^{(m)}
		-E(\sum_{l=1}^q \bc^{(l)\top} \bD\bc^{(l)} +
		2 \sum_{l \neq m} \bc^{(l)\top} \bD\bc^{(m)})
		\Big|>u/3\Big)\\
		&\le \sum_l P\{|\bc^{(l)\top} \bD\bc^{(l)} -E(\bc^{(l)\top} \bD\bc^{(l)})|>2u/3 q(q+1)\}\\
		& +  \sum_{l \neq m} P\{|\bc^{(l)\top} \bD\bc^{(m)} -E(\bc^{(l)\top} \bD\bc^{(m)})|>2u/3 q(q+1)\}.
	\end{align*}
	Recall that $\bc_t^{(l)}$ is defined as $\vec(\one_{N_1} \circ \cdots \circ \bbeta_{X_l, t}^0 \circ \cdots \one_{N_q}) \in \mR^{\prod_l N_l}$ in the notations,
	and denote $\bc^{(l)} = (\bc_t^{(l)\top}: 0 \le t \le T)^\top \in \mR^{(\prod_l N_l)(T+1)}$.
	Also denote the $C_u u = 2u /3q(q+1)$.
	Due to the bounded assumption for parameters in Assumption \ref{assum:para_space}, we treat $\bc_t^{(l)}$ as $\wt\x_t^{(l)\eta}$ in  Lemma \ref{lem:xDz},
	which indicates that
	\begin{align*}
		& P\{|\bc^{(l)\top} \bD\bc^{(l)} -E(\bc^{(l)\top} \bD\bc^{(l)})|>C_u u\}
		\\
		& =P\Big\{\Big|\sum_{l_1, s_2=0}^T\{\bc_{l_1}^{(l)\top} \bD_{l_1 s_2}\bc_{s_2}^{(l)} -E(\bc_{l_1}^{(l)\top}
		\bD_{l_1 s_2}\bc_{s_2}^{(l)})\} \Big|>C_u u\Big\}\\
		&\le2\exp\Big(-\frac{1}{C}\min\Big(
		\frac{(C_u u)^2}{\|\mD^{(l)}\|_F^2},
		\frac{C_u u}{\|\mD^{(l)}\|}
		\Big)\Big) \le2\exp\{-c_1\min(
		u^2T,uT^{1/2}
		)\},
	\end{align*}
	where $\mD^{(l)}$ is given in Lemma \ref{lem:xDz} and
	the last step is due to Lemma \ref{lem:D_upper}.
	Similar treatment leads to
	\begin{align*}
		P\{|\bc^{(l)\top} \bD\bc^{(m)} -E(\bc^{(l)\top} \bD\bc^{(m)})|>C_u u\} \le 2\exp\{-c_1\min(
		u^2T,uT^{1/2})\}
	\end{align*}
	Combining the above results, we get
	\begin{align*}
		P\Big(\Big|\frac{1}{T}\sum_t\bw^\top\mY_t^c\mY_t^{c\top}\bw
		-\bw^\top\bGamma^c\bw|>u/3\Big)
		\le 2 q^2\exp\{-c_1\min(Tu^2,T^{1/2}u)\},
	\end{align*}
	where $c_1$ is a constant.

	\noindent
	{\bf 2. Proof of (\ref{eq:Y_t_ce})}
	
	We can obtain that
		$\frac{1}{T}\sum_t\bw^\top\mY_t^c\mY_t^{e\top}\bw = \sum_{l_1 = 0}^T\sum_{s_2= 0}^T \bc_{l_1}^\top\bD_{l_1s_2}\E_{s_2}=
		\bc^\top\bD\E$,
	where $\E=(\E_0\trans,\dots, \E_t\trans)\trans  \in \mR^{(\prod_l N_l) (T+1)}$.
	Obviously,
	\begin{align*}
		P\Big(\Big|\frac{1}{T}\sum_t\bw^\top\mY_t^c\mY_t^{e\top}\bw
		|>u/6\Big)=P\Big(\Big|\sum_l \bc^{(l)\top} \bD\E \Big|>u/6\Big)\le \sum_l P\{  |\bc^{(l)\top} \bD\E |> u/(6q)\}.
	\end{align*}
	Treating $\bc_t^{(l)}$, which is defined as $\vec(\one_{N_1} \circ \cdots \circ \bbeta_{X_l,t}^0 \circ \cdots \circ \one_{N_q})$,
	as $\wt\x_t^{(l)\eta}$ in  Lemma \ref{lem:xDz}, we get
	\begin{align*}
		& P\{|\bc^{(l)\top} \bD\E |>u/(6q)\} =P\{|\sum_{l_1,s_2=0}^T|\bc_{l_1}^{(l)\top} \bD_{l_1 s_2}\E_{s_2}|>u/(6q)\}\\
		&\le
		2\exp\Big(-\frac{1}{C}\min\Big(\frac{u^2}{\|\wt\mD^{(l)}\|_F^2}, \frac{u}{\|\wt\mD^{(l)}\|}\Big)\Big) \le
		2\exp\{-c_2 \min(Tu^2,T^{1/2}u)\},
	\end{align*}
	where $\wt\mD^{(l)}$ is given in Lemma \ref{lem:xDz} and
	the last step is due to Lemma \ref{lem:D_upper}.
	Therefore, we get
	\begin{align*}
		P\Big(\Big|\frac{1}{T}\sum_t\bw^\top\mY_t^c\mY_t^{e\top}\bw
		|>u/6\Big)
		\le 2q\exp\{-c_2 \min(Tu^2,T^{1/2}u)\},
	\end{align*}
	where $c_2$ is a positive constant.

	\noindent
	{\bf 3. Proof of (\ref{eq:Y_t_e2})}

	Similar to the expression of (\ref{eq:YcAYc}), we can obtain that
	\begin{align*}
		\frac{1}{T}\sum_t\bw^\top\mY_t^e\mY_t^{e\top}\bw
		=
		\sum_{l_1 = 0}^T\sum_{s_2= 0}^T \E_{l_1}^\top\bD_{l_1s_2}\E_{s_2}=
		\E^\top\bD\E,
	\end{align*}
	where $\E=(\E_0\trans, \dots, \E_T\trans)\trans \in \mR^{\prod_l N_l}$.
	Since $\E$ follows the $K$-convex concentration property,
	following Lemma \ref{lem:xAx_convex},
	we  have
	\begin{align*}
		&P\Big(\Big|\frac{1}{T}\sum_t\bw^\top\mY_t^e\mY_t^{e\top}\bw
		-\bw^\top\bGamma^e \bw|>u/3\Big)
		=P\Big(\Big|\E^\top\bD\E - E(\E^\top\bD\E)\Big|> u/3\Big)\\
		&\le
		2\exp\Big(-\frac{1}{C}\min\Big(\frac{u^2/9}{K^4\|\bD\|_F^2 },
		\frac{u/3}{K^2\|\bD\|}\Big)\Big)
		\le
		2\exp\Big(-c_3\min\Big(Tu^2,
		T^{1/2}u\Big)\Big),
	\end{align*}
	where $c_3$ is a finite constant and the last step is due to
	(\ref{D_F_bound}) of Lemma \ref{lem:D_upper}.
	
\end{proof}

	\begin{lemma}\label{lem:Sij_concent}
		Under Assumptions \ref{assum:para_space}--\ref{assum:station}, denote $d_{i_1 \cdots i_q} \defeq d_{i_1 \cdots i_q}(\bTheta_{i_1 \cdots  i_q}, \bTheta_{i_1 \cdots  i_q}^0)$.
		{Recall that $S_{i_1 \cdots i_q}(\bTheta_{i_1 \cdots i_q}) = Q_{i_1 \cdots i_q}(\bTheta_{i_1 \cdots i_q})  - Q_{i_1 \cdots i_q}(\bTheta_{i_1 \cdots i_q}^0) $.}
		we have
		\begin{align}
			& P\Big\{ \sup_{\bTheta_{i_1 \cdots  i_q}} T^{-1} \frac{ |S_{i_1 \cdots  i_q} (\bTheta_{i_1 \cdots  i_q}) - S^*_{i_1 \cdots  i_q} (\bTheta_{i_1 \cdots  i_q})| }{\sqrt{d_{i_1 \cdots  i_q}}} > x\Big\} \nonumber\\
			& \le C_1\exp\Big\{ -C_2 \min(T x^2, \sqrt{T} x) + C_3 m \Big\}, \label{eq:Sij_concent}
		\end{align}
		where $m = \sum_l (p_l + 1) + 1$.
		Furthermore, we have
			\begin{align}
				& \max_l \sup_{i_l} \sup_{d_{i_1 \cdots i_q} \le \omega^2} T^{-1} |S_{i_1 \cdots i_q}(\bTheta_{i_1 \cdots  i_q}) - S^*_{i_1 \cdots i_q}(\bTheta_{i_1 \cdots  i_q})| = O_p\big\{  \omega (\sum_l \log N_l)/\sqrt{T}  \big\}  \label{eq:Sij_order}\\
				& \sup_{d(\bTheta, \bTheta^0) \le \omega^2}  (\prod_l N_l T)^{-1} |S(\bTheta) - S^*(\bTheta)| = O_p\big\{  \omega (\sum_l \log N_l)/\sqrt{T}  \big\}.\label{eq:S_order}
			\end{align}
		
	\end{lemma}
	
	\begin{proof}
		
		\noindent
		{\bf 1. Proof of \eqref{eq:Sij_concent}.}
		
		We first write
		\begin{align}
			& S_{i_1 \cdots  i_q} (\bTheta_{i_1 \cdots  i_q}) - S^*_{i_1 \cdots  i_q} (\bTheta_{i_1 \cdots  i_q})\nonumber \\
			& =\Big[(\bTheta_{i_1 \cdots i_q} - \bTheta_{i_1 \cdots i_q}^0)^\top \cX_{i_1 \cdots i_q, t} \cX_{i_1 \cdots i_q, t}^\top(\bTheta_{i_1 \cdots i_q} - \bTheta_{i_1 \cdots i_q}^0) \nonumber\\
			& - E\{ (\bTheta_{i_1 \cdots i_q} - \bTheta_{i_1 \cdots i_q}^0)^\top \cX_{i_1 \cdots i_q, t} \cX_{i_1 \cdots i_q, t}^\top(\bTheta_{i_1 \cdots i_q} - \bTheta_{i_1 \cdots i_q}^0) \} \Big] \label{eq:S_concent_1}\\
			& + 2 \cX_{i_1 \cdots i_q, t}^\top(\bTheta_{i_1 \cdots i_q} - \bTheta_{i_1 \cdots i_q}^0) \ve_{i_1 \cdots i_q, t}.\label{eq:S_concent_2}
		\end{align}
		We use similar techniques in the proof of Lemma \ref{lem:Q_diff} to derive the concentration inequality.
		For example, we can apply \eqref{eq:X2_conv} on term \eqref{eq:S_concent_1} by noticing that $\|\bTheta_{i_1 \cdots i_q} - \bTheta_{i_1 \cdots i_q}^0\| / \sqrt{d_{i_1 \cdots i_q}} \\ \le cR$ for some constant $c$.
		Similarly, \eqref{eq:S_concent_2} can be shown by applying the proof for \eqref{eq:epsX_conv}.
		Hence, we can obtain the conclusion under the same Assumptions of Lemma \ref{lem:Q_diff}.

			\noindent
			{\bf 2. Proof of \eqref{eq:Sij_order} and \eqref{eq:S_order}.}

		For simplicity, denote $d_{i_1 \cdots i_q} \defeq d_{i_1 \cdots i_q}(\bTheta_{i_1 \cdots  i_q}, \bTheta_{i_1 \cdots  i_q}^0)$, and $S_{i_1 \cdots i_q}  \defeq S_{i_1 \cdots i_q}(\bTheta_{i_1 \cdots  i_q})$.
		Note that we have
		\begin{align*}
			& P\Big\{ \sup_{d_{i_1 \cdots i_q} \le \omega^2} |S_{i_1 \cdots i_q} - S_{i_1 \cdots i_q}^*| > x  \Big\} \\ & = P\Big\{ \sup_{d_{i_1 \cdots i_q} \le \omega^2} |S_{i_1 \cdots i_q} - S_{i_1 \cdots i_q}^*|/\omega > x /\omega \Big\} \\
			& \le P\Big\{ \sup_{d_{i_1 \cdots i_q} \le \omega^2} |S_{i_1 \cdots i_q} - S_{i_1 \cdots i_q}^*|/ \sqrt{d_{i_1 \cdots i_q}} > x /\omega \Big\} \\
			& \le P\Big\{ \sup_{\bTheta_{i_1 \cdots  i_q}}  |S_{i_1 \cdots i_q} - S_{i_1 \cdots i_q}^*|/ \sqrt{d_{i_1 \cdots i_q}} > x /\omega \Big\}  \\
			& \le C_1 \exp \big\{ -C_2 \min(T x^2/\omega^2, \sqrt{T} x/ \omega) + C_3 m \big\},
		\end{align*}
		where the last line uses \eqref{eq:Sij_concent}.
		By taking the union bound, we have
		\begin{align*}
			\max_l \sup_{i_l} \sup_{d_{i_1 \cdots i_q} \le \omega^2 } |S_{i_1 \cdots i_q} - S^*_{i_1 \cdots i_q}| = O_p\big\{ \omega (\sum_l \log N_l)/\sqrt{T} \big\}.
		\end{align*}
		Using similar procedure, we can obtain \eqref{eq:S_order} holds.
	
	\end{proof}

\bel\label{lem:S_Xtheta_tail}
Under Assumptions \ref{assum:para_space}--\ref{assum:station},
for vector $\bDelta \in \mR^{\sum_l (p_l+ 1)+1}$ in the parameter space,
we have
\begin{align}
	& P \Big\{\sup_{i_l}	\sup_{\| \bDelta\|^2 \le \omega^2}  (\prod_{m \neq l} N_m T)^{-1}
	\Big| \sum_{m \neq l} \sum_{i_m}  \sum_t
		\nonumber\\
	&  \hspace{12em} \Big\{\bDelta^\top \cX_{i_1 \cdots i_q,t} \cX_{i_1 \cdots i_q,t}^\top \bDelta  -  \big[\bDelta^\top E(\cX_{i_1 \cdots i_q,t} \cX_{i_1 \cdots i_q,t}^\top) \bDelta \big] \Big\}  \Big| \ge x \Big\} \nonumber \\
	& \le C_1 \exp\Big\{ -C_2 \min\Big( T x^2/\omega^2, \sqrt{T}x/\omega \Big) + C_3 m  + \sum_{l} \log N_l \Big\},\label{eq:S_Xtheta2_tail}\\
	& P \Big\{ \sup_{i_l} \sup_{\| \bDelta\|^2 \le \omega^2}  (\prod_{m \neq l} N_m T)^{-1} \Big|  \sum_{m \neq l} \sum_{i_m}  \sum_t \cX_{i_1 \cdots i_q,t}^\top \bDelta \ve_{i_1 \cdots i_q, t} \Big| \ge x \Big\} \nonumber\\
	& \le  C_1 \exp\Big\{ -C_2 \min\Big( T x^2/\omega^2, \sqrt{T}x/\omega \Big) + C_3 m  + \sum_{l} \log N_l \Big\},\label{eq:S_XthetaE_tail}
\end{align}
where $\cX_{i_1 \cdots i_q, t}$ is defined in equation \eqref{eq:X_i1iq}, 
$m = \sum_l (p_l + 1 )+1$, and $C_1, C_2, C_3$ are positive constants.
\eel
\begin{proof}
	We first derive \eqref{eq:S_Xtheta2_tail}.
	Note that
	\begin{align*}
		& P \Big\{\sup_{i_l}\sup_{\| \bDelta\|^2 \le \omega^2}\frac{1}{\prod_{m \neq l} N_m T}
		\Big| \sum_{m \neq l} \sum_{i_m}  \sum_t
		 \Big\{\bDelta^\top \cX_{i_1 \cdots i_q,t} \cX_{i_1 \cdots i_q,t}^\top \bDelta \\
		 & \hspace{1em}  -  \big[\bDelta^\top E(\cX_{i_1 \cdots i_q,t} \cX_{i_1 \cdots i_q,t}^\top) \bDelta \big] \Big\}  \Big| \ge x \Big\} \\
		& \le P \Big\{\sup_{i_l} \sup_{\| \bDelta\|^2 \le \omega^2}\frac{1}{\prod_{m \neq l} N_m}
		 \sum_{m \neq l} \sum_{i_m} \\
		 & \hspace{1em}
	 \frac{T^{-1} \Big|  \sum_t \bDelta^\top \cX_{i_1 \cdots i_q,t} \cX_{i_1 \cdots i_q,t}^\top \bDelta - \big[\bDelta^\top E(\cX_{i_1 \cdots i_q,t} \cX_{i_1 \cdots i_q,t}^\top) \bDelta \big] \Big|}{\|\bDelta\|} \|\bDelta\| \ge x \Big\} \\
		& \le P \Big\{\sup_{i_l} \sup_{\| \bDelta\|^2 \le \omega^2} \\
		& \hspace{1em}  \sqrt{ \sum_{m \neq l} \sum_{i_m}
			\frac{T^{-2} \Big|  \sum_t \bDelta^\top \cX_{i_1 \cdots i_q,t} \cX_{i_1 \cdots i_q,t}^\top \bDelta  -  \big[\bDelta^\top E(\cX_{i_1 \cdots i_q,t} \cX_{i_1 \cdots i_q,t}^\top) \bDelta \big] \Big|^2 }{\|\bDelta\|^2 } } \|\bDelta\| \ge x \Big\} \\
		& = P \Big\{\sup_{i_l} \sup_{\| \bDelta\|^2 \le \omega^2}  \sum_{m \neq l} \sum_{i_m} \\
		& \hspace{1em}
			\frac{T^{-2} \Big|  \sum_t \bDelta^\top \cX_{i_1 \cdots i_q,t} \cX_{i_1 \cdots i_q,t}^\top \bDelta  -  \big[\bDelta^\top E(\cX_{i_1 \cdots i_q,t} \cX_{i_1 \cdots i_q,t}^\top) \bDelta \big] \Big|^2 }{\|\bDelta\|^2 }  \|\bDelta\|^2 \ge x^2 \Big\} \\
		& \le  P \Big\{\sup_{i_l} \sup_{\| \bDelta\|^2 \le \omega^2}  \sum_{m \neq l} \sum_{i_m} \\
		&  \hspace{1em}
		\frac{T^{-2} \Big|  \sum_t \bDelta^\top \cX_{i_1 \cdots i_q,t} \cX_{i_1 \cdots i_q,t}^\top \bDelta  -  \big[\bDelta^\top E(\cX_{i_1 \cdots i_q,t} \cX_{i_1 \cdots i_q,t}^\top) \bDelta \big] \Big|^2 }{\|\bDelta\|^2 }  \omega^2 \ge x^2 \Big\} \\
		& \le \sum_l \sum_{i_l = 1}^{N_l} P \Big\{  \sup_{\| \bDelta\|^2 \le \omega^2} \\
		&  \hspace{1em} \frac{T^{-2} \Big|  \sum_t \bDelta^\top \cX_{i_1 \cdots i_q,t} \cX_{i_1 \cdots i_q,t}^\top \bDelta  -  \big[\bDelta^\top E(\cX_{i_1 \cdots i_q,t} \cX_{i_1 \cdots i_q,t}^\top) \bDelta \big] \Big|^2 }{\|\bDelta\|^2 }  \omega^2 \ge x^2  \Big\} \\
		& \le  \sum_l \sum_{i_l = 1}^{N_l} P \Big\{ \sup_{\bDelta} 	\frac{T^{-1} \Big|  \sum_t \bDelta^\top \cX_{i_1 \cdots i_q,t} \cX_{i_1 \cdots i_q,t}^\top \bDelta  -  \big[\bDelta^\top E(\cX_{i_1 \cdots i_q,t} \cX_{i_1 \cdots i_q,t}^\top) \bDelta \big] \Big| }{\|\bDelta\|} \omega \ge x  \Big\} \\
		& \le C_1 \exp\Big\{ -C_2 \min(T x^2/\omega^2, \sqrt{T} x/\omega) + C_3 m + \log(\prod_l N_l) \Big\},
	\end{align*}
	where $m = \sum_l (p_l+1)+1$,
	the third line holds due to $\sum_i (a_i^2 b_i^2) \le (\sum_i a_i^2) (\sum_i b_i^2)$.
	The last line is obtained by applying \eqref{eq:X2_conv} and using that $\|\bDelta\|_{\max} / \|\bDelta\| \le c R$.
	The proof of \eqref{eq:S_XthetaE_tail} can be finished in a similar scheme, while applying \eqref{eq:epsX_conv}.
\end{proof}

\bel\label{lem:YMY}
Under Assumption \ref{assum:mixing} and \ref{assum:station},
let $\bM = \wt\bW^\top \wt\bW\in \mR^{(\prod_l N_l)\times (\prod_l N_l)}$ be a symmetric matrix.
Here $\wt\bW\in \mR^{m \times (\prod_l N_l)}$ is elementwisely positive and assume
$n^{-1}\one_{\prod_l N_l}^\top \wt \bW^\top \wt \bW\one_{\prod_l N_l} = O(1)$, where $n$ is the order of $N_l, ~ l \in [q]$.
Then we have
\begin{align*}
	P\Big(\Big|\frac{1}{n T}\sum_t\mY_t^\top\bM\mY_t - n^{-1}\tr(\bGamma\bM)\Big|>u\Big)\le 2(q^2+q+1)\exp\Big(-c\min(Tu^2,T^{1/2}u)\Big),
\end{align*}
and we have $ n^{-1}\tr(\bM\bGamma)\le c$ for a positive constant $c$.
\eel

\begin{proof}
	By (\ref{eq:Yt_expan}), we have
		$\frac{1}{T}\sum_t\mY_t^\top \bM\mY_t =
		\frac{1}{T}\sum_t\mY_t^{c\top}\bM\mY_t^c +
		\frac{2}{T}\sum_t\mY_t^{e\top}\bM\mY_t^c +
		\frac{1}{T}\sum_t\mY_t^{e\top}\bM\mY_t^e$.
	The proof follows the proof of Lemma \ref{lem:concenX}.
	The difference is that we need to replace the $\bA$ matrix with $\bM/n$.
	Since the procedure is the same we omit the details here.
	
	Furthermore, we have
		$n^{-1}\tr(\bM\bGamma)\le n^{-1}\tr(\bM\one_{\prod_l N_l}\one_{\prod_l N_l}^\top c_\Gamma)\le
		c_\Gamma\one_{\prod_l N_l}^\top\bM\one_{\prod_l N_l}/n = O(1)$,
	where the first inequality is obtained by Lemma \ref{lem:10}.
\end{proof}

\bel\label{lem:Y4_bound}
Under Assumption \ref{assum:mixing}, we have
$\max_l \max_{i_l} E(Y_{i_1 \cdots i_q, t}^4) \le c$,
where $c$ is a positive constant.
\eel

\begin{proof}
	Let $\bw = \e_{i_q}^{(N_q)} \otimes \cdots \otimes  \e_{i_1}^{(N_1)} \in \mR^{\prod_l N_l}$,
	where $\e_{i_l}^{(N_l)} \in \mR^{N_l}$ is a vector whose $i_l$th element being equal to 1 while others being equal to 0.
	Then we have
	$Y_{i_1 \cdots i_q, t} = \bw^\top \mY_t$.
	By (\ref{eq:Yt_expan}), we have
	\begin{align*}
		Y_{i_1 \cdots i_q, t}^2& =\bw^\top\mY_t\mY_t^\top\bw =
		\bw^\top\mY_t^c\mY_t^{c\top}\bw +
		2\bw^\top\mY_t^c\mY_t^{e\top}\bw +
		\bw^\top\mY_t^e\mY_t^{e\top}\bw
		\\
		& \le 2\bw^\top\mY_t^c\mY_t^{c\top}\bw +
		2\bw^\top\mY_t^e\mY_t^{e\top}\bw.
	\end{align*}
	Thus,
		$Y_{i_1 \cdots i_q, t}^4=(\bw^\top\mY_t\mY_t^\top\bw )^2
		\le 4(\bw^\top\mY_t^c\mY_t^{c\top}\bw)^2 +
		4(\bw^\top\mY_t^e\mY_t^{e\top}\bw)^2$.
	It suffices to investigate the above two terms respectively.

	\noindent
	{\bf 1. Proof of $E\{(\bw^\top\mY_t^c\mY_t^{c\top}\bw)^{2}\}<\infty$}
	
	First, let $\bA = \bw\bw^\top \in \mR^{\prod_l N_l \times \prod_l N_l}$, then by (\ref{eq:Yt_expan}) and the
	definition in Lemma \ref{lem:D_upper},
	\begin{align}
		&\bw^\top\mY_t^c\mY_t^{c\top}\bw
		= \mY_t^{c\top}
		\A\mY_t^c =
		\sum_{s_1 = 0}^t \sum_{s_2 = 0}^t \bc_{s_1}^\top (\bB_0^{t-s_1})^\top \A\bB_0^{t-s_2}\bc_{s_2}\nonumber\\
		& =\sum_{s_1 = 0}^t\sum_{s_2= 0}^t
		\bc_{s_1}^\top\L_{ t,s_1s_2}\bc_{s_2} =
		\sum_{s_1 = 0}^t\sum_{s_2= 0}^t
		(\sum_l \bc_{s_1}^{(l)\top})\L_{ t,s_1s_2}(\sum_l \bc_{s_2}^{(l)}) \nonumber\\
		&\le 2 \sum_l  \sum_{s_1 = 0}^t\sum_{s_2= 0}^t
		\bc_{s_1}^{(l)\top}\L_{ t,s_1s_2}\bc_{s_2}^{(l)}
		=2  \sum_l \sum_{s_1 = 0}^t\sum_{s_2=
			0}^t\bbeta_{X_l,s_1}^{(l)0\top}
		\mL_{t,s_1s_2}^{(l)} \bbeta_{X_l,s_2}^{(l)0}.\label{eq:wYcYcw}
	\end{align}
	where $\mL_{t,s_1s_2}^{(l)} = ( \one_{N_1}^\top \otimes \cdots \otimes \one_{N_{l-1}}^\top \otimes \bI_{N_l} \otimes \one_{N_{l+1}}^\top \otimes \cdots \otimes \one_{N_q}^\top ) \L_{t,s_1s_2}
	(\one_{N_1} \otimes \cdots \otimes \one_{N_{l-1}} \otimes \bI_{N_l} \otimes \one_{N_{l+1}} \otimes \cdots \otimes \one_{N_q} )\in \mR^{N_l \times N_l}$ as defined in
	Lemma \ref{lem:D_upper}.

	Further, let $\mL_{t}^{(l)} = (\wt \mL_{t,s_1s_2}^{(l)}: 0\le s_1,s_2\le T)$, where
	$\wt \mL_{t,s_1s_2}^{(l)} = \zero$ for $s_1>t$ or $s_2>t$, and
	$\wt \mL_{t,s_1s_2}^{(l)} = \mL_{t,s_1s_2}^{(l)}$ otherwise,
	as defined in Lemma \ref{lem:D_upper}.
	Let $\bbeta_{X_l}^{(l)0} = (\bbeta_{X_l,s}^{(l)0\top}: 0\le s\le T)^\top$.
	Then we have
		$\sum_{s_1 = 0}^t\sum_{s_2= 0}^t\bbeta_{X_l,s_1}^{(l)0\top} \mL_{ t,s_1s_2}^{(l)} \bbeta_{X_l,s_2}^{(l)0} = \bbeta_{X_l}^{(l)0\top} \mL_{t}^{(l)} \bbeta_{X_l}^{(l)0}$,
	which implies that
		$(\bw^\top\mY_t^c\mY_t^{c\top}\bw)^2\le 4 \sum_l (\bbeta_{X_l}^{(l)0\top} \mL_{t}^{(l)} \bbeta_{X_l}^{(l)0})^2$.
	Let $\mL_{t}^{(l)} = \sum_{i_l} \lambda_{i_l} \u_{i_l} \u_{i_l}^\top$ be the
	eigen-decomposition of $\mL_{t}^{(l)}$, where $\lambda_{i_l}$ and $\u_{i_l}$
	are the $i_l$th eigenvalue and eigenvector respectively.
	One can verify that $\mL_{t}^{(l)}$ is semi-definite, then we have all $\lambda_{i_l} \ge 0$.
	Then we have
	\begin{align*}
		(\bbeta_{X_l}^{(l)0\top} \mL_{t}^{(l)} \bbeta_{X_l}^{(l)0})^2 = \sum_{i_{l1}, i_{l2}}\lambda_{i_{l1}}\lambda_{i_{l2}}  \left(\bbeta_{X_l}^{(l)0\top}\u_{i_{l1}}\u_{i_{l1}}^\top \bbeta_{X_l}^{(l)0}\right)
		\left(\bbeta_{X_l}^{(l)0\top}\u_{i_{l2}}\u_{i_{l2}}^\top \bbeta_{X_l}^{(l)0}\right).
	\end{align*}
	Note that
	\begin{align*}
		&E\left\{\left(\bbeta_{X_l}^{(l)0\top}\u_{i_{l1}}\u_{i_{l1}}^\top \bbeta_{X_l}^{(l)0}\right)
		\left(\bbeta_{X_l}^{(l)0\top}\u_{i_{l2}}\u_{i_{l2}}^\top \bbeta_{X_l}^{(l)0}\right)\right\}
		=
		E\left(\langle\u_{i_{l1}},\bbeta_{X_l}^{(l)0} \rangle^2\langle\u_{i_{l2}},\bbeta_{X_l}^{(l)0} \rangle^2\right)\\
		&\le(1/2)  E\left(\langle\u_{i_{l1}},\bbeta_{X_l}^{(l)0} \rangle^4\right)+(1/2)E\left(\langle \u_{i_{l2}},\bbeta_{X_l}^{(l)0}\rangle^4\right)\\
		&= 2\int_0^\infty t^3 P\left\{|\langle \u_{i_{l1}},\bbeta_{X_l}^{(l)0}\rangle|>t\right\}dt +
		2\int_0^\infty t^3 P\left\{|\langle\u_{i_{l2}},\bbeta_{X_l}^{(l)0} \rangle|>t\right\}dt\\
		&\le 8 \int_0^\infty t^3 \exp(-t^2/K^2)dt\defeq c_K,
	\end{align*}
	where $c_K$ is a constant related to $K$, and it is the constant defined in Definition \ref{def:convex_concen}.
	Here the last inequality is obtained by noting that $\langle \u_{i_l},
	\bbeta_{X_l}^{(l)0}\rangle$ is a $1$-Lipschitz convex function of vector $\x$ with
	mean zero and $\x$ is $K$-convex defined in Definition \ref{def:convex_concen}.
	Therefore we have
	\begin{align*}
		E\left\{(\bbeta_{X_l}^{(l)0\top} \mL_{t}^{(l)} \bbeta_{X_l}^{(l)0})^2\right\} \le c_K\sum_{i_{l1}, i_{l2}}\lambda_{i_{l1}}\lambda_{i_{l2}}
		= c_K\tr(\mL_{t}^{(l)})^2 <\infty,
	\end{align*}
	where the last inequality is obtained by (\ref{eq:L12_tr_bound}) of
	Lemma \ref{lem:D_upper}.

	\noindent
	{\bf 2. Proof of $E\{(\bw^\top\mY_t^e\mY_t^{e\top}\bw)^{2}\}<\infty$}
	
	By the definition in Lemma \ref{lem:D_upper}, we have
	\begin{align*}
		\bw^\top\mY_t^e\mY_t^{e\top}\bw
		= \mY_t^{e\top}
		\A\mY_t^e =
		\sum_{s_1 = 0}^t\sum_{s_2 = 0}^t \E_{s_1}^\top (\bB_0^{t-s_1})^\top \A\bB_0^{t-s_2}\E_{s_2} =\sum_{s_1 = 0}^t\sum_{s_2= 0}^t
		\E_{s_1}^\top\L_{t,s_1s_2}\E_{s_2}
	\end{align*}
	Define $\wt \L_{t} = (\wt \L_{t,s_1s_2}: 0\le s_1,s_2\le T)$, where
	$\wt \L_{t,s_1s_2} = \zero$ for $s_1>t$ or $s_2>t$, and
	$\wt \L_{t,s_1s_2} = \L_{t,s_1s_2}$ otherwise.
	Then we have $ \bw^\top\mY_t^e\mY_t^{e\top}\bw = \E^\top \wt\L_{t} \E$,
	where $\E = (\E_0^\top, \cdots, \E_t^\top)^\top$.
	Similar to the proof given in Step 1, we can verify
		$E\{(\bw^\top\mY_t^e\mY_t^{e\top}\bw)^2\}=
		E\{(\E^\top \wt\L_{t} \E)^2\}\le c \tr(\wt \L_{t})^2
		{\le c \tr(\L_t)^2}
		<\infty$,
	where the second last inequality used the definition of $\L_t$
	in  Lemma
	\ref{lem:D_upper} and the
	last inequality is obtained by (\ref{eq:L12_tr_bound}) of Lemma
	\ref{lem:D_upper}.
\end{proof}

\bel\label{lem:Q_star_diff}
Under Assumption \ref{assum:tau_min}, it holds that
\begin{align}
	&\tau_{\min} \|\bTheta_{i_1 \cdots i_q} - \bTheta_{i_1 \cdots i_q}^0\|^2\le \frac{1}{T}\Big\{
	Q_{i_1 \cdots i_q}^*(\bTheta_{i_1 \cdots i_q})- Q_{i_1 \cdots i_q}^*(\bTheta_{i_1 \cdots i_q}^0)
	\Big\}\le  \tau_{\max} \|\bTheta_{i_1 \cdots i_q} - \bTheta_{i_1 \cdots i_q}^0\|^2\label{eq:Qij_star_diff}\\
	&\tau_{\min} d(\bTheta, \bTheta^0)\le \frac{1}{(\prod_l N_l) T}\Big\{
	Q^*(\bTheta)- Q^*(\bTheta^0)
	\Big\}\le  \tau_{\max} d(\bTheta, \bTheta^0).\label{eq:Q_star_diff}\\
	& \tau_{\min} d_{i_l}(\bTheta_{\cdot i_l \cdot}, \bTheta_{\cdot i_l \cdot}^0)\le\frac{1}{(\prod_{m \neq l}N_l)T}
	\Big\{Q_{i_l}^*(\bxi_{g_{i_l}^{(l)}}^{(l)} ;  \bxi_{g^{-(l)}}^{-(l)} ,\mG_{-l})
	-Q_{i_l}^*(\bxi_{g_{i_l}^{(l)0}}^{(l)0} ;  \bxi_{g^{-(l)}}^{-(l)0},\mG_{-l}^0))
	\Big\} \nonumber\\
	&\le  \tau_{\max} d_{i_l}(\bTheta_{\cdot i_l \cdot}, \bTheta_{\cdot i_l \cdot}^0),\label{eq:Qj_star_diff}
\end{align}
where $\tau_{\min}$ is defined in Assumption \ref{assum:tau_min}, $\tau_{\max}$ is defined in notation Section \ref{sec:notation}, and $d_{i_l}(\cdot, \cdot)$ is defined in \eqref{eq:def_dj}.
\eel

\begin{proof}
	We have
	\begin{align*}
		Q(\bTheta) &  = \sum_{i_1, \cdots, i_q}Q_{i_1 \cdots i_q}(\bTheta_{i_1 \cdots i_q}) = \sum_l \sum_{i_l  =1}^{N_l} \sum_{t = 1}^T
		\big(\ve_{i_1 \cdots i_q, t} +   \cX_{i_1 \cdots i_q,t}^\top\bTheta_{i_1 \cdots i_q}^0- \cX_{i_1 \cdots i_q,t}^\top\bTheta_{i_1 \cdots i_q}\big)^2\\
		& = \sum_l \sum_{i_l =1}^{N_l} \sum_{t = 1}^T\Big\{\ve_{i_1 \cdots i_q,t}^2 +
		2\ve_{i_1 \cdots i_q,t}\cX_{i_1 \cdots i_q,t}^\top(\bTheta_{i_1 \cdots i_q}^0- \bTheta_{i_1 \cdots i_q} )\\
		& +
		(\bTheta_{i_1 \cdots i_q}-\bTheta_{i_1 \cdots i_q}^0)^\top\cX_{i_1 \cdots i_q,t}\cX_{i_1 \cdots i_q,t}^\top (\bTheta_{i_1 \cdots i_q}-\bTheta_{i_1 \cdots i_q}^0)
		\Big\}.
	\end{align*}
	We have
		$\frac{1}{T}\Big\{
		Q_{i_1 \cdots i_q}^*(\bTheta_{i_1 \cdots i_q})- Q_{i_1 \cdots i_q}^*(\bTheta_{i_1 \cdots i_q}^0)
		\Big\} = (\bTheta_{i_1 \cdots i_q}-\bTheta_{i_1 \cdots i_q}^0)^\top\bSigma_{i_1 \cdots i_q} (\bTheta_{i_1 \cdots i_q}-\bTheta_{i_1 \cdots i_q}^0)$.
	By using Assumption \ref{assum:tau_min} and Lemma \ref{lem:tau_max},
	(\ref{eq:Qij_star_diff})
	can be obtained.

	Next, it holds
	\begin{align*}
		& \frac{1}{(\prod_l N_l)T}\Big\{Q^*(\bTheta) - Q^*(\bTheta^0)\Big\}\\
		& =
		\frac{1}{(\prod_l N_l)T} \sum_l \sum_{i  = i_l}^{N_l} \sum_{t = 1}^T (\bTheta_{i_1 \cdots i_q}-\bTheta_{i_1 \cdots i_q}^0)^\top\bSigma_{i_1 \cdots i_q} (\bTheta_{i_1 \cdots i_q}-\bTheta_{i_1 \cdots i_q}^0)\\
		& = \frac{1}{\prod_l N_l} \sum_l \sum_{i_l = 1}^{N_l} (\bTheta_{i_1 \cdots i_q}-\bTheta_{i_1 \cdots i_q}^0)^\top\bSigma_{i_1 \cdots i_q} (\bTheta_{i_1 \cdots i_q}-\bTheta_{i_1 \cdots i_q}^0).
	\end{align*}
	We have
	\begin{align*}
		&\frac{1}{\prod_l N_l} \sum_l \sum_{i_l = 1}^{N_l}  (\bTheta_{i_1 \cdots i_q}-\bTheta_{i_1 \cdots i_q}^0)^\top\bSigma_{i_1 \cdots i_q} (\bTheta_{i_1 \cdots i_q}-\bTheta_{i_1 \cdots i_q}^0)\\
		& \ge \frac{\tau_{\min}}{\prod_l N_l}\sum_l \sum_{i_l = 1}^{N_l} \Big\|\bTheta_{i_1 \cdots i_q}-\bTheta_{i_1 \cdots i_q}^0\Big\|^2
		= \tau_{\min} d(\bTheta, \bTheta^0),\\
		& \frac{1}{\prod_l N_l} \sum_l \sum_{i_l = 1}^{N_l}  (\bTheta_{i_1 \cdots i_q}-\bTheta_{i_1 \cdots i_q}^0)^\top\bSigma_{i_1 \cdots i_q} (\bTheta_{i_1 \cdots i_q}-\bTheta_{i_1 \cdots i_q}^0)\\
		& \le
		\frac{\tau_{\max}}{\prod_l N_l}\sum_l \sum_{i_l = 1}^{N_l} \Big\|\bTheta_{i_1 \cdots i_q}-\bTheta_{i_1 \cdots i_q}^0\Big\|^2
		= \tau_{\max} d(\bTheta, \bTheta^0)
	\end{align*}
	by using Assumption \ref{assum:tau_min} and Lemma \ref{lem:tau_max}.
	This proves (\ref{eq:Q_star_diff}).
	
	Similar arguments as that of the proof for (\ref{eq:Q_star_diff})
	at a fixed $i_l$ leads to (\ref{eq:Qj_star_diff}).
\end{proof}

\bel\label{lem:sig_h_bound}
Suppose $G_l \ge G_{l,0}$ for all $l \in [q]$, where $G_{l,0}$ is the true number of groups. In addition, assume Assumptions \ref{assum:para_space}--\ref{assum:station} hold.
Define
\begin{align*}
	\sigma_l(g^{(l)})& = \argmin_{\wt g^{(l)}  \in [G_l]}
	\|\wh \btheta_{\wt g^{(l)} }^{(l)} - \btheta_{g^{(l)} }^{(l)0}\|^2
	+\frac{1}{\prod_{m \neq l} N_m}  \sum_{m \neq l} \sum_{i_m} \Big|\wh \alpha_{\wt g^{(l)} \wh g_{\i_{-l}}^{-(l)}} - \alpha_{ g^{(l)} g_{\i_{-l}}^{-(l)0}}^0 \Big|^2.
\end{align*}
Then we have
\begin{align*}
	&\max_{g^{(l)} \in [G_{l,0}]}\Big\{
	\|\wh \btheta_{\sigma_l (g^{(l)} )}^{(l)} - \btheta_{g^{(l)}}^{(l)0}\|^2 +\frac{1}{\prod_{m \neq l} N_l} \sum_{m \neq l} \sum_{i_m}  \Big|\wh \alpha_{ \sigma_l(g^{(l)})  \wh g_{\i_{-l}}^{-(l)} } - \alpha_{g^{(l)} g_{\i_{-l}}^{-(l)0}}^0 \Big|^2
	\Big\} \\
	& \le \max_{g^{(l)} \in [G_{l,0}]}
	\frac{N_l}{N_{l g^{(l)}}} d(\wh \bTheta, \bTheta^0)  =O_p\Big(T^{-1}( \sum_l \log N_l)^2 \Big).\nonumber
\end{align*}

\eel

\begin{proof}
	We have
	\begin{align*}
		&\|\wh \btheta_{\sigma_l (g^{(l)} )}^{(l)} - \btheta_{g^{(l)}}^{(l)0}\|^2
		+\frac{1}{\prod_{m \neq l} N_l} \sum_{m \neq l} \sum_{i_m}  \Big|\wh \alpha_{\sigma_l(g^{(l)})  \wh g_{\i_{-l}}^{-(l)} } - \alpha_{g^{(l)}   g_{\i_{-l}}^{-(l)0} }^0 \Big|^2\\
		&= \frac{1}{N_{l g^{(l)}}}\sum_{i_l} I(g_{i_l}^{(l)0} = g^{(l)})
		\Big\{\|\wh \btheta_{\sigma_l (g_{i_l}^{(l)0})}^{(l)} - \btheta_{g_{i_l}^{(l)0}}^{(l)0}\|^2 \\
		& +\frac{1}{\prod_{m \neq l} N_l} \sum_{m \neq l} \sum_{i_m} \Big|\wh \alpha_{ \sigma_l(g_{i_l}^{(l)0})   \wh g_{\i_{-l}}^{-(l)} } - \alpha_{ g_{i_l}^{(l)0}  g_{\i_{-l}}^{-(l)0} }^0 \Big|^2\Big\}\\
		& \le \frac{1}{N_{l g^{(l)}}}\sum_{i_l} I(g_{i_l}^{(l)0} = g^{(l)})
		\Big\{\|\wh \btheta_{\wh g_{i_l}^{(l)}}^{(l)} - \btheta_{g_{i_l}^{(l)0}}^{(l)0}\|^2 + \frac{1}{\prod_{m \neq l} N_l} \sum_{m \neq l} \sum_{i_m} \Big|\wh \alpha_{ \wh g_{i_l}^{(l)}  \wh g_{\i_{-l}}^{-(l)} } - \alpha_{ g_{i_l}^{(l)0}  g_{\i_{-l}}^{-(l)0} }^0 \Big|^2 \Big\} \\
		& \le \frac{N_l}{N_{l g^{(l)}}}
		d(\wh \bTheta, \bTheta^0).
	\end{align*}
	Then by taking $\max_{g^{(l)} \in [G_{l,0}]}$ of both sides, we have
	\begin{align*}
		&\max_{g^{(l)} \in[G_{l,0}]}
		\left(\|\wh \btheta_{\sigma(g^{(l)})}^{(l)} - \btheta_{g^{(l)}}^{(l)0}\|^2+\frac{1}{\prod_{m \neq l}}\sum_{m \neq l}\sum_{i_m} |\wh \alpha_{\sigma(g^{(l)}) \wh g_{\i_{-l}}^{-(l)}} - \alpha_{g^{(l)} g_{\i_{-l}}^{-(l)0}}^0|^2
		\right)\\
		& \le \frac{N_l}{\min_{g^{(l)}} N_{lg}} d(\wh \bTheta, \bTheta^0)  =  O_p \Big(
		T^{-1}(\sum_l \log N_l)^2 \Big)
	\end{align*}
	by Theorem \ref{thm:pseudo_dist}.
	
\end{proof}

\bel\label{lem:h_consistency_prepare}
Under Assumptions \ref{assum:para_space}--\ref{assum:group_ratio}, and suppose $G_l \ge G_{l,0}$ for all $l \in [q]$.
For any $\bxi$, let $\wh g_{i_l}^{(l)}(\bxi)$ denote the membership obtained with \eqref{eq:g_i}.
Define $\bTheta = (\bTheta_{i_1 \cdots i_q}: i_l \in [N_l])$ with $\bTheta_{i_1 \cdots i_q} = ( \btheta^{(1)\top}_{\wh g_{i_1}^{(1)}(\bxi)}, \cdots, \btheta^{(q)\top}_{\wh g_{i_q}^{(q)}(\bxi)}, \alpha_{\wh g_{i_1}^{(1)}(\bxi) \cdots \wh g_{i_q}^{(q)}(\bxi)})^\top$.
Assume we have
\beq
 d(\bTheta, \bTheta^0) ={O_p(T^{-1}(\sum_l \log  N_l)^2) }= o_p(c_\gap),\label{eq:Theta_cond}
\eeq
as $\min_l N_l \to \infty$.
Let $\bxi^{(l)} = (\btheta^{(l)\top}, (\vec(\balpha_{\cdot  g^{(l)} \cdot})^\top: g^{(l)} \in [G_l]))^\top = \big( (\btheta_{ g^{(l)}}^{(l)}: g^{(l)} \in [G_l])^\top, (\vec(\balpha_{\cdot  g^{(l)} \cdot})^\top: g^{(l)} \in [G_l]) \big)^\top$. Define
\begin{align*}
	& \mM^{(l)} { (\bxi_{g^{(l)}}^{(l)}, \bxi_{g^{(l)0}}^{(l)0}; \mG_{-l}, \mG_{-l}^0)} = \|\btheta_{g^{(l)}}^{(l)} - \btheta_{g^{(l)0}}^{(l)0} \|^2 \\
	&+
	\frac{1}{\prod_{m \neq l} N_l} \sum_{m \neq l} \sum_{i_m = 1}^{N_m}  \big|\alpha_{g_{i_l}^{(1)} \cdots g_{i_{l-1}}^{(l-1)} g^{(l)} g_{i+1}^{(l+1)} \cdots g_{i_q}^{(q)}} -
	\alpha_{g_{i_1}^{(1)0} \cdots g_{i_{l-1}}^{(l-1)0} g^{(l)0} g_{i+1}^{(l+1)0} \cdots g_{i_q}^{(q)0}}^0 \big|^2\\
	& \defeq \|\btheta_{g^{(l)}}^{(l)} - \btheta_{g^{(l)0}}^{(l)0} \|^2
	+ \frac{1}{\prod_{m \neq l} N_l} \sum_{m \neq l} \sum_{i_m = 1}^{N_m}
	|\alpha_{g^{(l)}g_{\i_{-l}}^{-(l)}} -
	\alpha_{g^{(l)0}g_{\i_{-l}}^{-(l)0}}|^2
\end{align*}
with $\i_{-l} = (i_m: m\ne l)^\top$ and $g_{\i_{-l}}^{-(l)} = (g_{i_m}^{(m)}: m \neq l)^\top \in \mR^{q-1}$.
Further denote
\begin{align}
	d_S^{(l)}(\bxi^{(l)}, \bxi^{(l)0}; \mG_{-l}, \mG_{-l}^0) = 	\max\Big\{& \max_{g^{(l)0}\in [G_{l,0}]} \min_{g^{(l)} \in [G_l]}
	\Big( \mM^{(l)} { (\bxi_{g^{(l)}}^{(l)}, \bxi_{g^{(l)0}}^{(l)0}; \mG_{-l}, \mG_{-l}^0)} \Big),\nonumber\\
	&  \max_{g^{(l)} \in [G_l]}\min_{g^{(l)0}\in [G_{l,0}]}
	\Big(\mM^{(l)} { (\bxi_{g^{(l)}}^{(l)}, \bxi_{g^{(l)0}}^{(l)0}; \mG_{-l}, \mG_{-l}^0)}\Big)\Big\},\label{eq:distance_set}
\end{align}
where $\mG_{-l} = \{\mG_m: m \neq l\} = \{( g_{i_m}^{(m)}: 1 \le i_m \le N_m)^\top : m \neq l\}$.
In addition, define $\mN_\eta^{(l)}  = \{{\bxi:}
d_S^{(l)}(\bxi^{(l)}, \bxi^{(l)0}; \wh\mG_{-l}(\bxi), \mG_{-l}^0)<\eta\}$ given $\bxi$.
Correspondingly, denote
\begin{align}
	\mA^{(l)}_\eta(\bxi, g^{(l)0}, {\mG_{-l}^0}) &= \Big\{g^{(l)} \in [G_l]:
	\|\btheta_{g^{(l)}}^{(l)} - \btheta_{g^{(l)0}}^{(l)0} \|^2 \nonumber\\
	& +
	\frac{1}{\prod_{m \neq l} N_m} \sum_{m \neq l}\sum_{i_m} |\alpha_{g^{(l)}\wh g_{\i_{-l}}^{-(l)}(\bxi)} -
	\alpha_{g^{(l)0}g_{\i_{-l}}^{-(l)0}}^0| \le \eta \Big\}.\label{eq:A_eta}
\end{align}

Then the following conclusions hold:\\
(i) For all $ \bxi \in \mN^{(l)}_\eta$ with $ \eta<(c_\pi)^{q-1} c_\gap/4$, we have
$ \{\mA^{(l)}_\eta(\bxi, g^{(l)0},{ \mG_{-l}^0}), ~g^{(l)0} \in [G_{l,0}]\}$ is a partition of $[G_l]$;\\
(ii)
Define the event $\bOmega = \{\wh g_{i_l}^{(l)}(\bxi) \in \mA^{(l)}_\eta(\bxi, g_{i_l}^{(l)0}, \mG_{-l}^0),~\forall
i_l \in [N_l]\}$, where $\bxi$ satisfying that $\bxi \in \mN^{(l)}_\eta$, and
$\eta \le \tau_{\min}c_\gap
(c_\pi)^{q-1}/ \{8(\tau_{\min}+\tau_{\max})\}$.
Here, $\tau_{\max}$ is given in the notation Section \ref{sec:notation}.
Then we have
	$P(\bOmega^c) \le C \exp\Big(-c_1 T^{1/2}c_\gap +c_2m + \sum_l \log N_l\Big)$,
where $C, c_1, c_2$ are positive constants.
\eel

\begin{proof}
	\noindent
	{\bf 1. Proof of (i)}
	
	By the definition of $\mN^{(l)}_\eta$ and $\mA^{(l)}_\eta(\bxi, g^{(l)}, \mG_{-l}^0)$, we have
	$\cup_{g^{(l)} = 1}^{G_{l,0}}\mA^{(l)}_\eta(\bxi, g^{(l)}, \mG_{-l}^0) = [G_l]$.
	Then it remains to show that $\mA^{(l)}_\eta(\bxi, g^{(l)}, \mG_{-l}^0)$ is a partition of $[G_l]$.
	That is $\mA^{(l)}_\eta(\bxi, g_1^{(l)}, \mG_{-l}^0)\cap \mA^{(l)}_\eta(\bxi, g_2^{(l)}, \mG_{-l}^0) = \emptyset$
	for any $g_1^{(l)} \ne g_2^{(l)}$.
	
	We prove by contradiction. Suppose there exists $g_{12}^{(l)} \in [G_l]$ so that
	$g_{12}^{(l)}\in \mA^{(l)}_\eta(\bxi, g_1^{(l)}, \mG_{-l}^0)\cap \mA^{(l)}_\eta(\bxi, g_2^{(l)}, \mG_{-l}^0)$
	for $g_1^{(l)} , g_2^{(l)}\in [G_{l,0}]$ and $g_1^{(l)} \ne g_2^{(l)}$.
	{Denote $g^{-(l)} = (g^{(m)}: m \neq l)$, and correspondingly denote the event $\{g_{\i_{-l}}^{-(l)0} = g^{-(l)}\}$ as $\{ g_{i_m}^{(m)0} = g^{(m)}: m \neq l \}$.}
	In this case we have
	\begin{align*}
		c_\gap &\le \{\|\btheta_{g_1^{(l)}}^{(l)0}  - \btheta_{g_2^{(l)}}^{(l)0}\|^2 +\max_{g^{-(l)} \in [G_{-l}^0]} |\alpha_{g_1^{(l)} g^{-(l)}}^0 - \alpha_{g_2^{(l)} g^{-(l)}}^0 |^2\}\\
		&\le {2}\|\btheta_{g_1^{(l)}}^{(l)0}   -\btheta_{g_{12}^{(l)}}^{(l)}  \|^2+2
		\|\btheta_{g_2^{(l)}}^{(l)0} - \btheta_{g_{12}^{(l)}}^{(l)} \|^2\\
		& +\max_{g^{-(l)} \in [G_{-l}^0]} \left(\frac{1}{\prod_{m \neq l} (\pi_{g^{(m)}, N_m}^{(m)} N_m)} \sum_{m \neq l}\sum_{i_m} \big\{ |\alpha_{g_1^{(l)} g^{-(l)}}^0 - \alpha_{g_2^{(l)} g^{-(l)}}^0 |^2  I(g_{\i_{-l}}^{-(l)0} = g^{-(l)}) \big\}\right)\\
		&=  {2}\|\btheta_{g_1^{(l)}}^{(l)0}   -\btheta_{g_{12}^{(l)}}^{(l)}  \|^2+2
		\|\btheta_{g_2^{(l)}}^{(l)0} - \btheta_{g_{12}^{(l)}}^{(l)} \|^2\\
		& +\max_{g^{-(l)} \in [G_{-l}^0]} \left(\frac{1}{\prod_{m \neq l} (\pi_{g^{(m)}, N_m}^{(m)} N_m)} \sum_{m \neq l}\sum_{i_m} \big\{ |\alpha_{g_1^{(l)} g_{\i_{-l}}^{-(l)0}}^0 - \alpha_{g_2^{(l)} g_{\i_{-l}}^{-(l)0}}^0 |^2  I(g_{\i_{-l}}^{-(l)0} = g^{-(l)}) \big\}\right)\\
		&\le  {2}\|\btheta_{g_1^{(l)}}^{(l)0}   -\btheta_{g_{12}^{(l)}}^{(l)}  \|^2+2
		\|\btheta_{g_2^{(l)}}^{(l)0} - \btheta_{g_{12}^{(l)}}^{(l)} \|^2 \\
		&+ \max_{g^{-(l)} \in [G_{-l}^0]} \left(\frac{1}{\prod_{m \neq l} (\pi_{g^{(m)}, N_m}^{(m)} N_m)} \sum_{m \neq l}\sum_{i_m} |\alpha_{g_1^{(l)} g_{\i_{-l}}^{-(l)0}}^0 - \alpha_{g_2^{(l)} g_{\i_{-l}}^{-(l)0}}^0 |^2 \right)\\
		&= {2}\|\btheta_{g_1^{(l)}}^{(l)0}   -\btheta_{g_{12}^{(l)}}^{(l)}  \|^2+2
		\|\btheta_{g_2^{(l)}}^{(l)0} - \btheta_{g_{12}^{(l)}}^{(l)} \|^2 \\
		& + \max_{g^{-(l)} \in [G_{-l}^0]} \left(\frac{1}{\prod_{m \neq l} (\pi_{g^{(m)}, N_m}^{(m)} N_m)}\right) \sum_{m \neq l}\sum_{i_m} |\alpha_{g_1^{(l)} g_{\i_{-l}}^{-(l)0}}^0 - \alpha_{g_2^{(l)} g_{\i_{-l}}^{-(l)0}}^0 |^2 \\
		& \le  {2}\|\btheta_{g_1^{(l)}}^{(l)0}   -\btheta_{g_{12}^{(l)}}^{(l)}  \|^2+2
		\|\btheta_{g_2^{(l)}}^{(l)0} - \btheta_{g_{12}^{(l)}}^{(l)} \|^2 \\
		& +\frac{1}{\min_{g^{-(l)} \in [G_{-l}^0]} (\prod_{m \neq l}   \pi_{g^{(m)},N_m}^{(m)} N_m)}\sum_{m \neq l}\sum_{i_m} |\alpha_{g_1^{(l)} g_{\i_{-l}}^{-(l)0}}^0 - \alpha_{g_2^{(l)} g_{\i_{-l}}^{-(l)0}}^0 |^2  \\
		& \le\frac{2}{\min_{g^{-(l)} \in [G_{-l}^0]} (\prod_{m \neq l} \pi_{g^{(m)},N_m}^{(m)})} \times  \Big( \|\btheta_{g_1^{(l)}}^{(l)0}   -\btheta_{g_{12}^{(l)}}^{(l)}  \|^2+
		\|\btheta_{g_2^{(l)}}^{(l)0} - \btheta_{g_{12}^{(l)}}^{(l)} \|^2\\
		& + \frac{1}{\prod_{m \neq l} N_m} \sum_{m \neq l} \sum_{i_m} |\alpha_{g_1^{(l)} g_{\i_{-l}}^{-(l)0}}^0 -\alpha_{g_{12}^{(l)}  \wh g_{\i_{-l}}^{-(l)}(\bxi)}|^2 \\
		& + \frac{1}{\prod_{m \neq l} N_m} \sum_{m \neq l} \sum_{i_m}  |\alpha_{g_2^{(l)} g_{\i_{-l}}^{-(l)0}}^0 -\alpha_{g_{12}^{(l)}  \wh g_{\i_{-l}}^{-(l)}(\bxi)} |^2\Big)\\
		&  \le \frac{4 \eta}{\min_{g^{-(l)} \in [G_{-l}^0]} (\prod_{m \neq l} \pi_{g^{(m)},N_m}^{(m)})}.
	\end{align*}
	The last line contradicts the definition of $\eta$ that $\eta <c_\gap (c_{\pi})^{q-1}/4$ as $N_l \to \infty$. Therefore, there does not exist a $g_{12}^{(l)} \in [G_l]$ such that
	$g_{12}^{(l)}\in \mA^{(l)}_\eta(\bxi, g_1^{(l)}, \mG_{-l}^0)\cap \mA^{(l)}_\eta(\bxi, g_2^{(l)}, \mG_{-l}^0)$
	for any $g_1^{(l)}, g_2^{(l)}\in [G_{l,0}]$ and $g_1^{(l)} \ne g_2^{(l)}$, which suggests that $\mA^{(l)}_\eta(\bxi, g_1^{(l)}, \mG_{-l}^0)\cap \mA^{(l)}_\eta(\bxi, g_2^{(l)}, \mG_{-l}^0)=\emptyset$ for any $g_1^{(l)} \ne g_2^{(l)}$. This completes the proof of part (i).

	\noindent
	{\bf 2. Proof of (ii)}
	
	For the notation simplicity, we use $\wh g_{i_l}^{(l)}$ to replace $\wh g_{i_l}^{(l)}(\bxi)$, where $\bxi$ satisfies that $\bxi \in \mN^{(l)}_\eta$ {and also \eqref{eq:Theta_cond}}.
	By the definition of $\wh g_{i_l}^{(l)}$, we have
	\begin{align*}
		I(\wh g_{i_l}^{(l)}= g^{(l)})\le I\left(
		Q_{i_l} (\bxi_{g^{(l)}}^{(l)}; \bxi_{g^{-(l)}}^{-(l)} , \wh \mG_{-l}(\bxi) )<
		Q_{i_l} (\bxi_{\wt g^{(l)}}^{(l)};  \bxi_{g^{-(l)}}^{-(l)}, \wh \mG_{-l}(\bxi) )
		\right)
	\end{align*}
	for any $\wt g^{(l)} \ne g^{(l)}$.
	Therefore, for $\wt g_{i_l}^{(l)} \in \mA^{(l)}_\eta(\bxi, g_{i_l}^{(l)0},  \mG_{-l}^0 )$, we have
	\begin{align*}
		&I\left(\wh g_{i_l}^{(l)} \not\in  \mA^{(l)}_\eta(\bxi, g_{i_l}^{(l)0},  \mG_{-l}^0 ) \right)
		= \sum_{g^{(l)} = 1}^{G_l} I\left(g^{(l)}\not\in \mA^{(l)}_\eta(\bxi, g_{i_l}^{(l)0},  \mG_{-l}^0 ) \right)I(\wh g_{i_l}^{(l)} = g^{(l)})\\
		&\le  \sum_{g^{(l)} = 1}^{G_l} I\left(g^{(l)}\not\in \mA^{(l)}_\eta(\bxi, g_{i_l}^{(l)0},  \mG_{-l}^0 ) \right)
		I\left(Q_{i_l} (\bxi_{g^{(l)}}^{(l)};  \bxi_{g^{-(l)}}^{-(l)}, \wh \mG_{-l}(\bxi) )<
		Q_{i_l} (\bxi_{\wt g_{i_l}^{(l)}}^{(l)};  \bxi_{g^{-(l)}}^{-(l)}, \wh \mG_{-l}(\bxi) ) \right) \\
		& \defeq \sum_{g^{(l)} = 1}^{G_l} { W_{i_l g^{(l)}}(\bxi).}
	\end{align*}
	For all $g^{(l)} \not\in \mA^{(l)}_\eta(\bxi, g_{i_l}^{(l)0}, \mG_{-l}^0)$ and $g^{(l)} \in[G_l]$,
	due to (i), there exists a $g_{j_l}^{(l)0} \ne g_{i_l}^{(l)0}$,
	such that $g^{(l)} \in \mA^{(l)}_\eta(\bxi, g_{j_l}^{(l)0}, \mG_{-l}^0)$.
	Then we have
	\begin{align*}
		&\left\|\btheta_{g_{i_l}^{(l)0}}^{(l)0} - \btheta_{g^{(l)}}^{(l)}\right\|^2+
		\frac{1}{\prod_{m \neq l} N_m} \sum_{m \neq l} \sum_{i_m} |\alpha_{g^{(l)} \wh g_{\i_{-l}}^{-(l)}(\bxi)} - \alpha_{ g_{i_l}^{(l)0} g_{\i_{-l}}^{-(l)0}}^0|^2\\
		&\ge \frac{1}{2}
		\left\|\btheta_{g_{i_l}^{(l)0}}^{(l)0}  - \btheta_{g_{j_l}^{(l)0}}^{(l)0} \right\|^2+
		\frac{1}{2 \prod_{m \neq l} N_m} \sum_{m \neq l} \sum_{i_m}  |\alpha_{g_{j_l}^{(l)0} g_{\i_{-l}}^{-(l)0}}^0 - \alpha_{g_{i_l}^{(l)0} g_{\i_{-l}}^{-(l)0}}^0|^2 \\
		& - \left\{\left\| \btheta_{g_{j_l}^{(l)0}}^{(l)0}  - \btheta_{g^{(l)}}^{(l)}\right\|^2+
		\frac{1}{\prod_{m \neq l} N_m} \sum_{m \neq l} \sum_{i_m} |\alpha_{ g^{(l)} \wh g_{\i_{-l}}^{-(l)}(\bxi)} - \alpha_{ g_{j_l}^{(l)0} g_{\i_{-l}}^{-(l)0}}^0 |^2\right\} \ge c_\gap (c_\pi)^{q-1}/2 - \eta
	\end{align*}
	by Assumption \ref{assum:group_diff} when $N_l \to \infty$.
	By Lemma \ref{lem:Q_star_diff}, it holds for any $g^{(l)} \not\in \mA^{(l)}_{\eta}(\bxi, g_{i_l}^{(l)0}, \mG_{-l}^0)$,
	\begin{align}
		\frac{1}{(\prod_{m \neq l} N_m)T}\big\{Q_{i_l}^*(\bxi_{g^{(l)}}^{(l)}; \bxi_{g^{-(l)}}^{-(l)}, \wh\mG_{-l}(\bxi)) - Q_{i_l}^*(\bxi_{g_{i_l}^{(l)0}}^{(l)};  \bxi_{g^{-(l)}}^{-(l)0}, \mG_{-l}^0)\big\}
		\ge \tau_{\min}(c_\gap (c_\pi)^{q-1}/2 - \eta).\label{eq:Q_star_j_diff1}
	\end{align}
	On the other hand, for any $\wt g_{i_l}^{(l)}\in \mA^{(l)}_\eta(\bxi, g_{i_l}^{(l)0}, \mG_{-l}^0)$,
	it holds
	\begin{align}
		&\frac{1}{(\prod_{m \neq l} N_m)T}Q_{i_l}^*(\bxi_{\wt g_{i_l}^{(l)}}^{(l)};  \bxi_{g^{-(l)}}^{-(l)}, \wh \mG_{-l}(\bxi)) -
		\frac{1}{(\prod_{m \neq l} N_m) T}Q_{i_l}^*(\bxi_{g_{i_l}^{(l)0}}^{(l)0}; \bxi_{g^{-(l)}}^{-(l)0}, \mG_{-l}^0)\nonumber\\
		&\le \tau_{\max}\left\{ \frac{1}{\prod_m N_m} \sum_{m \neq l} \sum_{i_m} \left\|\btheta_{\wh g_{i_m}^{(m)}(\bxi)}^{(m)} - \btheta_{g_{i_m}^{(m)0}}^{(m)0}\right\|^2+
		\left\|\btheta_{g_{i_l}^{(l)0}}^{(l)0} - \btheta_{\wt g_{i_l}^{(l)}}^{(l)}\right\|^2 \right. \nonumber\\
		& + \left.
		\frac{1}{\prod_{m \neq l} N_m} \sum_{m \neq l}\sum_{i_m} |\wh \alpha_{ \wt g_{i_l}^{(l)} \wh g_{\i_{-l}}^{-(l)}(\bxi) } - \alpha_{g_{i_l}^{(l)0} g_{\i_{-l}}^{-(l)0}}^0|^2\right\}\nonumber\\
		&\le \tau_{\max}\big\{d(\bTheta, \bTheta^0)+\eta\big\}\le \tau_{\max} \big\{C T^{-1} (\sum_l \log N_l)^2 +\eta\big\}.\label{eq:Q_star_j_diff2}
	\end{align}
	in probability tending to 1 by \eqref{eq:Theta_cond}, where $C$ is a positive constant.
	Combining (\ref{eq:Q_star_j_diff1}) and (\ref{eq:Q_star_j_diff2}), we have
	\begin{align*}
		&\frac{1}{(\prod_{m \neq l} N_m)T}Q_{i_l}^*(\bxi_{g^{(l)}}^{(l)}; \bxi_{g^{-(l)}}^{-(l)}, \wh\mG_{-l}(\bxi)) -
		\frac{1}{(\prod_{m \neq l} N_m)T}Q_{i_l}^*(\bxi_{\wt g_{i_l}^{(l)}}^{(l)}; \bxi_{g^{-(l)}}^{-(l)}, \wh \mG_{-l}(\bxi))\\
		&\ge
		\tau_{\min}\{c_\gap (c_\pi)^{q-1}/2 - \eta\} - \tau_{\max}
		\{\eta+  C T^{-1} (\sum_l \log N_l)^2 \}
		\defeq \epsilon_\eta
	\end{align*}
	when $N_l \to \infty$.
	Note that $\tau_{\min}$ and $\tau_{\max}$ are both bounded
	positive constants due to Assumption \ref{assum:tau_min} and Lemma
	\ref{lem:tau_max}.
	This leads to
	\begin{align}
		& W_{i_l g^{(l)}}(\bx) \nonumber\\
		& = I\left(g^{(l)}\not\in \mA^{(l)}_\eta(\bxi, g_{i_l}^{(l)0}, \mG_{-l}^0) \right)
		I\left(Q_{i_l} (\bxi_{g^{(l)}}^{(l)}; \bxi_{g^{-(l)}}^{-(l)}, \wh \mG_{-l}(\bxi) )<
		Q_{i_l} (\bxi_{\wt g_{i_l}^{(l)}}^{(l)}; \bxi_{g^{-(l)}}^{-(l)}, \wh \mG_{-l}(\bxi) ) \right) \nonumber\\
		&\le I\left(g^{(l)}\not\in \mA^{(l)}_\eta(\bxi, g_{i_l}^{(l)0}, \mG_{-l}^0) \right)\nonumber\\
		&
		I\Big(
		\frac{1}{(\prod_{m \neq l} N_m) T}Q_{i_l} (\bxi_{\wt g_{i_l}^{(l)}}^{(l)};  \bxi_{g^{-(l)}}^{-(l)}, \wh \mG_{-l}(\bxi) ) -
		\frac{1}{(\prod_{m \neq l} N_m)  T}Q_{i_l} (\bxi_{g^{(l)}}^{(l)};  \bxi_{g^{-(l)}}^{-(l)}, \wh \mG_{-l} (\bxi)) \nonumber\\
		&
		+\frac{1}{(\prod_{m \neq l} N_m)  T}Q_{i_l}^*(\bxi_{g^{(l)}}^{(l)};  \bxi_{g^{-(l)}}^{-(l)}, \wh\mG_{-l}(\bxi))-
		\frac{1}{(\prod_{m \neq l} N_m)  T}Q_{i_l}^*(\bxi_{\wt g_{i_l}^{(l)}}^{(l)};  \bxi_{g^{-(l)}}^{-(l)} , \wh \mG_{-l}(\bxi))\ge \epsilon_\eta\Big) \nonumber\\
		&\le 2 I\left(\sup_{i_1, \cdots, i_q}\sup_{\|\bTheta_{i_1, \cdots i_q}\|_{\max} <R}
		\Big|\frac{1}{T}Q_{i_1\cdots i_q}(\bTheta_{i_1\cdots i_q}) - \frac{1}{T}Q_{i_1\cdots i_q}^*(\bTheta_{i_1\cdots i_q})\Big| \ge \epsilon_\eta/2\right).\label{eq:W_il}
	\end{align}
	Consequently, we have
	\begin{align}
		&P\left\{\sup_{1\le i_l \le N_l} I\left(\wh g_{i_l}^{(l)} \not\in  \mA^{(l)}_\eta(\bxi, g_{i_l}^{(l)0}, \mG_{-l}^0) \right)>0
		\right\}\le \sum_{g^{(l)} = 1}^{G_l} P\left\{\sup_{1\le i_l \le N_l} W_{i_l g^{(l)}}(\bxi)>0
		\right\}\nonumber\\
		&\le \sum_{g^{(l)} = 1}^{G_l}  P\left\{\sup_{i_1, \cdots, i_q}\sup_{\|\bTheta_{i_1, \cdots i_q}\|_{\max} <R}
		\Big|\frac{1}{T}Q_{i_1\cdots i_q}(\bTheta_{i_1\cdots i_q}) - \frac{1}{T}Q_{i_1\cdots i_q}^*(\bTheta_{i_1\cdots i_q})\Big| \ge \epsilon_\eta/2
		\right\}\nonumber\\
		& \le G_l (\prod_l N_l) \exp\Big\{-c_1' \min(T\epsilon_\eta^2,T^{1/2}\epsilon_\eta)+c_2m \Big\} \nonumber\\
		& = G_l \exp\Big\{-c_1' \min(T\epsilon_\eta^2,T^{1/2}\epsilon_\eta)+c_2m + \sum_l \log N_l\Big\} \label{eq:g_il_indicator}
	\end{align}
	by using (\ref{eq:Q_ij_diff}) of Lemma \ref{lem:Q_diff}.
	Finally, noting that
	$ c_\gap \gg T^{-1}(\sum_l \log N_l)^2$ {and we have $\eta < \{ \tau_{\min}c_\gap
		(c_\pi)^{q-1}/4-  \tau_{\max} C T^{-1} (\sum_l \log N_l)^2 \}/(\tau_{\min}+\tau_{\max})$, then we can obtain $\epsilon_\eta > \tau_{\min} c_\gap (c_\pi)^{q-1}/4$ as $\min_l N_l,T\to \infty$.}
	This results in
	\begin{align*}
		P\left\{\sup_{1\le i_l \le N_l} I\left(\wh g_{i_l}^{(l)} \not\in  \mA^{(l)}_\eta(\bxi, g_{i_l}^{(l)0}, \mG_{-l}^0) \right)>0
		\right\} \le C \exp\Big\{-c_1 T^{1/2}c_{\gap} +c_2m + \sum_l \log N_l\Big\}.
	\end{align*}
	This finishes the proof of (ii).
	
\end{proof}

\section{General Technical Lemmas}\label{sec:general_lemma}

\bel\label{hw_ine}
({\sc  Hanson-Wright Inequality})
Define the $\phi_2$-norm of variable $X$ as
\beq
\|X\|_{\phi_2} = \inf\Big\{t>0: E\exp\Big(\frac{X^2}{t^2}\Big)\le 2\Big\}.\nonumber
\eeq
Let $X_1,\cdots, X_n$ be independent variables satisfying $E(X_i) = 0$,
$E(X_i^2) = \sigma_i^2$, and $\|X_i\|_{\phi_2}\le M<\infty$
and $\A$ be a symmetric $n\times n$ matrix.
Define $\bx = (X_1,\cdots, X_n)^\top \in \mR^n$.
For any $t>0$, we have
\begin{align*}
	P\Big(|\bx^\top \A \bx - \tr(\A\bSigma_x)|>t\Big)\le 2\exp\Big\{-c\min\Big(
	\frac{t^2}{M^4\|\A\|_F^2}, \frac{t}{M^2\sigma_1(\A)}\Big)\Big\},
\end{align*}
where $\bSigma_x = \diag\{\sigma_1^2,\cdots,
\sigma_n^2\}$, $\sigma_1(\A)$ is  the
maximum singular value of $\A$,
$\|\bA \|_F$ denotes the Frobenius norm of matrix $\bA$,
and $c>0$ is a constant.
\eel

\begin{proof}
	The proof is given in \cite{hanson1971bound}.
\end{proof}

\bel\label{lem:xAx_convex}
let $\bx$ be a mean zero random vector in $\mR^n$ satisfying
$K$-convex concentration property according to Definition \ref{def:convex_concen}.
Then for any $\bA\in \mR^{n\times n}$ and $t>0$, it holds
\begin{align*}
	P\Big(\Big|\bx^\top\bA\bx - E(\bx^\top\bA\bx)\Big|\ge t\Big)\le
	2\exp\Big(-\frac{1}{C}\min\Big(\frac{t^2}{K^4\|\bA\|_F^2 },
	\frac{t}{K^2\|\bA\|}\Big)\Big).
\end{align*}
\eel

\begin{proof}
	The proof can be found in Theorem 2.5 of \cite{adamczak2015note}.
\end{proof}

\bel\label{lem:covx}
If $\bx\in \mR^n$ satisfies the $K$-convex concentration property according to
Definition \ref{def:convex_concen} with $E(\bx) = \zero$,
then we have $\|\cov(\bx)\|\le 2K^2$.
\eel

\begin{proof}
	For any unit vector $\bu\in \mR^n$ with $\|\bu\| = 1$,
	$\langle \bu, \bx\rangle$ is a $1$-Lipschitz convex function of $\bx$.
	Then we have
	\beq
	\bu^\top\cov(\bx)\bu = E\langle \bu, \bx\rangle^2 =
	2\int_0^\infty tP(|\langle \bu, \bx\rangle|>t)dt \le 4
	\int_0^\infty t\exp(-t^2/K^2)dt = 2K^2.\nonumber
	\eeq
	This implies $\|\cov(\bx)\|\le 2K^2$.

\end{proof}

{\bel\label{lem:xDz}
	Let $\{\bD_{s_1 s_2}\in \mR^{(\prod_l N_l)\times (\prod_l N_l)}: s_1, s_2\in \{0, \dots, T\}$
	be a sequence of matrices.
	Define the following three terms,
	$\mD_{s_1s_2}^{(l)} = (\one_{N_1}^\top \otimes \cdots \otimes  \one_{N_{l-1}}^\top \otimes \bI_{N_l} \otimes \one_{N_{l+1}}^\top \otimes \cdots \otimes  \one_{N_q}^\top) \bD_{s_1s_2}
	(\one_{N_1} \otimes \cdots \otimes  \one_{N_{l-1}} \otimes \bI_{N_l} \otimes \one_{N_{l+1}} \otimes \cdots \otimes  \one_{N_q})\in \mR^{N_l \times N_l}$,
	$\mD_{s_1s_2}^{(lm)} = (\one_{N_1}^\top \otimes \cdots \otimes  \one_{N_{l-1}}^\top \otimes \bI_{N_l} \otimes \one_{N_{l+1}}^\top \otimes \cdots \otimes  \one_{N_q}^\top) \bD_{s_1s_2}
	(\one_{N_1} \otimes \cdots \otimes  \one_{N_{m-1}} \otimes \bI_{N_m} \otimes \one_{N_{m+1}} \otimes \cdots \otimes  \one_{N_q})\in \mR^{N_l \times N_m}$,
	and $\wt\mD_{s_1s_2}^{(l)} = (\one_{N_1}^\top \otimes \cdots \otimes  \one_{N_{l-1}}^\top \otimes \bI_{N_l} \otimes \one_{N_{l+1}}^\top \otimes \cdots \otimes  \one_{N_q}^\top) \bD_{s_1s_2} \in \mR^{N_l \times (\prod_l N_l)}$.
	Correspondingly, let $\mD^{(l)} = (\mD_{s_1s_2}^{(l)}: s_1,s_2\in \{0, \dots, T\})\in
	\mR^{(N_l{(T+1)})\times (N_l{(T+1)})}$,
	let $\mD^{(lm)} = (\mD_{s_1s_2}^{(lm)}: s_1,s_2\in \{0, \dots, T\})\in
	\mR^{(N_l{(T+1)})\times (N_m{(T+1)})}$,
	and let $\wt\mD^{(l)} = (\wt\mD_{s_1s_2}^{(l)}: s_1, s_2 \in \{0, \cdots, T\}) \in \mR^{N_l(T+1) \times (\prod_l N_l)(T+1)}$.
	Define $\wt\x_t^{(l)\eta} = \one_{N_1} \circ \cdots \circ \bx_t^{(l) \eta} \circ \cdots \circ \one_{N_q} \in \mR^{\prod_l N_l}$, where $\bx_t^{(l) \eta}$ is defined in Assumption \ref{assum:mixing}.
	Further note that $\x^{(l)\eta} = (\x_t^{(l)\eta}: 0 \le t \le T) \in \mR^{N_l (T+1)}$.
	Then we have
	\begin{align}
		&
		P\Big\{\Big|\sum_{s_1,s_2 = 0}^T \wt\x_t^{(l)\eta \top}  \bD_{s_1s_2} \wt\x_t^{(l)\eta}  -
		\sum_{s_1,s_2 = 0}^T E(\wt\x_t^{(l)\eta \top}  \bD_{s_1s_2} \wt\x_t^{(l)\eta})
		\Big|\ge u\Big\}\nonumber\\
		&\le 2\exp\Big(-\frac{1}{C}\min\Big(
		\frac{u^2}{\|\mD^{(l)}\|_F^2},
		\frac{u}{\|\mD^{(l)}\|}
		\Big)\Big), \label{eq:xDx_concen}\\
		&P\Big\{\Big|\sum_{s_1,s_2 = 0}^T \wt\bx_{s_1}^{(l)\eta\top} \bD_{s_1s_2} \wt\bx_{s_2}^{(m)\eta}
		\Big|>u\Big\}\le 2\exp\Big(-\frac{1}{C}\min\Big(\frac{u^2}{\|\mD^{(lm)}\|_F^2}, \frac{u}{\|\mD^{(lm)}\|}\Big)\Big),
		\label{eq:xDz_concen}\\
		& P\Big\{\Big|\sum_{s_1,s_2 = 0}^T \wt\bx_{s_1}^{(l)\eta\top} \bD_{s_1s_2} \E_{s_2}
		\Big|>u\Big\} \le
		2\exp\Big(-\frac{1}{C}\min\Big(\frac{u^2}{\|\wt\mD^{(l)}\|_F^2}, \frac{u}{\|\wt\mD^{(l)}\|}\Big)\Big).
		\label{eq:xDE_concen}
	\end{align}
	\eel
	
	\begin{proof}
		\noindent
		{\bf 1. Proof of (\ref{eq:xDx_concen})}
		
		Note that we have
			$
			\sum_{s_1,s_2 = 0}^T \wt\x_{s_1}^{(l)\eta \top}  \bD_{s_1s_2} \wt\x_{s_2}^{(l)\eta}
			= \sum_{s_1,s_2 = 0}^T \bx_{s_1}^{(l)\eta\top}
			\mD_{s_1s_2}^{(l)} \bx_{s_2}^{(l)\eta} = \bx^{(l)\eta\top }\mD^{(l)} \bx^{(l)\eta}.
			$
		Here $\bx^{(l)\eta}$ satisfies the $K'$-convex concentration property
		defined in
		Definition \ref{def:convex_concen} by Assumption \ref{assum:mixing}
		for a constant $K'$.
		Therefore by Lemma \ref{lem:xAx_convex} the concentration
		inequality in (\ref{eq:xDx_concen}) is directly obtained.

		\noindent
		{\bf 2. Proof of (\ref{eq:xDz_concen})}

		Note that
			$\sum_{s_1,s_2 = 0}^T \bx_{s_1}^{(l)\eta\top} \bD_{s_1s_2}
			\bx_{s_2}^{(m)\eta}
			= \sum_{s_1,s_2 = 0}^T \bx_{s_1}^{(l)\eta\top}
			\mD_{s_1s_2}^{(lm)}\bx_{s_2}^{(m)\eta} = \bx^{(l)\eta\top }\mD^{(lm)} \bx^{(m)\eta}.$
		Define $\bh^\eta = (\bx^{(l)\eta\top}, \bx^{(m)\eta\top})^\top$,
		and $\mD = (\zero,\mD^{(lm)};\mD^{(lm)\top}, \zero)$.
		Then we have $\bx^{(l)\eta\top }\mD^{(lm)} \bx^{(m)\eta} =
		\bh^{\eta\top}\mD\bh^\eta/2$.
		By Lemma \ref{lem:xAx_convex} and Assumption \ref{assum:mixing}  we obtain that
		\begin{align*}
			P\Big(\Big|\bx^{(l)\eta\top }\mD^{(lm)} \bx^{(m)\eta}\Big|>u\Big) &\le
			2\exp\Big(-\frac{1}{C_1}\min\Big(\frac{u^2}{\|\mD\|_F^2}, \frac{u}{\|\mD\|}\Big)\Big)\\
			&\le
			2\exp\Big(-\frac{1}{C}\min\Big(\frac{u^2}{\|\mD^{(lm)}\|_F^2}, \frac{u}{\|\mD^{(lm)}\|}\Big)\Big).
		\end{align*}

		\noindent
		{\bf 3. Proof of (\ref{eq:xDE_concen})}
		
		Denote $\E = (\E_t^\top: 0 \le t \le T)^\top \in \mR^{(\prod_l N_l) (T+1)}$.
		Note that
			$\sum_{s_1,s_2 = 0}^T \wt\bx_{s_1}^{(l)\eta\top} \bD_{s_1s_2} \E_{s_2}
			= \bx^{(l)\eta \top} \wt\mD^{(l)}\E,$
		then we follow the proof of (\ref{eq:xDz_concen}) and obtain
		the result.
		
	\end{proof}

	\bel\label{lem:D_upper}
	Let $\bD_{s_1s_2} = 1/T\sum_{t = \max\{1, s_1,s_2\}}^T
	(\bB_0^{t-s_1})^\top \bw^{(l)} \bw^{(l)\top}\bB_0^{t-s_2}  \defeq
	1/T\sum_{t = \max\{1, s_1,s_2\}}^T \\ \L^{(l)}_{t,s_1s_2}$,
	where $\bw^{(l)} = \e_{k_1}^{(N_1)} \otimes \cdots \otimes \bw^{(l)}_{i_l \cdot} \otimes \cdots \otimes \e_{k_q}^{(N_q)} \in \mR^{\prod_l N_l}$ for any $l \in [q]$.
	Here $\bw_{l i_l \cdot} = (w_{l i_l j}: j \in [N_l])^\top \in \mR^{N_l}$, $w_{l i_l j} \ge 0$ and $\|\bw_{l i_l \cdot}\|_1 = 1$.
	In addition, let
	$\mD^{(l)}$, $\mD^{(lm)}$ and $\wt\mD^{(l)}$ be defined as in Lemma \ref{lem:xDz}.
	Furthermore, let $\L_t^{(l)} = (\L^{(l)}_{t,s_1s_2}:0\le s_1,s_2\le t)$ and
	$\mL^{(l)}_{t, s_1s_2} = (\one_{N_1}^\top \otimes \cdots \otimes \one_{N_{l-1}}^\top \otimes \bI_{N_l} \otimes \one_{N_{l+1}}^\top \otimes \cdots \otimes \one_{N_q}^\top) \L^{(l)}_{t,s_1s_2}
	(\one_{N_1} \otimes \cdots \otimes \one_{N_{l-1}} \otimes \bI_{N_l} \otimes \one_{N_{l+1}} \otimes \cdots \otimes \one_{N_q})\in \mR^{N_l \times N_l}$.
	Define $\mL_{t}^{(l)} = (\wt \mL^{(l)}_{t,s_1s_2}: 0\le s_1,s_2\le T)$ for $l \in [q]$, where
	$\wt \mL_{t,s_1s_2}^{(l)} = \zero$ for $s_1>t$ or $s_2>t$, and
	$\wt \mL_{t,s_1s_2}^{(l)} = \mL_{t, s_1s_2}^{(l)}$ otherwise.
	Then we have for any $l \in [q]$,
	\begin{align}
		&\|\mD^{(l)}\|_F^2 \le c_1T^{-1}, ~~~\|\mD^{(l)}\| \le c_1^{1/2}T^{-1/2},\label{D_r_F_bound}\\
		&\|\mD^{(lm)}\|_F^2 \le c_1T^{-1}, ~~~\|\mD^{(lm)}\| \le c_1^{1/2}T^{-1/2},\label{D_rc_F_bound}\\
		&\|\bD\|_F^2 \le c_2T^{-1}, ~~~\|\bD\| \le c_2^{1/2}T^{-1/2},\label{D_F_bound}\\
		&\|\wt\mD^{(l)}\|_F^2 \le c_2T^{-1}, ~~~\|\wt\mD^{(l)}\| \le c_2^{1/2}T^{-1/2},\label{D_r1_F_bound}\\
		&\max_l \tr(\mL_{t}^{(l)})<\infty,~~~
		\tr(\L_t^{(l)})<\infty,\label{eq:L12_tr_bound}
	\end{align}
	where $\bD$ is defined in \eqref{def_D}, and $c_1$, $c_2$ are two positive constants.
	\eel
	
	\begin{proof}
		\noindent
		{\bf 1. Proof of (\ref{D_r_F_bound})--(\ref{D_rc_F_bound})}
		
		Note that by Lemma \ref{lem:Bn_max} we have
		\begin{align*}
			&\|\mD^{(l)}\|_F^2 = \sum_{s_1,s_2 = 0}^T \|\mD_{s_1s_2}^{(l)}\|_F^2 \\
			&=  \sum_{s_1,s_2 = 0}^T
			\Big\|(\one_{N_1}^\top \otimes \cdots \otimes \one_{N_{l-1}}^\top \otimes \bI_{N_l} \otimes \one_{N_{l+1}}^\top \otimes \cdots \otimes \one_{N_q}^\top) \bD_{s_1s_2}\\
			&
			(\one_{N_1} \otimes \cdots \otimes \one_{N_{l-1}} \otimes \bI_{N_l} \otimes \one_{N_{l+1}} \otimes \cdots \otimes \one_{N_q})\Big\|_F^2\\
			& \le  \sum_{s_1,s_2 = 0}^T
			\Big(\one_{\prod_l N_l}^\top |\bD_{s_1s_2}|_e
			\one_{\prod_l N_l}\Big)^2\\
			&  \le \frac{1}{T^2}
			\sum_{s_1,s_2 = 0}^T
			\Big( \sum_{t = \max\{1, s_1,s_2\}}^T\kappa_{\max}^{2t-s_1-s_2}
			\one_{\prod_l N_l}^\top |\bw^{(l)}|_e|\bw^{(l)\top}|_e
			\one_{\prod_l N_l}\Big)^2\\
			& =\frac{1}{T^2}
			\sum_{s_1,s_2 = 0}^T
			\Big( \sum_{t = \max\{1, s_1,s_2\}}^T\kappa_{\max}^{2t-s_1-s_2}\Big)^2
			\le \frac{1}{T^2}
			\sum_{s_1,s_2 = 0}^T\Big( \frac{1}{1-\kappa_{\max}^2}\kappa_{\max}^{|s_1-s_2|}\Big)^2 \le c_1T^{-1}
		\end{align*}
		where $c_1$ is a finite constant.
		Furthermore, noting that
		$\|\mD^{(l)}\|\le \|\mD^{(l)}\|_F \le c_1^{1/2}T^{-1/2}$, we have the upper bound
		for $\|\mD^{(l)}\|$.
		Subsequently, the upper bounds for
		$\|\mD^{(lm)}\|_F^2$ and
		$\|\mD^{(lm)}\|$ can be established using the same technique.

		\noindent
		{\bf 2. Proof of (\ref{D_F_bound})}

		First, for any two matrices $\bM_1, \bM_2$,
		\begin{align*}
			|\bM_1\bM_2^\top|_e\cle \max_{i,j}|\bM_{1i\cdot}\bM_{2j\cdot}^\top| \one\one^\top
			\le \max_{i,j}\|\bM_{1i\cdot}\|\|\bM_{2j\cdot}\| \one\one^\top.
		\end{align*}
		When $\|\bM_1\|_{\max}\le 1$
		and $\|\bM_2\|_{\max}\le 1$, we further have
		$\|\bM_{1i\cdot}\|\|\bM_{2j\cdot}\|\le
		\|\bM_1\|_\infty \|\bM_2\|_\infty$
		Hence
		$
		|\bM_1\bM_2^\top|_e\cle \|\bM_1\|_\infty \|\bM_2\|_\infty
		\one\one^\top$,
		and therefore
		\beq\label{eq:useful}
		\bw^{(l)\top}\bM_1\bM_2^\top\bw^{(l)} \le \|\bM_1\|_\infty \|\bM_2\|_\infty \bw^{(l)\top}\one\one^\top\bw^{(l)}
		= \|\bM_1\|_\infty \|\bM_2\|_\infty.
		\eeq
		Note that Lemma \ref{lem:Bn_max} implies
		$\|\bB_0^k\|_{\infty}\le\kappa_{\max}^n<1$ for any $k>0$, hence
		\begin{align*}
			\|\bD\|_F^2& = \sum_{s_1,s_2 = 0}^T \|\bD_{s_1s_2}\|_F^2
			\\
			&=  \frac{1}{T^2}\sum_{s_1,s_2 = 0}^T\sum_{t_1,t_2 = \max\{1, s_1, s_2\}}^T
			\bw^{(l)\top} \bB_0^{t_1 - s_2} (\bB_0^{t_2 - s_2})^\top \bw^{(l)}
			\bw^{(l)\top} \bB_0^{t_2 - s_1} (\bB_0^{t_1 - s_1})^\top\bw^{(l)}\\
			&
			\le  \frac{1}{T^2}\sum_{s_1,s_2 = 0}^T\sum_{t_1,t_2 = \max\{1, s_1, s_2\}}^T\Big(\|\bB_0^{t_1 - s_2}\|_\infty \|\bB_0^{t_2 -s_2}\|_\infty
			\|\bB_0^{t_2 - s_1}\|_\infty \|\bB_0^{t_1 - s_1}\|_\infty\Big)\\
			&\le \frac{1}{T^2}\sum_{s_1,s_2 = 0}^T\sum_{t_1,t_2 = \max\{1, s_1, s_2\}}^T
			\kappa_{\max}^{2t_1+2t_2-2s_1-2s_2} \\
			& \le \frac{1}{T^2(1-\kappa_{\max}^2)^2}\sum_{s_1,s_2 = 0}^T
			\kappa_{\max}^{2|s_1-s_2|}\le c_2T^{-1}
		\end{align*}
		where $c_2$ is a finite constant and we used Lemma \ref{lem:Bn_max} in the
		second inequality.

		\noindent
		{\bf 3. Proof of (\ref{D_r1_F_bound})}
		
		We have
		\begin{align*}
			& \|\wt\mD^{(l)}\|_F^2 = \sum_{s_1,s_2 = 0}^T \|\wt\mD_{s_1s_2}^{(l)}\|_F^2 =  \sum_{s_1,s_2 = 0}^T
			\Big\|(\one_{N_1}^\top \otimes \cdots \otimes \one_{N_{l-1}}^\top \otimes \bI_{N_l} \otimes \one^\top_{N_{l+1}} \otimes \cdots \otimes \one_{N_q}^\top) \bD_{s_1s_2}\Big\|_F^2\\
			& = \frac{1}{T^2} \sum_{s_1,s_2 = 0}^T
			\sum_{t_1,t_2 = \max\{1, s_1, s_2\}}^T
			\bw^{(l)\top} \bB_0^{t_1 - s_2}
			\bB_0^{t_2- s_2\top} \w^{(l)}\w^{(l)\top} \bB_0^{t_2- s_1}\\
			&(\one_{N_1} \one_{N_1}^\top \otimes \cdots \otimes \one_{N_{l-1}} \one_{N_{l-1}}^\top \otimes \bI_{N_l} \otimes \one_{N_{l+1}} \one_{N_{l+1}}^\top \otimes \cdots \otimes \one_{N_q} \one_{N_q}^\top)
			(\bB_0^{t_1 - s_1})^\top \bw^{(l)}\\
			& \le \frac{1}{T^2} \sum_{s_1,s_2 = 0}^T
			\sum_{t_1,t_2 = \max\{1, s_1, s_2\}}^T
			\|\bB_0^{t_1 - s_2}\|_\infty
			\|\bB_0^{t_2- s_2\top}\|_\infty \\
			& \hspace{13em}
			\w^{(l)\top} |\bB_0^{t_2- s_1}|_e
			(\1_{\prod_l N_l}\one_{\prod_l N_l}\trans)
			|\bB_0^{t_1 - s_1}|_e^\top \bw^{(l)}\\
			&\le  \frac{1}{T^2}\sum_{s_1,s_2 = 0}^T\sum_{t_1,t_2 = \max\{1, s_1,
				s_2\}}^T\|\bB_0^{t_1 - s_2}\|_\infty \|\bB_0^{t_2 -s_2}\|_\infty
			\|\bB_0^{t_2 - s_1}\|_\infty \|\bB_0^{t_1 - s_1}\|_\infty
			\bw^{(l)\top} \one\one^\top\bw^{(l)}\\
			&= \frac{1}{T^2}\sum_{s_1,s_2 = 0}^T\sum_{t_1,t_2 = \max\{1, s_1,
				s_2\}}^T\|\bB_0^{t_1 - s_2}\|_\infty \|\bB_0^{t_2 -s_2}\|_\infty
			\|\bB_0^{t_2 - s_1}\|_\infty \|\bB_0^{t_1 - s_1}\|_\infty\\
			&\le \frac{1}{T^2}\sum_{s_1,s_2 = 0}^T\sum_{t_1,t_2 = \max\{1, s_1, s_2\}}^T
			\kappa_{\max}^{2t_1+2t_2-2s_1-2s_2}\\
			& \le \frac{1}{T^2(1-\kappa_{\max}^2)^2}\sum_{s_1,s_2 = 0}^T
			\kappa_{\max}^{2|s_1-s_2|}\le c_2T^{-1}
		\end{align*}
		In addition, it further implies that
		$\|\wt\mD^{(l)}\|\le c_2^{1/2}T^{-1/2}$.

		\noindent
		{\bf 4. Proof of \eqref{eq:L12_tr_bound}}

		Note that by Lemma \ref{lem:Bn_max} we have
		\begin{align*}
			&\tr(\mL_{t}^{(l)}) = \sum_{s = 0}^t \tr(\mL_{t,ss}^{(l)}) \\
			& =  \sum_{s = 0}^t\tr
			\Big(( \one_{N_1}^\top \otimes \cdots \otimes \one_{N_{l-1}}^\top \otimes \bI_{N_l}\otimes \cdots \otimes \one_{N_q}^\top ) \L_{t,ss}^{(l)}
			( \one_{N_1} \otimes \cdots \otimes \one_{N_{l-1}} \otimes \bI_{N_l}\otimes \cdots \otimes \one_{N_q} )\Big)\\
			& \le  \sum_{s = 0}^t
			\Big(\one_{\prod_l N_l}^\top |\L_{t,ss}^{(l)}|_e
			\one_{\prod_l N_l}\Big) \le
			\sum_{s = 0}^t
			\Big( \kappa_{\max}^{2t-2s}
			\one_{\prod_l N_l}^\top |\bw^{(l)}|_e|\bw^{(l)\top}|_e
			\one_{\prod_l N_l}\Big)\le
			\frac{1}{1-\kappa_{\max}^2}<\infty.
		\end{align*}
		Consequently the first inequality of (\ref{eq:L12_tr_bound}) holds.
		Next we note
		\begin{align*}
			\tr(\L_t^{(l)}) &\le  \sum_{s = 0}^t
			(\one_{\prod_l N_l}^\top |\L_{t,ss}^{(l)}|_e
			\one_{\prod_l N_l})\\
			&\le
			\sum_{s = 0}^t
			( \kappa_{\max}^{2t-2s}
			\one_{\prod_l N_l}^\top |\bw^{(l)}|_e|\bw^{(l)\top}|_e
			\one_{\prod_l N_l})\le
			{1}/({1-\kappa_{\max}^2})<\infty,
		\end{align*}
		then the second inequality holds.	
	\end{proof}

	\begin{lemma}\label{lem:Bn_max}
		
		Under Assumption \ref{assum:station}, we have
		$\|\bB_0^n\|_\infty
		\le\kappa_{\max}^n$ and
		$|\bB_0|_e^n\one_{\prod_l N_l}
		\cle\kappa_{\max}^n\one_{\prod_l N_l}$.
	\end{lemma}
	
	\begin{proof}
		Note that we have
		\begin{align*}
			&|\bB_0|_e\one_{\prod_l N_l} \cle \sum_l \one_{N_1} \otimes \cdots \one_{N_{l-1}} \otimes \L_{l,0} \one_{N_l} \otimes \one_{N_{l+1}} \otimes \cdots \otimes \one_{N_q} + |\vec(\mA_0)|_e \cle
			\kappa_{\max}\one_{\prod_l N_l}
		\end{align*}
		by Assumption \ref{assum:station}.
		As a result, we have
		$\|\bB_0^n\|_\infty \le \||\bB_0|_e^n\|_\infty = \||\bB_0|_e^n\one_{\prod_l N_l}\|_{\max}\le \kappa_{\max}^n$.
	\end{proof}

	\bel\label{lem:10}
	Under Assumptions \ref{assum:mixing} and \ref{assum:station}, we
	have $\|\bGamma\|_{\max}\le c_{\Gamma}$,
	where $c_\Gamma>0$ is a constant.
	\eel
	
	\begin{proof}
		Define $\bh_t = \c_t + \E_t$ and then $\bh_t$ follows
		$K'$-convex concentration property by Assumption \ref{assum:mixing}
		for some constant $K'$.
		In addition, let $\cH_t = (\bh_t^\top, \bh_{t-1}^\top,\cdots, \bh_0^\top)^\top$,
		$\cB = (\bI_{\prod_l N_l}, \bB_0, \bB_0^2, \cdots, \bB_0^t)$.
		Note that we have
		\begin{align*}
			\bGamma = \cov(\mY_t) &= \sum_{k_1, k_2=0}^t \bB_0^{k_1}
			\cov\Big\{\bh_{t-k_1},\bh_{t-k_2}\Big\}\bB_0^{k_2\top} = \cB \cov(\cH_t)\cB^\top.
		\end{align*}
		Then we have $\|\bGamma\|_{\max}= \max_{i,j}|\e_i^\top \bGamma\e_j|
		=\max_{i}|\e_i^\top \bGamma\e_i|
		\le \|\cov(\cH_t)\| \max_i |\e_i^\top \cB\cB^\top\e_i|$,
		where $\e_i \in \mR^{\prod_l N_l}$ is a vector with its $i$th element bing equal to 1 while others being 0.
		By Lemma \ref{lem:covx},
		it holds $\|\cov(\cH_t)\| \max_i |\e_i^\top
		\cB\cB^\top\e_i|\le  2K^{*2} |\e_i^\top \cB\cB^\top\e_i|$
		for some constant $K^*$.
		Next, we have $\e_i^\top \cB\cB^\top\e_i =\sum_{k = 0}^t
		\e_i^\top\bB_0^k\bB_0^{k\top}\e_i\le \sum_{k = 0}^t
		\|\bB_0^k\|_\infty^2\le \sum_{k = 0}^t \kappa_{\max}^{2k} \le
		1/(1-\kappa_{\max}^{2})<\infty$ by Lemma \ref{lem:Bn_max}
		and Assumption \ref{assum:station}. Thus the result holds with $c_\Gamma=1/(1-\kappa_{\max}^{2})$.
	\end{proof}

	\bel\label{lem:tau_max}
	Under Assumptions \ref{assum:mixing} and \ref{assum:station},
	we have $ \tau_{\max} = \max_{i_1, \cdots, i_q}\lambda_{\max}(\bSigma_{i_1 \cdots i_q})<\infty$.
	\eel
	
	\begin{proof}
		Recall that we have defined
		\begin{align*}
			\cX_{i_1 \cdots i_q, t} \defeq
			\big(& \sum_{k = 1}^{N_1} w^{(1)}_{ i_1 k}Y_{k i_2 \cdots i_q, (t-1)}, \bx_{i_1t}^{(1)\top} , \cdots, \\
			&
			\sum_{k = 1}^{N_q} w^{(q)}_{i_q k} Y_{i_1 \cdots i_{(q-1)} k, (t-1)} , \bx_{i_q t}^{(q)\top},
			Y_{i_1 \cdots i_q, (t-1)}\big)^\top\in \mR^{\sum_l (p_l+1) + 1}.
		\end{align*}
		Note that there are $(2q+1)$ elements in $\cX_{i_1 \cdots i_q, t}$, denote the $m$th element as $\cX_{i_1 \cdots i_q, t}^{m}, 1 \le m \le (2q+1)$.
		{Specifically,
			$\cX_{i_1 \cdots i_q, t}^{m} = \sum_{k = 1}^{N_1} w^{(l)}_{i_l k}Y_{i_1 \cdots i_{l-1} k i_{l+1} \cdots i_q, (t-1)}$ when $m$ takes $\{1, 3, \cdots, 2q-1\}$, $\cX_{i_1 \cdots i_q, t}^{m} = \bx_{i_l t}^{(l)\top}$ when $m$ takes $\{2, 4, \cdots, 2q\}$, and $\cX_{i_1 \cdots i_q, t}^{2q+1} = Y_{i_1 \cdots i_q, (t-1)}$.}
		We write
		$\bfeta = (\eta_1, \eta_2, \eta_3, \eta_4, \cdots, \eta_{2q+1})^\top$
		for any vector $\bfeta\in \mR^{\sum_l (p_l+1)+1}$.
		{Here the even elements, i.e., $\eta_2, \eta_4, \cdots, \eta_{2q}$ are vectors with dimension $p_2, p_4, \cdots, p_{2q}$, while other elements are scalar.}
		Then we have
		\begin{align*}
			\lambda_{\max}(\bSigma_{i_1 \cdots i_q}) &  = \sup_{\|\bfeta\|=1}
			\sum_{k_1,k_2 = 1}^{2q+1} \eta_{k_1}^\top \cov(\cX_{i_1 \cdots i_q, t}^{k_1}, \cX_{i_1 \cdots i_q, t}^{k_2})\eta_{k_2}
			\\
			& \le (2q+1) \sup_{\|\bfeta\|=1}\sum_{k=1}^{2q+1} \eta_k^\top \var(\cX_{i_1 \cdots i_q, t}^k)\eta_k.
		\end{align*}
		Note that $\cX_{i_1 \cdots i_q, t}^k$ has two typical forms.
		The first can be written
		as $\bw^\top \mY_t$ ($k = 1,3,5, \cdots$) by taking $\bw$ as a vector with
		non-negative elements and $\|\bw\|_1 = 1$.
		The second is $\bx_{i_l t}^{(l)}$, which
		represents the covariate information.
		The first form yields
		$\eta_{k} \var(\bw^\top \mY_t)\eta_{k}
		= \eta_{k}^2\bw^\top \bGamma \bw\le
		\|\bGamma\|_{\max}|\bw^\top \one\one^\top \bw| =
		\|\bGamma\|_{\max}<c_\Gamma$, where $c_\Gamma$ is defined in
		Lemma \ref{lem:10}.
		The second form yields
		$\eta_k^\top \var(\bx_{i_lt}^{(l)})\eta_k\le \var(\bx_{t}^{(l) \eta}) <2K^2$
		due to Assumption \ref{assum:mixing} and Lemma \ref{lem:covx}.
		Thus, $\tau_{\max}\le (2q+1) ((q+1) c_\Gamma+ q K^2)<\infty$.
	\end{proof}

\section{Additional Simulation Studies}\label{sec:add_simu}

\subsection{Group Number Selection Consistency}\label{subsec:simu_G_consist}
	
	To examine the finite sample performance of the group selection consistency, we conduct experiment when $q=2$.
	The data generating mechanism is the same as in Section \ref{subsec:model_set} in the main text.
	Specifically, we define
	\begin{align*}
		\varrho(G_1) = R^{-1} \sum_{r=1}^R I(\wh G_1^{(r)} = G_1), ~~~ \varrho(G_2) = R^{-1} \sum_{r=1}^R I(\wh G_2^{(r)} = G_2),
	\end{align*}
	where $\wh G_1^{(r)}$ and $\wh G_2^{(r)}$ are estimated group numbers in the $r$th replicate.
	Hence, $\varrho(G_1)$ and $\varrho(G_2)$ evaluate the proportion of the correctly group numbers for $G_1$ and $G_2$.
	The results are shown in
	Table \ref{table:select_G}, from which one could see that as the sample size increases, the correct group numbers selection percentage is closer to 1, which can illustrate the group numbers selection consistency in Theorem \ref{thm:select_GH} from the finite sample experiment.
	
	\begin{table}[]
		\centering
		\caption{The proportion of selected group numbers $G_1$ and $G_2$ in $R = 500$ replicates under different settings.}\label{table:select_G}
		\scalebox{0.9}{
			\begin{tabular}{c|c|c|cc|cc|cc}
				\hline
				\multirow{2}{*}{$N_1$} & \multirow{2}{*}{$N_2$} & \multirow{2}{*}{$T$} & \multirow{2}{*}{$G_1$} & \multirow{2}{*}{$G_2$} & \multicolumn{2}{c|}{Scenario 1 (SBM)} & \multicolumn{2}{c}{Scenario 2 (Power-Law)} \\ \cline{6-9}
				&                        &                      &                      &                      & $\varrho(G_1)$      & $\varrho(G_2)$      & $\varrho(G_1)$         & $\varrho(G_2)$        \\ \hline
				\multirow{6}{*}{100}   & \multirow{6}{*}{80}    & \multirow{3}{*}{20}  & 2                    & 2                    & 0.966             & 0.966             & 0.838                & 0.838               \\
				&                        &                      & 3                    & 3                    & 0.034             & 0.034             & 0.162                & 0.162               \\
				&                        &                      & 4                    & 4                    & 0.000                 & 0.000                 & 0.000                    & 0.000                   \\ \cline{3-9}
				&                        & \multirow{3}{*}{40}  & 2                    & 2                    & 0.134             & 0.134             & 0.000                    & 0.000                   \\
				&                        &                      & 3                    & 3                    & 0.866             & 0.866             & 1.000                    & 1.000                   \\
				&                        &                      & 4                    & 4                    & 0.000                 & 0.000                 & 0.000                    & 0.000                   \\ \hline
				\multirow{6}{*}{200}   & \multirow{6}{*}{150}   & \multirow{3}{*}{20}  & 2                    & 2                    & 0.756             & 0.756             & 0.010                 & 0.010                \\
				&                        &                      & 3                    & 3                    & 0.244             & 0.244             & 0.990                 & 0.990                \\
				&                        &                      & 4                    & 4                    & 0.000                 & 0.000                 & 0.000                    & 0.000                   \\ \cline{3-9}
				&                        & \multirow{3}{*}{40}  & 2                    & 2                    & 0.016              & 0.016              & 0.000                    & 0.000                   \\
				&                        &                      & 3                    & 3                    & 0.984              & 0.984              & 1.000                    & 1.000                   \\
				&                        &                      & 4                    & 4                    & 0.000                 & 0.000                 & 0.000                    & 0.000                   \\ \hline
			\end{tabular}
		}
	\end{table}

		\subsection{Random Initialization}\label{subsec:random_init}
	
	In the {\sc Step 2} of Algorithm \ref{alg:init} in Appendix \ref{sec:init}, we use the $k$-means clustering for initialization of the group memberships.
	Alternatively, we also consider using random initialization in {\sc Step 2}, and the estimation results are shown in Table \ref{tbl:random_init}.
	We take $q =2$ as an example, and set $N_1 = 100, N_2 = 80$.
	The time lengths vary in $\{40, 80, 150, 200\}$.
	All the evaluation metrics are the same as Section \ref{subsec:model_set} in the main text.	
	The results demonstrate that $k$-means initialization yields superior parameter estimation accuracy compared to random initialization under limited temporal lengths.
	Notably, as the temporal length increases, the RMSE gap between the two initialization methods narrows significantly, i.e., random initialization based procedure converges to the similar performance as the $k$-means initialization based procedure.
	This convergence pattern is consistently observed in mis-classification rates 
	(the last two columns in Table \ref{tbl:random_init}).
	Critically, $k$-means initialization maintains robust performance regardless of sample size,
	while random initialization achieves comparable finite-sample performance when sufficient data is available.
	
	

	\begin{sidewaystable}[]
		\centering
		\caption{RMSEs of estimated parameters and mis-clustering rates when $G_{1,0} = G_{2,0} =3$ with 300 replicates.
			The Initialization methods include random and $k$-means initialization.}	\label{tbl:random_init}
		\scalebox{1}{
			\begin{tabular}{c|c|c|c|c|c|ccccc|cc}
				\hline
				$G_1$              & $G_2$              & $N_1$                & $N_2$               & Initialization             & $T$ & $\wh\blambda^{(1)}$ & $\wh\blambda^{(2)}$ & $\wh\bzeta^{(1)}$ & $\wh\bzeta^{(2)}$ & $\wh\balpha$ & $\wh\eta_1$ & $\wh\eta_2$ \\
				\hline
				\multirow{8}{*}{3} & \multirow{8}{*}{3} & \multirow{8}{*}{100} & \multirow{8}{*}{80} & \multirow{4}{*}{$K$-means} & 40  & 0.0100              & 0.0071              & 0.0140            & 0.0094            & 0.0153       & 0.0313      & 0.0001      \\
				&                    &                      &                     &                            & 80  & 0.0054              & 0.0046              & 0.0068            & 0.0067            & 0.0092       & 0.0031      & 0.0001      \\
				&                    &                      &                     &                            & 150 & 0.0040              & 0.0038              & 0.0047            & 0.0060            & 0.0068       & 0.0000      & 0.0001      \\
				&                    &                      &                     &                            & 200 & 0.0034              & 0.0032              & 0.0041            & 0.0043            & 0.0057       & 0.0000      & 0.0001      \\ \cline{5-13}
				&                    &                      &                     & \multirow{4}{*}{Random}    & 40  & 0.0157              & 0.0135              & 0.0207            & 0.0135            & 0.0356       & 0.0673      & 0.0124      \\
				&                    &                      &                     &                            & 80  & 0.0089              & 0.0105              & 0.0112            & 0.0093            & 0.0208       & 0.0296      & 0.0044      \\
				&                    &                      &                     &                            & 150 & 0.0050              & 0.0039              & 0.0067            & 0.0053            & 0.0084       & 0.0117      & 0.0013      \\
				&                    &                      &                     &                            & 200 & 0.0043              & 0.0040              & 0.0054            & 0.0054            & 0.0083       & 0.0078      & 0.0036      \\ \hline
		\end{tabular}}
	\end{sidewaystable}

	\subsection{Robustness Evaluation}\label{subsec:simu_robust}
	
	To show the robustness of our proposed model,
	we add several simulations in this section, including model mis-specification and heavy-tail noise settings.

	\subsubsection{Model Mis-specification}\label{subsec:mis_spec}

	In this subsection,
	we consider two model-specification settings.
	In both settings, we set $(N_1, N_2) \in \{ (100, 80), (200, 150) \}$, and the training time length $T \in \{20, 40, 80, 120\}$.

		\noindent
	{\bf 1. Mis-specified network structures.}
	
	First, we consider the case that the network weighting matrices $\{\W^{(l)}\}$ are misspecified when $q=2$.	
	First,
	we generate the ``correct'' networks $\A^{(l)}$ by setting $P(a_{ij,l} = 1) = 1/N_l$ and set $a_{ii} = 0$ by convention.
	Then, we simulate mis-specified networks by setting $P(a^\prime_{ij,l} = 1) = 1/(10 N_l)$ for those $a^\prime_{ij, l} = 0, ~ i \neq j$.
Consequently, we observe more links than the correctly specified networks.
	Denote the mis-specified networks by $\A_{\text{mis}}^{(1)}$ and $\A_{\text{mis}}^{(2)}$.
	Correspondingly, denote the row-normalized weighting matrices as $\W_{\text{mis}}^{(1)}$ and $\W_{\text{mis}}^{(2)}$.
	Our data is generated by model \eqref{eq:model00} with the ``correct'' weighting matrices $\W^{(1)}$ and $\W^{(2)}$, while the estimation is conducted using mis-specified weighting matrices $\W_{\text{mis}}^{(1)}$ and $\W_{\text{mis}}^{(2)}$.

			\noindent
	{\bf 2. Non-linear covariates.}
	
	Second, we consider the case when covariate structure is mis-specified.
	We generate data using the following model,
	\begin{align}
		Y_{ij, t} &=\lambda_{g_i^{(1)}}^{(1)} \sum_{k = 1}^{N_1} \frac{a^{(1)}_{ik}}{n_{1i}} Y_{kj, (t-1)}
		+ \lambda_{g_j^{(2)}}^{(2)} \sum_{k = 1}^{N_2} \frac{a^{(2)}_{kj}}{n_{2j}}Y_{ik, (t-1)} \nonumber\\
		&\hspace{5em} + \alpha_{g_i^{(1)} g_j^{(2)}}Y_{ij, (t-1)} + f(\bx_{it}^{(1)})^\top \bzeta_{g_i^{(1)}}^{(1)} +
		f(\bx_{jt}^{(2)})^\top \bzeta_{g_j^(2)}^{(2)} +\ve_{ij, t},\label{eq:model_quad_cov}
	\end{align}
	where $f(\bv) = (f(v_1), \cdots, f(v_{p_l}))$ with $f(v_m) = 0.1 v_m^3 + \sin(0.1 v_m \pi)$.
	Then we estimate the model using our proposed algorithm \ref{alg:gmnar_q2}, mis-specifying the true covariates $f(\bx_{it}^{(1)})$ and $f(\bx_{jt}^{(2)})$ with $\bx_{it}^{(1)}$ and $\bx_{jt}^{(2)}$.

	The data generating scheme for other parts of the model is the same as Section \ref{subsec:model_set} in the main text.
	To evaluate the performance of the two mis-specification settings,
	we calculate the out-of-sample mean square prediction error (MSPE) for the fitted response $\wh Y_{ij, t}$, respectively.
	Specifically, we set the subsequent $T_{test} = T/2$ samples after the training set as the testing set.
	Denote the predicted response for the testing set as $\wh Y_{ij, t}^{\textup{te}}$.
	Correspondingly, calculate the MSPE as
	\begin{align*}
		\textup{MSPE} = (N_1 N_2 T_{test})^{-1} \sum_{t=T_{train} + 1}^{T_{test}} \sum_{i,j} (\wh Y_{ij, t}^{\textup{te}} - Y_{ij, t})^2.
	\end{align*}
	Then we calculate the relative mean square prediction error as
	 $\textup{ReMSPE}= \textup{MSPE}/\textup{MSPE}_0$,
	where the baseline $\textup{MSPE}_0 = (N_1 N_2 T_{train})^{-1} \sum_{t = 1}^{T_{train}} (Y_{ij, t} - \mu_{ij, train})^2$, and $\mu_{ij, train} = T_{train}^{-1} \sum_{t=1}^{T_{train}} Y_{ij, t}$.
	For comparison, we calculate the ReMSPEs for the true models, denoted as $\text{ReMSPE}_{\text{true}}$.
	Specifically, we use $\W^{(1)}$ and $\W^{(2)}$ as the true networks in the first scenario, and use $f(\bx_{it}^{(1)})$, $f(\bx_{jt}^{(2)})$ as the true covariates in the second scenario.
	The results are shown in Table \ref{tbl:model_mis}.
	Across both scenarios, ReMSPEs for misspecified models exhibit a pronounced decreasing trend as temporal length increases.
	Notably, prediction errors under misspecification are close to those of true models, demonstrating the robust performance of the proposed GTNAR framework against model deviations.

\begin{table}
	\centering
	\caption{ReMSPEs of the mis-specified model and the true model in two scenarios.}
	\label{tbl:model_mis}
\begin{tabular}{ccc|cc|cc}
	\hline
	\multirow{2}{*}{$N_1$} & \multirow{2}{*}{$N_2$} & \multirow{2}{*}{$T$} & \multicolumn{2}{c|}{\textbf{Changing Networks}}                  & \multicolumn{2}{c}{\textbf{Nonlinear Covariates}}                \\ \cline{4-7}
	&                        &                      & $\text{ReMSPE}_{\textup{mis}}$ & $\text{ReMSPE}_{\textup{true}}$ & $\text{ReMSPE}_{\textup{mis}}$ & $\text{ReMSPE}_{\textup{true}}$ \\ \hline
	\multirow{4}{*}{100}   & \multirow{4}{*}{80}    & 20                   & 0.7097                         & 0.7082                          & 0.9190                         & 0.8924                          \\
	&                        & 40                   & 0.6845                         & 0.6833                          & 0.8758                         & 0.8497                          \\
	&                        & 80                   & 0.6445                         & 0.6435                          & 0.8589                         & 0.8339                          \\
	&                        & 120                  & 0.6290                         & 0.6282                          & 0.8514                         & 0.8260                          \\ \hline
	\multirow{4}{*}{200}   & \multirow{4}{*}{150}   & 20                   & 0.7309                         & 0.7300                          & 0.9257                         & 0.8985                          \\
	&                        & 40                   & 0.6926                         & 0.6921                          & 0.8792                         & 0.8531                          \\
	&                        & 80                   & 0.6741                         & 0.6736                          & 0.8560                         & 0.8305                          \\
	&                        & 120                  & 0.6415                         & 0.6410                          & 0.8489                         & 0.8236                          \\ \hline
\end{tabular}
\end{table}

	\subsubsection{Heavy-tail Random Noise}\label{subsec:robust_subgaussian}
	
In this subsection, we generate heavy-tailed data to validate the robustness of our proposed methodology.
The experimental design specifies $q=2$ with $G_{1,0} = G_{2,0} = 3$ underlying groups, with network structures generated from Stochastic Block Model (SBM) networks {as in Section \ref{subsec:model_set}}.
We conduct $R = 300$ replicates {for a reliable evaluation}.
For each replicate, the idiosyncratic noise $\ve_{i_1 i_2,t}$ follows an independent and identical $t$-distribution with {5 degrees of freedom} (i.e., $\ve_{i_1 i_2,t} \stackrel{\text{i.i.d.}}{\sim} t(5)$). All other simulation parameters {align with} Section \ref{subsec:model_set} of the main text.
The results are shown in Table \ref{tbl:simu_t_noise}.
As both the sample sizes ($N_1, N_2$) and the time length ($T$) increase, the RMSEs for all parameters decrease significantly. Our proposed estimators closely approach their oracle counterparts, particularly at larger sample sizes.
Notably, the group membership errors ($\widehat{\eta}_1, \widehat{\eta}_2$) converge rapidly to zero, reaching near-zero values at $T=40$ for $N_1 \geq 200$, which confirms the high accuracy of the proposed methodology.
This demonstrates the GTNAR method's robustness in maintaining finite sample performance despite heavy-tailed noise setting.

\begin{sidewaystable}
	\centering
	\caption{RMSEs of estimated parameters when $G_{1,0} = G_{2,0} = 3$ with noise distribution being $t(5)$. The experiments are conducted with 300 replications. The performances are evaluated for different sample sizes $N_1, N_2$ and the time length $T$. Results of the oracle scenario (given true group memberships $\mG_1^0, \mG_2^0$) are provided. The corresponding CPs are shown in the parenthesis.}
	\label{tbl:simu_t_noise}
	\scalebox{0.9}{
	\begin{tabular}{ccc|ccccc|ccccc|cc}
		\hline
		$N_1$                & $N_2$                & $T$ & $\wh\blambda^{(1)}$                                       & $\wh\blambda^{(2)}$                                       & $\wh\bzeta^{(1)}$                                         & $\wh\bzeta^{(2)}$                                         & $\wh\balpha$                                              & $\wh\blambda^{(1)\text{or}}$                              & $\wh\blambda^{(2)\text{or}}$                              & $\wh\bzeta^{(1)\text{or}}$                                & $\wh\bzeta^{(2)\text{or}}$                                & $\wh\balpha^{\text{or}}$                                  & $\wh\eta_1$ & $\wh\eta_2$ \\ \hline
		\multirow{2}{*}{100} & \multirow{2}{*}{80}  & 20  & \begin{tabular}[c]{@{}c@{}}0.0212\\ (0.7378)\end{tabular} & \begin{tabular}[c]{@{}c@{}}0.0243\\ (0.8078)\end{tabular} & \begin{tabular}[c]{@{}c@{}}0.0582\\ (0.5885)\end{tabular} & \begin{tabular}[c]{@{}c@{}}0.0325\\ (0.8322)\end{tabular} & \begin{tabular}[c]{@{}c@{}}0.0716\\ (0.6544)\end{tabular} & \begin{tabular}[c]{@{}c@{}}0.0127\\ (0.9400)\end{tabular} & \begin{tabular}[c]{@{}c@{}}0.0123\\ (0.9500)\end{tabular} & \begin{tabular}[c]{@{}c@{}}0.0167\\ (0.9500)\end{tabular} & \begin{tabular}[c]{@{}c@{}}0.0169\\ (0.9422)\end{tabular} & \begin{tabular}[c]{@{}c@{}}0.0203\\ (0.9467)\end{tabular} & 0.2332      & 0.0569      \\
		&                      & 40  & \begin{tabular}[c]{@{}c@{}}0.0113\\ (0.8656)\end{tabular} & \begin{tabular}[c]{@{}c@{}}0.0077\\ (0.9489)\end{tabular} & \begin{tabular}[c]{@{}c@{}}0.0223\\ (0.8185)\end{tabular} & \begin{tabular}[c]{@{}c@{}}0.0118\\ (0.9463)\end{tabular} & \begin{tabular}[c]{@{}c@{}}0.0189\\ (0.8726)\end{tabular} & \begin{tabular}[c]{@{}c@{}}0.0087\\ (0.9478)\end{tabular} & \begin{tabular}[c]{@{}c@{}}0.0077\\ (0.9511)\end{tabular} & \begin{tabular}[c]{@{}c@{}}0.0116\\ (0.9541)\end{tabular} & \begin{tabular}[c]{@{}c@{}}0.0118\\ (0.9470)\end{tabular} & \begin{tabular}[c]{@{}c@{}}0.0139\\ (0.9507)\end{tabular} & 0.0642      & 0           \\ \hline
		\multirow{2}{*}{200} & \multirow{2}{*}{150} & 20  & \begin{tabular}[c]{@{}c@{}}0.0109\\ (0.7933)\end{tabular} & \begin{tabular}[c]{@{}c@{}}0.0065\\ (0.9333)\end{tabular} & \begin{tabular}[c]{@{}c@{}}0.0248\\ (0.7430)\end{tabular} & \begin{tabular}[c]{@{}c@{}}0.0088\\ (0.9430)\end{tabular} & \begin{tabular}[c]{@{}c@{}}0.0178\\ (0.8189)\end{tabular} & \begin{tabular}[c]{@{}c@{}}0.0067\\ (0.9389)\end{tabular} & \begin{tabular}[c]{@{}c@{}}0.0064\\ (0.9356)\end{tabular} & \begin{tabular}[c]{@{}c@{}}0.0085\\ (0.9511)\end{tabular} & \begin{tabular}[c]{@{}c@{}}0.0088\\ (0.9444)\end{tabular} & \begin{tabular}[c]{@{}c@{}}0.0102\\ (0.9448)\end{tabular} & 0.0829      & 0.0001      \\
		&                      & 40  & \begin{tabular}[c]{@{}c@{}}0.0054\\ (0.9133)\end{tabular} & \begin{tabular}[c]{@{}c@{}}0.0043\\ (0.9356)\end{tabular} & \begin{tabular}[c]{@{}c@{}}0.0074\\ (0.9256)\end{tabular} & \begin{tabular}[c]{@{}c@{}}0.0060\\ (0.9563)\end{tabular} & \begin{tabular}[c]{@{}c@{}}0.0075\\ (0.9381)\end{tabular} & \begin{tabular}[c]{@{}c@{}}0.0049\\ (0.9367)\end{tabular} & \begin{tabular}[c]{@{}c@{}}0.0043\\ (0.9333)\end{tabular} & \begin{tabular}[c]{@{}c@{}}0.0060\\ (0.9485)\end{tabular} & \begin{tabular}[c]{@{}c@{}}0.0060\\ (0.9563)\end{tabular} & \begin{tabular}[c]{@{}c@{}}0.0069\\ (0.9519)\end{tabular} & 0.0068      & 0.0001      \\ \hline
		\multirow{2}{*}{300} & \multirow{2}{*}{250} & 20  & \begin{tabular}[c]{@{}c@{}}0.0053\\ (0.9133)\end{tabular} & \begin{tabular}[c]{@{}c@{}}0.0039\\ (0.9356)\end{tabular} & \begin{tabular}[c]{@{}c@{}}0.0085\\ (0.9256)\end{tabular} & \begin{tabular}[c]{@{}c@{}}0.0053\\ (0.9563)\end{tabular} & \begin{tabular}[c]{@{}c@{}}0.0077\\ (0.9381)\end{tabular} & \begin{tabular}[c]{@{}c@{}}0.0042\\ (0.9367)\end{tabular} & \begin{tabular}[c]{@{}c@{}}0.0039\\ (0.9333)\end{tabular} & \begin{tabular}[c]{@{}c@{}}0.0055\\ (0.9485)\end{tabular} & \begin{tabular}[c]{@{}c@{}}0.0053\\ (0.9563)\end{tabular} & \begin{tabular}[c]{@{}c@{}}0.0064\\ (0.9519)\end{tabular} & 0.0153      & 0           \\
		&                      & 40  & \begin{tabular}[c]{@{}c@{}}0.0027\\ (0.9600)\end{tabular} & \begin{tabular}[c]{@{}c@{}}0.0027\\ (0.9467)\end{tabular} & \begin{tabular}[c]{@{}c@{}}0.0039\\ (0.9478)\end{tabular} & \begin{tabular}[c]{@{}c@{}}0.0038\\ (0.9437)\end{tabular} & \begin{tabular}[c]{@{}c@{}}0.0044\\ (0.9463)\end{tabular} & \begin{tabular}[c]{@{}c@{}}0.0027\\ (0.9600)\end{tabular} & \begin{tabular}[c]{@{}c@{}}0.0027\\ (0.9467)\end{tabular} & \begin{tabular}[c]{@{}c@{}}0.0039\\ (0.9478)\end{tabular} & \begin{tabular}[c]{@{}c@{}}0.0038\\ (0.9437)\end{tabular} & \begin{tabular}[c]{@{}c@{}}0.0044\\ (0.9467)\end{tabular} & 0           & 0           \\ \hline
	\end{tabular}}
\end{sidewaystable}

	\subsection{Computational Cost}\label{subsec:compute_cost}


	As we comment in Remark \ref{rmk:alg1_reduce_dim}, the memberships update equation \eqref{eq:update_g} can be computed in parallel for each inner-layer subject ($i_l$ in	the $l$th layer).
	Therefore, we implement the algorithm by a multi-core computational scheme.
	For each layer, the group memberships estimation is conducted in R version 4.4.3, running on a Apple M3 platform.
	Parallel processing implements 4 CPU cores using the {\sf doParallelSNOW} package (version 1.0.17).
	Total available hardware resources included 8 physical cores supporting 8 concurrent threads.
	We report the average computational cost of 10 replicated experiments in Figure \ref{fig:compute_time}.
	The results demonstrate the computational efficiency of the proposed methodology under varying network sizes and time lengths.
	The initialization stage (orange dashed line) exhibits near-constant time complexity with stable
	execution times (minimal fluctuation as $T$ increases), indicating
	negligible scaling overhead in initialization.
	 The iterative estimation phase (blue solid line) shows linear scaling with time length $T$.
	 Notably, even at the maximum network size configuration ($N_1 = 300, N_2 = 250$), total computational costs remain no more than 20 seconds across all time horizons,
	 which highlights the GTNAR method's suitability for large-scale applications despite increasing data dimensionality.

	\begin{figure}
		\centering
		\includegraphics[width=0.32\textwidth]{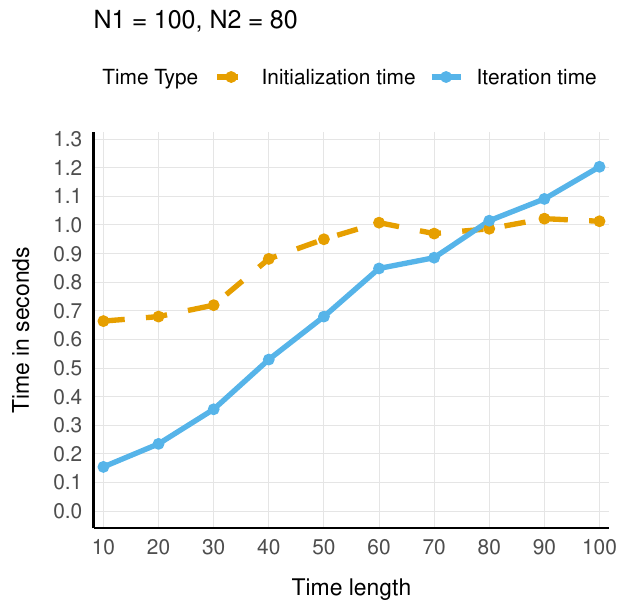}
		\includegraphics[width=0.32\textwidth]{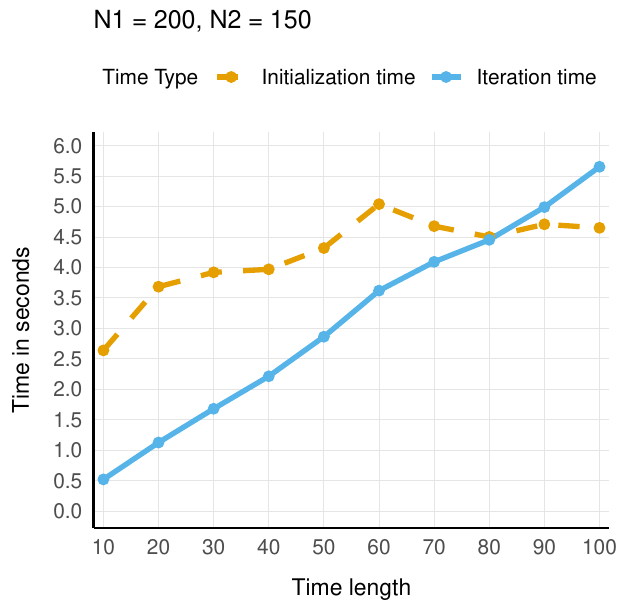}
		\includegraphics[width=0.32\textwidth]{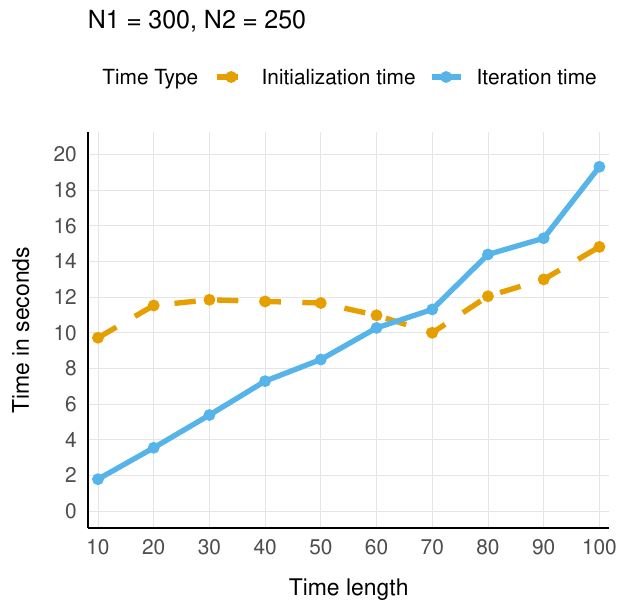}
		\caption{Computational costs (in second) of the initialization (orange dashed line) and iterative estimation (blue solid line) stage under different sample sizes and time lengths.}
		\label{fig:compute_time}
	\end{figure}

	\subsection{Numerical Convergence Analysis}\label{subsec:converge}

	In this subsection, we take $q=2$ for example to evaluate the numerical convergence rate of our proposed algorithm.
	Specifically, denote the loss function in the $k$th iteration as $Q^{[k]}(\wh\bTheta^{[k]})$, where $\wh\bTheta^{[k]}$ is the estimators in the $k$th iteration.
	The convergence criterion is set as $|Q^{[k+1]}(\wh\bTheta^{[k+1]})- Q^{[k]}(\wh\bTheta^{[k]})| \le 10^{-6}$.
	We first show the steps of iteration required for algorithm convergence and the corresponding total loss defined in equation \eqref{eq:Q_obj_simple} in each iteration step in Figure \ref{fig:converge_step}.
	Specifically, we set the network sizes $(N_1, N_2) \in \{ (100, 80), (200, 150) \}$ and the time lengths $T \in \{20, 40\}$.
	We repeat the experiments for 50 replicates in each sample size setting.
	For a baseline comparison, we calculate the oracle loss by using the oracle estimators, which are obtained by setting the group memberships as true values.
	Figure \ref{fig:converge_step} demonstrates the highly efficient convergence of our proposed iterative algorithm to the oracle loss across diverse sample size configurations.
	Notably, either larger network sizes ($N_1 =200, N_2 =150$) or longer time periods $(T=40)$ enhance convergence speed.
	
	\begin{figure}
		\centering
		\includegraphics[width=0.4\textwidth]{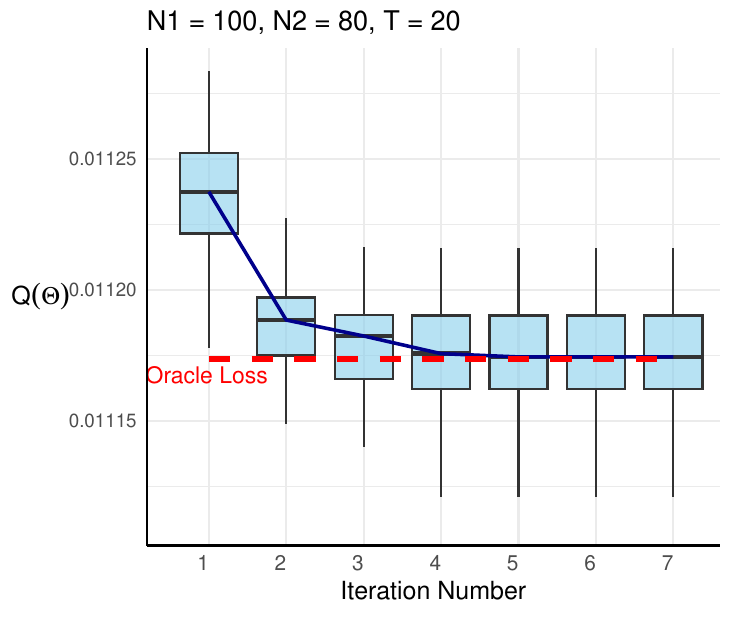}
		\includegraphics[width=0.4\textwidth]{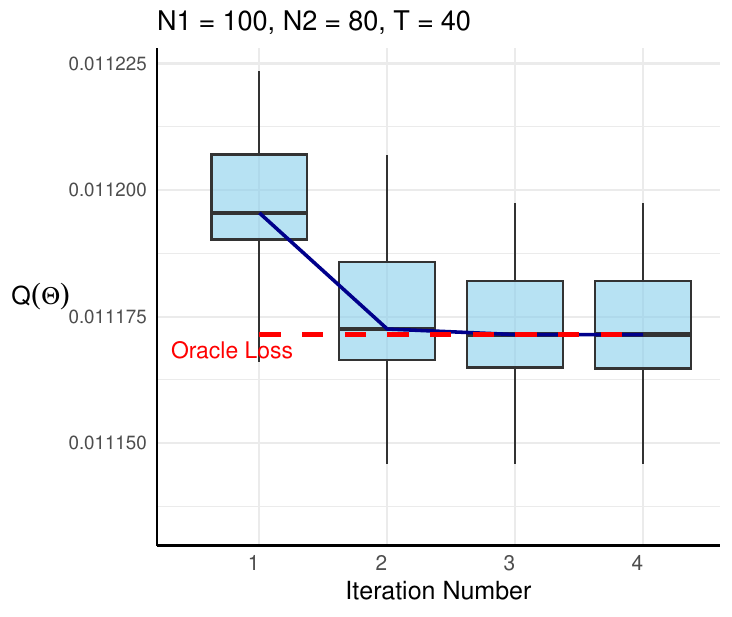}
		\includegraphics[width=0.4\textwidth]{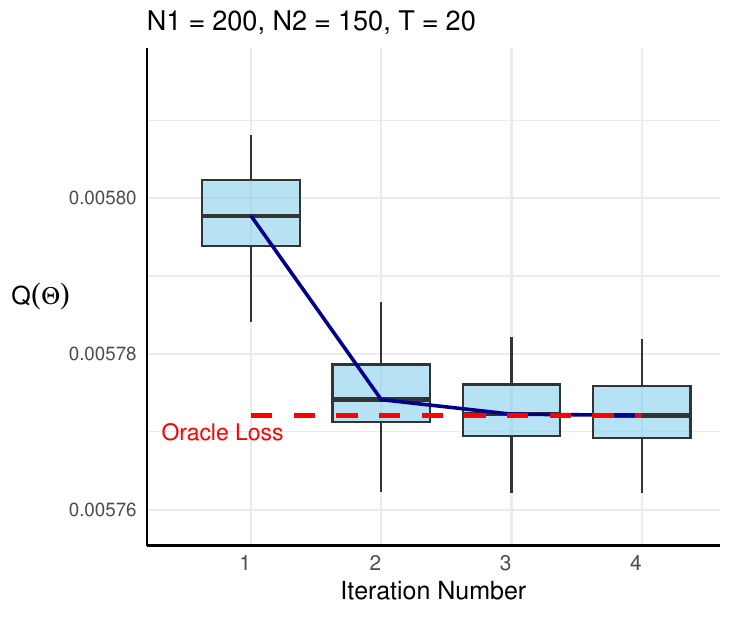}
		\includegraphics[width=0.4\textwidth]{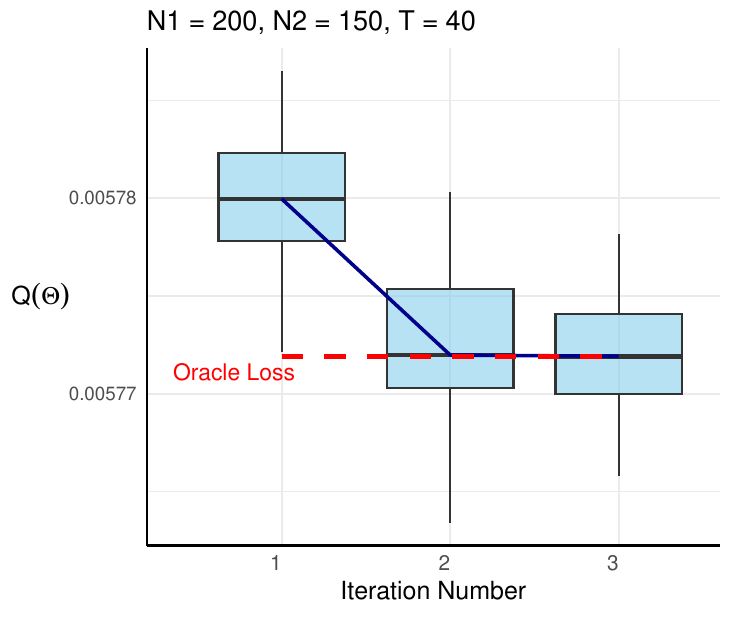}
		\caption{Total loss in each iteration step and the number of iterations required for algorithm convergence in different network sizes and time lengths. Each box shows the total loss of $R=50$ replicates. The red line shows the oracle loss.}
		\label{fig:converge_step}
	\end{figure}
	
	We further show the average and the maximum number of iterations.
	Fix network sizes as $N_1 = 100, N_2 = 80$, we show the results as the time length $T$ grows from 10 to 100;
	and fix the time length $T = 20$, we increase the network sizes throughout $(N_1, N_2) \in \{ (30,20), (50, 40), (80, 60), (100, 80), (120, 100), (160, 120), (200, 150), (250, 180), (300, 250) \}$.\\
	The results are shown in Figure \ref{fig:converge_step_vary}.
	The average iteration number demonstrates consistent monotonic decline as sample sizes increase, which highlights the algorithm's intrinsic efficiency.
	While the maximum iteration number exhibits fluctuations at minimal network sizes, it undergoes rapid stabilization followed by progressive decline as sample sizes grow.

	\begin{figure}
		\centering
		\includegraphics[width=0.48\textwidth]{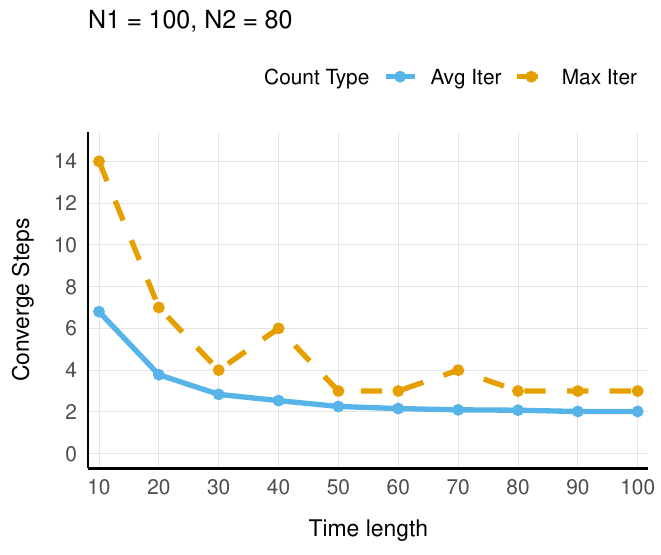}
		\includegraphics[width=0.43\textwidth]{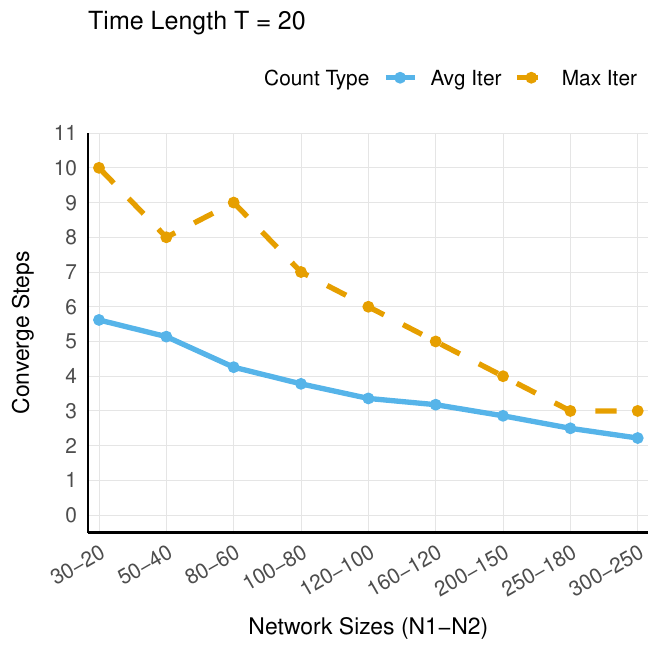}
		\caption{Average and maximum number of iterations required for convergence under different sample sizes throughout 50 replicates. The orange dashed line shows the maximum number, while the blue solid lines shows the average one.}
		\label{fig:converge_step_vary}
	\end{figure}
}
	
		\subsection{Simulation Results when $q=3$}\label{subsec:simu_q3}

	In this subsection, we conduct experiment when $q=3$ to show the finite sample performance of our proposed method.
	We set the true group numbers in the three dimensions as $G_{1,0} = 3, G_{2,0} = 3, G_{3, 0} = 2$.
	The true parameters are provided in Table \ref{tbl:true_param_3d}.
	We consider the networks generated from stochastic block models.
	The network sizes are set as $(N_1, N_2, N_3) \in \{ (20, 20, 20), (30, 30, 30) \}$, and the time length is set to be $T \in \{ 10, 40, 80\}$.
	The network and the data generation schemes are the same as those in Section \ref{subsec:model_set} in the main text.
	The experiments are repeated for $R = 300$ times.

	We first evaluate estimation accuracy under correctly specified group numbers, with comprehensive results presented in Table \ref{tbl:simu_q3}.
	We observe three key patterns.
	First, the mis-classification rates decline rapidly to zero as sample sizes increase, confirming precise recovery of group memberships.
	Second, the RMSEs decrease monotonically with increasing network sizes or time length, demonstrating consistent estimation efficiency.
	Third, the coverage probabilities converge to the nominal 95\% confidence level under larger samples, validating asymptotic normality of the estimators. Collectively, these findings meet theoretical results for statistical consistency.

	\begin{table}[]
		\centering
		\caption{True parameters when $G_{1,0} = 3, G_{2,0} = 3, G_{3, 0} = 2$.}
		\label{tbl:true_param_3d}
		\begin{tabular}{ccc}
			\hline
			\multicolumn{1}{c|}{$G_{1,0}=3$}                                                                                               & \multicolumn{1}{c|}{$G_{2,0} = 3$}                                                                                            & $G_{3,0} = 2$                                                                    \\ \hline
			\multicolumn{1}{c|}{$\lambda_{g^{(1)}}^{(1)}$}                                                                                 & \multicolumn{1}{c|}{$\lambda_{g^{(2)}}^{(2)}$}                                                                                & $\lambda_{g^{(3)}}^{(3)}$                                                        \\
			\multicolumn{1}{c|}{(-0.1, 0.2, 0.3)}                                                                                          & \multicolumn{1}{c|}{(0.15, 0.2, 0.4)}                                                                                         & (-0.2, 0.25)                                                                     \\ \hline
			\multicolumn{1}{c|}{$\bzeta_{g^{(1)}}^{(1)}$}                                                                                  & \multicolumn{1}{c|}{$\bdelta_{g^{(2)}}^{(2)}$}                                                                                & $\bdelta_{g^{(3)}}^{(3)}$                                                        \\
			\multicolumn{1}{c|}{$\left(\begin{array}{ccc}0.2 & 0.25 & -0.3 \\ 0.15 & 0.35 & -0.35 \\0.24& 0.30& -0.32 \end{array}\right)$} & \multicolumn{1}{c|}{$\left(\begin{array}{ccc}0.25 & -0.3 & 0.35 \\ 0.2 & -0.25 & 0.32 \\ 0.1& -0.2 & 0.2 \end{array}\right)$} & $\left(\begin{array}{ccc}-0.1& 0.2 & 0.4 \\ 0.3 & 0.1 & -0.32\end{array}\right)$ \\ \hline
			\multicolumn{3}{c}{$\balpha_{\cdot \cdot 1}$~~~~~~~~~ ~~~ ~~~ ~~~ ~~~ ~~~ $\balpha_{\cdot \cdot 2}$}                                                                                                                                                                                                                                                                                            \\
			\multicolumn{3}{c}{$\left(\begin{array}{ccc}0.2 & 0.25 & -0.3 \\ 0.15 & 0.35 & -0.35 \\0.24& 0.30& -0.32 \end{array}\right)$~~~$\left(\begin{array}{ccc}0.2 & 0.25 & -0.3 \\ 0.15 & 0.35 & -0.35 \\0.24& 0.30& -0.32 \end{array}\right)$}                                                                                                         \\ \hline
		\end{tabular}
	\end{table}

	\begin{sidewaystable}[]
		\caption{RMSEs of estimated parameters when $G_{1,0} = G_{2,0} = 3$ and $G_{3,0} = 2$ with 300 replications. The performances are evaluated for different sample sizes $N_1, N_2, N_3$ and the time length $T$.
			The corresponding CPs are shown in the parenthesis.}\label{tbl:simu_q3}
		\begin{tabular}{ccc|c|cccccccccc}
			\hline
			$N_1$               & $N_2$               & $N_3$               & $T$ & $\wh\blambda^{(1)}$                                       & $\wh\blambda^{(2)}$                                       & $\wh\blambda^{(3)}$                                       & $\wh\bzeta^{(1)}$                                         & $\wh\bzeta^{(2)}$                                         & $\wh\bzeta^{(3)}$                                         & $\wh\balpha$                                              & $\wh\eta_1$ & $\wh\eta_2$ & $\wh\eta_3$ \\ \hline
			\multirow{5}{*}{20} & \multirow{5}{*}{20} & \multirow{5}{*}{20} & 10  & \begin{tabular}[c]{@{}c@{}}0.0165\\ (0.9256)\end{tabular} & \begin{tabular}[c]{@{}c@{}}0.0141\\ (0.9411)\end{tabular} & \begin{tabular}[c]{@{}c@{}}0.0097\\ (0.9367)\end{tabular} & \begin{tabular}[c]{@{}c@{}}0.0207\\ (0.9226)\end{tabular} & \begin{tabular}[c]{@{}c@{}}0.0204\\ (0.9315)\end{tabular} & \begin{tabular}[c]{@{}c@{}}0.0128\\ (0.9272)\end{tabular} & \begin{tabular}[c]{@{}c@{}}0.0666\\ (0.9222)\end{tabular} & 0.0083      & 0           & 0           \\
			&                     &                     & 40  & \begin{tabular}[c]{@{}c@{}}0.0066\\ (0.9356)\end{tabular} & \begin{tabular}[c]{@{}c@{}}0.0067\\ (0.9378)\end{tabular} & \begin{tabular}[c]{@{}c@{}}0.0044\\ (0.9533)\end{tabular} & \begin{tabular}[c]{@{}c@{}}0.0093\\ (0.9411)\end{tabular} & \begin{tabular}[c]{@{}c@{}}0.0096\\ (0.9452)\end{tabular} & \begin{tabular}[c]{@{}c@{}}0.0059\\ (0.9372)\end{tabular} & \begin{tabular}[c]{@{}c@{}}0.0263\\ (0.9422)\end{tabular} & 0           & 0           & 0           \\
			&                     &                     & 80  & \begin{tabular}[c]{@{}c@{}}0.0025\\ (0.9311)\end{tabular} & \begin{tabular}[c]{@{}c@{}}0.0024\\ (0.9467)\end{tabular} & \begin{tabular}[c]{@{}c@{}}0.0024\\ (0.9383)\end{tabular} & \begin{tabular}[c]{@{}c@{}}0.0065\\ (0.9459)\end{tabular} & \begin{tabular}[c]{@{}c@{}}0.0065\\ (0.9519)\end{tabular} & \begin{tabular}[c]{@{}c@{}}0.0042\\ (0.9478)\end{tabular} & \begin{tabular}[c]{@{}c@{}}0.0158\\ (0.9478)\end{tabular} & 0           & 0           & 0           \\ \hline
			\multirow{5}{*}{30} & \multirow{5}{*}{30} & \multirow{5}{*}{30} & 10  & \begin{tabular}[c]{@{}c@{}}0.0105\\ (0.9422)\end{tabular} & \begin{tabular}[c]{@{}c@{}}0.0082\\ (0.9400)\end{tabular} & \begin{tabular}[c]{@{}c@{}}0.0057\\ (0.9383)\end{tabular} & \begin{tabular}[c]{@{}c@{}}0.0130\\ (0.9196)\end{tabular} & \begin{tabular}[c]{@{}c@{}}0.0106\\ (0.9370)\end{tabular} & \begin{tabular}[c]{@{}c@{}}0.0068\\ (0.9378)\end{tabular} & \begin{tabular}[c]{@{}c@{}}0.0435\\ (0.9269)\end{tabular} & 0.0078      & 0           & 0           \\
			&                     &                     & 40  & \begin{tabular}[c]{@{}c@{}}0.0034\\ (0.9489)\end{tabular} & \begin{tabular}[c]{@{}c@{}}0.0033\\ (0.9567)\end{tabular} & \begin{tabular}[c]{@{}c@{}}0.0023\\ (0.9583)\end{tabular} & \begin{tabular}[c]{@{}c@{}}0.0050\\ (0.9526)\end{tabular} & \begin{tabular}[c]{@{}c@{}}0.0049\\ (0.9556)\end{tabular} & \begin{tabular}[c]{@{}c@{}}0.0033\\ (0.9411)\end{tabular} & \begin{tabular}[c]{@{}c@{}}0.0133\\ (0.9459)\end{tabular} & 0           & 0           & 0           \\
			&                     &                     & 80  & \begin{tabular}[c]{@{}c@{}}0.0013\\ (0.9478)\end{tabular} & \begin{tabular}[c]{@{}c@{}}0.0013\\ (0.9422)\end{tabular} & \begin{tabular}[c]{@{}c@{}}0.0012\\ (0.9650)\end{tabular} & \begin{tabular}[c]{@{}c@{}}0.0034\\ (0.9489)\end{tabular} & \begin{tabular}[c]{@{}c@{}}0.0037\\ (0.9507)\end{tabular} & \begin{tabular}[c]{@{}c@{}}0.0024\\ (0.9461)\end{tabular} & \begin{tabular}[c]{@{}c@{}}0.0084\\ (0.9463)\end{tabular} & 0           & 0           & 0           \\ \hline
		\end{tabular}
	\end{sidewaystable}
	
	Next, we consider the setting of unspecified group numbers.
	We implement the QIC introduced in Section \ref{sec:group_number} in the main text, with $\kappa$ set as the same with Section \ref{subsec:model_set}.
	The results are shown in Table \ref{tbl:simu_q3_qic}.
	When the group numbers are under-specified (i.e., $G_{l} = 2, ~ l=1,2,3$), the node-wise RMSEs are large and do not decrease as the sample sizes increase.
	On the contrary, when the group numbers are correctly- or over-specified, the node-wise RMSEs show a decreasing pattern as the theoretical results show.
	These findings are consistent with those for two dimensional tensor in Section \ref{subsec:simu_res} of the main text.
	
	
	\begin{sidewaystable}[]
		\caption{Simulation results with pre-specified group numbers as well as the QIC selection group numbers $\wh G_l, ~l = 1,2, 3$. The true group numbers are set as $G_{1,0} = G_{2,0}= 3$ and $G_{3,0}=2$. The node-wise RMSEs of different estimators are denoted as $\wh\blambda^{(l)}_{\text{all}}, \bzeta^{(l)}_{\text{all}}, \balpha_{\text{all}}$ for $l=1,2,3$.}\label{tbl:simu_q3_qic}
		\begin{tabular}{c|c|c|c|ccc|ccccccc|ccc}
			\hline
			$N_1$               & $N_2$               & $N_3$               & $T$                 & $G_1$   & $G_2$   & $G_3$   & $\wh\blambda^{(1)}_{\text{all}}$ & $\wh\blambda^{(2)}_{\text{all}}$ & $\wh\blambda^{(3)}_{\text{all}}$ & $\wh \bzeta^{(1)}_{\text{all}}$ & $\wh \bzeta^{(2)}_{\text{all}}$ & $\wh \bzeta^{(3)}_{\text{all}}$ & $\wh\balpha_{\text{all}}$ & $\wh\xi_1$ & $\wh\xi_2$ & $\wh\xi_3$ \\ \hline
			\multirow{5}{*}{20} & \multirow{5}{*}{20} & \multirow{5}{*}{20} & \multirow{5}{*}{40} & \multicolumn{3}{c|}{Oracle} & 0.0037                           & 0.0038                           & 0.0031                           & 0.0052                          & 0.0046                          & 0.0040                          & 0.0058                    & -          & -          & -          \\
			&                     &                     &                     & 2       & 2       & 2       & 0.0559                           & 0.0805                           & 0.0551                           & 0.0361                          & 0.0468                          & 0.0051                          & 0.1400                    & 0.2500     & 0.2080     & 0          \\
			&                     &                     &                     & 3       & 3       & 2       & 0.0037                           & 0.0038                           & 0.0031                           & 0.0052                          & 0.0046                          & 0.0040                          & 0.0174                    & 0.1403     & 0.0142     & 0          \\
			&                     &                     &                     & 4       & 4       & 4       & 0.0047                           & 0.0048                           & 0.0047                           & 0.0061                          & 0.0060                          & 0.0063                          & 0.0123                    & 0          & 0          & 0          \\
			&                     &                     &                     & $\wh G_1$ & $\wh G_2$ & $\wh G_3$ & 0.0037                           & 0.0038                           & 0.0031                           & 0.0052                          & 0.0046                          & 0.0040                          & 0.0058                    & -          & -          & -          \\ \hline
			\multirow{5}{*}{20} & \multirow{5}{*}{20} & \multirow{5}{*}{20} & \multirow{5}{*}{80} & \multicolumn{3}{c|}{Oracle} & 0.0014                           & 0.0017                           & 0.0017                           & 0.0034                          & 0.0034                          & 0.0026                          & 0.0037                    & -          & -          & -          \\
			&                     &                     &                     & 2       & 2       & 2       & 0.1140                           & 0.0571                           & 0.0580                           & 0.0343                          & 0.0213                          & 0.0041                          & 0.1447                    & 0.2449     & 0.1959     & 0          \\
			&                     &                     &                     & 3       & 3       & 2       & 0.0037                           & 0.0028                           & 0.0028                           & 0.0040                          & 0.0038                          & 0.0026                          & 0.0065                    & 0          & 0          & 0          \\
			&                     &                     &                     & 4       & 4       & 4       & 0.0040                           & 0.0033                           & 0.0038                           & 0.0046                          & 0.0042                          & 0.0043                          & 0.0099                    & 0          & 0          & 0          \\
			&                     &                     &                     & $\wh G_1$ & $\wh G_2$ & $\wh G_3$ & 0.0014                           & 0.0017                           & 0.0017                           & 0.0034                          & 0.0034                          & 0.0026                          & 0.0037                    & -          & -          & -          \\ \hline
		\end{tabular}
	\end{sidewaystable}

\subsection{Comparison with Existing Methods}

		\subsubsection{Comparing with Estimation Using A Series of Matrices Models}\label{subsec:compare_2d}
	
	To enhance the difference between our proposed GTNAR model and a series of matrix models, we conduct an experiment when $q=3$ to show the fundamental role of the intrinsic inner-tensor data information.
	Specifically, we generate data using the same setting as in Appendix \ref{subsec:simu_q3}.
	The total time length is set to be $T_{train} + T_{test}$, and we use the first $T_{train}$ as training set, while leaving the following $T_{test}$ as the testing set.
	By implementing Algorithm \ref{alg:gmnar}, we obtain the in-sample and out-of-sample predicted values $\wh \cY_t^{\text{GTNAR}}$.
	For comparison, we divide the data by the third dimension into $N_3$ parallel data slices. Each slice contains a matrix-valued time series.
	By algorithm \ref{alg:gmnar_q2}, we can estimate the two-dimensional parameters for each series slice.
	Take the $i_3$ slice for example,
	we can calculate the in-sample fitted matrix $\wh\bY_{\cdot \cdot i_3, t}, (t \in [T_{train}])$ and out-of-sample prediction values $\wh\bY_{\cdot \cdot i_3, t}, (t = T_{train}+1, \cdots T_{train} + T_{test})$ for each slice.
	Then we aggregate the fitted and predicted values in all $N_3$ slices, obtaining the in-sample and out-of-sample predicted tensor $\wh\cY_t^{\text{MatSlice}}$.
	We call this method by ``Matrix Slices'' method.
	Using the same metrics as in Appendix \ref{subsec:mis_spec}, we calculate ReMSPE for both methods regarding in-sample and out-of-sample performance.
	The experiments are both repeated for $R = 100$ times.
	The results are shown in Table \ref{tbl:compare_2d}.
	GTNAR exhibits consistent improvement in both in-sample and out-of-sample prediction accuracy as sample size increases.
	In contrast, Matrix Slices shows no systematic improvement trend, with error magnitudes actually increasing substantially at larger $T$.
	Compared with the Matrix Slices method, our proposed GTNAR shows a superior prediction performance.
	These results show that the GTNAR fundamentally differs from simple matrix slicing approaches.
	The error gaps reveals critical information loss inherent in slice-based approximations when the inherent data is generated from the tensor-valued model.

	\begin{table}[]
		\centering
		\caption{In-sample and out-of-sample ReMSPE of GTNAR and the matrix slices estimation methods.}
		\label{tbl:compare_2d}
		\begin{tabular}{ccc|c|cc|cc}
			\hline
			\multirow{2}{*}{$N_1$} & \multirow{2}{*}{$N_2$} & \multirow{2}{*}{$N_3$} & \multirow{2}{*}{$T$} & \multicolumn{2}{c|}{\textbf{GTNAR}}                                             & \multicolumn{2}{c}{\textbf{Matrix Slices}}                                      \\ \cline{5-8}
			&                        &                        &                      & {$\text{ReMSPE}_{\textup{tr}}$} & {$\text{ReMSPE}_{\textup{te}}$} & {$\text{ReMSPE}_{\textup{tr}}$} & {$\text{ReMSPE}_{\textup{te}}$} \\ \hline
			\multirow{3}{*}{20}    & \multirow{3}{*}{20}    & \multirow{3}{*}{20}    & 10                   & 0.6739                                 & 0.6549                                 & 0.8836                                 & 1.0319                                 \\
			&                        &                        & 40                   & 0.5882                                 & 0.5871                                 & 0.7692                                 & 0.8003                                 \\
			&                        &                        & 80                   & 0.0234                                 & 0.0049                                 & 2.4769                                 & 4.3342                                 \\ \hline
			\multirow{3}{*}{30}    & \multirow{3}{*}{30}    & \multirow{3}{*}{30}    & 10                   & 0.6261                                 & 0.6173                                 & 0.8789                                 & 1.0281                                 \\
			&                        &                        & 40                   & 0.4710                                 & 0.4660                                 & 0.9787                                 & 1.7442                                 \\
			&                        &                        & 80                   & 0.0066                                 & 0.0006                                 & 4.8092                                 & 2.9876                                 \\ \hline
		\end{tabular}
	\end{table}

\subsubsection{Comparing with Other Methods}\label{subsec:simu_compare}

In this section, we conduct the prediction and compare the accuracy with a number of competing methods.
Denote $T_{train}$ and $T_{test}$ as the time length for training set and the testing set.
We generate the simulation data when $q=2$, and the generation scheme is the same as that in Section \ref{subsec:model_set} in the main text.
We set the sample sizes as $(N_1, N_2, T_{train}) \in \{ (30, 20, 20), (50, 40, 30), (80, 60, 40), (100, 80, 50), (120, 100, 60), (160, 120, 70), (200, 150, 80),\\ (250, 180, 90)\}$.
For each sample size setting, we set the following $T_{test} = 0.5 T_{train}$ as the test set.
To evaluate the prediction accuracy, we repeat the experiments for each setting for $R=50$ times, and 
we compute average relative mean square prediction error (ReMSPE).
Specifically, first calculate mean square prediction error as
\begin{align*}
	\textup{MSPE}^r = (N_1 N_2 T_{test})^{-1} \sum_{t = t_{test, 0}}^{t_{test, 0} + T_{test}} \sum_{i, j} (\wh Y_{ij, t}^r - Y_{ij, t}^r)^2,
\end{align*}
where $\wh Y_{ij, t}^r$ is the predicted {response} in the $r$th replicate, {and $t_{test, 0} = T_{train} + 1$ is the starting time point.}
Then, calculate
\begin{align*}
	\textup{MSPE}^r_0 = (N_1 N_2 T_{test})^{-1} \sum_{t = t_{test, 0}}^{t_{test, 0} + T_{test}} \sum_{i, j} (Y_{ij, t}^r - \wh \mu_{ij, train}^r)^2,
\end{align*}
where $\wh \mu_{ij, train}^r = T_{trian}^{-1} \sum_{t = 1}^{T_{train}} Y_{ij, t}^r$ as the mean response in the training set.
Subsequently, we calculate the average ReMSPE as $\textup{ReMSPE} = R^{-1} \sum_r \textup{MSPE}^r/ \textup{MSPE}_0^r$.

\begin{figure}
	\centering
	\includegraphics[width=0.5\textwidth]{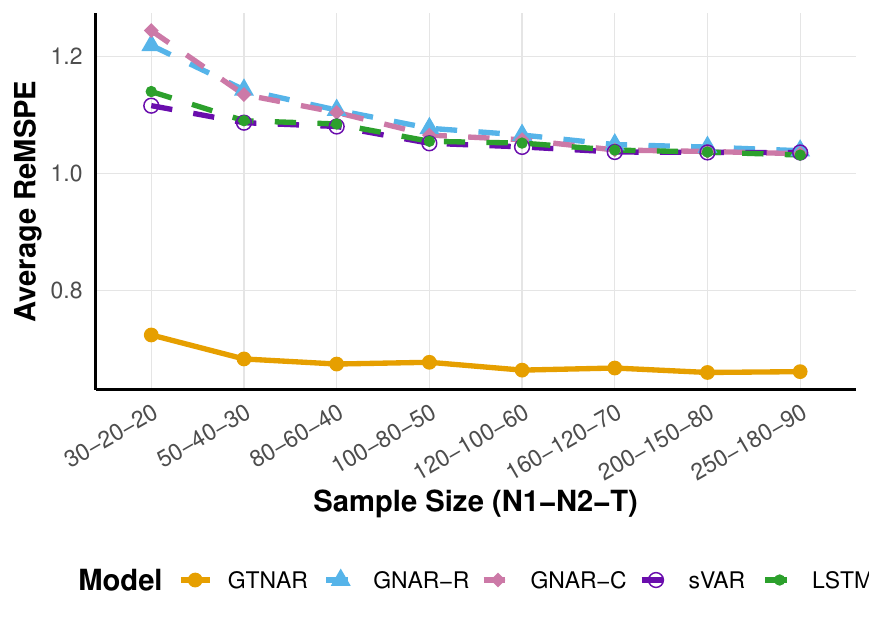}
	\caption{Average ReMSPEs of GTNAR, GNAR-R, GNAR-C, sVAR and LSTM under different sample sizes.}
	\label{fig:simu_compare_line}
\end{figure}

Specifically, we compare the prediction accuracy estimated by the proposed algorithm with that estimated by
the sparse VAR method (sVAR, \cite{nicholson2020high}), the multilinear tensor regression model (BiTR, \cite{hoff2015multilinear}), the group NAR model (GNAR, \cite{zhu2023simultaneous}), and the deep learning method for time series, LSTM \citep{hochreiter1997long}.
For the sVAR model, we first vectorize the matrix response $\Y_t$ to be in
the vector form as $\vec(\Y_t)$. Subsequently, we apply the sVAR method in the \textsf{bigtime} package to $\{\vec(\Y_t)\}$ to obtain the model estimation and prediction result.
For BiTR model, we use $\Y_t$ as the response variable, and set the explanatory variables as the concatenation by the row network term $\bW^{(1)} \bY_{t-1}$, the and column network term $\bY_{t-1} \bW^{(2)}$, the lag term $\bY_{t-1}$ and the covariates $\bX_{t}^{(1)}$, $\bX_{t}^{(2)}$.
	We apply the alternating least squares algorithm proposed by \cite{hoff2015multilinear} to estimate the model.
For the GNAR model proposed by \cite{zhu2023simultaneous}, we note that it
	only involves a single network 
	and can only be applied for the vector time series data. Therefore we make prediction with their model form by involving one network matrix a time.
	To involve the user network (i.e., row network), we apply the GNAR model in the following form
	\begin{align}
		Y_{ij, t} = \sum_{k = 1}^{N_1} \beta_{g_i^{(1)} g_k^{(1)}} w_{1, ik} Y_{kj,t-1} + \nu_{g_i^{(1)}} Y_{ij, t-1} + \bx_i^{(1)\top} \bzeta_{g_i^{(1)}} + \ve_{ij,t},\label{eq:zhu_matrix}
	\end{align}
	and estimate the model parameters and make prediction for each $j\in [N_2]$.
	The model prediction performance is denoted as GNAR-R in Table \ref{tbl:pred_compare}.
	Similarly, to incorporate the spatial network (i.e., column network), we estimate the following GNAR model as
	\begin{align*}
		 Y_{ij, t} = \sum_{k=1}^{N_2} \beta_{g_k^{(2)} g_j^{(2)} } w_{2, kj} Y_{ik, t-1} + \nu_{g_j^{(2)}} Y_{ij, t-1} + \bx_j^{(2)\top} \bzeta_{g_j^{(2)}} + \ve_{ij, t},
	\end{align*}
	for each $i\in [N_1]$, whose prediction performance is denoted as GNAR-C in Figure \ref{fig:simu_compare_line}.
	For the above two models, we use the open source code provided by \cite{zhu2023simultaneous} to obtain the result.
	Lastly, we apply the LSTM method  \citep{hochreiter1997long}
	to our data $\{ (\bx_{it}^{(1)} \in \mR^{p_1}, \bx_{jt}^{(2)} \in \mR^{p_2}, Y_{ij, t} ): i \in [N_1], j \in [N_2], t \in [T]\}$.
	During training, we set the learning rate as 0.05, the dimension of hidden layers as 64, and set the number of iteration as 100.
	The momentum coefficient is set as 0.5.

We remark that due to the suboptimal estimation accuracy of BiTR, which exhibits orders of magnitude differences compared to others, we present the ReMSPEs of GTNAR, GNAR-R, GNAR-C, sVAR and LSTM in Figure \ref{fig:simu_compare_line} for visual clarity. 
The results for BiTR are provided in Table \ref{tbl:simu_compare} for detailed examination.
Based on the simulation results, when the data is generated using the GTNAR model, the prediction accuracy of all alternative methods proves suboptimal. 
Even the LSTM model, which is recognized for its strong fitting capabilities by its non-linear structure, fails to achieve satisfactory prediction accuracy. 
Furthermore, in Appendix \ref{subsec:predict_compare}, we conduct rolling-window prediction comparisons across six methods using the Yelp dataset, which also demonstrate the superiority of our proposed approach.

\begin{table}[]
	\centering
	\caption{Average ReMSPEs of GTNAR and BiTR under different sample sizes.}
	\label{tbl:simu_compare}
	\begin{tabular}{cc|c|cc}
		\hline
		$N_1$ & $N_2$ & $T_{train}$ & \textbf{GTNAR} & \textbf{BiTR} \\ \hline
		30    & 20    & 20          & 0.7237            & 94.8849       \\
		50    & 30    & 30          & 0.6827           & 160.6119      \\
		80    & 60    & 40          & 0.6740       & 219.2674      \\
		100   & 80    & 50          & 0.6770      & 138.1804      \\
		120   & 100   & 60          & 0.6636          & 105.4973      \\
		160   & 120   & 70          & 0.6672           & 139.3628      \\
		200   & 150   & 80          & 0.6596            & 101.1558      \\
		250   & 180   & 90          & 0.6610         & 28.5175       \\ \hline
	\end{tabular}
\end{table}

\section{Additional Yelp Data Analysis}\label{sec:add_yelp}

\subsection{Additional Results for Yelp Data Analysis}\label{subsec:yelp_append_des}

\begin{figure}[]
	\begin{center}
		\includegraphics[width=\textwidth]{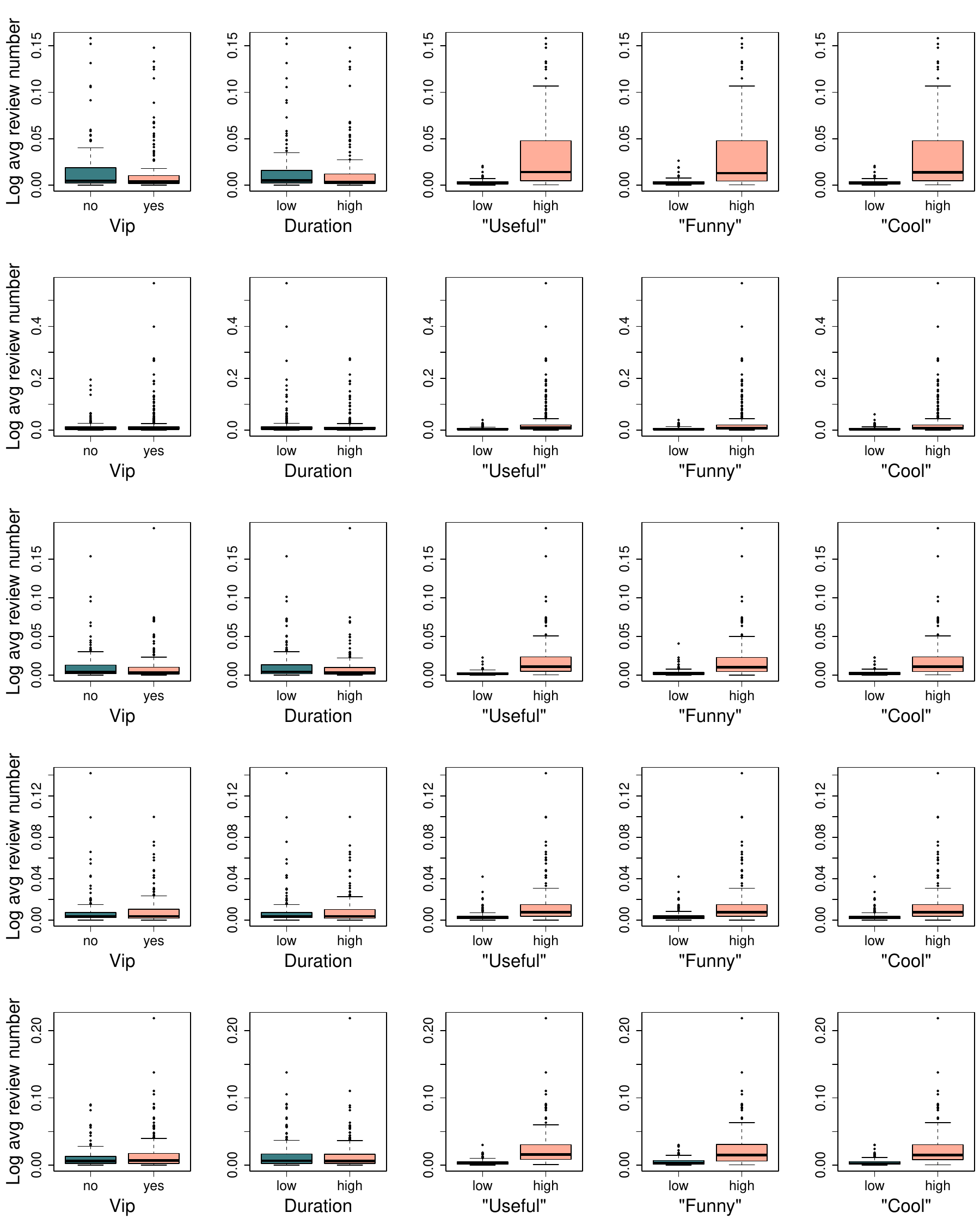}
		\caption{\small Boxplots for response variable with regards to the users' covariates in Charlotte, Las Vegas, Phoenix, Scottsdale, and Toronto (displayed in each line). Specifically, the y-axis is the log-transformed average review number given by each user throughout the time span. The x-axis shows the corresponding high or low level for each covariate, separated by the median for continuous variables. 
		}
		\label{fig:box}
	\end{center}
\end{figure}

We visualize the relationship between these user-related covariates and the response variable in Figure \ref{fig:box}. 
The plot reveals that users who receive more tags for their reviews tend to be motivated to contribute more reviews.
Notably, VIP users in Scottsdale and Toronto tend to write more reviews, whereas VIP users in Charlotte exhibit comparatively less activity.
The estimation results for city Scottsdale and Toronto are provided in Table \ref{tbl:real_res_2_yelp}.

	\begin{sidewaystable}[]
	\centering
	\caption{Estimation results for Scottsdale and Toronto. The $p$-values are shown in the parenthesis.}
	\label{tbl:real_res_2_yelp}
	\scalebox{0.8}{
		\begin{tabular}{c|cccc|cccc}
			\hline
			Parameters                                    & \multicolumn{4}{c|}{Scottsdale ($N_1 = 391, ~ N_2 = 60$)}                                                                                                                                                                                                                                                                           & \multicolumn{4}{c}{Toronto ($N_1 = 462, ~ N_2 = 56$)}                                                                                                                                                                                                                                                                             \\ \hline
			\multirow{2}{*}{}                             & \multicolumn{2}{c|}{$\lambda_{g^{(1)}}^{(1)}$}                                                                                                                               & \multicolumn{2}{c|}{$\lambda_{g^{(2)}}^{(2)}$}                                                                                                          & \multicolumn{2}{c|}{$\lambda_{g^{(1)}}^{(1)}$}                                                                                                                              & \multicolumn{2}{c}{$\lambda_{g^{(2)}}^{(2)}$}                                                                                                          \\ \cline{2-9}
			& \begin{tabular}[c]{@{}c@{}}0.067\\ (\textless{}0.001)\end{tabular}  & \multicolumn{1}{c|}{\begin{tabular}[c]{@{}c@{}}0.011\\ (\textless{}0.001)\end{tabular}}  & \begin{tabular}[c]{@{}c@{}}0.069\\ (\textless{}0.001)\end{tabular}  & \begin{tabular}[c]{@{}c@{}}0.239\\ (\textless{}0.001)\end{tabular} & \begin{tabular}[c]{@{}c@{}}0.013\\ (\textless{}0.001)\end{tabular} & \multicolumn{1}{c|}{\begin{tabular}[c]{@{}c@{}}0.037\\ (\textless{}0.001)\end{tabular}}  & \begin{tabular}[c]{@{}c@{}}0.305\\ (\textless{}0.001)\end{tabular} & \begin{tabular}[c]{@{}c@{}}0.107\\ (\textless{}0.001)\end{tabular} \\ \hline
			& \multicolumn{2}{c|}{$\bzeta_{g^{(1)}}^{(1)}$}                                                                                                                                & \multicolumn{2}{c|}{$\bzeta_{g^{(2)}}^{(2)}$}                                                                                                         & \multicolumn{2}{c|}{$\bzeta_{g^{(1)}}^{(1)}$}                                                                                                                               & \multicolumn{2}{c}{$\bzeta_{g^{(2)}}^{(2)}$}                                                                                                         \\
			Intercept                                     & \begin{tabular}[c]{@{}c@{}}-0.002\\ (0.175)\end{tabular}            & \multicolumn{1}{c|}{\begin{tabular}[c]{@{}c@{}}0.002\\ (\textless{}0.001)\end{tabular}}  & \begin{tabular}[c]{@{}c@{}}-0.001\\ (\textless{}0.001)\end{tabular} & \begin{tabular}[c]{@{}c@{}}0.003\\ (\textless{}0.001)\end{tabular} & \begin{tabular}[c]{@{}c@{}}-0.009\\ (0.001)\end{tabular}           & \multicolumn{1}{c|}{\begin{tabular}[c]{@{}c@{}}0.009\\ (\textless{}0.001)\end{tabular}}  & \begin{tabular}[c]{@{}c@{}}0.016\\ (\textless{}0.001)\end{tabular} & \begin{tabular}[c]{@{}c@{}}-0.001\\ (0.274)\end{tabular}           \\
			$\zeta_{\text{dur}}^{(1)}$ / $\zeta_{\text{star}}^{(2)}$ & \begin{tabular}[c]{@{}c@{}}0.009\\ (\textless{}0.001)\end{tabular}  & \multicolumn{1}{c|}{\begin{tabular}[c]{@{}c@{}}-0.003\\ (\textless{}0.001)\end{tabular}} & \begin{tabular}[c]{@{}c@{}}0.002\\ (\textless{}0.001)\end{tabular}  & \begin{tabular}[c]{@{}c@{}}0.004\\ (\textless{}0.001)\end{tabular} & \begin{tabular}[c]{@{}c@{}}$10^{-4}$\\ (0.453)\end{tabular}            & \multicolumn{1}{c|}{\begin{tabular}[c]{@{}c@{}}-0.006\\ (\textless{}0.001)\end{tabular}} & \begin{tabular}[c]{@{}c@{}}0.003\\ (0.209)\end{tabular}            & \begin{tabular}[c]{@{}c@{}}0.003\\ (\textless{}0.001)\end{tabular} \\
			$\zeta_{\text{vip}}^{(1)}$ / $\zeta_{\text{num}}^{(2)}$  & \begin{tabular}[c]{@{}c@{}}0.002\\ (0.037)\end{tabular}             & \multicolumn{1}{c|}{\begin{tabular}[c]{@{}c@{}}0.001\\ (\textless{}0.001)\end{tabular}}  & \begin{tabular}[c]{@{}c@{}}0.003\\ (\textless{}0.001)\end{tabular}  & \begin{tabular}[c]{@{}c@{}}0.002\\ (\textless{}0.001)\end{tabular} & \begin{tabular}[c]{@{}c@{}}0.001\\ (\textless{}0.001)\end{tabular} & \multicolumn{1}{c|}{\begin{tabular}[c]{@{}c@{}}0.001\\ (0.155)\end{tabular}}             & \begin{tabular}[c]{@{}c@{}}0.003\\ (0.009)\end{tabular}            & \begin{tabular}[c]{@{}c@{}}0.005\\ (\textless{}0.001)\end{tabular} \\
			$\zeta_{\text{use}}^{(1)}$                          & \begin{tabular}[c]{@{}c@{}}0.227\\ (\textless{}0.001)\end{tabular}  & \multicolumn{1}{c|}{\begin{tabular}[c]{@{}c@{}}0.023\\ (\textless{}0.001)\end{tabular}}  &                                                                     &                                                                    & \begin{tabular}[c]{@{}c@{}}0.032\\ (\textless{}0.001)\end{tabular} & \multicolumn{1}{c|}{\begin{tabular}[c]{@{}c@{}}0.069\\ (\textless{}0.001)\end{tabular}}  &                                                                    &                                                                    \\
			$\zeta_{\text{fun}}^{(1)}$                          & \begin{tabular}[c]{@{}c@{}}-0.360\\ (\textless{}0.001)\end{tabular} & \multicolumn{1}{c|}{\begin{tabular}[c]{@{}c@{}}-0.025\\ (\textless{}0.001)\end{tabular}} &                                                                     &                                                                    & \begin{tabular}[c]{@{}c@{}}-0.002\\ (0.401)\end{tabular}           & \multicolumn{1}{c|}{\begin{tabular}[c]{@{}c@{}}0.013\\ (\textless{}0.001)\end{tabular}}  &                                                                    &                                                                    \\
			$\zeta_{\text{cool}}^{(1)}$                         & \begin{tabular}[c]{@{}c@{}}0.176\\ (\textless{}0.001)\end{tabular}  & \multicolumn{1}{c|}{\begin{tabular}[c]{@{}c@{}}0.028\\ (\textless{}0.001)\end{tabular}}  &                                                                     &                                                                    & \begin{tabular}[c]{@{}c@{}}0.002\\ (0.483)\end{tabular}            & \multicolumn{1}{c|}{\begin{tabular}[c]{@{}c@{}}0.041\\ (\textless{}0.001)\end{tabular}}  &                                                                    &                                                                    \\ \hline
			\multirow{3}{*}{}                             & \multicolumn{4}{c|}{$\balpha^\top \in \mR^{G_1 \times G_2}$}                                                                                                                                                                                                                                                       & \multicolumn{4}{c}{$\balpha^\top \in \mR^{G_1 \times G_2}$}                                                                                                                                                                                                                                                      \\
			& \multicolumn{2}{c}{\begin{tabular}[c]{@{}c@{}}0.071\\ (\textless{}0.001)\end{tabular}}                                                                         & \multicolumn{2}{c|}{\begin{tabular}[c]{@{}c@{}}0.023\\ (\textless{}0.001)\end{tabular}}                                                  & \multicolumn{2}{c}{\begin{tabular}[c]{@{}c@{}}0.055\\ (\textless{}0.001)\end{tabular}}                                                                        & \multicolumn{2}{c}{\begin{tabular}[c]{@{}c@{}}0.290\\ (\textless{}0.001)\end{tabular}}                                                  \\
			& \multicolumn{2}{c}{\begin{tabular}[c]{@{}c@{}}0.235\\ (\textless{}0.001)\end{tabular}}                                                                         & \multicolumn{2}{c|}{\begin{tabular}[c]{@{}c@{}}0.051\\ (\textless{}0.001)\end{tabular}}                                                  & \multicolumn{2}{c}{\begin{tabular}[c]{@{}c@{}}0.033\\ (\textless{}0.001)\end{tabular}}                                                                        & \multicolumn{2}{c}{\begin{tabular}[c]{@{}c@{}}0.096\\ (\textless{}0.001)\end{tabular}}                                                  \\ \hline
		\end{tabular}
	}
\end{sidewaystable}

	\subsection{Ablation Experiment}\label{subsec:simu_ablation}

	In this subsection, we use ablation experiments to show the necessity of each component in our GTNAR model.
	We implement both the GTNAR model \eqref{eq:model00} and the Mixed GTNAR model \eqref{eq:model_int_2} to analyze the Yelp dataset.
	The evaluation metric is the similar as that in Appendix \ref{subsec:simu_compare} yet under a rolling window.
	Specifically, first calculate mean square prediction error as
	\begin{align*}
		\textup{MSPE} = (N_1 N_2 T_{test})^{-1} \sum_{t = t_{test, 0}}^{t_{test, 0} + T_{test}} \sum_{i, j} (\wh Y_{ij, t} - Y_{ij, t})^2,
	\end{align*}
	Then, calculate
	\begin{align*}
		\textup{MSPE}_0 = (N_1 N_2 T_{test})^{-1} \sum_{t = t_{test, 0}}^{t_{test, 0} + T_{test}} \sum_{i, j} (Y_{ij, t} - \wh \mu_{ij, train})^2,
	\end{align*}
	where $\wh \mu_{ij, train} = T_{trian}^{-1} \sum_{t = t_{test, 0} - T_{train}}^{t_{test, 0} -1} Y_{ij, t}$ as the mean response in the training set.
	Subsequently, we calculate the ReMSPE as $\textup{ReMSPE} = \textup{MSPE}/ \textup{MSPE}_0$.
	Recall the GTNAR model 
	\begin{align}
		\Y_t = \underbrace{ (\L^{(1)} \W^{(1)} ) \Y_{t-1} + \Y_{t-1} (\W^{(2)} \L^{(2)})}_{\text{network terms}} +\underbrace{ \A \odot \Y_{t-1}}_{\text{momentum term}} + \underbrace{\bbeta_{X_1, t}^{(1)} \one_{N_2}^\top + \one_{N_1} \bbeta_{X_2, t}^{(2)\top}}_{\text{covariates terms}} + \bE_t,\label{eq:model_abla}
	\end{align}
	For the ablation experiment, we remove the network terms, the lag term, and the covariates terms from GTNAR \eqref{eq:model_abla} and perform predictions, respectively.
	For performance evaluation, we predict the out-of-sample review number in the same rolling window settings, and calculate the ReMSPE as defined in Appendix \ref{subsec:simu_compare}.
	The prediction results are shown in Table \ref{tbl:ablation}, from which we can see that both GTNAR and Mixed GTNAR consistently achieve the highest prediction accuracy across most of the five evaluation testing time periods.
	This performance superiority is particularly evident in Charlotte, Las Vegas, and Scottsdale, where these models outperform all ablated versions at every starting time point.
	The ablation study reveals performance deterioration when each component is removed, validating the contribution of each model component (network terms, momentum term, and covariates terms) to GTNAR's robust predictive capability.
	
	\begin{table}[]
		\centering
		\caption{ReMSPEs of the ablation study in five cities under the rolling window setting.}
		\label{tbl:ablation}
		\begin{tabular}{c|c|ccccc}
			\hline
			&                     & \multicolumn{5}{c}{Starting time point}                                                                                                                       \\ \cline{3-7}
			&                     & 2010-Q1                       & 2010-Q2                       & 2010-Q3                       & 2010-Q4                       & 2011-Q1                       \\ \hline
			
			&\textbf{ GTNAR } &  {0.8386} &  {0.8132}&  \textbf{0.8361} &  \textbf{0.8646} &  \textbf{0.8951} \\
			&\textbf{Mixed GTNAR}&\textbf{0.8228}&\textbf{0.7725}&0.8382&0.9019&0.9629\\
			\cline{2-7}
			& no network terms    & 0.8421                        & 0.8140                    & 0.8392                      &0.8689                      & 0.8979                  \\
			& no momentum term         & {\color[HTML]{000000} 0.8717} & {\color[HTML]{000000} 0.8199} & {\color[HTML]{000000} 0.8629} & {\color[HTML]{000000} 0.8824} & {\color[HTML]{000000} 0.9624} \\
			\multirow{-4}{*}{Charlotte}  & no covariates terms & 0.8539                       & 0.8297                        & 0.8565                        & 0.8772                     & 0.9131                        \\ \hline
			& \textbf{ GTNAR }  & {0.8336} & {0.7811} & {0.8549} & {0.9102} & \textbf{0.9601} \\
			&\textbf{Mixed GTNAR}&\textbf{0.8228}&\textbf{0.7725}&\textbf{0.8382}&\textbf{0.9019}&0.9629\\
			\cline{2-7}
			& no network terms    & 0.8373                        & 0.8021                   & {0.8519}                    & 0.9237                    & 0.9784                    \\
			& no momentum term         & {\color[HTML]{000000} 0.8238} & {\color[HTML]{000000} 0.7913} & {\color[HTML]{000000} 0.8520} & {\color[HTML]{000000} 0.9145} & {\color[HTML]{000000} 0.9729} \\
			\multirow{-4}{*}{Las Vegas}  & no covariates terms & {\color[HTML]{000000} 0.8119} & 0.7918                   & {0.8395  }              & { 0.9070}                    & 0.9691                     \\ \hline
			& \textbf{ GTNAR }  & {1.1625} & \textbf{1.0784} & {0.9419} & \textbf{0.8730} & {0.8581} \\
			& \textbf{Mixed GTNAR}& 1.1406& 1.0888& 0.9571 & 0.8822 & 0.8582 \\
			\cline{2-7}
			& no network terms    & \textbf{1.1048} & {\color[HTML]{000000} 1.0793} & \textbf{0.9393} & {\color[HTML]{000000}0.8677} & {\color[HTML]{000000} 0.8616} \\
			& no momentum term         & {\color[HTML]{000000} 1.1426} & {\color[HTML]{000000} 1.0859} & {\color[HTML]{000000} 0.9424} & {\color[HTML]{000000} 0.8731} & \textbf{0.8554} \\
			\multirow{-4}{*}{Phoenix}    & no covariates terms & {\color[HTML]{000000} 1.1310	} & {\color[HTML]{000000} 1.1023} & {\color[HTML]{000000} 0.9542} & {\color[HTML]{000000} 0.8877} & {\color[HTML]{000000} 0.8664} \\ \hline
			& \textbf{ GTNAR }  & \textbf{0.8725} & \textbf{0.8588} & \textbf{0.8535} & \textbf{0.8757} & \textbf{0.8415} \\
			&\textbf{Mixed GTNAR}& 0.8986& 0.8757& 0.8629& 0.8859& 0.8581\\
			\cline{2-7}
			& no network terms    &0.8761                      & 0.8629                   & 0.8587                    &0.8771                     & 0.8441                      \\
			& no momentum term         & 0.8992                        & {\color[HTML]{000000} 0.8699} & {\color[HTML]{000000} 0.8545} & 0.9057                        & 0.8931                        \\
			\multirow{-4}{*}{Scottsdale} & no covariates terms & 0.8831                      & 0.8685                     & 0.8636                    & 0.8854                  & 0.8492                      \\ \hline
			&\textbf{ GTNAR }  & {1.5779} & \textbf{1.4670} & \textbf{1.3785} & \textbf{1.2241} & \textbf{1.0669} \\
			&\textbf{Mixed GTNAR} &1.5829& 1.5279& 1.3961& 1.2313& 1.0773\\
			\cline{2-7}
			& no network terms    & \textbf{1.5751  }                   & 1.4678                     & 1.3863                      & 1.2334                    & 1.0681                   \\
			& no momentum term         & {\color[HTML]{000000} 1.5783} & {\color[HTML]{000000} 1.4787} & 1.4079                        & {\color[HTML]{000000} 1.2362} & {\color[HTML]{000000} 1.0766} \\
			\multirow{-4}{*}{Toronto}    & no covariates terms & 1.5995                   & 1.4942                    &  1.4019                    & 1.2513                     &1.0877                       \\ \hline
		\end{tabular}
	\end{table}

	\subsection{Prediction Accuracy of Comparing Methods}\label{subsec:predict_compare}
	
	In this section, we conduct the prediction for the Yelp dataset and compare the accuracy with a number of competing methods.
	To illustrate the prediction performance of our proposed GTNAR model,
	we take a rolling window prediction setting.
	Specifically, we set the training length as $T_{train} = 25$ from the first data point, and set $T_{test} = 5$.
	Then we use a sliding window approach to evaluate the prediction performance by
	calculating the Root Mean Square Prediction Error (RMSPE) with moving one quarter in each step, which is calculated by 
	\begin{align*}
		\textup{RMSPE} = \Big\{(N_1 N_2 T_{test})^{-1} \sum_{t = t_{test, 0}}^{t_{test, 0} + T_{test}} \sum_{i, j} (\wh Y_{ij, t} - Y_{ij, t})^2 \Big\}^{1/2},
	\end{align*}
	where $\wh Y_{ij, t}$ is the predicted {response} from user $i$ to the district $j$ in the $t$th quarter, and $t_{test, 0}$ is the starting time point.
	The settings for the competing methods, namely, sVAR, BiTR, GNAR-R, GNAR-C and LSTM are the same as those in Appendix \ref{subsec:simu_compare}.

The prediction results are detailed in Table \ref{tbl:pred_compare}.
In comparison to the {sVAR, BiTR, GNAR-R and GNAR-C}, the GTNAR method demonstrates superior and more stable prediction accuracy across various staring points in all five cities.
{In comparison to} the LSTM model, we find that the prediction performances are comparable.
In particular,
for Charlotte, Las Vegas, Scottsdale and Toronto, the GTNAR method outperforms the LSTM with higher prediction accuracy.
For Phoenix, the prediction error of the LSTM method in the first start time point is slightly smaller than that of the GTNAR method.
We would like to remark that
other than our relative robust prediction performance, our modeling approach provides a
clearer model interpretation compared with complex nonlinear machine learning methods.

\begin{table}[htbp]
	\centering
		\caption{Prediction RMSPE in five cities under the rolling window setting.}
		\label{tbl:pred_compare}
		\begin{tabular}{c|c|ccccc}
			\hline
			\multirow{2}{*}{City} & \multirow{2}{*}{Model} & \multicolumn{5}{c}{{Start Time Point}} \\
			\cmidrule{3-7}
			& & {2010-Q1} & {2010-Q2} & {2010-Q3} & {2010-Q4} & {2021-Q1} \\
			\hline
\multirow{7}{*}{Charlotte} & \textbf{GTNAR} & 0.0848 & \textbf{0.0841} & \textbf{0.0845} & \textbf{0.0848 }&\textbf{ 0.0854} \\ & sVAR & 0.1145 & 0.1067 & 0.0999 & 0.1048 & 0.1061 \\ & BiTR & \textbf{0.0560} & 0.2329 & 0.7017 & 0.5040 & 0.6093 \\ & GNAR-R & 0.0964 & 0.1128 & 0.1227 & 0.0923 & 0.0949 \\ & GNAR-C & 2.6539 & 0.1023 & 0.2522 & 0.1080 & 0.0953 \\ & LSTM & 0.0866 & 0.0851 & 0.0863 & 0.0860 & 0.0871 \\\hline 
\multirow{7}{*}{LasVegas} & \textbf{GTNAR} & \textbf{0.0786} & \textbf{0.0760} & \textbf{0.0790} & \textbf{0.0814} & \textbf{0.0829} \\ & sVAR & 0.1118 & 0.1344 & 0.1300 & 0.1273 & 0.1442 \\ & BiTR & 1.5677 & 1.5457 & 1.6350 & 1.6465 & 1.2869 \\ & GNAR-R & 0.0862 & 0.1124 & 7.0599 & 0.6934 & 0.0849 \\ & GNAR-C & 43.9337 & 3.4841 & 37.4009 & 0.0943 & 0.2487 \\ & LSTM & 0.0789 & 0.0782 & 0.0790 & 0.0816 & 0.0837 \\\hline 
\multirow{7}{*}{Phoenix} & \textbf{GTNAR} & 0.0791 & \textbf{0.0762} & \textbf{0.0716} & \textbf{0.0690} & \textbf{0.0682} \\& sVAR & 0.0877 & 0.0881 & 0.0861 & 0.0930 & 0.0881 \\ & BiTR & 0.6108 & 1.1864 & 0.2036 & 0.2322 & 0.2437 \\ & GNAR-R & 0.0801 & 0.0796 & 0.0792 & 0.0736 & 0.0733 \\ & GNAR-C & 1.3056 & 0.0795 & 0.1082 & 0.0745 & 0.1303 \\ & LSTM & \textbf{0.0782} & 0.0790 & 0.0738 & 0.0700 & 0.0717 \\\hline 
\multirow{7}{*}{Scottsdale} & \textbf{GTNAR} & \textbf{0.0617} & \textbf{0.0614} & \textbf{0.0611} & \textbf{0.0619} & 0.0600 \\  & sVAR & 0.0772 & 0.0720 & 0.0715 & 0.0766 & 0.0730 \\& BiTR & 0.7890 & 2.0271 & 0.8091 & 0.8288 & \textbf{0.0446} \\ & GNAR-R & 0.0655 & 0.0639 & 0.0680 & 0.0638 & 0.0639 \\ & GNAR-C & 0.1423 & 0.0778 & 0.0656 & 0.0713 & 0.1265 \\ & LSTM & 0.0632 & 0.0624 & 0.0616 & 0.0624 & 0.0606 \\\hline 
\multirow{7}{*}{Toronto} & \textbf{GTNAR} & \textbf{0.0989} & \textbf{0.0972} & \textbf{0.0958 }& \textbf{0.0915 }& \textbf{0.0865 }\\ & sVAR & 0.1117 & 0.1106 & 0.1105 & 0.1067 & 0.1083 \\& BiTR & 1.3064 & 0.9224 & 0.6444 & 1.2483 & 0.9182 \\ & GNAR-R & 0.1013 & 0.1006 & 0.1013 & 0.1022 & 0.0951 \\ & GNAR-C & 5.1932 & 3.0433 & 6.2586 & 18.4431 & 0.4216 \\ & LSTM & 0.0999 & 0.0985 & 0.0969 & 0.0932 & 0.0877 \\\hline
		\end{tabular}
\end{table}

\section{Addtional Trading Data Application}\label{sec:add_realdata}

\subsection{Data Description}\label{subsec:trading_data}

In this section, we embark on an analysis of the monthly import and export volumes of goods among countries/regions, spanning the period from 1993 to 2022, comprising a total of $T = 360$ months.  The source for our multilateral trading data is the IMF-DOTS database \citep{imf2017}. To enhance the stationarity of the time series, we define the response variable, denoted by $Y_{ij,t}$, as the first-order difference in export volume, which is referred to as the ``increased export volume (IEV)'' in the subsequent discussions. IEV quantifies the change in export volume from the $i$th country/region to the $j$th country/region between the $t$th month and the preceding $(t-1)$th month.

\begin{figure}[htpb!]
\begin{center}
	\includegraphics[width=0.7\textwidth]{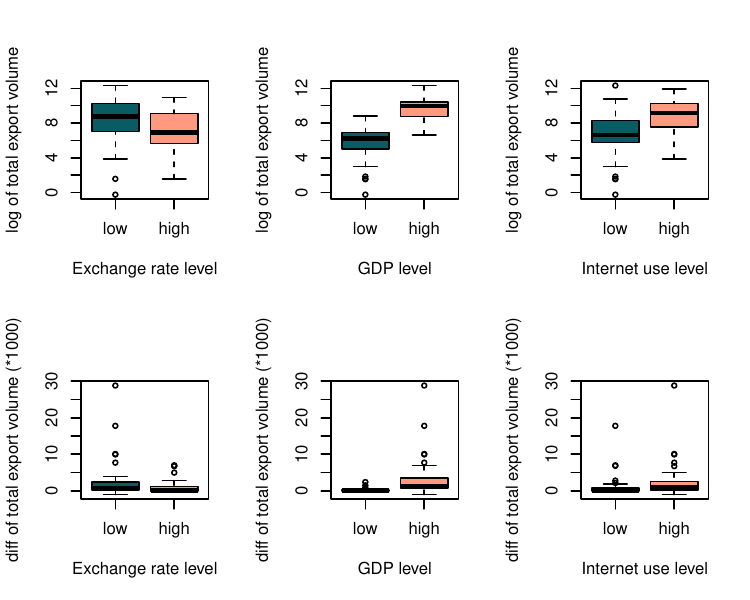}
	\caption{\small Top panels: the relationship between the total export volume in 2022 and the levels of three covariates in 2021 (the categorization into "low" and "high" levels is determined by the medians). Bottom panels:  the relationship between the first-order difference in exports between 2022 and 2021 and the levels of the same three covariates in 2021.
	}
	\label{fig:box_trade}
\end{center}
\end{figure}

{\bf Covariates.} In our analysis, we consider three essential covariates for each country or region.
Firstly, the annual gross domestic product ($x_{it, \text{gdp}}^{(1)} / x_{jt, \text{gdp}}^{(2)}$) serves as a measure of the overall size of the economy.
{Secondly, we incorporate the difference of log-annual average exchange rate to US dollars ($ x_{it, \text{exc}}^{(1)} / x_{jt, \text{exc}}^{(2)}$) in consecutive years, which influences the actual costs of exporting goods.
The influence from exchange rate toward trading pattern is underscored in prior research \citep{hooper1978effect, auboin2013relationship}.}
Lastly, we include the proportion of the population with internet access ($ x_{it, \text{net}}^{(1)} / x_{jt, \text{net}}^{(2)}$), a metric that captures the effects of the rapid development of e-commerce and its effect on cross-regional commodity circulation, as discussed in studies such as those by \cite{freund2004effect} and \cite{terzi2011impact}.
This covariate data has been sourced from both the World Bank Database$^1$ and the IMF-DOTS database, encompassing the years 1992 to 2021.
Figure \ref{fig:box_trade} depicts the accumulated response for export volume in the year 2022 versus the three covariates of export countries in 2021, which suggests that a lower exchange rate, higher GDP, or a higher proportion of internet users in 2021 are associated with potentially higher export volumes in 2022.

{\bf Spatial Networks.} The countries under study are interconnected through various international relationships. In this analysis, our focus is on the spatial network spillover effects, a phenomenon well-documented in the literature~\citep{redding2004economic,YIN2020102112}.
After filtering out countries/regions with missing data, we have a total of $N_1 = N_2 = 82$ observed units. To construct the adjacency matrices $\bA^{(1)}$ and $\bA^{(2)}$, we reference neighboring countries and territories for each country and region, as listed on Wikipedia$^2$\footnote[2]{https://en.wikipedia.org/wiki/List\_of\_countries\_and\_territories\_by\_number\_of\_land\_borders}. In these adjacency matrices, an element $a^{(1)}_{ij} = 1$ ($a^{(2)}_{ij} = 1$) signifies the presence of at least one land border between the two regions. Consequently, our response variable, $Y_{ij,t}$, is indexed by two networks, the exporter network $\bA^{(1)}$  and the importer network $\bA^{(2)}$. In particular, we note that although these two networks share the same network nodes and topology, they exert different influences on the response variable, accounting for export and import spatial spillover effects on IEV, respectively.

\subsection{Estimating Results}\label{subsec:trading_res}
\begin{figure}[htpb!]
\begin{center}
	\includegraphics[width=0.7\textwidth]{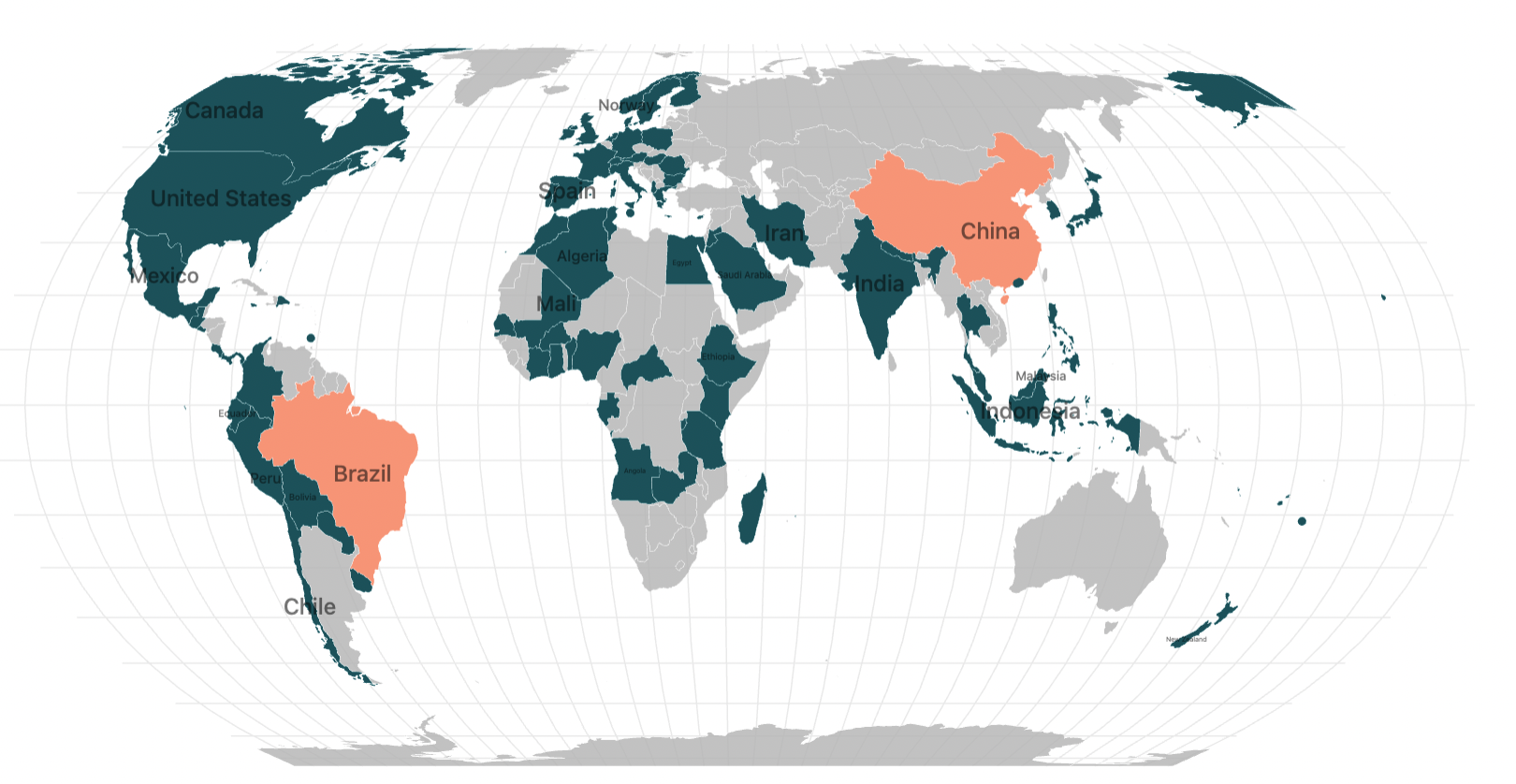}
	\includegraphics[width=0.7\textwidth]{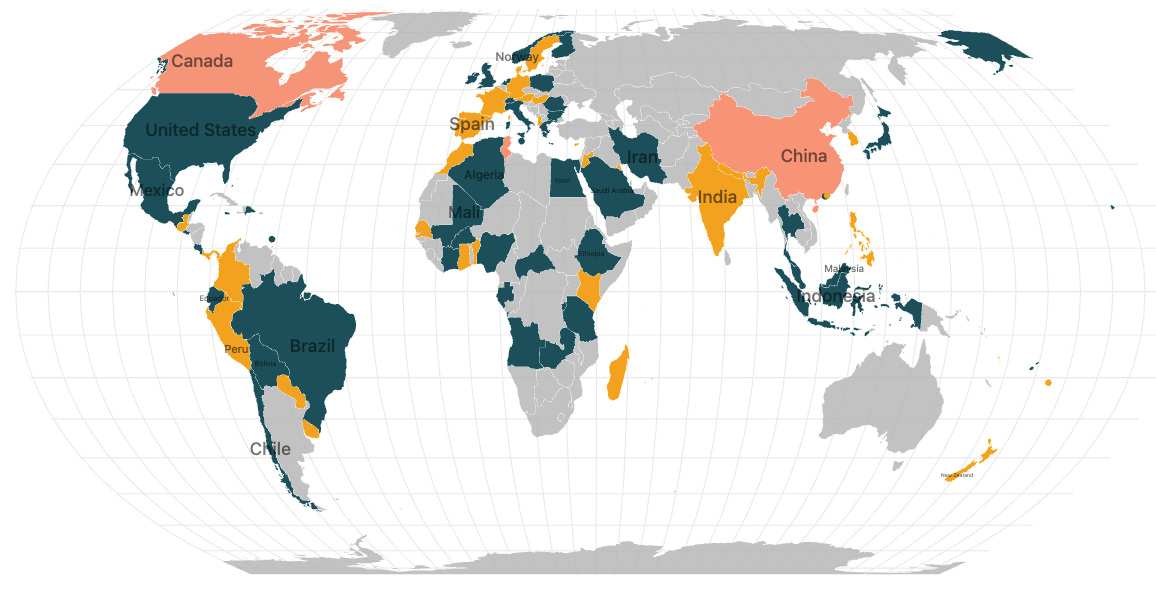}
	\caption{\small The group patterns of export (the top) and import (the bottom) markets.
		The Group 1 and 2 in the export market are marked with dark green and pink.
		The Group 1, 2, 3  in the import market are marked with dark green, pink and orange, respectively.
		The gray color refers to the regions excluded in the model dataset.
	}
	\label{fig:world_map_group}
\end{center}
\end{figure}

We next implement the GMNAR model to investigate the dynamic trading patterns around the world. The QIC suggests two groups for the exporter network ($\bA^{(1)}$) nodes and  three groups for the importer network ($\bA^{(2)}$) nodes. The group patterns are color-coded in Figure \ref{fig:world_map_group} and estimated model parameters are summarized in Table \ref{tbl:real_res_2}.
We summarize the estimated results as follows.

{\bf Self-momentum effects.}
Table \ref{tbl:real_res_2} reveals that the self-momentum effects $\wh\balpha$ are consistently negative for all cases, with the exception of $\wh\alpha_{22}$. One possible explanation for the negative auto-correlation is that the demand for exported goods from specific countries within the importer network tends to remain relatively stable over the years. Consequently, a significant increase in imports in the previous year may indicate a reduction in the importation of these goods in the following year. The sole exception is $\wh\alpha_{22}>0$, representing the self-momentum of exported goods from group 2 of the exporter network (i.e., mainland China and Brazil) to group 2 of the importer network (i.e., Canada, mainland China and Tunisia). This implies that trade between these countries has consistently grown at an accelerated speed from 1992 to 2021. For instance,
mainland China's exports to the Canada have increased at annualized rates of 10.8\% and 14.7\%, respectively,
while Brazil's exports to mainland China and Canada have grown at rates of 17.6\% and 8.41\%\footnote[3]{https://oec.world/en/profile/bilateral-country}. Such observations align well with the estimation by the GTNAR method.

{\bf Spatial exporter network effects.} Table \ref{tbl:real_res_2} indicates that the export network effects ($\wh\lambda_{g^{(1)}}^{(1)}$'s) are negative for both groups, but the second group, consisting of Brazil and mainland China, exhibits a substantially larger magnitude. The negative exporter network effects imply that a country's export is negatively influenced when neighboring countries experience significant increases in their exports, particularly pronounced in the case of large developing countries like Brazil and mainland China. This observation suggests that countries in neighboring regions tend to be competitors rather than collaborators in the export market.

{\bf Spatial importer network effects.} In Table \ref{tbl:real_res_2}, we also observe that the import network effects $\wh\lambda_{g^{(2)}}$ are positive in the first two groups (i.e., $\wh\lambda_1^{(2)} = 0.042$ and $\wh\lambda_2^{(2)} = 0.032$), but negative in the third group (i.e., $\wh\lambda_3^{(2)} = -0.005$). The positive values of $\wh\lambda_{g^{(2)}}$ for the first two groups indicate that if neighboring countries import more goods, it positively influences their own imports.
This implies a positive spillover effect of the import market for regions within the first two groups (such as mainland China, the United States and Canada).
Conversely, the third group predominantly comprises developing countries in South America and Africa, which are shown to be less affected by their neighbors' imports.

{\bf Covariates effects.}
In Table \ref{tbl:real_res_2}, we can observe distinct effects on IEV (International Export Volume) resulting from the three covariates. Firstly, it is notable that for the second group of exporters, $\wh\zeta_{\text{exc},2}^{(1)} = -1.710$, signifying a significant negative correlation. This suggests that higher export costs lead to reduced IEV in Brazil and mainland China. Conversely, for the second group of importers, $\wh\zeta_{\text{exc},2}^{(2)} = 0.035$, which is a positive value and aligns with expectations, as higher exchange rates generally correspond to increased import volumes. Secondly, both $\wh\zeta_{\text{gdp}}^{(1)}$ and $\wh\zeta_{\text{gdp}}^{(2)}$ display positive effects on IEV across all countries. This aligns with existing research indicating that domestic economic development bolsters international trade \citep{bernard1999exceptional, bernard2004some, berman2010financial}. Thirdly, we can observe that $\wh\zeta_{\text{net},1}^{(1)} = 0.228$, suggesting that a higher level of Internet access in most exporting countries leads to increased IEV. However, it's worth noting that countries in the second export group (Brazil and mainland China), as well as the second importer group (mainland China and Canada), exhibit negative coefficients concerning Internet population percentage. This discrepancy may be attributed to the rapid growth of internet users in China, which has outpaced the growth in export volume from 1993 to 2022.

\begin{table}[]
\centering
\caption{Estimation results for trading data analysis. The $p$-values are shown in the parenthesis.}
\label{tbl:real_res_2}
\scalebox{1}{
	\begin{tabular}{c|ccccc}
		\hline
		Parameters                               & \multicolumn{2}{c|}{$\lambda_{g^{(1)}}^{(1)}$}                                                                                                                               & \multicolumn{3}{c}{$\lambda_{g^{(2)}}^{(2)}$}                                                                                                                                                                                \\ \hline
		& \begin{tabular}[c]{@{}c@{}}-0.003\\ (\textless{}0.001)\end{tabular} & \multicolumn{1}{c|}{\begin{tabular}[c]{@{}c@{}}-0.512\\ (\textless{}0.001)\end{tabular}} & \begin{tabular}[c]{@{}c@{}}0.042\\ (\textless{}0.001)\end{tabular} & \begin{tabular}[c]{@{}c@{}}0.032\\ (\textless{}0.001)\end{tabular} & \begin{tabular}[c]{@{}c@{}}-0.005\\ (\textless{}0.001)\end{tabular} \\ \hline
		& \multicolumn{2}{c|}{$\bzeta_{g^{(1)}}^{(1)}$}                                                                                                                                  & \multicolumn{3}{c}{$\bzeta_{g^{(2)}}^{(2)}$}                                                                                                                                                                               \\
		Intercept                                & \begin{tabular}[c]{@{}c@{}}-0.011\\ (0.960)\end{tabular}            & \multicolumn{1}{c|}{\begin{tabular}[c]{@{}c@{}}2.072\\ (0.002)\end{tabular}}             & \begin{tabular}[c]{@{}c@{}}-0.369\\ (0.112)\end{tabular}           & \begin{tabular}[c]{@{}c@{}}2.674\\ (\textless{}0.001)\end{tabular} & \begin{tabular}[c]{@{}c@{}}-0.244\\ (0.329)\end{tabular}            \\
		$\bzeta_{\text{exc}}^{(1)}$/$\bzeta_{\text{exc}}^{(2)}$ & \begin{tabular}[c]{@{}c@{}}-0.120\\ (0.484)\end{tabular}            & \multicolumn{1}{c|}{\begin{tabular}[c]{@{}c@{}}-1.710\\ (0.036)\end{tabular}}            & \begin{tabular}[c]{@{}c@{}}-0.104\\ (0.549)\end{tabular}           & \begin{tabular}[c]{@{}c@{}}0.035\\ (\textless{}0.001)\end{tabular} & \begin{tabular}[c]{@{}c@{}}-1.399\\ (0.027)\end{tabular}            \\
		$\bzeta_{\text{gdp}}^{(1)}$/$\bzeta_{\text{gdp}}^{(2)}$ & \begin{tabular}[c]{@{}c@{}}0.693\\ (\textless{}0.001)\end{tabular}  & \multicolumn{1}{c|}{\begin{tabular}[c]{@{}c@{}}2.544\\ (\textless{}0.001)\end{tabular}}  & \begin{tabular}[c]{@{}c@{}}1.046\\ (\textless{}0.001)\end{tabular} & \begin{tabular}[c]{@{}c@{}}1.015\\ (\textless{}0.001)\end{tabular} & \begin{tabular}[c]{@{}c@{}}1.681\\ (\textless{}0.001)\end{tabular}  \\
		$\bzeta_{\text{net}}^{(1)}$/$\bzeta_{\text{net}}^{(2)}$ & \begin{tabular}[c]{@{}c@{}}0.228\\ (0.022)\end{tabular}             & \multicolumn{1}{c|}{\begin{tabular}[c]{@{}c@{}}-2.962\\ (0.010)\end{tabular}}            & \begin{tabular}[c]{@{}c@{}}0.029\\ (0.838)\end{tabular}            & \begin{tabular}[c]{@{}c@{}}-1.314\\ (0.001)\end{tabular}           & \begin{tabular}[c]{@{}c@{}}0.206\\ (0.186)\end{tabular}             \\ \hline
		\multirow{4}{*}{}                        & \multicolumn{5}{c}{$\balpha^\top \in \mR^{G_1 \times G_2}$}                                                                                                                                                                                                                                                                                                                             \\
		& \multicolumn{2}{c}{\begin{tabular}[c]{@{}c@{}}-0.431\\ (\textless{}0.001)\end{tabular}}                                                                        & \multicolumn{3}{c}{\begin{tabular}[c]{@{}c@{}}-0.494\\ (\textless{}0.001)\end{tabular}}                                                                                                                       \\
		& \multicolumn{2}{c}{\begin{tabular}[c]{@{}c@{}}-0.287\\ (\textless{}0.001)\end{tabular}}                                                                        & \multicolumn{3}{c}{\begin{tabular}[c]{@{}c@{}}0.021\\ (\textless{}0.001)\end{tabular}}                                                                                                                        \\
		& \multicolumn{2}{c}{\begin{tabular}[c]{@{}c@{}}-0.315\\ (\textless{}0.001)\end{tabular}}                                                                        & \multicolumn{3}{c}{\begin{tabular}[c]{@{}c@{}}-0.311\\ (\textless{}0.001)\end{tabular}}                                                                                                                       \\ \hline
	\end{tabular}
}
\end{table}

\end{document}